\documentclass[rmp,
               aps,
               reprint,
               longbibliography,
               floatfix]{revtex4-2}

\usepackage[utf8]{inputenc}
\usepackage[english]{babel}
\usepackage{amssymb,amsthm,amsmath,amstext,amsbsy,amsopn}
\usepackage{bbm}
\usepackage{nicefrac}
\usepackage{slashed}
\usepackage{pstricks}
\usepackage{graphicx}
\usepackage[colorlinks]{hyperref}
\usepackage{leftidx}
\usepackage{environ}
\usepackage{suffix}
\usepackage{mathtools}
\usepackage{mathrsfs}
\usepackage{xspace}
\usepackage{overpic}
\usepackage{tensor}
\usepackage{microtype}
\usepackage{soul}
\usepackage{listings}
\usepackage{enumitem}
\usepackage{orcidlink}

\definecolor{aeired}{RGB}{173, 0, 0}
\definecolor{aeiblu}{RGB}{1, 67, 95}
\hypersetup{linkcolor=aeired}
\hypersetup{citecolor=aeiblu}
\hypersetup{urlcolor=aeiblu}

\def\gammab{\hat{\gamma}}

\def\psib{\bar{\psi}}

\newcommand{\cc}{{\rm c}}
\newcommand{\dd}{{\rm d}}
\newcommand{\ii}{{\rm i}}
\newcommand{\mm}{{\rm m}}
\newcommand{\dV}{{\rm d}^{4}x \, \sqrt{-g} \,}

\def\Veff{V_{\rm eff}}

\def\subdef{{\mbox{\tiny DEF}}}
\def\DAlembert{\Box}

\def\eff{{\rm eff}}   
\def\pnt{{}^{\ast}RR} 

\newcommand{\msun}{{\rm M}_{\odot}}

\usepackage{acronym}
\def\newacronym#1#2#3{\gdef#1{\gdef#1{#2\xspace}#3 (#2)\xspace}}
\newacronym{\EFT}{EFT}{effective field theory}
\newacronym{\LVK}{LVK}{LIGO-Virgo-Kagra}
\newacronym{\NICER}{NICER}{Neutron Star Interior Composition Explorer}
\newacronym{\APR}{APR}{Akmal, Pandharipande, and Ravenhall}

\newacronym{\sm}{SM}{standard model}
\newacronym{\bhs}{BHs}{black holes}
\newacronym{\bh}{BH}{black hole}
\newacronym{\ns}{NS}{neutron star}
\newacronym{\eob}{EOB}{effective-one-body}
\newacronym{\gr}{GR}{general relativity}
\newacronym{\gw}{GW}{gravitational wave}
\newacronym{\snr}{SNR}{signal-to-noise ratio}
\newacronym{\qnm}{QNM}{quasinormal mode}
\newacronym{\pn}{PN}{post-Newtonian}
\newacronym{\bbh}{BBH}{black-hole binary}
\newacronym{\bns}{BNS}{neutron-star binary}
\newacronym{\DEF}{DEF}{Damour-Esposito-Far\`ese}
\newacronym{\ADM}{ADM}{Arnowitt-Deser-Misner}
\newacronym{\isco}{ISCO}{innermost stable circular orbit}
\newacronym{\sGB}{sGB}{scalar-Gauss-Bonnet}
\newacronym{\TMST}{TMST}{tensor-multiscalar theories}
\newacronym{\qpos}{QPOs}{quasiperiodic oscillations}
\newacronym{\EOS}{EOS}{equations of state}
\newacronym{\WEP}{WEP}{weak equivalence principle}

\begin{document}

\title{Spontaneous scalarization}

\author{Daniela.~D.~Doneva\,\orcidlink{0000-0001-6519-000X}}
\email{daniela.doneva@uni-tuebingen.de}
\affiliation{Theoretical Astrophysics, Eberhard Karls University of T\"ubingen, 72076 T\"ubingen, Germany\looseness=-3}
\affiliation{INRNE - Bulgarian Academy of Sciences, 1784  Sofia, Bulgaria\looseness=-3}

\author{Fethi M. Ramazano\u{g}lu\,\orcidlink{0000-0003-3075-1457}}
\email{framazanoglu@ku.edu.tr}
\affiliation{Department of Physics, Ko\c{c} University,
Rumelifeneri Yolu, 34450 Sariyer, Istanbul, Turkey\looseness=-3}

\author{Hector O. Silva\,\orcidlink{0000-0002-0066-9471}}
\email{hector.silva@aei.mpg.de}
\affiliation{Max Planck Institute for Gravitational Physics (Albert Einstein Institute),
Am M\"uhlenberg 1, D-14476 Potsdam, Germany\looseness=-3}

\author{Thomas P. Sotiriou\,\orcidlink{0000-0002-9089-4866}}
\email{Thomas.Sotiriou@nottingham.ac.uk}
\affiliation{Nottingham Centre of Gravity and School of Mathematical Sciences and School of Physics and Astronomy, University of Nottingham, University Park, Nottingham, NG7 2RD, United Kingdom\looseness=-4}

\author{Stoytcho~S.~Yazadjiev\,\orcidlink{0000-0002-1280-9013}}
\email{yazad@phys.uni-sofia.bg}
\affiliation{Theoretical Astrophysics,Eberhard Karls University of T\"ubingen, 72076 T\"ubingen, Germany\looseness=-3}

\affiliation{Department of Theoretical Physics, Faculty of Physics, Sofia University, 1164 Sofia, Bulgaria\looseness=-3}

\affiliation{Institute of Mathematics and Informatics, Bulgarian Academy of Sciences, Academy Georgi Bonchev Street 8, 1113 Sofia, Bulgaria\looseness=-4}

\begin{abstract}
Scalarization is a mechanism that endows strongly self-gravitating bodies, such as neutron stars and black holes, with a scalar-field configuration. It resembles a phase transition in that the scalar configuration appears only when a certain quantity that characterizes the compact object, for example, its compactness or spin, is beyond a threshold. A critical and comprehensive review of scalarization, including the mechanism itself, theories that exhibit it, its manifestation in neutron stars, black holes and their binaries, potential extension to other fields, and a thorough discussion of future perspectives, is provided.
\end{abstract}

\maketitle

\tableofcontents

\section{Introduction}
\label{sec:intro}
\vphantom{\bh \ns \gw }

Exploring the nature of gravity in the strong curvature regime has seen a recent surge of interest.
This is expected to intensify, as it is driven by current and future observations of compact objects: black holes (\bh{s}) and neutron stars (\ns{s}).
In particular, gravitational waves (\gw{s}) produced by coalescing compact binaries have by now been routinely detected by the \LVK Collaboration~\cite{LIGOScientific:2018mvr,LIGOScientific:2020ibl,LIGOScientific:2021djp}.
These observations enabled us to probe the highly dynamical and strong-field regime of \gr for the first time. They have enabled one to perform new tests of \gr and to constrain modifications thereof in a hitherto unexplored regime.
Future spaceborne and ground-based \gw observatories have testing \gr and the \sm among their key priorities~\cite{Barausse:2020rsu,LISA:2022kgy,Sathyaprakash:2019yqt,Kalogera:2021bya}.
At the same time, there is a new suite of electromagnetic observations that probe \ns{s} with
unprecedented sensitivity and timing resolution~\cite{Gendreau2012:SPIE,Arzoumanian2014:SPIE,Gendreau2017:NatAst}.
On other fronts, the precision timing of binary pulsars has continually improved (\textcite{Kramer:2021jcw}), measurements of the motion of stars at the Galactic Center are becoming more precise~\cite{GRAVITY:2018ofz,Do:2019txf,GRAVITY:2020gka}, and we have witnessed breakthroughs in supermassive \bh imaging~\cite{EventHorizonTelescope:2019dse}.

An interesting prospect is that these observations may reveal the existence of some otherwise
elusive new fundamental fields, which could be an ingredient of new physics beyond the \sm
or beyond \gr; see~\textcite{Clifton:2011jh},~\textcite{Yunes:2013dva},~\textcite{Berti:2015itd},
and~\textcite{Barack:2018yly}.
For such fields to have remained undetected, there has to exist a mechanism to
suppress them when gravity is weak. For scalar fields, which are ubiquitous in extensions of the \sm and of \gr, a possible realization of such a mechanism was first proposed by \textcite{Damour:1993hw} and named \emph{spontaneous scalarization}. They showed that a specific type of nonminimal coupling between scalar field and gravity (or matter, after a field redefinition) leads to a theory that is indistinguishable from \gr in weak-field gravitational experiments and yet predicts order unity deviations from general-relativistic expectations in the strong-gravity regime of \ns{s}.

As today, the first model of spontaneous scalarization came about at a time in which gravitational experiments were producing new data from a then unexplored regime of gravity: the slow-velocity but strong-field regime of binary pulsars discovered by~\textcite{Hulse:1974eb}; see~\textcite{Damour:2014tpa}.
This discovery inaugurated a new arena to test \gr and its contenders~\cite{Taylor:1982zz,Damour:1991rd,Taylor:1993zz}.
Meanwhile, slow velocities and weak-field Solar System tests had reached an accuracy that made it questionable whether viable theories that predict deviations from \gr that are measurable with binary pulsars can exist~\cite{Will:2014kxa}. Spontaneous scalarization settled this question and provided further motivation for the use of binary-pulsar observations to test \gr.
By now these observations have ruled out the \DEF scalarization model~\cite{Antoniadis:2013pzd,Kramer:2021jcw,Zhao:2022vig}.

In recent years, however, spontaneous scalarization has received renewed interest.
This is due to the realization that vacuum \bh solutions of \gr can also scalarize when the scalar field or fields couple suitably to the spacetime curvature~\cite{Doneva:2017bvd,Silva:2017uqg}. This development also showed that the earlier \DEF model is part of a much broader class of theories \cite{Andreou:2019ikc} that exhibit what resembles a phase transition in the strong field: once a quantity that describes a compact object, such as its compactness \cite{Damour:1993hw,Doneva:2017bvd,Silva:2017uqg} or spin \cite{Dima:2020yac}, exceeds a certain threshold, the scalar field switches from a trivial constant configuration to a nontrivial one, and large deviations from \gr appear. Conversely, one can think of this as deviations from \gr getting severely ``screened'' as soon as one crosses the same threshold in the opposite direction. It is this phase transition behavior that distinguishes scalarization from other models in which deviation from \gr is induced and controlled by coupling to curvature, such as more general scalar-tensor theories with linear \cite{Yunes:2011we,Sotiriou:2013qea,Sotiriou:2014pfa} or exponential~\cite{Kanti:1995vq} couplings to the Gauss-Bonnet invariant.

Since the advent of \gw astronomy, the broader class of theories that exhibit spontaneous scalarization have played a role similar to the one that the \DEF model has played for binary-pulsar observations. These theories provide a putative explanation of why we have not detected new fundamental fields with existing observations, but we might still uncover them with high precision observations of astrophysical systems with specific characteristics.

The aim of this review is to summarize, in a unified manner, the current status of this field.
In Sec.~\ref{sec:theory_bg}, we start by providing the theoretical background of the scalarization mechanism and its various subcases, following a pedagogical, rather than historical, approach.
Next we discuss in more detail the literature on scalarization, first of \ns{s} in Sec.~\ref{sec:ns_scalarization} and then of \bh{s} in Sec.~\ref{sec:bh_scalarization}.
In so doing, we present the state of the art of our understanding of the consequences of scalarization for various situations of observational interest.
In Sec.~\ref{sec:other}, we discuss attempts to generalize scalarization to other field types.
In Sec.~\ref{sec:conclusion}, we outline open issues and summarize future perspectives in the field.
Unless stated otherwise, we use geometrical units $G = 1 = c$
and employ the mostly plus metric signature convention.

\section{Theoretical background: mechanism and theories}
\label{sec:theory_bg}
\subsection{Spontaneous scalarization mechanism}
\label{sec:scalarization_mechanism}

\subsubsection{Tachyonic instability and nonlinear quenching}
\label{tachyonquenched}

Before we discuss spontaneous scalarization in the context of gravity, it is
instructive to review the dynamics of a real scalar field $\varphi$ with a
quartic self-interaction in Minkowski spacetime.
The Lagrangian for this field is
\begin{equation}
\label{phi4L}
{\cal L}=\tfrac{1}{2} \eta^{\mu\nu} \partial_\mu \varphi\partial_\mu \varphi+V(\varphi),
\end{equation}
where $\eta_{\mu\nu}$ is the Minkowski metric
\begin{equation}
\label{eq:phi4P}
V(\varphi)=\tfrac{1}{2}\mu^2\varphi^2+\tfrac{1}{4}\lambda \varphi^4,
\end{equation}
$\mu$ is the bare mass, and $\lambda$ is a coupling constant.
The scalar then satisfies the following field equation:
\begin{equation}
\label{eq:phi4}
\Box_{\eta} \, \varphi - \mu^2\varphi - \lambda \varphi^3=0,
\end{equation}
where $\Box_{\eta} = \eta^{\mu\nu} \partial_{\mu} \partial_{\nu}$ is the
flat-spacetime d'Alembertian.
$\varphi=0$ is a solution of this equation.
Consider now small perturbations $\delta \varphi$ around $\varphi=0$.
By linearizing~Eq.~\eqref{eq:phi4}, we find that $\delta \varphi$ obeys
\begin{equation}
\label{eq:phi4lin}
\Box_{\eta} \, \delta\varphi-\mu^2\delta\varphi=0.
\end{equation}
The corresponding dispersion relation is $\omega^2=k^2+\mu^2$, where $\omega$
is the frequency and $k$ is the wavenumber.
If $\mu^2>0$, the relevant solutions to this equation are plane waves and the perturbations
decay. If instead $\mu^2<0$, one encounters a \emph{tachyonic} instability and
the perturbations with small wave number exhibit exponential growth.

This exponential growth seems catastrophic at first sight, but it does not have to be.
As $\varphi$ grows, the previously used linear approximation will quickly become
invalid, and the nonlinear self-interaction $\lambda \varphi^3$ will become
important. It will be this interaction that will determine the end point of the
instability.
Assume that $\lambda>0$ (and $\mu^2<0$), in which case the potential has the
well-known ``Mexican hat'' shape. Equation~\eqref{eq:phi4} will then admit a second
solution with constant $\varphi$, which we denote as $\varphi_{\rm min}$, as it
will correspond to the minimum of the potential. Equation~\eqref{eq:phi4} implies that
$\varphi^2_{\rm min}=-\mu^2/\lambda$.
Thus, the tachyonic instability simply drives the scalar field away from the
unstable local maximum of the potential and toward a stable minimum. This is
sometimes referred to as tachyon condensation and is associated to a phase
transition of the system.

The key message from this simple example is that linearized perturbations
around the unstable maximum capture the onset of the tachyonic instability but
they are oblivious to the shape of the rest of the potential and hence
cannot determine the end point.
Nonlinear interactions, represented in this specific case by the $\varphi^4$ term
in the potential~\eqref{eq:phi4P}, eventually quench the instability and
drive the field to a different, stable configuration.

\subsubsection{Tachyonic instability in curved spacetime}
\label{sec:tachcurved}

In Sec.~\ref{tachyonquenched} we considered a scalar0field that exhibited a tachyonic
instability in flat spacetime.
The generalization to curved spacetime is simple. If we promote the Minkowski
metric to some general curved background described by a metric $g_{\mu\nu}$,
Eq.~\eqref{eq:phi4lin} becomes
\begin{equation}
\label{phi4lincurv}
\Box \, \delta\varphi-\mu^2\delta\varphi=0,
\end{equation}
where $\Box = g^{\mu\nu}\nabla_\mu \nabla_\nu$, with $\nabla_{\mu}$ the covariant derivative.
The key difference here is that in curved space $\mu^2<0$ is no longer
sufficient for having a tachyonic instability.

To see this, we take $g_{\mu\nu}$ to be the Schwarzschild metric, which
can describe either a nonrotating \bh or the exterior spacetime of a nonrotating
\ns in \gr.
The line element is
\begin{align}
\dd s^2=-\left(1-\frac{2M}{r}\right) \, \dd t^2+ \left(1-\frac{2M}{r}\right)^{-1} \, \dd r^2+r^2 \, \dd \Omega^2,
\nonumber \\
\end{align}
where $\dd \Omega^2= \dd \theta^2 + \sin^2\theta \, \dd \phi^2$
and $M$ is the mass of the compact object.
Because the spacetime is static and spherically symmetric, we can decompose the
scalar perturbation $\delta \varphi$ into spherical harmonics $Y_{\ell m} (\theta, \phi)$ and assume a
harmonic time dependence
\begin{equation}
\delta \varphi = \sum_{\ell m}\frac{\psi_{\ell m}(r)}{r}
Y_{\ell m}(\theta,\phi) e^{- i \omega t},
\end{equation}
and by substitution into Eq.~\eqref{phi4lincurv} we obtain a Schr\"odinger-like
equation
\begin{equation}
\frac{\dd^2 \psi_{\ell m}}{\dd r_{\ast}^2}
+ \left[ \omega^2 - V_{\rm eff}(r) \right] \psi_{\ell m} = 0,
\label{eq:eom_pert_schw}
\end{equation}
where we introduced the tortoise coordinate $r_\ast$ defined as $\dd r/\dd
r_{\ast} = 1-2M/r$ and $V_{\rm eff}$ is an effective potential given by
\begin{equation}
V_{\rm eff} = \left( 1 - \frac{2M}{r} \right)
\left[
\frac{\ell(\ell+1)}{r^2} + \frac{2M}{r^3} + \mu^2
\right],
\label{eq:eff_potential}
\end{equation}
which encodes information about the background, curved spacetime.
To have an instability in a Schwarzschild \bh spacetime it is sufficient but not
necessary that
\begin{equation}
    \int_{-\infty}^{\infty} \dd r_{\ast} \, V_{\rm eff}(r)  \leqslant  0,
\label{eq:bound_state_criteria}
\end{equation}
where $r_\ast=-\infty$ corresponds to the horizon radius in tortoise coordinates. For $M=0$ (flat spacetime) this condition is always satisfied when $\mu^2<0$ and Eq.~\eqref{eq:eom_pert_schw} yields the same dispersion relation
that we discussed in Sec.~\ref{sec:scalarization_mechanism}.
The situation is different for $M\neq 0$. Although it is not immediately obvious from upon inspection of Eq.~\eqref{eq:eff_potential}, it turns out that  $\mu^2$ would have to be sufficiently negative for  the tachyonic instability to occur.
The main lesson here is that, in curved spacetime, the threshold for the
tachyonic instability to happen depends on the spacetime; we will
return to how one can determine it later.
Note that, although we previously used Schwarzschild spacetime as an example, one
can rederive Eq.~\eqref{eq:eom_pert_schw} for a general static, spherically
symmetric background provided that $r_\ast$ is chosen appropriately.

\subsubsection{Scalarization and gravity}
\label{sec:theory_sca_and_grav}

We have thus far considered a scalar field with a negative bare mass squared,
which is not well motivated.
However, fields can acquire an effective mass squared $\mu_{\eff}^2$
in specific situations due to their coupling to other fields.
As an example, consider a scalar field that is nonminimally coupled to gravity
and a term $\varphi^2 R$ is present in the action, where $R$ is the Ricci
scalar.
We then expect a contribution proportional to $\varphi R$ to the scalar's field
equation; hence, the Ricci scalar contributes to the field's effective
mass, that is, $\mu_\eff^2 \propto \mu^2 + R$.
Assume that the scalar field has no bare mass (i.e.,~$\mu = 0$) and that the coupling
to $R$ is the only contribution to its effective mass $\mu_\eff$.
In flat spacetime, scalar-field perturbations would then be massless, whereas
in curved spacetimes (with $R\neq 0$) they would be massive and in general
also position dependent.
Moreover, the sign of $\mu^2_{\rm eff}$ would be
controlled by the sign of $R$ in this case. Hence, in some situations it would
be possible for $\mu^2_{\rm eff}$ to become sufficiently negative in some
spacetime region and cause the scalar field to become tachyonically unstable
despite this being impossible in flat spacetime.

Just as in the flat-spacetime example of the scalar with negative $\mu^2$ and
quartic interactions in Sec.~\ref{tachyonquenched}, this tachyonic instability
does not have to be catastrophic. It can simply signal that the scalar needs to
transition to a different configuration once curvature exceeds some threshold.
The instability implies that the scalar field will grow, nonlinearities will
become important, and, if they can quench the instability, then one can end up
with a stable, different configuration, for both the scalar field and the
spacetime.

This is precisely the idea behind spontaneous scalarization,\footnote{To our knowledge, the
expression ``spontaneous scalarization'' was first used in print by~\textcite{Damour:1996ke}.} first proposed in
\textcite{Damour:1993hw}. In a given generalized scalar-tensor theory, a
configuration with a constant scalar field and a metric that solves Einstein's
equations describes all gravitating systems except some that exhibit strong
gravity. In the latter case, curvature becomes significant enough to render
the constant scalar configurations tachyonically unstable.
The tachyonic instability is eventually quenched by nonlinear effects, and
there is a stable configuration with a nontrivial scalar and a spacetime
that is no longer a solution to Einstein's equations.

\subsubsection{Strong-field phase transitions and weak-field screening}

We have not yet shown that the mechanism of spontaneous scalarization, as
described heuristically earlier, can be at play within a consistent gravity
theory: we do so in Sec.~\ref{models}.
Nonetheless, assuming that the proposal can be successfully implemented in some
model, the following key observations can already be made:
\begin{itemize}
    \item Scalarization is a sharp transition to a new configuration that can
        differ significantly from the \gr configuration for the same object,
        even when one is very near the threshold of the tachyonic instability.
        This is intuitive when one thinks of scalarization as a linear
        instability quenched by nonlinearity: even for a mild instability
        (large timescale) the scalar has to grow significantly for nonlinear
        effects to become important and manage to stop further growth.
    \item The onset of instability can be controlled by curvature
        couplings. In Sec.~\ref{sec:theory_sca_and_grav}, we considered as an example a coupling between the
        scalar field and the Ricci scalar $R$, but one can envisage couplings with other curvature
        invariants, as we see in in Sec.~\ref{models}. Hence, there
        can be models in which scalarization will occur only in the strong-field regime
        (where curvature can become large), while objects that exhibit weak gravity will show no
        deviation from \gr (because the curvature is small).
        Combined with the previous point, this suggests that spontaneous
        scalarization can be thought of as a strong-field phase transition,
        whereby a field that is dormant in the weak field transitions to a
        nontrivial configuration in the strong field. Alternatively, one can
        think of scalarization in the reverse way: as a screening mechanism
        that forces a scalar field to transition to a trivial configuration in
        the weak field and hence explain why this field has managed to remain
        undetected so far.
    \item The previous argument is based on the rather naive expectation that
        curvature invariants are a good measure of how strong the gravitational
        interaction is. As an example of the failure of this expectation, recall
        that for a Schwarzschild \bh the Ricci scalar is zero; however, other
        curvature scalars, such as the Kretschmann scalar $R_{\mu\nu\rho\sigma} \, R^{\mu\nu\rho\sigma}$,
        are nonzero.
        In general, curvature invariants  can have a
        complicated dependence on the properties of compact objects. Hence, more
        work is needed to understand what controls the threshold of the
        tachyonic instability and the onset of scalarization. This is
        addressed in Sec.~\ref{types}.
\end{itemize}

\subsection{Models of scalarization}
\label{models}

\subsubsection{Tachyonic instability and the minimal action}
\label{sec:minimal_action}

We now turn our attention to the gravity theories that can exhibit spontaneous
scalarization.
As discussed, at the perturbative level the hallmark
of spontaneous scalarization is a tachyonic instability.
This begs the following question: Can we construct a minimal gravity theory which
can have scalar-field perturbations that are tachyonically unstable?
To do so, consider a gravity theory with a metric $g_{\mu\nu}$ and a scalar
field $\varphi$. Assume that the theory is such that the following hold:
\begin{enumerate}[label=\textsc{A.\arabic*}]
    \item\label{itm:a1_grsol} Spacetimes that are solutions of Einstein's equations,
        potentially with a cosmological constant, and a constant scalar field
        are admissible solutions of this theory as well.
    \item\label{itm:a2_quad} All terms in the action are at least quadratic in $\varphi$.
    \item\label{itm:a3_2ndordereom} The field equations are second-order partial differential equations.
\end{enumerate}
Under these requirements, the equation governing the dynamics of scalar
perturbations $\delta \varphi$ on \gr spacetimes can be cast in the form
\begin{equation}
\label{genLinEq}
g_{\rm eff}^{\mu\nu}\nabla^{(0)}_\mu\nabla^{(0)}_\nu\delta\varphi - \mu_{\rm eff}^2 \, \delta\varphi+{\rm NLC}=0,
\end{equation}
where $g^{\mu\nu}_{\rm eff}$, $\nabla^{(0)}$ and $\mu_{\rm eff}^2$ are all computed in the
background spacetime $g^{(0)}_{\mu\nu}$, and NLC denotes nonlinear corrections.
In Eq.~\ref{genLinEq} $g^{\rm eff}_{\mu\nu}$ is an effective metric that can differ from
$g^{(0)}_{\mu\nu}$ for certain types of nonminimal couplings between the metric and
the scalar field and $\mu_{\rm eff}^2$ may contain not only a bare mass term but also
other contributions.

If one neglects nonlinearities and assumes that $g_{\rm eff}$ is nondegenerate and
has a Lorentzian signature, Eq.~\eqref{genLinEq} becomes a curved-spacetime
version of Eq.~\eqref{eq:phi4lin}.
This means that one can identify all theories with a single scalar field that are
expected to lead to spontaneous scalarization by considering which couplings
between a scalar and the metric can contribute to $g_{\rm eff}$ and $\mu_{\rm eff}^2$
while still satisfying the aforementioned assumptions \ref{itm:a1_grsol}--\ref{itm:a3_2ndordereom}.
The benefit of taking into account all possible such terms is that it
would allow one to fully explore the onset of scalarization and identify a
class of gravity theories that result in a scalarized
spacetime.

Assumption~\ref{itm:a2_quad} appears to be essential to avoid having a source term in Eq.~\eqref{genLinEq}. We return to this shortly and show that it is a redundant assumption. But we first consider assumption~\ref{itm:a3_2ndordereom}. It  ensures that there are no unwanted degrees of freedom, as
would generically be the case if the equations contained higher-order derivatives (exceptions can exist, most notably in cases where field redefinitions can reduce the differential order of the equations).
This assumption does limit the possibilities of the terms that one can consider.
For example, a coupling term of the type $\varphi^2
\, R^{\mu\nu\lambda\sigma}R_{\mu\nu\lambda\sigma}$ in the action would
contribute to $\mu_{\rm eff}^2$ but leads to higher-order field equations (in
the absence of suitable counterterms).
To deal with this potential pitfall, one can follow the lines
of~\textcite{Andreou:2019ikc} and start with the Horndeski action
\cite{Horndeski:1974wa,Deffayet:2009mn}, also known as generalized scalar-tensor
theory,\footnote{See also~\textcite{Motohashi:2018wdq} for a classification
of a broader class of scalar-tensor theories according to their \bh solutions,
including those of \gr.}
\begin{equation}
\label{eq:Action_Horndeski}
S=\frac{1}{16 \pi G}\sum_{i=2}^{5}\int \dV \mathcal{L}_i + S_{\mm}[\Psi_{\mm};\, g_{\mu\nu}]
\end{equation}
and
\begin{subequations} \label{eq:Lfunctions}
\begin{align}
\label{L2}
\mathcal{L}_2 &=  G_2(\varphi,X),\\
\label{L3}
\mathcal{L}_3 &= -G_3(\varphi,X) \, \DAlembert \varphi,\\
\label{L4}
\mathcal{L}_4 &=  G_4(\varphi,X)R+G_{4X} \left[(\DAlembert \varphi)^2-(\varphi_{\mu\nu})^2 \right],\\
\nonumber
\mathcal{L}_5 &= G_5(\varphi,X) G^{\mu\nu} \varphi_{\mu\nu} \nonumber  - \tfrac{1}{6} \, G_{5X} \left[ \left(\DAlembert \varphi \right)^3 - 3 \, \DAlembert \varphi \, (\varphi_{\mu\nu})^2 \right.\\
              &\quad \left. + 2 \, (\varphi_{\mu\nu})^3\right],
\label{L5}
\end{align}
\end{subequations}
where $G_{2}$, $G_{3}$, $G_{4}$, and $G_{5}$ are arbitrary functions of the scalar field $\varphi$ and
its kinetic term $X=-\nabla_\mu \varphi \nabla^\mu\varphi/2$.
In Eq.~\eqref{eq:Lfunctions} $G_{iX}=\partial G_{i}/\partial X$ ($i = 4$ and $5$), $G^{\mu\nu}$ is the Einstein tensor, and the
notation $\varphi_{\mu\nu} = \nabla_{\mu} \nabla_{\nu} \varphi$ was introduced so, for example,
$(\varphi_{\mu\nu})^2 = \varphi_{\mu\nu} \varphi^{\mu\nu} = \nabla_\mu\nabla_\nu\varphi \, \nabla^\mu\nabla^\nu\varphi$.
Finally, $S_{\mm}$ is the matter action, with matter fields collectively denoted by $\Psi_{\mm}$.
This is the most general action for a metric and a scalar field that leads to
second-order field equations in four dimensions upon direct variation; see~\textcite{Kobayashi:2019hrl} for a review.
We assume for the moment that matter couples minimally to the metric only.
This means that the choice of fields $g_{\mu\nu}$ and $\varphi$
correspond to the so-called Jordan frame; we return to this issue later.

Imposing assumptions~\ref{itm:a1_grsol} and~\ref{itm:a2_quad} on the action~\eqref{eq:Action_Horndeski} places
restrictions on the $G_{i}$ functions, as we later see.
We refer to \textcite{Andreou:2019ikc} for a detailed discussion.
For our purposes, it is sufficient to say that, by perturbing around an
arbitrary spacetime that is assumed to be a solution of Einstein's equations
with a constant scalar field, we can identify all of the terms that
contribute, at the linear level, to $g^{\mu\nu}_{\rm eff}$ and $\mu^2_{\rm eff}$, as
defined in Eq.~\eqref{genLinEq}.
These terms amount to the following action:
\begin{align}\label{eq:ActionCaseI}
    S_{\rm min} & = \frac{1}{16 \pi G} \int \dV \left[R-\frac{1}{2}(\gamma_{1}+\gamma_{2}R)\nabla_\mu \varphi\nabla^\mu \varphi  \right.
                \nonumber \\
                &\left.\quad + \, \gamma_{2}R_{\mu\nu}\nabla^\mu\varphi\nabla^\nu\varphi - \frac{1}{2} \mu_\varphi^2\varphi^2 - \frac{1}{4}\beta\varphi^2 R + \frac{1}{2}\alpha\,\varphi^2\,\mathscr{G} \right.
                \nonumber \\
                &\left.\quad - \, 2 \Lambda \right] + S_{\mm}[\Psi_{\mm};\, g_{\mu\nu}],
\end{align}
where $\mathscr{G}$ is the Gauss-Bonnet invariant, which is defined in terms of the Riemann
tensor and its familiar contractions as
\begin{equation}
\mathscr{G} = R^{\mu\nu\rho\sigma}R_{\mu\nu\rho\sigma}
- 4 R^{\mu\nu} R_{\mu\nu}
+ R^2,
\label{eq:def_gauss_bonnet}
\end{equation}
and where $\alpha$, $\beta$, $\gamma_i$, and $\mu^2_\varphi$ can be expressed
in terms of the $G_i$ functions and their derivatives evaluated in the
background configuration~\cite{Andreou:2019ikc}.\footnote{We are not following the notation of
\textcite{Andreou:2019ikc}, but have instead adapted it to match that of some
of the specific models that we later study. While expected, it is nontrivial
to show how the Gauss-Bonnet invariant emerges from Eq.~\eqref{eq:Action_Horndeski}.
This was first shown by~\textcite{Kobayashi:2011nu} at the level of the field equations,
and by~\textcite{Langlois:2022eta} at the level of the action.}
We refer to this action, in a slight abuse of terminology, as the
minimal action for scalarization, in the sense that it contains all of the
terms that contribute to the \emph{onset of scalarization} manifesting as a tachyonic
instability. As such, it can be used to study and understand what triggers
scalarization and to determine the relevant instability thresholds.

Before we go further, we examine what happens if we decide to drop
assumption~\ref{itm:a2_quad} altogether, but still impose assumptions~\ref{itm:a1_grsol} and~\ref{itm:a3_2ndordereom}.
Working along the same lines as before, one arrives at a different set of
terms composing the action
\begin{align}\label{eq:ActionCaseII}
    S'_{\rm min} & = \frac{1}{16 \pi G} \int \dV \left[R-\frac{1}{2}\frac{\gamma'_{1}+\gamma'_{2}R}{\varphi}\nabla_\mu \varphi\nabla^\mu \varphi \right. \nonumber \\
                 &\left. \quad + \, \frac{\gamma'_{2}}{\varphi}R_{\mu\nu}\nabla^\mu\varphi\nabla^\nu\varphi +\tau\varphi +\eta\,\varphi R +\lambda\,\varphi\,\mathscr{G} \right. \nonumber \\
                 &\left. \quad - \, 2 \Lambda \right] + S_{\mm}[\Psi_{\mm}; g_{\mu\nu}].
\end{align}
As before, $\tau$, $\eta$, $\lambda$, and $\gamma'_i$ can be expressed in terms
of the $G_i$ functions and their derivatives evaluated in the background
configuration. It might seem counterintuitive that an action containing terms
linear in $\varphi$ leads to perturbation equations that are in the form of
Eq.~\eqref{genLinEq}, which has no source terms. This is due to the presence of
terms that are nonanalytic in $\varphi$.

The first impression may be that abandoning assumption~\ref{itm:a2_quad} has given rise to a second
minimal action for scalarization. However, action \eqref{eq:ActionCaseII}
is just a field redefinition away from action \eqref{eq:ActionCaseI}.
Indeed, one can start with Eq.~\eqref{eq:ActionCaseII}, introduce the redefinition\footnote{Here and elsewhere in the text, when two actions are related by a field redefinition we do
not relabel the field in order to keep the notation lighter.} $\varphi \to \varphi^2$, and obtain action \eqref{eq:ActionCaseI}, with the following mapping of parameters: $\gamma_1=4 \gamma'_1$, $\mu^2_\varphi=-4\tau$, $\beta=-\eta$, and $\alpha=\lambda$ \cite{Andreou:2019ikc}.
This equivalence demonstrates (i) that assumption~\ref{itm:a2_quad} is redundant and
(ii) that up to field redefinitons Eq.~\eqref{eq:ActionCaseI} is
sufficient to capture all terms that contribute to the onset of
scalarization and satisfy assumptions~\ref{itm:a1_grsol} and~\ref{itm:a3_2ndordereom}.

One can see by inspection that the $\gamma_i$ terms in
Eq.~\eqref{eq:ActionCaseI} will contribute to $g^{\mu\nu}_{\rm eff}$, while the
rest of the terms will contribute to $\mu^2_{\rm eff}$. Hence, if the latter
vanish, the former cannot trigger scalarization by themselves, as the effective mass would vanish.
Nevertheless, the $\gamma_i$ terms will affect the threshold of the tachyonic
instability we associate with scalarization; cf.~the discussion about the
tachyonic instability in curved spacetime in Sec.~\ref{sec:tachcurved}.
Additionally, $\mu^2_\varphi$ corresponds to the bare mass of the scalar field,
so it is expected to be positive. We then conclude that the terms that are expected to trigger
scalarization in the strong-field regime are only the couplings of $\varphi$ to
$R$ and $\mathscr{G} $. In fact, we see shortly that these are indeed the terms
present in the known models of scalarization.

To summarize, the minimal action \eqref{eq:ActionCaseI} can be used to study
the onset of spontaneous scalarization triggered by a nonminimal coupling to
gravity.
It contains all possible terms that contribute to the associated tachyonic
instability at the linearized level, so it could be used to study the threshold
and onset of this instability in full generality. As previously discussed, as the
instability progresses it is expected to be quenched nonlinearly, and the
end point will be a scalarized configuration.
The terms in Eq.~\eqref{eq:ActionCaseI} can contribute nonlinearly as well,
but one could add a plethora of other nonlinear interactions, ranging from
scalar self-interactions, for example, $\varphi^4$, to nonminimal coupling terms that
do not contribute to linear perturbations around curved spacetime with constant
scalar, for instane, $\varphi^4 \, \mathscr{G} $.
That is, one can start with Eq.~\eqref{eq:ActionCaseI}, or even a subset of
terms therein, and construct different scalarization models. Models that differ
only by terms that are not in Eq.~\eqref{eq:ActionCaseI} will have the same
behavior regard to the onset of scalarization, and hence the
configurations that one expects to not scalarize, but they can differ in
the properties of scalarized solutions \cite{Andreou:2019ikc,Silva:2018qhn,Macedo:2019sem,Minamitsuji:2019iwp}.
This is further discussed alogn with specific examples in Secs.~\ref{sec:bh_scalarization_sgb} and
\ref{sec:bhs_stability_nr}.

Before proceeding to discuss more specific known models, we return
to the issue of the coupling to matter. We have thus far assumed that matter couples
minimally to the metric only. This assumption is sufficient to ensure that the
theory is compatible with the \WEP~\cite{Will:2018bme}.
To satisfy the \WEP it is sufficient to have matter
couple minimally to some metric, but this does not need to be the same metric
(or choice of other fields) for which the theory has second-order field
equations.
However, it is known that a disformal transformation~\cite{Bekenstein:1992pj}
of the form
\begin{equation}\label{eq:disformal}
g_{\mu\nu} \rightarrow C(\varphi)\left[g_{\mu\nu}+D(\varphi)\nabla_\mu\varphi\nabla_\nu\varphi\right]
\end{equation}
leaves the Horndeski action~\eqref{eq:Action_Horndeski} formally
invariant~\cite{Bettoni:2013diz,Zumalacarregui:2013pma}. It was shown
by \textcite{Andreou:2019ikc} that coupling matter minimally to a metric that
is related to $g_{\mu\nu}$ by such a disformal transformation as done, \textcite{Minamitsuji:2016hkk},
amounts to a redefinition of $\gamma_2$ in the linearized theory around spacetimes that are solutions of Einstein's equations.
Hence, such a coupling would be redundant when
the onset of scalarization is studied using the minimal action~\eqref{eq:ActionCaseI}.
One could also entertain the idea of coupling matter to some composite metric
$\bar{g}_{\mu\nu}$ that depends on both $g_{\mu\nu}$ and $\varphi$ in a
different manner than in Eq.~\eqref{eq:disformal}.
In such a case, it is likely that assumption~\ref{itm:a3_2ndordereom} would be violated and one
would have to start the analysis presented here with a generalization of the
action~\eqref{eq:ActionCaseI}.

\subsubsection{Damour--Esposito-Far\`ese Model}
\label{sec:theory_def_model}

As mentioned, the concept of scalarization as we described it was
first discussed by \textcite{Damour:1993hw}. They considered the theory
\begin{align}
S &= \frac{1}{16\pi G_{*}} \int\dV \left[R -
2 \nabla_{\mu}\varphi \nabla^{\mu}\varphi
\right]
\nonumber \\
&\quad + S_{\mm}[\Psi_{\mm};\, {\cal A}^{2}(\varphi)g_{\mu\nu}].
\label{eq:action_st}
\end{align}
Equation~\eqref{eq:action_st} is said to be written in the Einstein frame, which means that,
contrary to our previous assumptions and conventions, the scalar field is coupled
minimally to gravity and has a canonical kinetic term.
The coupling with matter field $\Psi_{\mm}$
is through the function ${\cal A}^{2}(\varphi)$.
In Eq.~\eqref{eq:action_st} $G_{\ast}$ carries a subscript, as it is not generally equal to $G$ used
thus far. Variation of the action~\eqref{eq:action_st} with respect to $\varphi$
yields the field equation
\begin{equation}
\label{DEFscalareq}
    \Box \varphi= -4\pi G_{*}\alpha(\varphi) T,
\end{equation}
where
\begin{equation}\label{eq:coupling_STT}
    \alpha(\varphi) = {\dd \ln {\cal A}(\varphi)} / {\dd \varphi}
\end{equation}
and $T = g_{\mu\nu} \, T^{\mu\nu}$ is the trace of the matter energy-momentum tensor in the Einstein frame
defined as $T^{\mu\nu} = 2 (- g)^{-1/2} \delta S_{\mm} / \delta g_{\mu\nu}$.
We see that $\alpha(\varphi)$ controls the coupling strength between the scalar field and matter.

If $\alpha(\varphi_0)=0$ for some
constant scalar-field value $\varphi_0$, the constant scalar configuration
with $T\neq 0$ will be an admissible solution of the theory.
It then follows from the generalized Einstein's equations
\begin{equation}\label{eq:DEF_EinsteinEq}
    R_{\mu\nu} = 2 \nabla_{\mu} \varphi \nabla_{\nu} \varphi + 8 \pi G_{\ast} (T_{\mu\nu} - \tfrac{1}{2} g_{\mu\nu} \, T)
\end{equation}
that these will be solutions of \gr since the first term on the right-hand side vanishes.

At the same time, if we perturb Eq.~\eqref{DEFscalareq} linearly in $\varphi$ in a fixed
background metric that is a solution of \gr and compare with Eq.~\eqref{genLinEq} we find that $\beta(\varphi_0) = (\dd
\alpha/\dd \varphi)_{\varphi = \varphi_0} $ and $T$ determine the value and
sign of the effective mass square of the perturbations, namely,
\begin{equation}
    \mu_{\rm eff}^2 = -4\pi G_{*} \beta(\varphi_0) T.
    \label{eq:mueffsqr_def}
\end{equation}
For stars, one generally has $T < 0$. Hence, for
a negative sign of $\beta(\varphi_0)$ and the right magnitude of both quantities, the scalar can develop a tachyonic instability around a spacetime that describes stars in \gr, as previously discussed and
as studied in detail by~\textcite{Harada:1997mr}.\footnote{Exceptions exist [see \textcite{Mendes:2014vna}], as discussed in Sec.~\ref{sec:NS_DEF_model}.}
It was shown by~\textcite{Novak:1998rk} that this instability is quenched by
nonlinearities and that the outcome is a \ns with a nontrivial scalar-field configuration.
These scalarized \ns{s} were shown by~\textcite{Damour:1993hw} to have
properties, such as their masses $M$ and radii $R$, that can be dramatically
different from their \gr counterparts.

In much of the literature considering scalarization, the function $\alpha(\varphi)$ is taken to have the
form\footnote{\textcite{Damour:1993hw} also studied the case in which ${\cal A} = \cos(\sqrt{6} \varphi)$
and hence $\alpha = - \sqrt{6} \, \tan(\sqrt{6} \varphi) \approx - 6 \varphi + 12 \varphi^3 + \dots$, which
includes higher powers in the scalar-field-matter interaction series~\eqref{eq:DEF_alpha} for
$\alpha_0 = 0$ and $\beta_0 = - 6$.}
\begin{equation}\label{eq:DEF_alpha}
  \alpha = \alpha_0 + \beta_0 \varphi \equiv \alpha_{\subdef}\,,
\end{equation}
where $\alpha_0$ and $\beta_0$ are dimensionless constants. Sometimes this
choice, rather that the more general action of Eq.~\eqref{eq:action_st}, is
referred to as the \DEF model.
The constant $\alpha_0$ is then assumed to vanish to allow for constant
$\varphi$ solutions [cf.~Eq.~\eqref{DEFscalareq}], or it is assumed to be small.
In the latter case, all stars will carry some nontrivial scalar field, but by
tuning down $\alpha_0$ any deviation from \gr would be undetectable until
scalarization kicks in.
In its original formulation, the \DEF model did not include a bare mass or
self-interactions for the scalar, but a potential $V(\varphi)$ can be added to
the action and this option has been considered in the literature by
\textcite{Popchev2015,Chen:2015zmx,Ramazanoglu:2016kul}, as we see in
Sec.~\ref{sec:ns_scalarization}.

Thus far it appears that the DEF model is not covered by our minimal action
\eqref{eq:ActionCaseI}, because of our earlier assumption that the scalar does
not couple to the matter.
However, by defining ${\cal A}^2(\varphi) g_{\mu\nu}$ as a new metric and
rewriting the action \eqref{eq:action_st} in terms of that new metric, the
scalar field is no longer coupled to matter. This is referred to as the Jordan
frame. It was shown by \textcite{Andreou:2019ikc} that, at linearized
level and after a suitable scalar-field redefinition, the \DEF model is equivalent to
the action
\begin{align}\label{eq:ActionphisqR}
S &=\frac{1}{16 \pi G} \int \dd^4x \sqrt{-g} \left[\left(1-\frac{1}{4}\beta_0 \varphi^2\right) R-\frac{1}{2}\nabla_\mu \varphi\nabla^\mu \varphi \right]
 \nonumber \\
&\quad + S_{\mm}[\Psi_{\mm};\, g_{\mu\nu}].
\end{align}
Equation~\eqref{eq:ActionphisqR} is indeed a particular case of the action \eqref{eq:ActionCaseI} in which $\gamma_1=1$, $\gamma_2=\alpha=\Lambda=0$, and $\beta=\beta_0$.
Hence, the \DEF model, with what regards to the onset of the tachyonic instability
that leads to scalarization, is captured by the minimal action
\eqref{eq:ActionCaseI} and corresponds to one of the two couplings to curvature
that can trigger scalarization.

Before moving on, we mention that the original formulation of the
\DEF model, which leads directly to Eq.~\eqref{DEFscalareq}, suggests that it
is the coupling to matter that controls and triggers scalarization. Indeed,
when $T=0$, as is the case for \bh{s}, Eq.~\eqref{DEFscalareq} becomes
$\Box \varphi= 0$ and admits only constant $\varphi$ solutions for stationary and asymptotically
flat configurations by virtue of a no-hair theorem by \textcite{Hawking:1972qk}
[this remains true when one includes a potential;
see \textcite{Sotiriou:2011dz}].
However, our previous analysis and the
correspondence between the \DEF model and action \eqref{eq:ActionphisqR} at the
linearized level makes it clear that the \DEF model is part of a broader class
of theories in which scalarization is present and controlled by the couplings
to curvature, rather than matter, and this observation has been crucial for the
development of models that exhibit \bh scalarization.
It is the fact that $R$ and $T$ are related through the trace of the theory's
generalized Einstein equation [cf.~Eq.~\eqref{eq:DEF_EinsteinEq}] that
allows for both interpretations in the \DEF model.

\subsubsection{Scalar-Gauss-Bonnet gravity}
\label{sec:theory_sgb}

It was first shown by \textcite{Doneva:2017bvd} and \textcite{Silva:2017uqg}
that theories described by the \sGB action
\begin{align}
\label{actionsgb}
    S &= \frac{1}{16 \pi G} \int \dd^4x \sqrt{-g} \left[ R - \tfrac{1}{2} \nabla_{\mu} \varphi \nabla^{\mu} \varphi + f(\varphi) \, \mathscr{G} \right]
    \nonumber \\
    &\quad + S_{\mm} [\Psi_{\mm};\, g_{\mu\nu}]
\end{align}
can exhibit \bh scalarization provided that $(\dd f/ \dd\varphi)_{\varphi = \varphi_0}=0$
for some constant $\varphi_0$. This is an existence condition for constant $\varphi$ configurations that are solutions of \gr.
As proven by \textcite{Silva:2017uqg}, \bh solutions of \gr are unique solutions
to the theory~\eqref{actionsgb} provided that $(\dd^2 f / \dd \varphi^2)_{\varphi = \varphi_0} \, \mathscr{G} < 0$.
To understand this, we can proceed as follows. By varying the action with respect to $\varphi$, we find
\begin{equation}
\label{eq:sgbscalar}
    \Box \varphi+ f_{,\varphi}(\varphi) \, \mathscr{G} = 0\,, \quad f_{,\varphi}(\varphi) = {\dd f} / {\dd\varphi}.
\end{equation}
Once again, we can consider linear perturbation of $\varphi$ on a fixed background and
compare with Eq.~\eqref{genLinEq}. We find that $(\dd^2
f/\dd\varphi^2)_{\varphi = \varphi_0} \, \mathscr{G} $ plays the role of an effective mass square for
the scalar perturbations,
\begin{equation}
    \mu^{2}_{\eff} = - (\dd^2 f / \dd \varphi^2)_{\varphi = \varphi_0} \, \mathscr{G}.
    \label{eq:mueffsqr_sgb}
\end{equation}
Hence, violating the condition $(\dd^2 f / \dd \varphi^2)_{\varphi = \varphi_0} \, \mathscr{G} < 0$  is
necessary, but not sufficient, to develop a tachyonic instability that can lead
to scalarization.

\textcite{Doneva:2017bvd} chose $f(\varphi)$ to be $f(\varphi) =
\lambda^{2}(1 - e^{-3/2 \varphi^2})/12$, whereas \textcite{Silva:2017uqg}
focused on  $f(\varphi)=\alpha\varphi^2/2$. Note that, in the linearized theory
around $\varphi=\varphi_0$ and provided that the condition $(\dd f/\dd\varphi)_{\varphi = \varphi_0}=0$ is satisfied, any choice of $f(\varphi)$ is equivalent to
$f(\varphi)=\alpha\varphi^2/2$.
Hence, all scalarization models described by action~\eqref{actionsgb} are
captured by the minimal action~\eqref{eq:ActionCaseI} and, in particular, by the
coupling between the scalar and the Gauss-Bonnet invariant in regard to
the onset of scalarization. However, different choices of $f(\varphi)$ will
exhibit different behavior in the nonlinear regime and hence scalarized \bh{s}
will generally have different properties.

Indeed, it was shown by \textcite{Blazquez-Salcedo:2018jnn} that the static,
spherically symmetric scalarized \bh{s} that were found by
\textcite{Silva:2017uqg} for $f(\varphi)=\alpha\phi^2/2$ are unstable against radial perturbations,
unlike their for counterparts for $f(\varphi) =
\lambda^{2}(1 - e^{-3/2 \varphi^2})/12$ found by \textcite{Doneva:2017bvd}.
\textcite{Silva:2018qhn} later demonstrated by examining the case $f(\varphi)=\alpha\varphi^2/2 + \xi
\varphi^4$ that it is indeed the nonlinearity in $\varphi$ that controls the
stability of scalarized \bh{s}. It was further shown by
\textcite{Macedo:2019sem} and~\textcite{Minamitsuji:2019iwp} that a quartic scalar-field
self-interaction would be sufficient to make scalarized \bh{s} stable
for the $f(\varphi)=\alpha\varphi^2/2$ case.
These results, which are discussed in Sec.~\ref{sec:bh_scalarization}, are
a clear demonstration that although the onset of scalarization can be described
fully using action \eqref{eq:ActionCaseI}, the end point of the tachyonic
instability and the properties of the scalarized configurations will depend to
the nonlinear interaction between the scalar and curvature and are thus model
dependent.

We remark that models described by action \eqref{actionsgb} lead to
scalarization of compact stars as well for certain regions of their parameter
spaces; see~\textcite{Doneva:2017duq} and \textcite{Silva:2017uqg}.

\subsubsection{The Ricci-Gauss-Bonnet model}
\label{sec:mixed_model}
As we have seen, the \DEF model and the \sGB models of scalarization correspond
respectively, to the $\varphi^2 \, R$ and $\varphi^2 \, \mathscr{G} $ terms (in
addition to $R$ and the canonical kinetic term for the scalar field) in the minimal
action~\eqref{eq:ActionCaseI}. We have also argued that these two terms
are the only terms that can trigger scalarization as a tachyonic instability
around a spacetime that is a solution of \gr.
These facts together suggest considering the
following action \cite{Antoniou:2020nax,Antoniou:2021zoy}:
\begin{align}\label{action:RGmixed}
    S &= \frac{1}{16 \pi G} \int \dV \left[R-\tfrac{1}{2}\nabla_\mu \varphi\nabla^\mu \varphi-\tfrac{1}{4}\beta\varphi^2 R
    \right. \nonumber \\
      &\quad \left. +\tfrac{1}{2}\alpha\,\varphi^2\,\mathscr{G}\right] + S_{\mm}[\Psi_{\mm};\, g_{\mu\nu}].
\end{align}
This theory is interesting from the perspective of an \EFT.
The previously considered terms can be seen as part of an \EFT in which the scalar
field enjoys reflection symmetry (i.e., invariance under $\varphi \to - \varphi$),
while shift symmetry (i.e., invariance under $\varphi \to \varphi + \rm{const.}$)
is broken only by the coupling to the curvature scalars.
This theory is not a complete \EFT, as there are other
operators compatible with these symmetries, such as $\varphi^4 \, R$ and
$G^{\mu\nu}\nabla_\mu \varphi\nabla_\nu \varphi$.
Nonetheless, the theory is phenomenologically interesting for various reasons.

To begin, it has \gr with a constant scalar field as a late-time cosmic attractor for
$\beta>0$ \cite{Antoniou:2020nax}. To appreciate why this is important, recall
that one can think of scalarization in terms of a tachyonic instability of compact object configurations that are solutions of \gr with
$\varphi=\varphi_0$.
Below the threshold of this instability, these configurations are expected to
be stable and exhibit no deviation from \gr. The attractive feature of
scalarization is that weakly gravitating systems will belong in this category
and hence scalarization can be a form of weak-field screening of the scalar
field. However, this argument assumes that $\varphi=\varphi_0$ everywhere in
the Universe and deviates from this value only due to scalarization. If there
were another reason for $\varphi$ to evolve away from $\varphi_0$, this would
make even weakly gravitating objects develop a nontrivial scalar configuration.
Cosmic evolution can indeed cause such evolution, as shown by~\textcite{Damour:1992kf}, to the extent that
weakly gravitating systems would become sufficiently scalarized to make the
\DEF model (\textcite{Anderson:2016aoi}) and \sGB scalarization models
(\textcite{Anson:2019uto,Franchini:2019npi}) fail weak-field and cosmological tests
of gravity without severely fine-tuning the initial conditions for cosmic
evolution.

The theory described by action \eqref{action:RGmixed} provides an elegant
solution to this problem \cite{Antoniou:2020nax}.
As pointed out by \textcite{Damour:1992kf}, the \DEF model has \gr as a cosmic
attractor for $\beta_0>0$, whereas scalarization requires $\beta_0$ to be
sufficiently negative.
It was already mentioned
that, at the linearized level around a \gr
background, the \DEF model is equivalent to action \eqref{eq:ActionphisqR}, so
it is reasonable to expect that the $\varphi^2 R$ term in action
\eqref{action:RGmixed} would tend to drive the scalar field to a constant in late-time cosmology.
Moreover, this term should be dominant over the $\varphi^2 \,
\mathscr{G}$ term at low curvatures, while the latter should dominate at high
curvatures and trigger scalarization.
It was indeed shown by \textcite{Antoniou:2020nax} that cosmic evolution
in the model of action \eqref{action:RGmixed} tracks the cosmic evolution of
\gr from radiation domination onward, and that $\varphi$ is driven to $\varphi_0$
rapidly during matter domination.

It was also shown by \textcite{Antoniou:2021zoy} and
\textcite{Ventagli:2021ubn} that for $\beta>0$ one can have a range of values
for $\alpha$ in which \bh{s} scalarize but \ns{s} do not. This is interesting
because the strongest constraints on scalarization thus far, and specifically the
\DEF model, are based on binary pulsars~\cite{Kramer:2021jcw,Zhao:2022vig}.
These constraints can be evaded if scalarization is limited to \bh{s}.
\textcite{Antoniou:2021zoy} also provided strong indication that
scalarized \bh{s} should be radially stable for $\beta>0$, which is not the
case for $\beta=0$. Both of these issues are discussed in
Sec.~\ref{sec:bh_scalarization}.

In summary, adding a $\beta \varphi^2 R$ term with $\beta>0$ to the simplest \sGB scalarization model addresses a series of concerns.
This term would be there anyway in an \EFT, as it has lower mass dimensions
than $\varphi^2 \, \mathscr{G}$. These considerations should act as a reminder that
scalarization theories are currently still toy models in need of a completion.

\subsubsection{Tensor-multiscalar theories}

Thus we have considered theories with a single field, but one could also study scalarization in models with multiple scalar fields.
These tensor-multiscalar theories were studied by~\textcite{Damour:1992we} and
in their simplest form they are described by the action
\begin{align}\label{ActionTMST}
    S &= \frac{1}{16\pi G_{*}}\int \dV [R - 2\gamma_{ab}(\boldsymbol{\varphi})\nabla_{\mu}\varphi^{a}\nabla^{\mu}\varphi^{b}   \nonumber \\
      &\quad - 4V(\boldsymbol{\varphi})] + S_{\mm}[\Psi_{\mm};\, {\cal A}^{2}(\boldsymbol{\varphi})g_{\mu\nu}],
\end{align}
where $\boldsymbol{\varphi}$ denotes a multiplet of $N$ scalar fields
$\{\varphi_1, \dots, \varphi_N\}$ and $\gamma_{ab}(\boldsymbol{\varphi})$ is a
target-space metric that mixes their kinetic terms.
Equation~\ref{ActionTMST} can be seen as a generalization of action~\eqref{eq:action_st} to
multiple fields.
We return to this theory in Sec.~\ref{sec:ns_scalarization}, as it was studied in the context of scalarized \ns{s}
by~\textcite{Horbatsch:2015bua} and~\textcite{Doneva:2020afj}.
However, we emphasize that one could consider further generalizations that
involve, for example, couplings to the Gauss-Bonnet invariant or further derivative
interactions between the scalar fields.

\subsection{Types of scalarization}
\label{types}

One can classify types of scalarization based on which property of the compact
object triggers scalarization and controls the threshold of the tachyonic
instability.
As we saw in our discussion of the model~\eqref{eq:ActionCaseI}, it is
the couplings between the scalar and curvature invariants that control the
onset of scalarization. Therefore, if one thinks of the onset of scalarization as an
instability around a spacetime of \gr, what controls the onset reduces to how
curvature invariants depend on the properties of the object that curves
spacetime.

\subsubsection{Induced by compactness}

For static, spherically symmetric \bh{s}, there is a straightforward answer: in
\gr the exterior is described by the Schwarzschild spacetime, so $R=0$ and
$\mathscr{G}=48 M^2 /r^6$, where $M$ is the mass of the \bh and $r$ is the
areal radial coordinate. Hence, for a given $M$, $\mathscr{G}$ scales
monotonically with $r$ and whether the scalar will develop a negative enough
effective mass square outside the horizon is controlled by $M$, with lighter \bh{s} being susceptible to the scalarization instability.
The opposite trend is present in \ns scalarization in the \DEF model: the trace of the energy-momentum tensor of the star is what controls the onset, and for most \EOS its magnitude shows a
steady increase as the radius decreases. The tachyonic instability is
stronger at the center of the star, where the density is higher, and this in turn correlates with the total mass of the star, with heavier stars scalarizing.

To unify these two pictures, one can observe that \bh{s} get more compact, in the sense of average mean density, as their mass decreases, whereas for \ns{s}
it is the other way around for most \EOS. Hence, how compact an object is plays a key role in scalarization and can determine whether the object will be affected by the tachyonic instability that triggers scalarization.

Although we colloquially refer to compactness as one of the triggers of scalarization in the rest of the review, there are some caveats that we ought to mention and on which we further elaborate. First, compactness for \bh{s} is often defined in the literature as the mass over the horizon radius. By that definition, unlike with the more colloquial meaning of the term, Schwarzschild \bh{s} of any mass have the same compactness (but not the same curvature near the horizon). Second, for \ns{s} the relation between compactness and scalarization is highly dependent on the \EOS, and there are cases where the tachyonic instability is not strongest at the center of the star. Finally, as we next discuss, other properties of compact objects also affect the value of curvature invariants, and hence can also control scalarization.

\subsubsection{Induced by spin}
\label{sec:incded_by_spin}

Rapidly rotating \ns{s} in the context of scalarization were first studied by
\textcite{Doneva:2013qva}. They showed that for \DEF-like models rapid
rotation can enhance scalarization by increasing the parameter space where
scalarization can occur.
Conversely, it is also known that larger spin leads to weaker scalarization for
\bh{s} in sGB gravity with $\dd^2f/\dd\phi^2 <0$ \cite{Cunha:2019dwb}.

In suitable theories, spin can in
itself induce a tachyonic instability that triggers scalarization~\cite{Dima:2020yac},
and spinning scalarized \bh{s} in these theories were
subsequently constructed explicitly by~\textcite{Herdeiro:2020wei} and~\textcite{Berti:2020kgk}.
The fact that spin can trigger \bh scalarization (rather than just controlling the
scalar charge) is important because it opens up the possibility that only
rapidly rotating \bh{s} carry scalar charge, irrespective of their mass.

These results, as well as further work on scalarized rotating compact objects
that is discussed in Secs.~\ref{sec:ns_scalarization}
and~\ref{sec:bh_scalarization}, show that spin, a ubiquitous
property of astrophysical objects, can play an
important role in the amount of scalar charge that a scalarized object carries.

\subsubsection{Induced by matter or coupling to other fields}
\label{sec:matter_induced_other_field}

Thus far we have linked scalarization to a tachyonic instability at the
perturbative level (although it is fundamentally a nonperturbative effect),
and we have focused on models in which that instability is controlled by
the nonminimal coupling to gravity.
However, as we saw, in the \DEF model one can  think of $\mu_{\rm eff}^2$
as being controlled either by the trace of the stress-energy tensor of matter
$T$ [see Eq.~\eqref{eq:mueffsqr_def}] or by the Ricci scalar $R$ [see
the action~\eqref{eq:ActionphisqR}]
The latter interpretation has the advantage of providing a unified framework of
scalarization of \bh{s} and stars as linked to a nonminimal coupling to
gravity, which is the perspective that we followed. However, the former
interpretation highlights that $\mu_{\eff}^2$ in Eq.~\eqref{genLinEq} could
instead be attributed to any type of coupling between $\varphi$ and another
matter field.

For example,~\textcite{Stefanov:2007eq} considered a scalar field coupled to
nonlinear electrodynamics, while~\textcite{Herdeiro:2018wub} focused on
Einstein-Maxwell-scalar theory with the addition of the coupling $e^{-\lambda
\varphi^2}F_{\mu\nu}F^{\mu\nu}$, where $F_{\mu\nu}$ is the usual Faraday tensor.
In both cases it was shown that electrically charged \bh{s} can develop scalar
hair through scalarization. Further work in this direction is summarized in
Sec.~\ref{sec:bh_scalarization_others}.

When $\mu_{\rm eff}^2$ in Eq.~\eqref{genLinEq} is thought of as being introduced by a
coupling to matter, one is also led to consider whether surrounding matter,
such as an accretion disk, a companion, or the Galaxy, could scalarize a \bh
even in models where \bh{s} cannot scalarize in vacuum. It was shown by~\textcite{Cardoso:2013fwa,Cardoso:2013opa}
that this can indeed occur in the \DEF model.

\subsubsection{Dynamical scalarization}
\label{sec:theory_dyn_sca}

Thus far we have discussed the scalarization of isolated compact objects, but what
happens when they form a binary?
As we alluded to, when we embed a \ns in an ambient scalar-field
environment [which is the case for the nonvanishing $\alpha_0$ in Eq.~\eqref{eq:DEF_alpha}], stars will always carry some small scalar charge
and they can still nonperturbatively develop large charge values when they
scalarize.
This is called induced scalarization~\cite{Salgado:1998sg},
and binaries provide a natural scenario for it to occur.
Imagine two \ns{s}, each with its own compactness, such that one is scalarized
and the other is not.
As the system inspirals, at some point the nonscalarized \ns will start
experiencing the presence of the scalar field sourced by its companion,
and induced scalarization will then take place~\cite{Barausse:2012da}.

Another, perhaps more dramatic, scenario is that of \emph{dynamical scalarization}.
In this case, two nonscalarized \ns{s} can become scalarized once their orbital
separation becomes sufficiently small. Qualitatively this can be quantified
by some measure of an ``effective compactness'' of the binary that, only for an isolated \ns,
can trigger scalarization once it reaches a certain threshold~\cite{Palenzuela:2013hsa,Shibata:2013pra,Taniguchi:2014fqa}.
In a quasicircular binary this effective compactness only increases (it scales inversely with the
orbital separation), but that does not have to be the case in an eccentric orbit.
In such cases, the effective compactness oscillates in time, being largest when the \ns{s} are closest.
This leads to a \emph{transient dynamical scalarization} of the system, whereby the two \ns{s}
continuously scalarize and descalarize as the system inspirals.

What about \bh{s}? In \sGB models the scalar field is sourced by the Gauss-Bonnet invariant,
and therefore a binary composed of two scalarized \bh{s} would in general result in an unscalarized
\bh remnant since the latter has a larger mass and therefore a smaller spacetime curvature.
That is, the system descalarizes~\cite{Silva:2020omi}. However, depending on the initial
nonscalarized \bh{s}' spins and masses, one can have cases where the remnant scalarizes due to its large spin, i.e., there can be \emph{dynamical spin-induced scalarization}~\cite{Elley:2022ept}.

As we have seen, compact binaries lead to new manifestations of scalarization.
We further discuss these in Secs.~\ref{sec:ns_scalarization} and~\ref{sec:bh_scalarization}.

\subsubsection{Beyond scalarization}
\label{sec:beyond_scalarization}

Thus far we have discussed scalarizations as a linear, tachyonic instability for a scalar field that is then quenched nonlinearly and leads to a nontrivial scalar configuration. There are many ways to extend this paradigm and yet keep the key outcome: to have fields that undergo what resembles a phase transition --- a (sharp change from a trivial to a nontrivial configuration) in the strong-field regime.

One direction is to generalize the mechanism to different fields, such as vectors, tensors, and spinors, generating models of \emph{spontaneous vectorization},
\emph{tensorization} or \emph{spinorization}~\cite{Ramazanoglu:2017xbl,Ramazanoglu:2017yun,Ramazanoglu:2018tig,Ramazanoglu:2018hwk}. Another approach would be to construct models in which the transition is triggered not by a tachyonic instability but by some other linear instability~\cite{Ramazanoglu:2017yun}. Both of these directions are discussed in Sec.~\ref{sec:other}.
A third direction comes from the possibility that the transition might not be triggered by a linear instability, but might instead be a fully nonlinear effect~\cite{Doneva:2021tvn}.
It was shown that if one chooses $f(\varphi)=\exp(\beta\varphi^4)$ in the action~\eqref{actionsgb}, one obtains a theory in which scalar perturbations are massless around Kerr \bh{s}, and hence there cannot be any linear tachyonic instability. Yet, stable scalarized solutions still exist for certain masses and spins.

\subsection{Quantum aspects and classical analogs}

While most of this review deals with scalarization from a classical
field theory perspective, it is noteworthy to comment on the quantum
aspects of this phenomenon.
In particular, \textcite{Lima:2010na} studied a quantum scalar field
nonminimally coupled to gravity living in the classical background of
a compact star spacetime (i.e., they worked in the semiclassical
approximation). They showed that some field modes can go through an exponential
growth causing the vacuum expectation value of the field operator
$\hat{\varphi}^2$ to grow, ultimately causing the vacuum
expectation value of the energy-momentum tensor $\hat{T}_{\mu\nu}$
of the field itself to grow in the same manner; see also~\textcite{Lima:2010xw}.
In this sense, one can say that classical curved spacetimes can ``awake the
vacuum state'' of quantum fields.
But what is the connection with spontaneous scalarization? The model studied
by~\textcite{Lima:2010na} is simply Eq.~\eqref{eq:ActionphisqR} for a
massive scalar field: the Jordan-frame equivalent of the \DEF model for a massive scalar.
In the parameter space spanned by the scalar-field--Ricci-scalar coupling
constant and stellar compactness relevant to our discussion,
the regions where the instability occurs agree with the classical prediction to where
scalarization should happen, as shown by~\textcite{Pani:2010vc}.
Thus, one can think of the classical ``scalar-field perturbations,'' which
we have so often spoken about
to be seeded by quantum field fluctuations~\cite{Landulfo:2014wra}.
We refer the interested reader
to~\textcite{Landulfo:2012nz,Lima:2013uya,Mendes:2013ija,Mendes:2014vna,Santiago:2015bve} for
other works on the semiclassical approach.

Finally, while our review focuses on astrophysical implications of
scalarization, we remark that the realization of this effect in condensed
matter systems has also been studied.
More specifically \textcite{Ribeiro:2019yzv} devised a classical analog that
exhibits this phenomenon based on the nonlinear optics of metamaterials, and this
could in principle be observed experimentally.

\section{Neutron star scalarization}
\label{sec:ns_scalarization}
Scalarization of compact objects was first considered in the context of \ns{s} by \textcite{Damour:1993hw}.
As discussed in Sec.~\ref{sec:theory_bg}, in this case the nonzero trace of the energy-momentum tensor of the nuclear matter acts as a source of the scalar field and evades the no-hair theorems existing for vacuum \bh{s} in certain scalar-tensor theories.
Since then, the scalarized \ns{s} in the \DEF model have attracted significant attention, and they are perhaps the most studied compact objects beyond \gr.
Over the years, \ns scalarization has also been examined in other scalar-tensor theories beyond the \DEF model.

In this section, we present the developments in the field over the past three
decades.
In Sec.~\ref{sec:NS_DEF_model}, we start with the \DEF model and its generalizations, placing a
special emphasis on the constraints coming from binary-pulsar observations.
In Sec.~\ref{sec:NS_DEF_DYN}, we proceed to the dynamics of such compact objects, both isolated
ones and those in binaries, discussing their stability in addition to \gw emission.
In Sec.~\ref{sec:NS_DEF_ASTRO}, we discuss the
various astrophysical implications of scalarized \ns{s}.
Finally, in Sec.~\ref{sec:ns_extended} we review
\ns{s} scalarization beyond the \DEF model.

\subsection{Equilibrium neutron stars in the Damour--Esposito-Far\`ese model}
\label{sec:NS_DEF_model}

In this section, we review the equilibrium properties of \ns{s} in the
\DEF theory, both in its original form and in extensions of the theory.
The strongest constraints on the former come from binary-pulsar constraints,
which we also discuss here.

\subsubsection{The original Damour--Esposito-Far\`ese model and binary-pulsar constraints}
\label{sec:DEF_model_equil}

\paragraph{Static neutron stars:}~As we saw in Sec.~\ref{sec:theory_def_model},
the effective mass squared of scalar-field perturbations in the \DEF model is given by
Eq.~\eqref{eq:mueffsqr_def}, namely,
\begin{equation*}
    \mu_{\rm eff}^2 = -4\pi G_{*} \beta(\varphi_0) T,
\end{equation*}
where we recall that $\beta(\varphi_0)$ is the derivative of the scalar-matter
coupling $\alpha(\varphi)$ evaluated at a constant background scalar-field value
$\varphi_0$.
Thus, the condition for scalarization $\mu_\eff^2 < 0$ can be satisfied when the trace of the energy-momentum tensor $T$ and $\beta(\varphi_0)$ have the
same sign.
For realistic \ns{s}, which are modeled as a perfect fluid with pressure $p$ and energy density $\varepsilon$,
we normally have
\begin{equation}
    T = 3 p - \varepsilon < 0,
\end{equation}
and hence scalarization can happen when $\beta(\varphi_0) < 0$.
In particular, for the coupling function~\eqref{eq:DEF_alpha} studied by~\textcite{Damour:1993hw},
we have $\beta(\varphi_0) = \beta_0$ being a constant.
Thus, most studies focus on the case where both $T$ and $\beta_0$
are negative, but we remark that at sufficiently high densities some \EOS{s}
predict a positive sign of $T$, resulting in scalarization when $\beta_0 > 0$;
see e.g.,~\textcite{Mendes:2014ufa,Podkowka:2018gib}.
For most \ns models, $p$ and $\varepsilon$ are related by a barotropic \EOS
$\varepsilon = \varepsilon(p)$.

\begin{figure*}[t]
\includegraphics[width=0.315\textwidth]{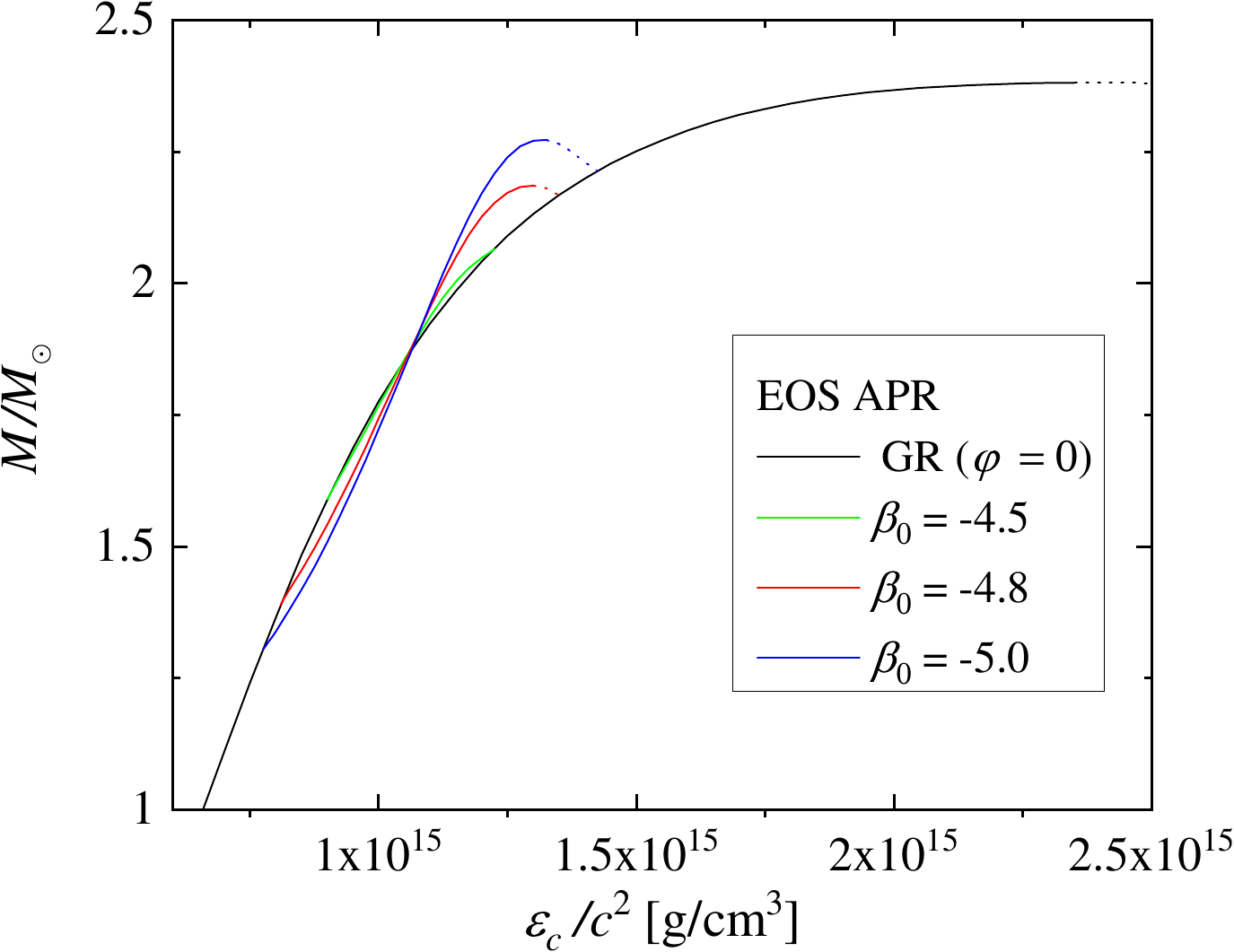}
\includegraphics[width=0.3\textwidth]{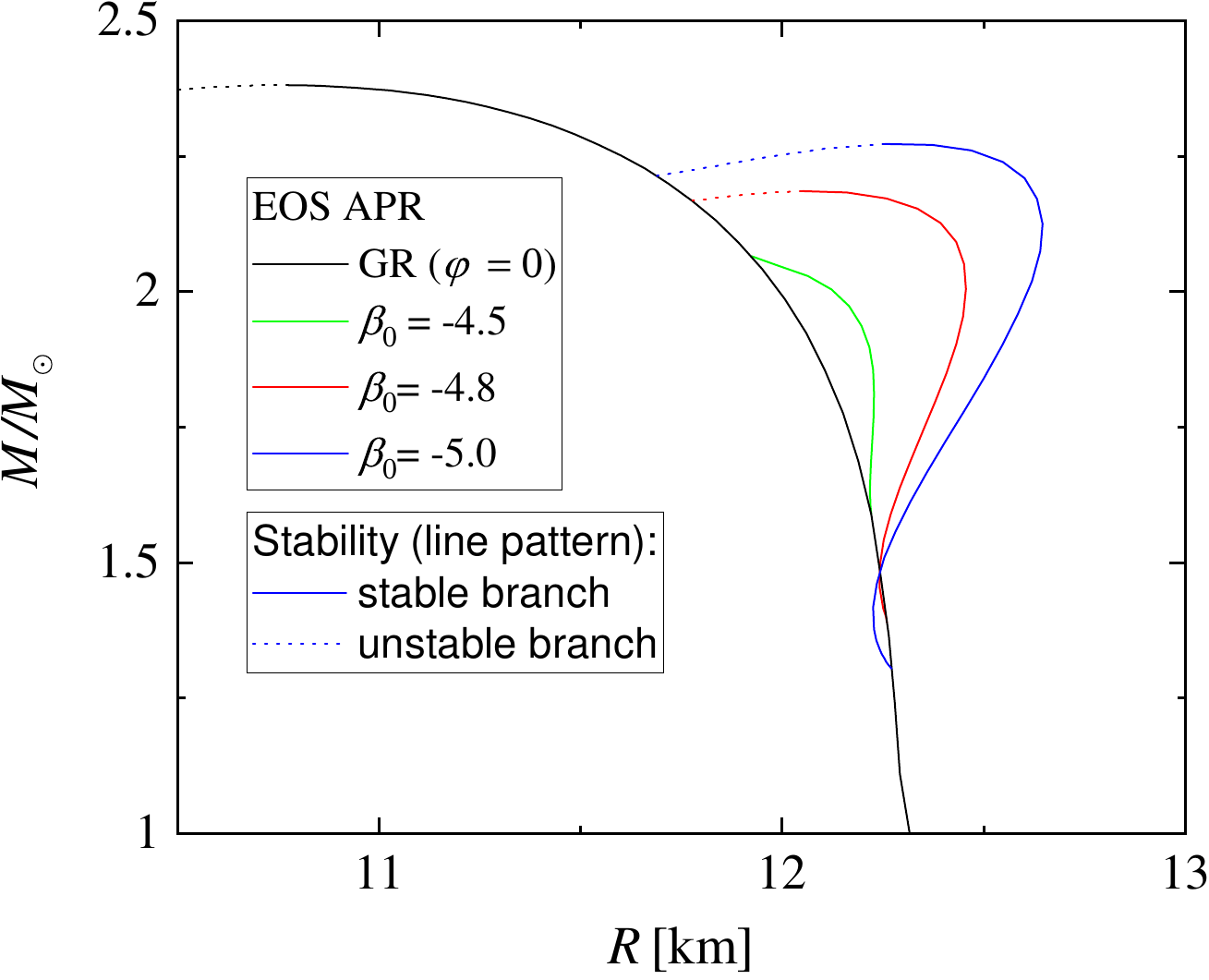}
\includegraphics[width=0.3\textwidth]{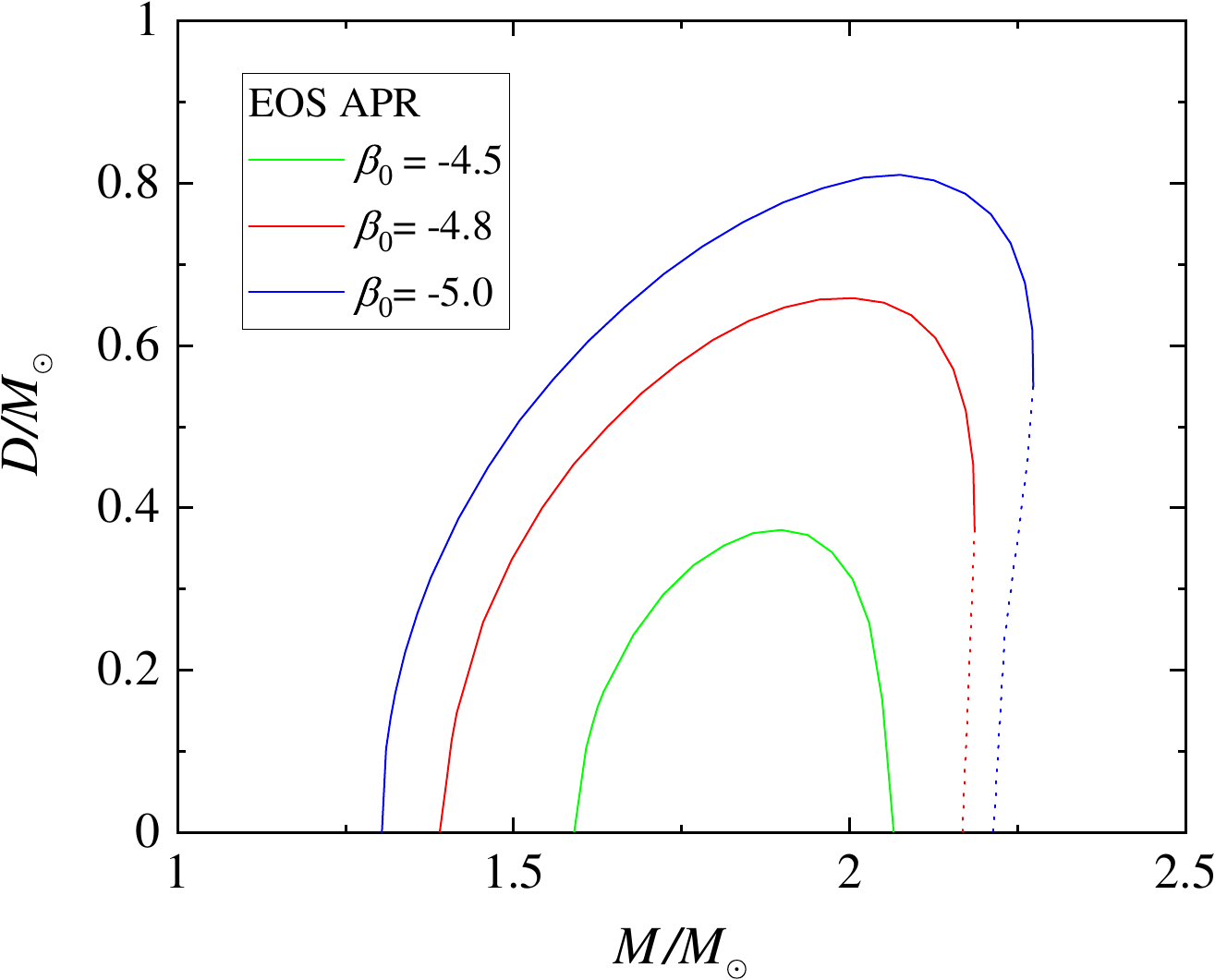}
\caption{
Some properties of \ns{s} in the \DEF model with $\alpha_0 = 0$, $\beta_0 < 0$,
and using the \APR \EOS.
We show the mass $M$ as a function of the central energy density $\varepsilon_\cc$ (left panel),
the mass $M$ as a function of the radius $R$ (middle panel), and
the scalar charge $D$ as a function of $M$ (right panel).
The solid black lines correspond to the \gr solutions, which are
also solutions in the \DEF model. We use differently colored lines to represent the
scalarized branches for $\beta_0 = -5.0,\, -4.8$, and $-4.5$.
\ns{s} past the maximum mass in the $M$-$\varepsilon_{\rm c}$ plane are unstable to radial oscillations.
These sequences of stars are shown as dotted lines.
}
\label{fig:DEF_NSmodels_Nonrot}
\end{figure*}

In Fig.~\ref{fig:DEF_NSmodels_Nonrot} we show some properties of nonrotating scalarized \ns{s}
for $\alpha_0 = 0$ and some illustrative values of $\beta_0 < 0$. Here in Sec.~\ref{sec:DEF_model_equil}, we consider the case of a vanishing scalar-field potential as in the original \DEF model. The influence of a nonzero potential is further discussed later. The \EOS is taken to be that
of \APR~\cite{Akmal:1998cf}.
In the left panel of Fig.~\ref{fig:DEF_NSmodels_Nonrot}, we show the Einstein-frame \ADM mass of a sequence of \ns solutions,
parametrized by the energy density at the center of the star $\varepsilon_{\cc}$.

When $\alpha_0 = 0$, \ns solutions of \gr are also solutions of the \DEF model;
they are indicated by solid black lines. These solutions are
characterized by having a zero scalar field.
We see that when a specific critical central energy-density value is reached
(say, $\varepsilon_{\cc, 1}$), the \gr sequence becomes unstable and a new branch of
stable solutions with a nontrivial scalar field (i.e., scalarized stars) bifurcates from it.
In our example, the value of $\varepsilon_{\cc, 1}$ depends only on $\beta_0$
and on the \EOS.
The scalarized branch merges again with
the \gr branch at a second bifurcation point at a larger energy density
$\varepsilon_{\cc, 2}$. Hence, scalarized \ns{s} exist only in the range
$\varepsilon_\cc \in [\varepsilon_{\cc, 1}, \, \varepsilon_{\cc, 2}]$.
We see that the larger $|\beta_0|$ becomes, the more dramatic the deviations
in the \ns mass relative to \gr are. In addition, the range of $\varepsilon_{\cc}$
in which scalarization can happen increases.
This is also shown in the middle panel of Fig.~\ref{fig:DEF_NSmodels_Nonrot}, where we plot the
\ADM mass as a function of the radius $R$, and in the right panel of Fig.~\ref{fig:DEF_NSmodels_Nonrot},
where we show the scalar charge $D$ as a function of the \ADM mass.
The scalar charge is defined in terms of an expansion at spatial infinity
of the scalar field $\varphi$, namely,
\begin{equation} \label{eq:scalar_charge}
    \varphi = \varphi_0 + D/r + {\mathcal O}\left(r^{-2}\right) ,
\end{equation}
where $\varphi_0$ is the cosmological background value of $\varphi$,
often assumed to be zero for simplicity.
The right panel of Fig.~\ref{fig:DEF_NSmodels_Nonrot} shows how $D$ has
a small magnitude near the bifurcation point where scalarization kicks in,
grows monotonically with $M$, and then approaches zero again onc e
$M$ approaches the mass corresponding to the second bifurcation point at $\varepsilon_{\cc, 2}$.

For which values of $\beta_0$ does scalarization occur?
Recall that the effective potential $V_\eff$ needs to be sufficiently negative in order to support at least one tachyonic mode; cf.~Eqs.~\eqref{eq:eff_potential} and~\eqref{eq:bound_state_criteria}.
For typical \ns densities, Eq.~\eqref{eq:mueffsqr_def} implies that scalarization exists for $\beta_0 \lesssim -4$. \textcite{Damour:1993hw} made this estimate using a Newtonian approximation and confirmed
it by integrating the fully relativistic equations of stellar equilibrium.
Subsequent works refined the threshold for scalarization to $\beta_0 \lesssim -4.35$ and also showed
that this bound is not very sensitive to the \EOS~\cite{Novak:1997hw,Silva:2014fca,Motahar:2017blm}.
We note that, similar to the \gr branch of \ns{s}, scalarized solutions in the original \DEF model
are stable up to the maximum mass of the corresponding branch,\footnote{For an
exception in the case of a massive scalar field, see Sec.~\ref{sec:ns_massive}.}
and the stable solutions are generally energetically favorable over the \gr
\ns{s}. This is further discussed in Sec.~\ref{sec:NS_dynamics}.

Let us briefly comment on the exact definition of mass in scalar-tensor
theories (shown in Fig.~\ref{fig:DEF_NSmodels_Nonrot} and elsewhere).
In contrast to \gr, the definition of mass in
scalar-tensor theories is subtle, due to the fact that
these theories violate the strong equivalence principle.
This results
in the appearance of different possible masses as a measure of the total energy
of the star~\cite{Lee:1974pt,Scheel:1994yr,Scheel:1994yn,Whinnett:1999ws,Yazadjiev:1999hy}.
These works showed that only the so-called tensor mass
has natural energylike properties.
For example, the tensor mass is positive definite, it decreases monotonically by the
emission of \gw{s} and it is well defined even in dynamical spacetimes~\cite{Lee:1974pt,Scheel:1994yr,Scheel:1994yn}.
In addition, only the tensor mass leads to a physically acceptable picture
since it peaks at the same point as the particle number, a property crucial for
the stability of the static stars~\cite{Whinnett:1999ws,Yazadjiev:1999hy}.
Therefore, the tensor mass, which is defined as the \ADM mass in the Einstein
frame, should be taken as the physical mass. As a matter of fact, though, for most of the commonly used coupling functions the Jordan-frame and the Einstein-frame \ns masses are identical.

After the discovery of the phenomenon, scalarization was examined for a larger
set of parameters and in more detail by~\textcite{Damour:1996ke}.
They showed that the presence of some externally imposed scalar-field background
$\varphi_0$, as well as considering $\alpha_0\ne 0$, smoothens the transition to
a scalarized state. This is what we described as induced scalarization in Sec.~\ref{sec:theory_dyn_sca}.
(Strictly speaking, we do not have pure scalarization when $\alpha_0\ne 0$, since \gr is
no longer a solution of the field equations.)
Spontaneous scalarization was further studied by
\textcite{Salgado:1998sg}, who considered the problem in the Jordan
frame and performed an approximate Newtonian analysis of the system.
They showed that scalarization can also be associated with the fact that the
effective gravitational constant in scalar-tensor theories decreases for large
scalar fields.
It was further argued by \textcite{Whinnett:2004rr} that scalarization leads to violation of the weak
energy condition in the inner regions of \ns{s}, which can cause instabilities.
It was later demonstrated by~\textcite{Salgado:2004tg} that this is not a general feature of scalar-tensor theories
and that there are subclasses of the theory where the weak energy condition is
easily satisfied.

Another consequence of scalarization is that for sufficiently large $\beta_0$ the maximum allowed mass for \ns{s} increases compared to \gr\footnote{This happens not only in scalar-tensor theories but also in other modified gravity theories. Upper bounds on the maximum mass of \ns{s} in \gr can be found under minimal assumptions on the \EOS using the approach of~\textcite{Rhoades:1974fn}. See~\textcite{Hartle:1978201} for an early account of applications of this method to other gravity theories.}, which can have various observational consequences. This problem was studied by~\textcite{Sotani:2017pfj}, who took the effects of various microphysics parameters into account. Empirical relations were derived for the maximum mass of scalarized \ns{s} that are parametrized with respect to the nuclear saturation parameters and the maximum sound velocity in the core.

Until now we have discussed works that considered negative $\beta_0$.
However, as we mentioned, when the trace of the energy-momentum tensor
is positive, scalarization can occur for $\beta_0>0$ as well, and this
scenario introduces qualitative differences relative to our story thus
far~\cite{Mendes:2014ufa}.
In particular, for the commonly used coupling function $\alpha(\varphi) = \beta_0 \varphi$ of the \DEF theory, scalarized stars are not stable for high values of $\beta_0$ ($\beta_0 \gg 1$), which is further discussed in Sec.~\ref{sec:ns_dynamics}.

The calculation of \ns parameters spanning the \DEF theory parameter space
provides a challenging technical task. In particular, tests of this theory
against binary-pulsar observations (described in Sec.~\ref{sec:ns_binary_pulsar}) require
knowledge of the scalar charges (and their derivatives with respect to the star's
mass) for a large catalog of \EOS{s}.
This calculation was performed most extensively in the works by~\textcite{Anderson:2019hio,Guo:2021leu},
who provide the results in tabulated form or through surrogate models.
\textcite{Yagi:2021loe} [see also~\textcite{Horbatsch:2010hj}] computed scalar charges in the \DEF model analytically using a combination of perturbative weak-field expansion and Pad\'e resummation
and found excellent agreement with numerical calculations.

Spontaneous scalarization for the case of several nearby compact objects was considered by~\textcite{Cardoso:2020cwo}.
Their analytical analysis showed that, even though an isolated body might be below the threshold for scalarization, a collection of such bodies could develop a nonzero scalar field while maintaining average compactness much below the scalarization limit.

We have already discussed spontaneous scalarization of \ns{s}, but other compact objects can also scalarize.
The case of \bh{s} is considered in Sec.~\ref{sec:bh_scalarization}, but as an additional nonvacuum example
we now discuss the case of boson stars [cf.~\textcite{Liebling:2012fv} for a review], which were also shown to scalarize~\cite{Whinnett:1999sc}. In such systems, one has a complex scalar field as a matter source for the boson star and an additional real scalar field responsible for scalarization similar to the \DEF model.
The dynamics of this process was examined by~\textcite{Alcubierre:2010ea} who showed that nonlinear development of the scalar field is observed in the absence of self-interactions in the complex scalar field.
\textcite{Ruiz:2012jt} studied spontaneous and induced scalarization starting with initial data corresponding to stable boson stars in \gr. They showed that a strong emission of scalar radiation occurs during the scalarization process.

\paragraph{Observational constraints from binary pulsars:}
\label{sec:ns_binary_pulsar}

To date binary pulsars have set the best constraints on scalarized \ns{s} in the \DEF models~\cite{Kramer:2021jcw,Zhao:2022vig}. This is because determining different characteristics of these systems through pulsar timing can be made extremely precise by accumulating yearslong observations.
One of such important observables is the rate at which the binary's orbit decreases by energy loss
through \gw emission.
In contrast to \gr, the \DEF model has an additional scalar degree of freedom that leads to a new channel of energy loss. Thus, the shrinking of the orbit should happen faster. The energy flux of the scalar-dipole radiation that gives the dominant contribution is given by~\cite{Damour:1992we},
\begin{equation}\label{eq:NS_scalar_dipole_radiation}
F^{\rm dipole}_{\rm scalar}= A \, (1+ q_1 q_2)^3 (q_2-q_1)^2 ,
\end{equation}
where $A$ is a function depending on the properties of the binary, such as its total and reduced masses, and its eccentricity; $q_2$ and $q_1$ are the scalar charges of each binary component normalized with respect to their
masses $q_{i} = D_{i}/M_{i}$ ($i = 1$ and $2$).

Binary-pulsar observations were considered in the context of scalarized \ns{s} for the first time by~\textcite{Damour:1996ke}. They calculated the gravitational form factors (also known as sensitivities) of slowly rotating \ns{s}, which form the set
of coupling constants appearing in the post-Keplerian description of the binary in scalar-tensor theory; \textcite{Horbatsch:2011nh} for a summary.
Only a few such binary systems were known at the time, and $\beta_0$ was constrained to be greater than $-5$ using polytropic \EOS{s}. These results were later refined to include realistic \EOS{s} and a limit of $\beta_0>-4.5$ was derived by \textcite{Damour:1998jk}. In addition, an estimate was made that it would be difficult for LIGO and VIRGO to improve $\beta_0$ bounds for most \EOS possibilities [some exceptions were pointed out by \textcite{Sampson:2014qqa} and \textcite{Shao:2017gwu}]. The reason is that even though the merger events observed using such \gw detectors can lead to much stronger scalar-dipole radiation, they are inferior in accuracy compared to the radio observations of binary pulsars, leading to weaker overall constraints. Next-generation detectors such as the Cosmic Explorer and the Einstein Telescope, though, will be able to improve the bounds on scalar-dipole radiation~\cite{Sampson:2014qqa,Shao:2017gwu}.

\begin{figure}[t]
\includegraphics[width=\columnwidth]{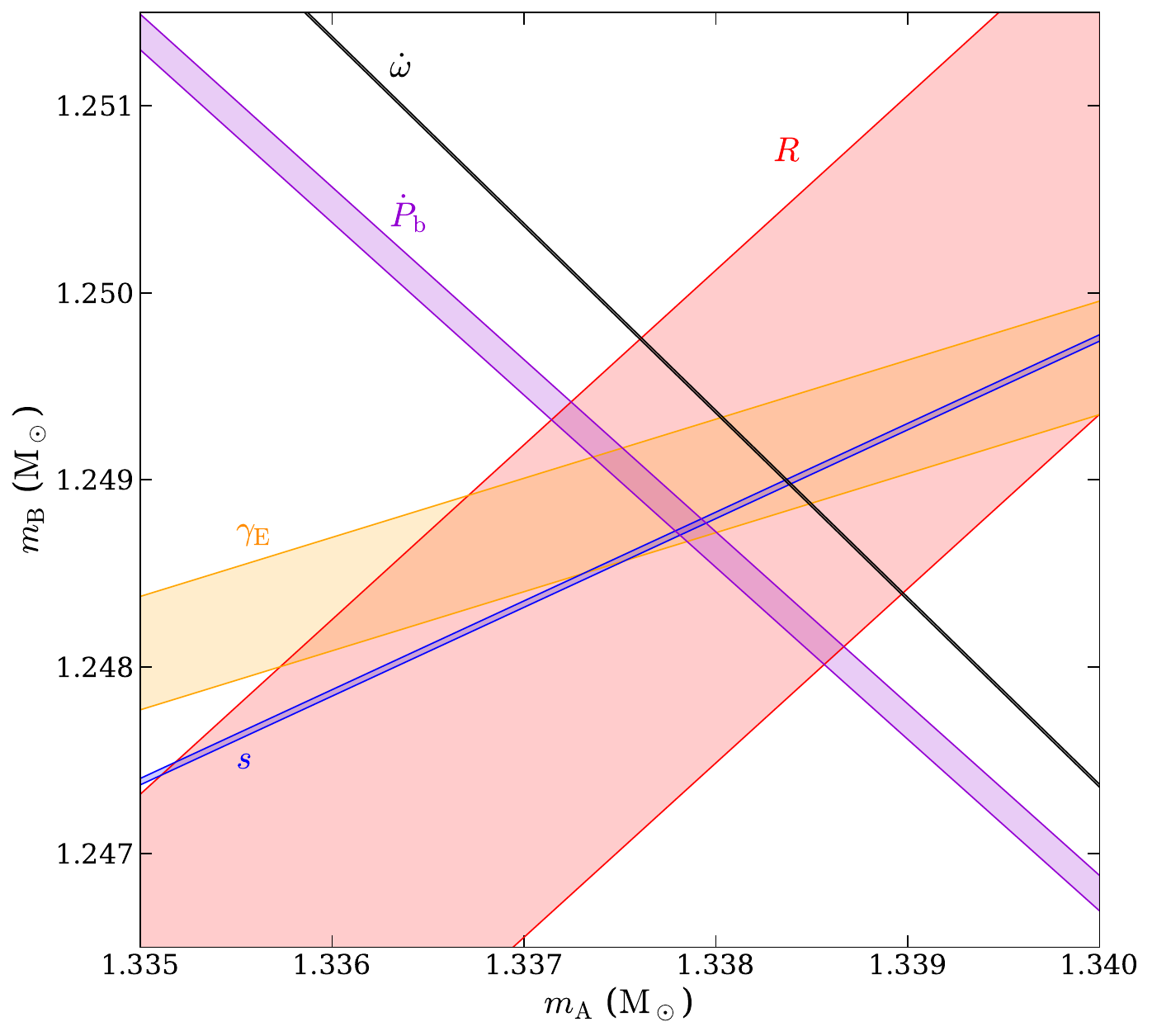}
\caption{Mass-mass diagram for the double pulsar PSR~J0737-3039A/B for the
\DEF model with $\alpha_0 = 5 \times 10^{-4}$, $\beta_0 = -4$, and
the assumption that the \ns{s} are described by \EOS MPA1.
Various measured post-Keplerian parameters are shown as different curves,
with the width indicating the measurement uncertainty in each parameter.
For the point $\alpha_0 = 5 \times 10^{-4}$ and $\beta_0 = -4$ of the
parameter space of the \DEF model to be consistent with observations, all
curves would have to intersect in a region of the mass-mass plane.
This is not the case, as can be seen with the $\dot{\omega}$ (periastron
advance) and $\dot{P}_{\rm b}$ (change of orbital period) curves.
For further details and the definitions of the other post-Keplerian parameters
see~\textcite{Kramer:2021jcw}.~From~\textcite{Kramer:2021jcw}.}
\label{fig:BinaryPulsarObservationsMAMBplane}
\end{figure}

An example of how constraints can be imposed on \DEF models is presented in Fig.~\ref{fig:BinaryPulsarObservationsMAMBplane}, which illustrates how
the model fails the double pulsar test for specific values of the theory parameters. In this case, the failure is due to the additional energy loss from scalar GWs predicted by the \DEF model, predominantly the dipolar contribution.

With the advances in observational astronomy, more pulsars in binary systems suitable for constraining the scalar-dipole radiation have been discovered, a complete and up-to-date list can be found in~\textcite{FreireWebSite}. Consequently, observational bounds on the scalar-tensor gravity parameter $\beta_0$ for such systems have been widely discussed in the literature~\cite{EspositoFarese:2004cc,Antoniadis:2013pzd,Freire:2012mg,Shibata:2013pra,Wex:2014nva,Shao:2017gwu,Voisin:2020lqi,Kramer:2021jcw,Chiba:2021rqa,Zhao:2022vig}.
The strongest current limit comes from \textcite{Zhao:2022vig}, who practically closed the scalarization window for the original \DEF model; i.e., the possibility for scalarization is ruled out in this theory.
As we later discuss, though, there are a number of other well-motivated models where scalarization is still possible or cannot be constrained at all by binary-pulsar observations. These include theories with a massive scalar field, tensor-multiscalar theories, or even the standard \DEF model when rapid rotation of \ns{s}, which enhances the effect of scalarization, is considered. Furthermore, the theoretical and numerical approaches developed for the study of the \DEF model are still applicable to these generalized theories in most situations. Thus, we spend considerable time here on the aspects of the \DEF model despite its original form being essentially ruled out.

Constraints on scalarization with $\beta_0>0$ using pulsar-timing observations were investigated by \textcite{Mendes:2019zpw}. Owing to the fact that the scalar charge is suppressed as $\beta_0$ increases while the range of masses allowing spontaneous scalarization decreases, it turns out that only weak constraints can be imposed by binary-pulsar observations in this part of parameter space.

\paragraph{Rotating scalarized neutron stars:}
Thus far we have commented only on static \ns models.
All observed \ns{s} are at least slowly rotating, and
some dynamical processes such as \ns mergers or stellar core collapse can
produce relatively long-lived rapidly rotating supramassive protoneutron
stars.
Hence, the inclusion of rotation in \ns physics is an inseparable part of the
goal to explore their astrophysical implications.

\textcite{Damour:1996ke} were the first to study slowly rotating scalarized \ns{s} to leading order in rotation frequency ${\mathcal O} (\Omega)$ using the formalism of~\textcite{Hartle:1967he,Hartle:1968si},
which allowed them to calculate the \ns moment of inertia; see also \textcite{Sotani:2012eb}.
It is interesting that in this case there is an exact analytical solution for the \ns exterior~\cite{Damour:1996ke}. Static and slowly rotating \ns{s} for a wide range of realistic \EOS{s}, including examples with hyperons or quark matter, were considered by \textcite{Motahar:2017blm}.
The extension to second order in the rotational frequency ${\mathcal O} (\Omega^2)$ was made by \textcite{Pani:2014jra}, who used the extension to calculate rotational corrections to the stellar radius and mass, and also its quadrupole moment.

\begin{figure}
	\centering
	\includegraphics[width=0.45\textwidth]{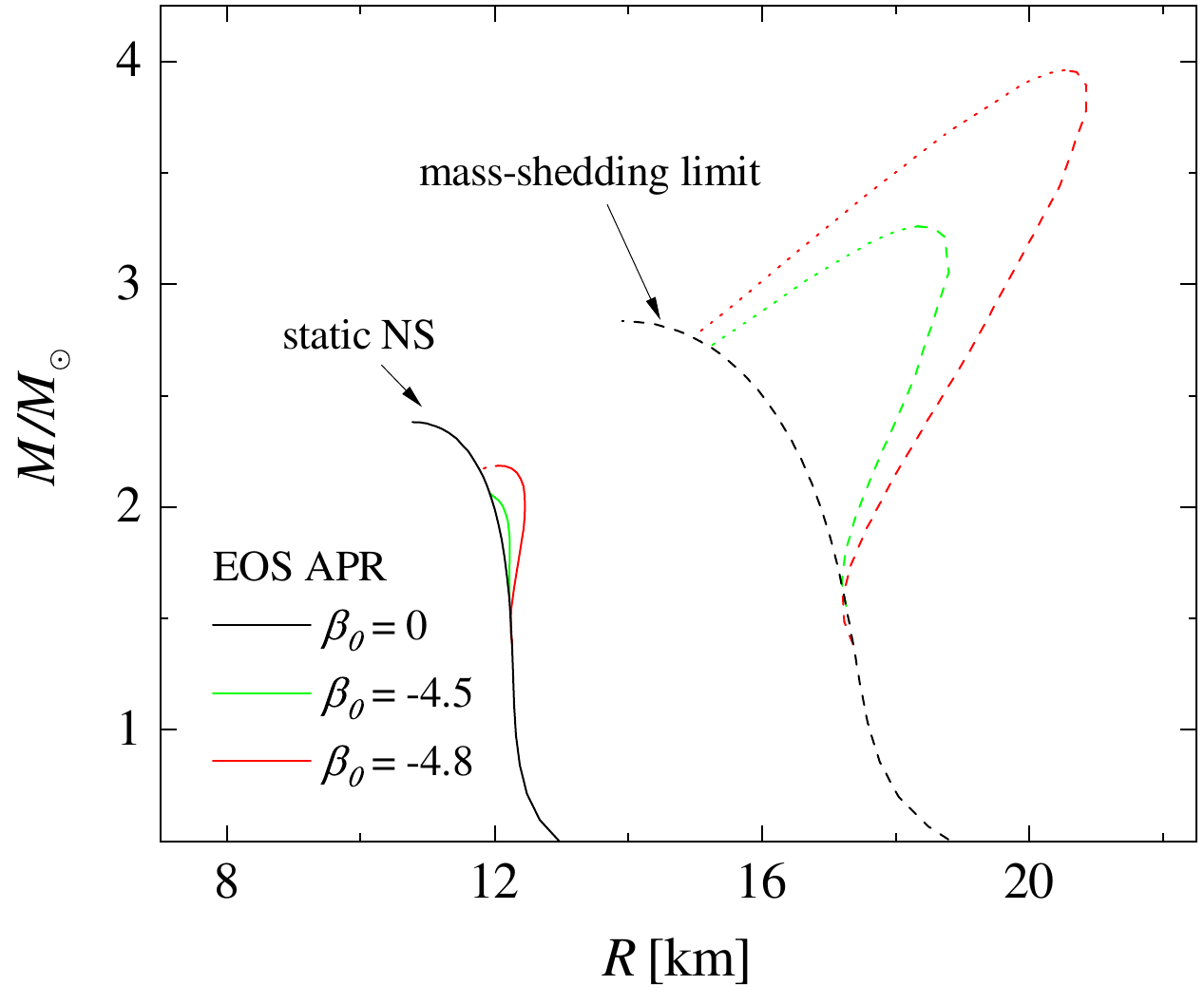}
\caption{Mass-radius relation for \ns{s} in the \DEF model for $\beta_0 = 0,\, -4.5$~and~$-4.8$
and employing the \APR \EOS. Nonrotating \ns{s} are shown as solid curves, whereas stars rotating at the mass-shedding limit are depicted as dashed lines. As in Fig.~\ref{fig:DEF_NSmodels_Nonrot}, stars unstable to radial oscillations
are shown as dotted lines. The radius of rotating stars refers to their equatorial radius.}
	\label{fig:DEF_NS_RNS}
\end{figure}

Rapidly uniformly rotating scalarized stars, without
approximation, were obtained by \textcite{Doneva:2013qva}.
They showed that for a fixed $\beta_0$ the maximum deviation from \gr that is
achieved at the mass-shedding limit is considerably larger than in the
static case, and the range of central energy densities where scalarization is
possible is significantly broadened.
This can be seen in Fig.~\ref{fig:DEF_NS_RNS}, where sequences of static
scalarized \ns{s} are compared to \ns{s} rotating at the Kepler limit for the
same values of $\beta_0$.

There are a number of factors leading to differences from the static case. The first, more intuitive one is that the rotational energy of the star also acts as a source for the scalar field, and thus can change the onset and degree of scalarization. Meanwhile, the rapidly rotating models tend to be less compact, which can reduce the degree of scalarization. The large deviations from \gr compared to the static case, conversely, are due mainly to the fact that scalarized stars can sustain much larger angular momentum before reaching the Kepler limit. This is a nonlinear effect that could not be normally caught in the slow-rotation approximation.

A natural consequence of the aforementioned rotation effects is that the minimum $|\beta_0|$ where scalarization is possible changes compared to the static case. Thus, for the same \EOS II~\cite{1985ApJ...291..308D} used by \textcite{Damour:1996ke}, scalarization happens for $\beta_0<-3.9$ in rotating stars \cite{Doneva:2013qva} compared to $\beta_0<-4.35$ in the static case \cite{Damour:1996ke}.
Therefore, one can conclude that, although binary-pulsar
observations seem to rule out \DEF scalarization for static or
slowly rotating \ns{s}~\cite{Zhao:2022vig}, there is still an observationally viable range of $\beta_0$ where only rapidly rotating \ns{s} can scalarize.
This is potentially relevant for binary mergers and stellar core collapse, where
such rapidly rotating \ns{s} can form.

A caveat in the previous argument is that one does not expect the star to rotate uniformly in these extremely dynamical events.
Such differential rotation was first studied by \textcite{Doneva:2018ouu}, who adopted a simple rotation law that can still capture some of the main properties of the merger remnants, especially a few tens of milliseconds after the merger of the binary~\cite{Bauswein:2017aur}.

When scalarization is considered, larger values of the maximum mass as well as
of the angular momentum can be achieved for supramassive \ns{s} compared to
\gr.
Moreover, the scalar field causes rapidly rotating models to be
less quasitoroidal than their general-relativistic counterparts.
This can have direct astrophysical implications, especially for binary \ns mergers, where the maximum possible mass and angular momentum that a \ns can sustain are crucial for determining the merger outcome and the lifetime of the merger remnant (in case a supramassive or hypermassive \ns forms\footnote{Supramassive \ns{s} do not have a stable static limit but are supported against collapse due to rapid rotation. Hypermassive \ns{s} do not have a stable uniformly rotating limit but are supported against collapse due to differential rotation \cite{Paschalidis:2016vmz}.});
see~\textcite{Bauswein:2013jpa,Takami:2014zpa,Weih:2017mcw,Bauswein:2017aur}.

\subsubsection{Massive scalar field}\label{sec:NS:MassiveSF}
\label{sec:ns_massive}

\begin{figure}
	\includegraphics[width=\columnwidth]{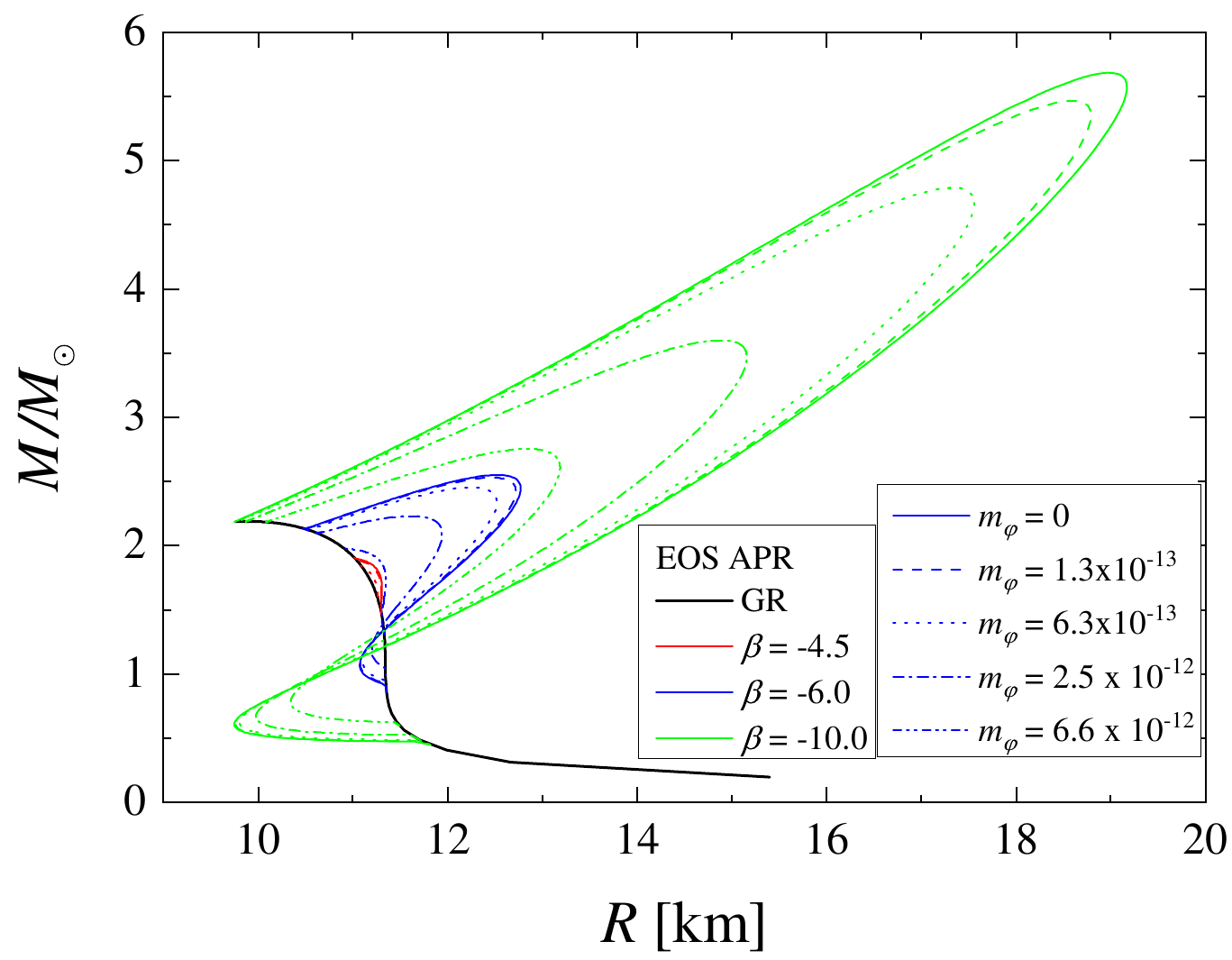}
\caption{Mass as a function of radius for the APR4 \EOS. The results for different values of the coupling constant $\beta_0$ and mass of the scalar field $\mu$, given in eV, are plotted. The potential is assumed to have the form $V(\varphi) = (\mu^2 / 2) \varphi^2$, with the scalar field defined though the action \eqref{actionsgb}. From \textcite{Yazadjiev:2016pcb}.}
	\label{fig:DEF_massive_SF}
\end{figure}

A key property of scalar-tensor theories that was neglected in the previously
discussed studies is the possibility of having a nonzero scalar-field potential.
The simplest case is to take a potential that leads to a nonzero scalar-field
mass $\mu$, but more complicated potentials, such as those with
self-interaction terms, can be considered as well; see Eq.~\eqref{eq:phi4P}.
Although this seems like a simple extension, it has a dramatic effect on the observational properties of \ns{s}, especially on the \gw emission. The reason lies in the different asymptotic behavior of the scalar field.  In the case of zero potential, the scalar field decreases as $D/r$ at infinity according to Eq.~\eqref{eq:scalar_charge}. This leads to a nonzero scalar charge $D$ and thus nonzero scalar-dipole radiation. In the presence of nonzero scalar-field mass $\mu$, though, the scalar field tends exponentially to zero after some characteristic distance related to its Compton wavelength $\lambda_\varphi = 2\pi/\mu$ as discussed by \textcite{Ramazanoglu:2016kul}. Hence, the scalar field is effectively confined to a characteristic radius and its scalar charge is zero. If the orbital separation between the two objects in a binary-pulsar system is much larger that $\lambda_\varphi$, the dynamics will not be directly altered by the scalar field and there is no significant emission of scalar-dipole radiation~\cite{Alsing:2011er}.
Since $\lambda_\varphi$ is controlled by $\mu$ in the simplest case of a scalar-field potential,
one can reconcile the \DEF model (for arbitrarily small
$\beta_0$) with binary pulsar constraints by giving the scalar field a mass
$\mu \gtrsim 10^{-16}$~eV.

Scalarized \ns{s} in massive scalar-tensor theories were first studied in the static case~\cite{Chen:2015zmx,Popchev2015,Ramazanoglu:2016kul} and later extended to slow~\cite{Yazadjiev:2016pcb} and rapid rotation~\cite{Doneva:2016xmf}.
The inclusion of a quartic self-interaction term to the potential was considered by \textcite{Staykov:2018hhc}.
These works showed that the mass of the scalar field and the self-interaction have similar effects on the scalar field around \ns{s} and that they both suppress scalarization.
The quartic interaction by itself cannot affect the range of central energy
densities where scalarized solutions exist, because it is a nonlinear
contribution to the linearized scalar-field equation of motion [recall
Eq.~\eqref{genLinEq}].
In contrast, the mass term shrinks the domain of existence of scalarized \ns{s}
and, for large enough masses, no scalarization is possible at all.
This is evident in Fig. \ref{fig:DEF_massive_SF}, where \ns mass is plotted as a function of its radius for different combinations of $\beta_0$ and
$\mu$.
The massive scalar-field solutions are confined between the zero scalar-field
mass models (the original \DEF models) and the \gr ones, corresponding loosely
speaking to $\mu \rightarrow \infty$.
Note that for $\mu \gtrsim 10^{-16}$~eV the solutions are almost
indistinguishable from the massless \DEF model. For this reason the latter
represents an upper limit on the possible deviations from \gr in massive
scalar-tensor theories.

The exponential asymptotic behavior of the massive scalar field brings
computational challenges to the construction of scalarized \ns{s}, which led
to new numerical approaches~\cite{Rosca-Mead:2020bzt} and also facilitated the construction of \ns{s} for highly negative $\beta_0$ and large scalar-field masses.
\textcite{Rosca-Mead:2020bzt} showed that for sufficiently negative $\beta_0$
qualitative changes in the strongly scalarized branch of solutions are
possible.
For example, the maximum of the scalar field can be located away from the stellar center. In their most extreme form, these solutions are composed of a highly compact \ns model surrounded by a scalar-field shell.
Also,~\textcite{Tuna:2022qqr} showed that some scalarized solutions in this
part of the $(\beta_0, \mu)$-parameter space gave indications of metastability:
they were stable to small perturbations but had lower binding energy
than their \gr counterparts.

An extension of these results to other forms of coupling functions and
scalar-field potentials is the asymmetron model~\cite{Chen:2015zmx,Morisaki:2017nit}. It is interesting because of the fact that the asymmetron model realizes proper cosmic evolution, and it can also account for the cold dark matter. \textcite{Chen:2015zmx,Morisaki:2017nit} focused especially on large scalar-field masses spanning several orders of magnitude and having a Compton wavelength shorter than 10km, which is the typical size of a neutron star.

\subsubsection{Incorporating further physics}
New aspects of the original \DEF model were recently studied with the inclusion of different physical details.
For instance,~\textcite{Silva:2014fca} studied the presence of anisotropic pressure of nuclear matter for both static and slowly rotating \ns{s}.
The motivation for this comes from the fact that some theoretical considerations,
for instance, with magnetic fields or within the Skyrme model (a low-energy \EFT of quantum chromodynamics)~\cite{Nelmes:2012uf}
suggest that at high densities the \ns \EOS might have a significant degree of anisotropy~\cite{Herrera:1997plx}.
In such a case, the effects of scalarization increase (decrease) when the
tangential pressure is larger (smaller) than the radial pressure. The threshold value of $\beta_0$ for the development of scalarization, which in the
isotropic case is $\beta_0<-4.35$, can be increased due to the
presence of anisotropy, thus widening the range of parameters in which
scalarization is possible.

Another astrophysically interesting extension of scalarization is to include
the magnetic fields. According to observations and modeling, \ns magnetic
field values can span from $10^8$ to $10^{12}$~G for standard ``old''  pulsars,
ranging from $10^{16}$~G at the surface of some magnetars to,
hypothetically, as high as $10^{17}-10^{18}$~G in the cores of newly formed
protoneutron stars.
Such strong magnetic fields impact the properties of scalarized \ns{s}, including their magnetic deformability, maximum mass, and range of scalarization, as studied by~\textcite{Soldateschi:2020hju}.
They found a magnetically induced spontaneous scalarization whose essence is the following:
strong toroidal magnetic fields can support descalarized configurations and, if
the star's magnetic field decreases during some nonideal magnetohydrodynamical process, the star can
undergo a rapid growth of the scalar field; i.e., it scalarizes.
The magnetic quadrupolar deformations of scalarized \ns{s} and the related
\gw{s} produced by rotating magnetars were studied by~\textcite{Soldateschi:2020zxb}

Another interesting extension to the standard \DEF model is related to
challenging the idea that the fundamental physics remains unchanged in the star’s
interior, which is a common assumption when a nuclear matter \EOS is constructed.
This was studied by \textcite{Coates:2016ktu}, who considered two models
in which the mass of the photon had a different value in the interior and the
vicinity of a compact star compared to the mass measured by experiments
performed in a weak-gravity regime. The first model is based on a Proca-like
mass with an effective mass term dependent on $\varphi$.
The second model can be thought of as a gravitational Higgs mechanism where the
Higgs potential is replaced by the scalar-gravity coupling. In both cases the
scalar field undergoes spontaneous scalarization, thus acquiring a nontrivial profile
if the compactness passes a certain threshold, providing a mass to the photon
by coupling to it in an appropriate manner. Although the focus of
\textcite{Coates:2016ktu} was on the electromagnetic field as a proof of
principle, these results can be extended to other fields of the standard model.
The signatures of such a gravitational Higgs mechanism on the
behavior of magnetic field of \ns{s} in Einstein-Maxwell theory was studied by~\textcite{Krall:2020kto}.

\subsection{Dynamics of scalarized neutron stars and binary mergers}
\label{sec:NS_DEF_DYN}

The dynamics of isolated \ns{s} can be studied by solving the full nonlinear field equations of scalar-tensor gravity,
which is often a challenging task. Instead, one usually first approaches the problem by linearizing the field equations around a background solution and then analyzing the resulting linearized dynamics. The study of nonlinear dynamics is then done when necessary and feasible.
We follow this sequence in the Secs.~\ref{sec:NS_dynamics}--\ref{sec:dyn_sca_nsm}. We then review what happens when \ns{s} in scalar-tensor theories are placed in binary systems.

\subsubsection{Linearized dynamics}\label{sec:NS_dynamics}

The studies of linearized dynamics concern the stability of scalarized stars and
the analysis of their \qnm spectrum. The latter involves the study of different classes
of \ns oscillation modes that are tied to the emission of \gw{s}.
See \textcite{Kokkotas:1999bd} for a review.

\paragraph{Stability:}
\label{sec:NS_Stability}

\begin{figure}
\includegraphics[width=\columnwidth]{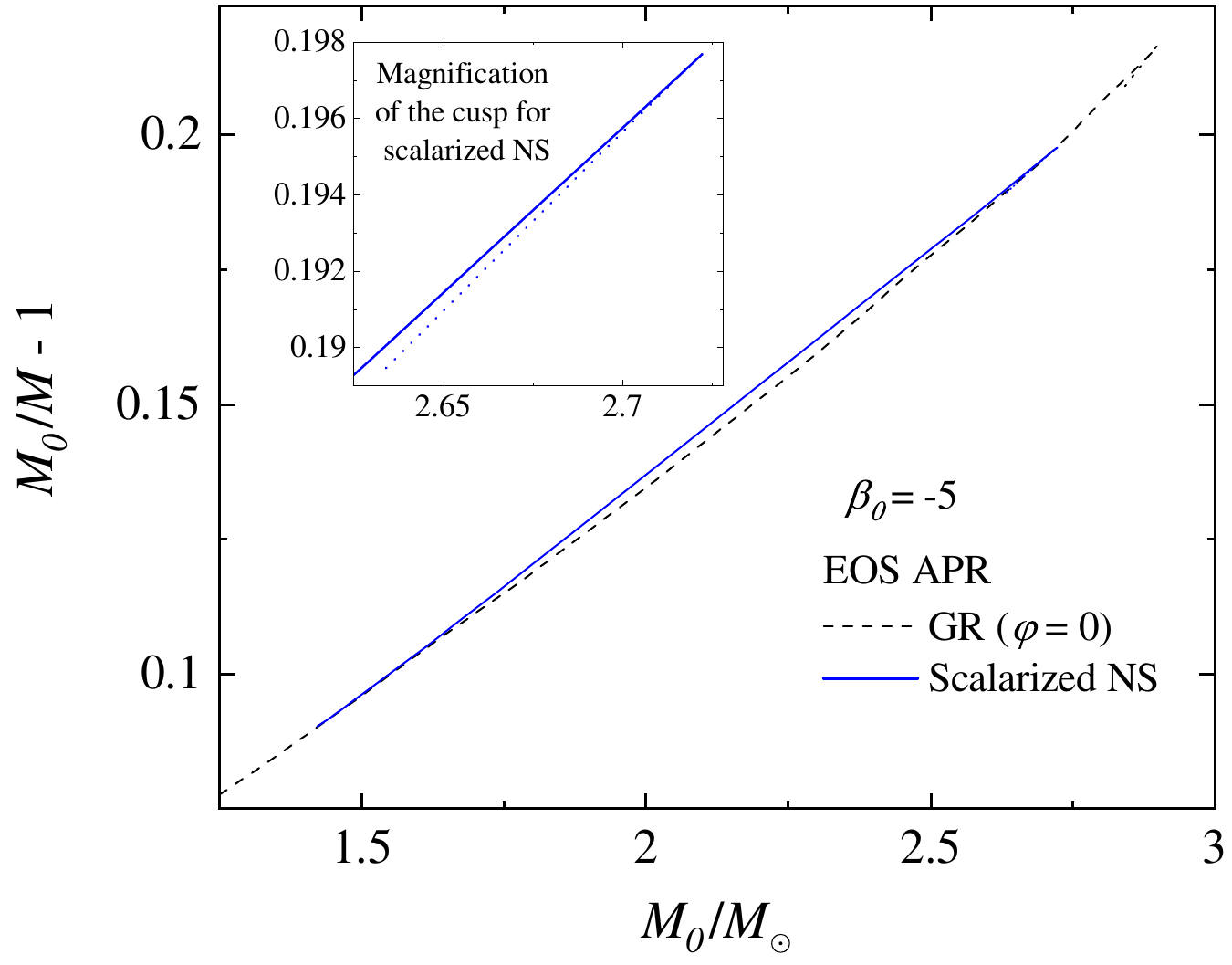}
\caption{Fractional binding energy $M_0 / M - 1$ as a function of the baryonic mass $M_0$ for scalarized \ns{s} in the \DEF model with $\beta_0 = - 5$ using the \APR \EOS. Inset: enlargement of the scalarized branch showing the formation of a cusp at the maximum of the mass in the upper-right corner.
}
\label{fig:DEF_NSstability_Nonrot}
\end{figure}

We have seen that scalarized \ns{s} coexist with their nonscalarized counterparts
as solutions within the \DEF model. Which of these branches of solutions is the one realizable in nature?
One way of answering this question consists of calculating the fractional binding
energy $M_0 / M - 1$, where $M_0$ is the star's baryonic mass.
Through this calculation,~\textcite{Damour:1993hw} showed
that scalarized \ns{s} are usually energetically favored over the \gr ones.
This is evident in Fig.~\ref{fig:DEF_NSstability_Nonrot}, where we plot the binding energy
as a function of the baryonic mass. At constant baryonic mass, the scalarized solutions (solid curve) have larger binding energies relative to the \gr solutions (dashed curve) and are thus energetically favorable.
In addition, we can see a cusp at the maximum
of the mass for both the scalarized and nonscalarized \ns{s}. This suggests a change of stability. Since the branches beyond the maximum of the mass have a lower binding energy (dotted curve in the inset), they are unstable.

The aforementioned stability analysis relied on the bulk properties of the star. A rigorous complementary approach
considers the linear perturbations of the star.
The first step in this direction was taken by~\textcite{Harada:1997mr}, who studied scalar-field perturbation in the background of a \ns with a zero (or constant) scalar field within the \DEF model.
\textcite{Harada:1997mr} studied the perturbation equations in the frequency domain and showed that the \gr solution becomes unstable after a specific critical central energy density.
This is the point where the scalarized solutions branch out from the \gr ones; see Fig.~\ref{fig:DEF_NSmodels_Nonrot}.
\textcite{Harada:1998ge} reached similar conclusions but worked in the context of catastrophe theory.
The radial stability of scalarized \ns{s} was also studied by~\textcite{Mendes:2018qwo}.
They considered metric and scalar-field perturbations for both signs of $\beta_0$.
They found that scalarized \ns{s} are stable against linear perturbations and
that instability takes place past the point of maximum mass.

\paragraph{Gravitational waves from perturbed neutron stars:}
\label{sec:NS_QNMs}
\begin{figure*}[th]
\includegraphics[width=0.9\textwidth]{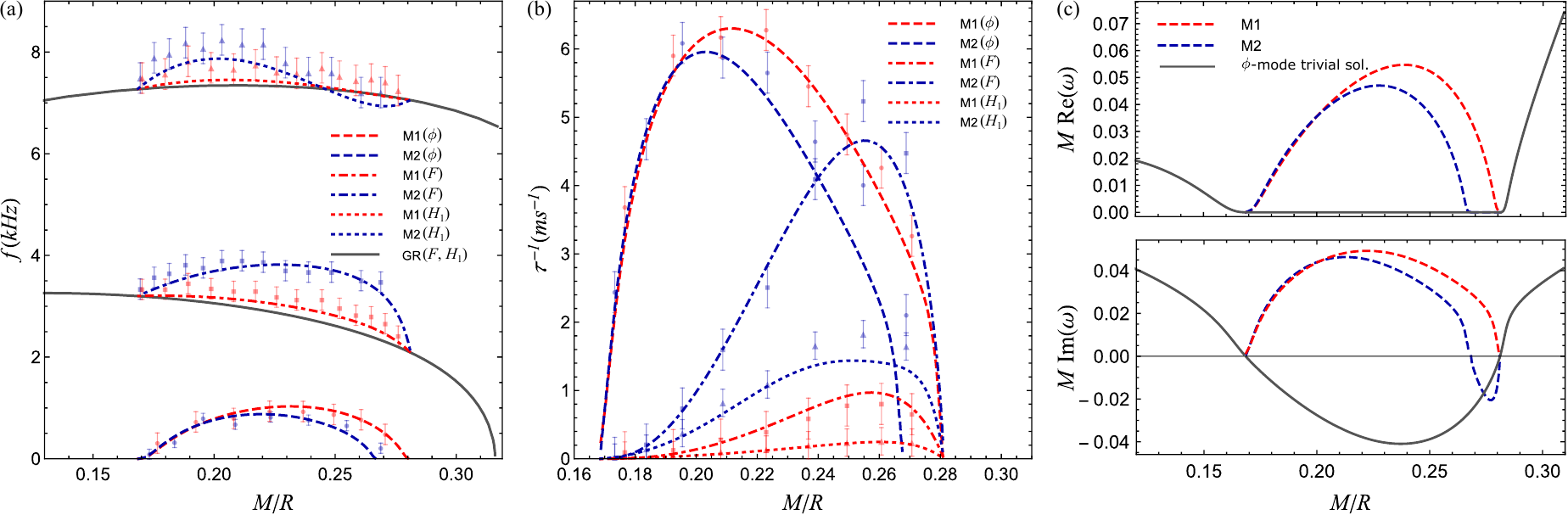}
\caption{(a) Frequency and (b) inverse damping time of the three lowest-frequency radial modes of stellar models in scalar-tensor theories for the coupling function ${\mathcal A}(\varphi) = \left[ \cosh \left(\sqrt{3} \beta_0 \varphi \right) \right]^{1/(3\beta_0)}$, denoted by M1, and ${\mathcal A}(\varphi) = e^{\beta_0 \varphi^2/2}$, denoted by M2 with $\beta_0 = -5$, as functions of compactness.
Lines (points) represent values computed with frequency-domain (time-domain) techniques. $F$ stands for the fundamental radial oscillation mode (with no nodes) and $H_1$ for its first overtone (with one node), while $\varphi$ denotes the fundamental scalar mode (with no nodes). (c) Depiction of the fundamental scalar mode showing its real and imaginary parts. The employed \EOS is a two-phase polytrope consisting of a stiff core and a soft crust. From \textcite{Mendes:2018qwo}.}
\label{fig:DEF_Mendes_RadialOscillation}
\end{figure*}

Linearized perturbations are also helpful for studying the \ns oscillation modes directly related to \gw emission.
We first note that \gw{s} in scalar-tensor gravity can carry additional polarizations compared to \gr. In particular, one can have breathing modes in addition to the standard ``plus'' and ``cross'' polarizations of \gr~\cite{Will:2018bme}.
Moreover, radial perturbations in scalar-tensor gravity can excite \gw{s}, contrary
to what happens in \gr. These perturbations source monopole scalar waves that result in a nonvanishing contribution to the perturbed Jordan-frame Riemann tensor (linearized around a Minkowski background) in the transverse-traceless gauge~\cite{Damour:1992we,Novak:1999jg}. For \DEF-like scalar-tensor theories,
this requires $\alpha_0 \neq 0$. This contribution is then linked to the existence of a breathing polarization mode of the \gw.
We can then conclude that radially oscillating scalarized \ns{s} in scalar-tensor theories with $\alpha_0 \neq 0$ will emit \gw{s}. Their amplitude is connected to $\alpha_0$ that controls the generation of tensorial waves from the dynamics of the scalar field.
Thus, larger $\alpha_0$ leads to stronger coupling and stronger excitation of the breathing modes.

The first study of radial oscillations of \ns{s} in the \DEF model was performed by~\textcite{Sotani:2014tua} in the Cowling approximation~\cite{Cowling:1941MNRAS,McDermott:1983ApJ}.
In the original GR version of this approximation, the spacetime is held fixed while only the fluid is perturbed.
In scalar-tensor gravity, the scalar-field perturbations are also often neglected.
Despite its limitations, the Cowling approximation can actually capture well the qualitative features of the neutron star oscillation spectrum.
The full problem (i.e., when both the metric and the scalar-field perturbations are taken into account) was addressed by~\textcite{Mendes:2018qwo}. They found a new family of modes named scalar modes that have no counterpart in \gr.
The results show that they have distinct frequencies and damping times compared to the fluid radial oscillation modes.
In Fig.~\ref{fig:DEF_Mendes_RadialOscillation} we show the frequencies and damping times of the fundamental fluid radial oscillation mode and its first overtone, as well as the fundamental scalar radial mode, as functions of the stellar compactness. We see in Fig.~\ref{fig:DEF_Mendes_RadialOscillation}{\textcolor{aeired}{(c)}} that scalarized \ns{s} become unstable at the maximum mass, and it is the scalar mode that is responsible for the instability. This contrasts with \gr, where it is the fundamental radial fluid mode that becomes unstable at the maximum of the mass.

Not only do the radial oscillations of scalarized \ns{s} differ significantly from those in the \gr case, but the nonradial modes that are related to the tensorial gravitational wave emission can be strongly influenced by the scalar field.
The behavior of nonradial oscillations modes, which are sources of the usual tensor polarizations of \gw{s}, can also be strongly affected by the scalar field.
These perturbations can be classified as ``polar'' or ``axial'' depending on how
they behave under parity transformations~\cite{Regge:1957td,Thorne:1967ApJ}.

In particular, scalar-field perturbations are of the polar type, meaning that
they couple only to polar perturbations of the fluid and metric. In \gr, these
perturbations are the most efficient \gw sources.
\textcite{Sotani:2004rq} were the first to study the nonradial oscillations of scalarized \ns{s}. They used the Cowling approximation, in which only polar-parity fluid perturbations are dynamical. This simplifies the problem considerably and \textcite{Sotani:2004rq} calculated the fundamental $f$-mode frequency and the pressure $p$-mode frequency.
In a follow-up work,~\textcite{Sotani:2005qx} went beyond the Cowling approximation and
derived equations for both axial and polar perturbations that included metric perturbations. They analyzed only the simpler axial perturbations (discussed later).
A study of polar perturbations became possible only after new techniques were
developed in \gr by~\textcite{Kruger:2019zuz,Kruger:2020ykw}. These techniques
were then used in scalar-tensor theory by~\textcite{Kruger:2021yay}, who also considered
self-interacting massive scalar fields.\footnote{See \textcite{Staykov:2015cfa,Blazquez-Salcedo:2020ibb} for a discussion of a case of massive scalar-tensor theory that is mathematically equivalent to $R^2$ gravity.}
\textcite{Kruger:2021yay} found that the scalar field leaves clear imprints on the oscillation frequencies of \ns{s}.
An extension of these results for rapidly rotating \ns{s}, but now back to the Cowling approximation, was made by \textcite{Yazadjiev:2017vpg}. They also studied the Chandrasekhar--Friedman-Schutz instability driven by rotation~\cite{Chandrasekhar:1992pr,Friedman:1978hf}.

\begin{figure}
\includegraphics[width=\columnwidth]{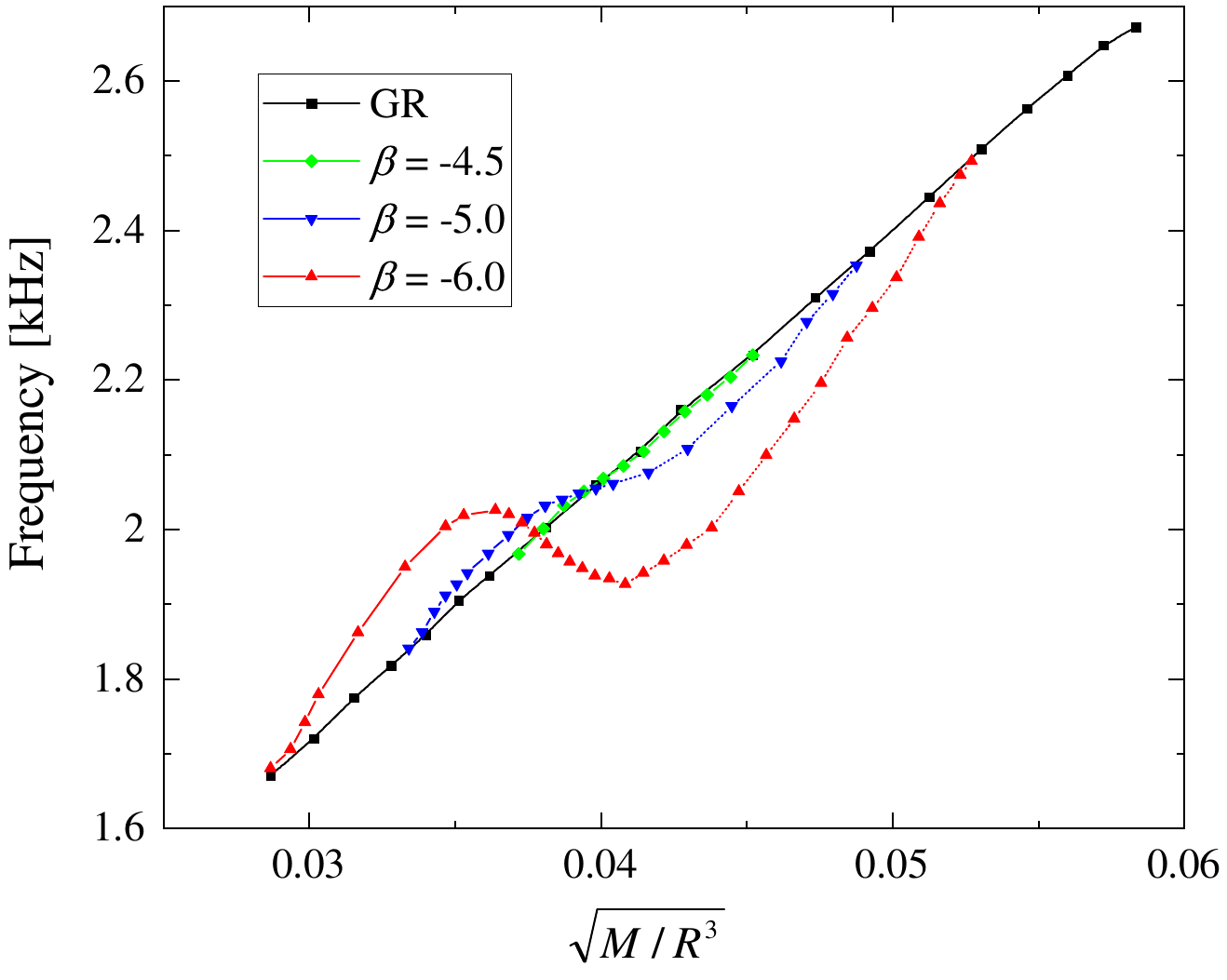}
\caption{$f$-mode frequencies (for $\ell = 2$) as a function of the average \ns density $\sqrt{M/R^3}$ for several values of $\beta$. The SLy \EOS is used~\cite{Douchin:2001sv}. From \textcite{Kruger:2021yay}.}
\label{fig:DEF_fmodes_l2}
\end{figure}

An interesting property of \ns oscillations in \gr is the almost linear
and \EOS-independent relation between the quadrupole ($\ell = 2$) $f$-mode frequency
and the average density of the \ns~\cite{Andersson:1997rn}.
\textcite{Sotani:2004rq} and~\textcite{Kruger:2021yay} showed that this linear scaling
is dramatically broken by scalarization.
We show this in Fig. \ref{fig:DEF_fmodes_l2}, where we also see that the deviations
from \gr increase with an increase of $|\beta_0|$.

Another class of modes that can be attributed, loosely speaking, to the ``oscillations'' of the spacetime itself are the axial spacetime $w$ modes. These modes are somehow easier to calculate (without approximations) because, in this case, the perturbations of the fluid and the scalar field are zero.
The axial modes of scalarized \ns{s} were considered for the first time by \textcite{Sotani:2005qx}, who calculated the frequencies and the damping times of different classes of $w$ modes.
Extensions of these results to a variety of realistic \EOS{s} including nuclear, hyperonic, and hybrid matter, were carried out by~\textcite{AltahaMotahar:2018djk}, while the case of massive self-interacting scalar fields was studied by~\textcite{AltahaMotahar:2019ekm}.
Numerical calculations showed that the effect of scalarization is stronger on the damping times than the effect on the frequencies, and in general the values of both are lower than their \gr values.
In addition, \EOS-independent relations between $w$-mode properties known to exist in \gr can also be obtained in scalar-tensor theory.

A class of \ns modes related to the crustal torsional oscillation was studied by \textcite{Silva:2014ora} for the \DEF model in the Cowling approximation. These oscillations probably follow the giant flares in soft gamma-ray repeaters~\cite{Israel:2005av,Strohmayer:2005ks,Strohmayer:2006py} and are associated with motions in the \ns crust. \textcite{Silva:2014ora} found that, for values of $\beta_0$ consistent with binary-pulsar constraints at the time, the effect of scalarization on the torsional oscillation frequencies is smaller than the uncertainties in the microphysics modeling of the crust.

\subsubsection{Nonlinear stability and collapse to a black hole}
\label{sec:ns_dynamics}
While the stability and oscillations of a scalarized \ns can be studied perturbatively, the formation of scalar hair starting from a \gr \ns solution is a fully nonlinear process.
More precisely the initial exponential growth of the scalar field starting from an unstable \gr solution can be modeled by linearized dynamics, but the subsequent saturation of the scalar field to an equilibrium value is a nonlinear phenomenon.

The transition from a nonscalarized to a scalarized \ns was first considered by~\textcite{Novak:1998rk} [see also \textcite{Degollado:2020lsa}], who studied the nonlinear evolution in spherical symmetry.
The numerical simulations of \textcite{Novak:1998rk} showed that the scalar field for an unstable \gr \ns first grows exponentially and then saturates, with the system saturating to an equilibrium scalarized end state.\footnote{A realistic astrophysical scenario for it is a low-mass \ns that cannot scalarize on its own. If this star accretes matter, its mass will gradually increase, eventually crossing the point of instability and then scalarizing}
A consequence of the work by~\textcite{Novak:1998rk} was the proof of nonlinear stability of scalarized \ns{s}. That is, the numerical simulations showed that the scalarized \ns{s} have stable evolution if the full nonlinear system of field equations (in spherical symmetry) is considered.
Although effort was not strictly dedicated to a stability analysis, a fully nonlinear evolution of scalarized \ns{s} was performed in a series of papers that we discuss later.
This provides strong support for the stability of these objects under the most general circumstances.

\textcite{Novak:1997hw} studied the collapse of a scalarized \ns to a \bh in spherical symmetry.
Since \bh no-hair theorems include \DEF models [see \textcite{Herdeiro:2015waa} for a review], the resulting \bh will be bald and the scalar field has to be radiated away
during collapse.\footnote{For a discussion of collapse in theories where \bh{s} can be endowed with a scalar field, see Sec.~\ref{sec:ns_extended}.}
This takes place in the form of scalar waves, in analogy to what we discussed in the case of radial \ns oscillations in Sec.~\ref{sec:NS_dynamics}.

Our previous discussion covered the $\beta_0<0$ case.
\textcite{Palenzuela:2015ima} and \textcite{Mendes:2016fby} studied the end state of the tachyonic instability in scalar-tensor theories for representative coupling functions with $\beta_0>0$ and realistic \EOS{s}.
This was done both through an energy balance analysis of the existing equilibrium configurations and by nonlinear dynamical simulations.
They found that (contrary to the $\beta_0<0$ case) the final state of the instability is highly sensitive to the details of the coupling function, varying from gravitational collapse to spontaneous scalarization.
They also found that in the original \DEF model [where  $\alpha(\varphi)=\beta_0 \varphi$ with $\alpha_0=0$] scalarized solutions can become unstable compared to the \gr ones when $\beta_0 \gg 1$.
However, stability can be recovered for all values of $\beta_0$ by considering different coupling functions. This is the case for coupling functions with bounded values, such as $\alpha(\varphi \to \infty) = \alpha_\infty$ for some constant $\alpha_\infty$.
This distinction could give rise to novel astrophysical tests for determining the detailed form of the coupling.

\subsubsection{Stellar core collapse}
Once we know that equilibrium scalarized \ns solutions exist in the \DEF model and they are stable against linear perturbations, the next step is to study how they are formed. Isolated \ns{s} can form after the collapse of the core of a massive star during a supernova explosion. A \ns on the other hand can collapse to a \bh if a threshold for the mass and the angular momentum is reached.
All these are highly dynamical nonlinear processes that can have strong observational signatures in both the electromagnetic and \gw signals.

The first study of a degenerate stellar core collapse (more specifically with white dwarf initial data) to a scalarized \ns through a bounce and the formation of a shock was performed by \textcite{Novak:1999jg}. The simulations were done in spherical symmetry, which allowed them to calculate only the resulting gravitational monopolar radiation. They found that the emitted breathing modes can be potentially detected by LIGO or VIRGO. Notably the emitted signal will be substantially different than the collapse of a \ns to a \bh, allowing one to distinguish between them.

\begin{figure}
\centering
\includegraphics[width=0.95\columnwidth]{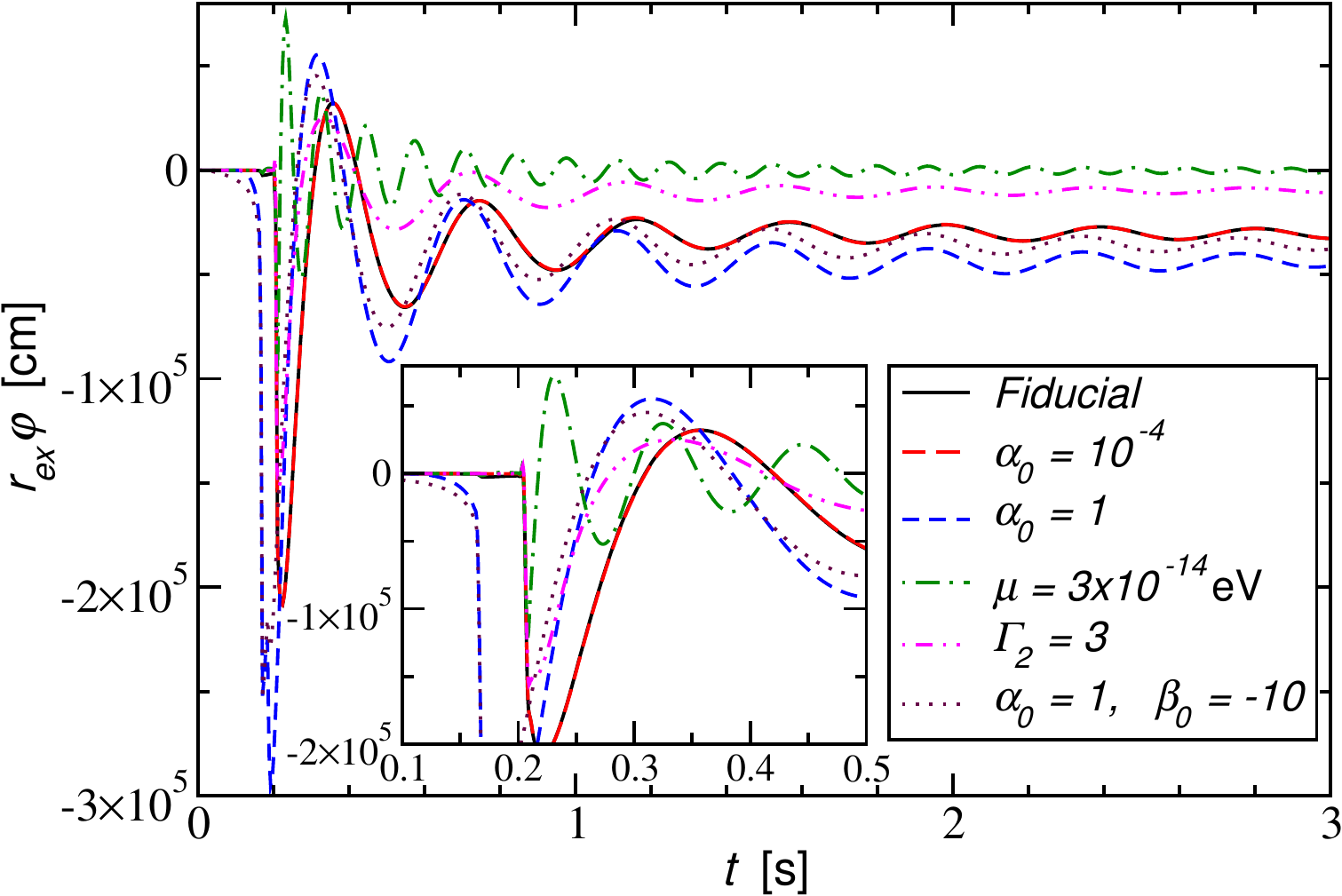}
\caption{Scalar wave $r \varphi$ during a stellar core collapse extracted at $5\times 10^4\,{\rm km}$. The legend lists deviations from the fiducial parameters  $\mu=10^{-14}\,{\rm eV},~\alpha_0=10^{-2},~\beta_0=-20,~ \Gamma_1=1.3,~\Gamma_2=2.5$, and $\Gamma_{\rm th}=1.35$, where $\mu$ is the scalar-field mass and $\Gamma_1$, $\Gamma_2$, and $\Gamma_{\rm th}$ are parameters of the polytropic \EOS and the thermal contribution. See \textcite{Sperhake:2017itk} for more details. From \textcite{Sperhake:2017itk}.}
\label{fig:NS_core_collapse}
\end{figure}

\citet{Gerosa:2016fri} studied the problem of spherically symmetric core collapse in the \DEF model in further detail. Two types of initial data were used: collapse of a stellar iron core and collapse of ``realistic'' \ns progenitors that were obtained from computations of stellar evolution. Depending on the theory parameters three possible outcomes of the core collapse are possible -- collapse to a \gr \ns, collapse to a scalarized \ns, or collapse to a short-lived protoneutron star followed by a nonscalarized \bh formation. It was in the last case that the most prominent \gw signal with a clear signature from the presence of nontrivial scalar fields was observed. While the fluid dynamics during the collapse is only weakly affected by the scalar field, the converse is not true: the scalar radiation depends strongly on the specifics of the matter collapse as well as on the choice of the coupling parameters $\alpha_0$ and $\beta_0$.

The inclusion of a scalar-field mass to the core-collapse simulations has some interesting consequences~\cite{Sperhake:2017itk}. The waveform $r\varphi$ during such events is shown in Fig. \ref{fig:NS_core_collapse}. The mass itself does not have a large influence on the dynamics, but it allows the theory to be reconciled with binary-pulsar observations for a much larger range of $\alpha_0$ and $\beta_0$, as discussed in Sec.~\ref{sec:NS:MassiveSF}. Thus, a dramatic increase in the radiated \gw signal observable even with the existing \LVK detectors was predicted.

A prominent feature is that we expect to receive an inverse chirp signal from such events that will last for years, with a near monochromatic signature on timescales of $\sim 1$ month. In fact, the inverse chirp signal is connected to the scalar field's mass and, consequently, to the dispersion relation. Thus, the scalar wave burst that is emitted during the collapse is effectively stretched in time with decreasing amplitude. Extensions of these results to the case of a self-interacting scalar-field potential were made by~\textcite{Cheong:2018gzn} and~\textcite{Rosca-Mead:2019seq}. Constraints on the theory based on the scalar-field evolution in the Einstein frame were imposed by~\textcite{Geng:2020slq}. The problem was considered in greater detail by~\textcite{Rosca-Mead:2020ehn}, who found the three possible scenarios of the collapse outcome in scalar-tensor theory with sufficiently negative $\beta_0$. These constitute the formation of a \bh following multiple \ns stages, the multistage formation of a strongly scalarized \ns, and the single-stage formation of a strongly scalarized NS.~\textcite{Rosca-Mead:2019seq} found that the resulting \gw signal can reach a signal-to-noise ratio of over $20$ for the existing \gw detectors, which has the potential to put strong constraints on the theory.

\subsubsection{Dynamical scalarization and neutron star mergers}
\label{sec:dyn_sca_nsm}

Another highly dynamical and nonlinear process that has important astrophysical, and especially \gw implications is binary \ns mergers.
This problem can be more challenging to solve than the stellar core collapse because it is not possible by construction to apply certain approximations, such as spherical symmetry.
That is why the binary merger dynamics in the \DEF model was addressed shortly thereafter, first by~\textcite{Barausse:2012da} and then by~\textcite{Shibata:2013pra}. The overall conclusions of both works are similar and they reside in the fact that, even if one or both of the \ns{s} are not scalarized before the merger, they can develop nonzero scalar fields during the inspiral. This was called dynamical scalarization.
The main advances made by~\textcite{Shibata:2013pra} constitute using several realistic \EOS with consistently derived bounds on the parameter $\beta_0$, as well as developing an initial data solver for scalarized binary \ns{s}. The overall results demonstrated a significant change of the inspiral \gw signal at the moment of dynamical scalarization and afterward. The reason is that the inspiral is accelerated due to the scalar-dipole radiation and the total number of \gw cycles is significantly decreased with respect to the general-relativistic case.

\begin{figure}
	\centering
	\includegraphics[width=0.9\columnwidth]{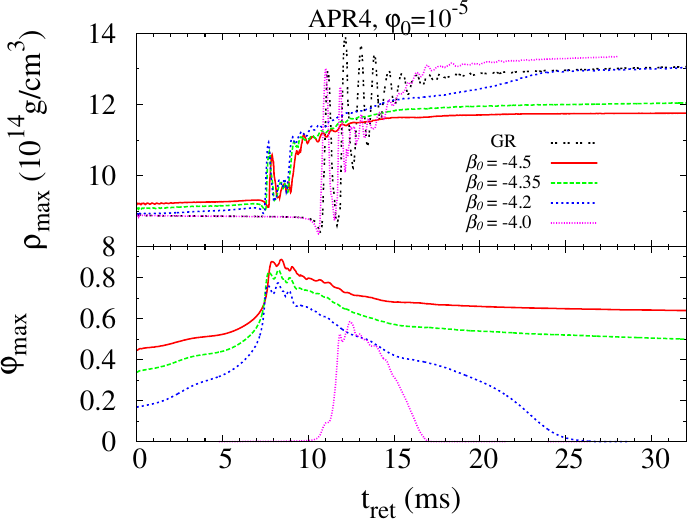}
	\includegraphics[width=0.9\columnwidth]{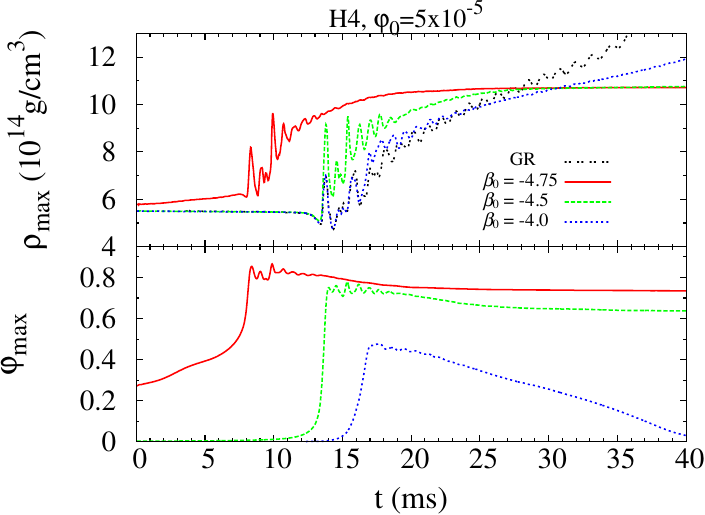}
\caption{Evolution of the maximum values of the rest-mass density $\rho_{\rm max}$ and scalar field $\varphi_{\rm max}$ during \ns merger for several models with a total mass $M = 2.7~\msun$ with the APR4 \EOS (top panel) and the H4 \EOS (bottom panel). The merger sets in at the time as the maximum density steeply increases. We note that for $\beta_0 \ge -4.2$ with the APR4 \EOS and for $\beta_0=-4.75$ with the H4 \EOS, the scalarization had already occurred at $t=0$. From \textcite{Shibata:2013pra}.}
	\label{fig:NS_merger}
\end{figure}

The actual merger and the postmerger phases were addressed only by~\textcite{Shibata:2013pra}, who showed that dynamical scalarization can happen not only in the inspiral phase but also during the merger since a massive compact star is formed in this process. The evolution during the \ns merger of the maximum values of the rest-mass density $\rho_{\rm max}$ and scalar field $\varphi_{\rm max}$ for several binary \ns{s} with total mass  $M = 2.7~\msun$ and several values of $\beta_0$ is shown in Fig. \ref{fig:NS_merger}. Two \EOS{s} are considered in the two panels and the cosmological value of the scalar field is taken to be $\varphi_0=10^{-5}$ in order to satisfy constraints from the binary-pulsar observations \cite{Freire:2012mg,Antoniadis:2013pzd}. Some of the binary \ns{s} are not scalarized at the beginning of the evolution (the models with  $\varphi_{\rm max}=0$ at the beginning), but the values of $\beta_0$ are chosen such that all of them develop scalar fields at a certain point of the evolution. The actual merger of the two \ns{s} is marked by the rapid increase of $\varphi_{\rm max}$. Afterward, either a hypermassive or supramassive \ns is formed or the merger remnant collapses to a bald \bh and the scalar field is radiated away.
In agreement with the studies of equilibrium differentially rotating \ns{s} discussed in Sec.~\ref{sec:DEF_model_equil}, the scalarized merger remnant can sustain a larger mass without collapsing to a \bh~\cite{Doneva:2018ouu}. This is evident in the bottom panel of Fig.~\ref{fig:NS_merger}, where collapse to a \bh is observed in pure \gr, while a supramassive \ns is formed after the merger in scalar-tensor gravity with small enough $\beta_0$.

The quasiperiodic oscillations of the merger remnant were examined by \textcite{Shibata:2013pra}, showing a clear distinction compared to the \gr case. Such oscillations were studied in a series of papers in \gr (see e.g. \textcite{Bauswein:2011tp,Bauswein:2012ya,Hotokezaka:2013iia,Bauswein:2014qla,Takami:2014zpa,Clark:2014wua,Takami:2014tva,Rezzolla:2016nxn,Maione:2016zqz}), mainly as a tool to determine the nuclear matter \EOS from the postmerger GW signal.
The observed differences with the scalarized case can potentially be used to discriminate between \gr and modified gravity theories. This is especially the case because scalarized merger remnants are produced only if specific initial conditions related to the mass of the merging compact objects and the specifics of the \EOS are met~\cite{Shibata:2013pra}.

\textcite{Taniguchi:2014fqa} took a different approach in which they calculated quasiequilibrium sequences of binary \ns{s} at different time instants instead of performing time evolution.
This approach is well known in \gr~\cite{Gourgoulhon:2000nn,Taniguchi:2003hx,Taniguchi:2010kj}, and it is assumed that the characteristic time of the system to settle to equilibrium is much smaller than the inspiral timescale. This is supposed to give a relatively accurate picture of the binary evolution even close to the merger. The results are in agreement with \textcite{Shibata:2013pra}, while the small deviations relative to~\textcite{Barausse:2012da} are probably due to the fact that \gr initial data were used in the latter.
\textcite{Taniguchi:2014fqa} showed that the absolute value of the binary binding energy is smaller than in \gr. In addition, the \gw cycles prior to the merger were significantly reduced compared to pure \gr once scalarization kicks in, and the effect is considerably stronger than the one due to tidal interactions.

The \pn approximation has also been a valued tool to model \ns inspiral
in the \DEF model. Initial work has focused on tensor-multiscalar theories
and simple scalar-tensor theories that do not allow scalarization~\cite{Damour:1992we,Damour:1995kt,Mirshekari:2013vb,Lang:2013fna}.
This approach cannot be immediately applied to the case of dynamical scalarization, since this is a nonperturbative effect that is absent from the weak-field regime.
\textcite{Palenzuela:2013hsa} were the first to address this issue. They used the
equations of motion at 2.5PN order, derived in scalar-tensor gravity by~\textcite{Mirshekari:2013vb}, modified in such a way that the changes in the stars’ scalar charges are taken into account. More specifically, they solved a system of nonlinear algebraic equations at each step of the orbital evolution to compute the scalar charges.
The resulting inspiral evolution was found to be in agreement with the numerical simulations of~\textcite{Barausse:2012da}. The approach of \textcite{Palenzuela:2013hsa}
is computationally inexpensive, and this allowed them the to study large portions of the parameter space, including (un)equal-mass and eccentric binaries. This approach can also be used to efficiently generate inspiral \gw templates in the \DEF model.
A further step forward taken by~\textcite{Sennett:2016rwa} introduced a methodology called post-Dickean expansion. Their main improvement with respect to \textcite{Palenzuela:2013hsa} is a 1PN extension of the feedback mechanism.
The post-Dickean expansion was compared against the quasiequilibrium calculations of~\textcite{Taniguchi:2014fqa}, and it was shown that this can accurately predict the onset and magnitude of dynamical scalarization.

\textcite{Sennett:2017lcx} took a different approach in the perturbative study of the inspiral that constitutes an analytical model of dynamical scalarization using an effective action. The motivation was to cure two deficiencies in the previous \pn studies.
In an effective action approach, the nonlinear scalarization process is reduced to a pair of cubic equations that have a closed-form solution depending on the binary separation when dynamical scalarization happens and on the magnitude of the developed scalar charge. This simplifies the problem relative to \textcite{Palenzuela:2013hsa} and \textcite{Sennett:2016rwa}.
In addition, the effective action approach allows one to construct a simple two-body Hamiltonian that can be used to compute the binary's binding energy.~\textcite{Sennett:2017lcx} used this Hamiltonian, in combination with Landau's theory of phase transitions, to interpret dynamical scalarization as a second-order phase transition.
\textcite{Khalil:2019wyy} extended these results to general theories admitting scalarization for either \bh{s} or \ns{s} and are valid for adiabatic (quasistationary) and quasicircular orbits.
\textcite{Khalil:2022sii} took it a step further, where these approximations were dropped and the dynamical evolution around the phase transition to the scalarized regime was studied. The results showed that in some cases assuming a quasistationary evolution might not be accurate enough even for quasicircular binaries.

Last, \textcite{Ponce:2014hha} studied the effect of the scalar field on the electromagnetic radiation emitted in the merger of magnetized \ns{s}. They found that deviations in the emitted electromagnetic flux due to scalarization are not negligible yet are challenging to measure.
However, if combined with \gw observations, constraints on scalar-tensor theory can in principle be placed, showing the usefulness of multimessenger astronomy.

\subsection{Astrophysical implications of scalarized neutron stars in the Damour--Esposito-Far\`ese model}
\label{sec:NS_DEF_ASTRO}

In this section, we further discuss the astrophysical implications of scalarized \ns{s}. Many aspects have already been covered, such as the binary-pulsar observations, \ns oscillations, stellar core collapse, and binary mergers, as they naturally appeared in the presentation. Here, we shed further light on the possible astrophysical implications of scalarization, trying to be as complete as possible in two main areas. We first discuss the astrophysical implications directly related to electromagnetic observables. Afterward, we address the problem of universal relations for scalarized \ns models.

\paragraph{Electromagnetic observations:}
Scalarized \ns{s} in the \DEF model and its extensions were studied in a variety of astrophysical scenarios in an attempt to probe the existence of the scalar field.
Since \ns{s} are often surrounded by accretion disks, it is natural to study the effect of a nontrivial scalar field on disk properties.
The simplest model is called the thin disk model, in which particles are assumed to move on geodesics around the central compact object.
Any small perturbation acting upon these particles will lead to oscillations around their equilibrium orbit with some characteristic epicyclic frequencies. Epicyclic and
orbital frequencies are thought to be, in one way or another, related to the interpretation
of different accretion disk properties. An example is the \qpos observed in the spectrum of low-mass x-ray binaries.
For this reason, some works calculated these frequencies in the spacetime of scalarized \ns{s}. \textcite{DeDeo:2004kk} did this for nonrotating \ns{s},~\textcite{Staykov:2019pwj} included slow rotation and also a mass to the scalar field, while \textcite{Doneva:2014uma} studied rapidly rotating \ns{s}.
The conclusion of these works is that, if one takes into account current observational constraints, only the cases of rapid rotation and massive scalar fields leave room
for a significant effect of the scalar field.
Yet, a question that deserves further work is how one can disentangle uncertainties in the \ns \EOS and in the disk model from modifications to \gr.

Another prospective way to infer information about accreting compact objects is through the shape of the Fe~$K\alpha$ fluorescent line at 6.4~keV. Since it is emitted from the inner regions of the accretion disk, it carries traces of the underlying spacetime geometry. This line can be observed for both accreting \ns{s} and \bh{s}, and the accuracy of the observations is expected to be improved through future x-ray missions such as Athena. \textcite{Bucciantini:2020owm} calculated the shape of this line by taking the light propagation around a scalarized \ns into account. They argued that the influence of \gr modification both on the intensity of the low-energy tails and on the position of the high-energy edge of the line are potentially observable in the future.

Accretion onto \ns{s} can also trigger a process called the gravitational phase transition, which was named in analogy with matter phase transitions from confined nuclear matter to deconfined quark matter~\cite{Kuan:2022oxs}. This process can happen when the maximum mass of the scalarized \ns{s} is smaller than the maximum mass of the zero scalar-field (\gr) solution. The idea then is that if a scalarized \ns close to this maximum mass accretes some matter, it may pass beyond the stability point. In \gr this would cause a collapse to a \bh. In scalar-tensor theory, the star will radiate its scalar hair and evolve toward the zero scalar-field (i.e., the~\gr) branch. A significant amount of \gw{s} can be produced in this process, and they will potentially be detectable with the next generation of \gw detectors.

The \ns surface as an emitter of x-ray radiation is also an important probe in strong-field gravity because the electromagnetic radiation is emitted from a region with large spacetime curvature.
In this regard, the simplest observable is the gravitational redshift of surface atomic lines. The redshift carries information about the \ns mass, radius, spin, and, in scalar-tensor gravity, the scalar field. \textcite{DeDeo:2003ju} showed that scalarization had a significant effect on the redshift, but only for negative enough values of $\beta_0$. Such values are already ruled out by binary-pulsar observations. One could entertain the idea that a small scalar-field mass could allow for large deviations in redshift relative to \gr, while reconciling the theory with binary-pulsar constraints~\cite{Popchev2015,Ramazanoglu:2016kul,Doneva:2016xmf}. A study of this problem has not yet been performed.

The  x-ray pulse profiles emitted by hot spots at \ns surfaces have also received considerable attention. Observations of these signals allow for a relatively clean inference of \ns masses and radii~\cite{Watts:2016uzu}. This potential was met with observations~\cite{Bogdanov:2019ixe,Miller:2019cac,Riley:2019yda} by the \NICER~\cite{Arzoumanian2014:SPIE}.
\textcite{Sotani:2017rrt} and~\textcite{Silva:2018yxz} developed pulse-profile models for \ns{s} in a massless \DEF model.~\textcite{Xu:2020vbs} considered the case of massive scalar fields and~\textcite{Hu:2021tyw}
studied the positive $\beta_0$ case. \textcite{Silva:2019leq} made the first study of what constraints can be placed on the \DEF model with pulse-profile observations and found they can be competitive with binary-pulsar observations. However, their work is rather simple and not at the level of realism found in the analysis of real data. This remains an important avenue for future work.

\begin{figure}
\includegraphics[width=0.95\columnwidth]{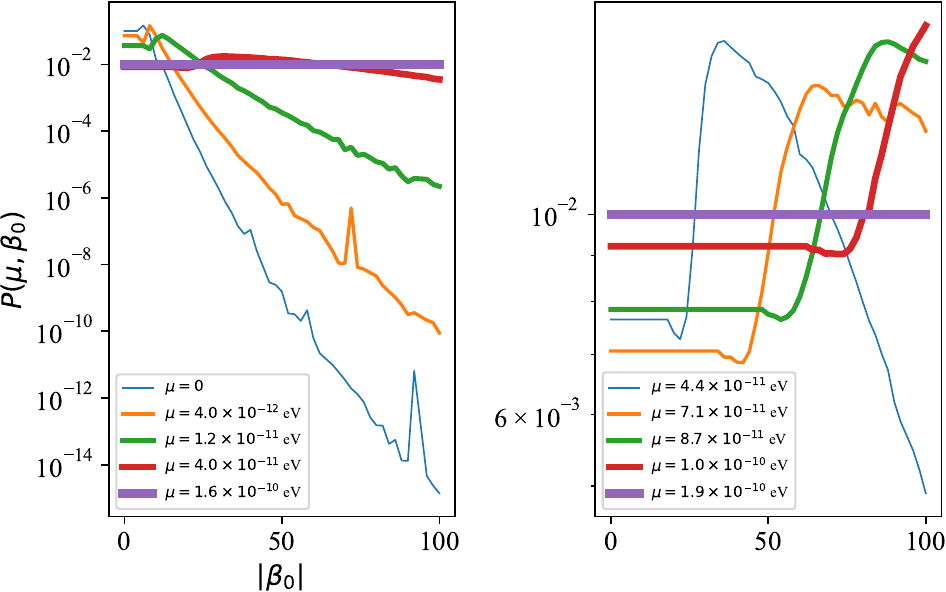}
\caption{The posterior probability density for  $\beta_0 \in [-100,\ 0]$ for various assumed values of the scalar-field mass $\mu$ in a massive version of the \DEF-model that was introduced by \textcite{Ramazanoglu:2016kul} (marginalized over various \EOS{s}). One can obtain the bound $\beta_0 \gtrsim -20$ for $\mu \lesssim 2\times 10^{-11}$~eV (left panel), but there is no effective bound for higher masses, at least not in the interval $\beta_0 \in [-100,\ 0]$ (right panel). No bound is possible for $\mu \gtrsim 10^{-10}$~eV using these data since scalarization does not occur for such high $\mu$ values within the considered $\beta_0$ interval. From \textcite{Tuna:2022qqr}.
}
\label{fig:mass_radius_constraint}
\end{figure}

At last, \textcite{Tuna:2022qqr} used \ns mass and radius measurements~\cite{Ozel:2015fia,Bogdanov:2016nle} to constrain the massive extensions of the \DEF model. They obtained a weak lower bound $\beta_0 \gtrsim -20$ for scalar-field masses $\mu \lesssim 2 \, \times10^{-11}$~eV; see Fig.~\ref{fig:mass_radius_constraint}. This is significant since no other bound is known for $\mu \gg 10^{-16}$~eV. These results show that large scalar masses enable agreement with observations, even for extremely negative $\beta_0$, and demonstrate the difficulty to constrain scalarization when the scalar field is massive.

\paragraph{Universal relations:}

One of the largest obstacles when using \ns observations to test modified theories of gravity is the uncertainty in the \ns \EOS, which remains unknown at high densities~\cite{Lattimer:2015nhk,Baym:2017whm}. In general modifications to \ns properties predicted by different \EOS{s} are degenerate with changes to the underlying gravity theory used to model these stars. One way to break this degeneracy is to consider relations between different \ns properties that depend weakly on the \EOS~\cite{Yagi:2016bkt,Doneva:2017jop}. In fact, in Sec.~\ref{sec:NS_QNMs} we already met one such example relating the \qnm frequencies and damping times to the mean density of the \ns{s}. Here we focus on other \EOS-independent relations connecting various \emph{equilibrium properties} of scalarized \ns{s}.

\begin{figure}
\centering
\includegraphics[width=\columnwidth]{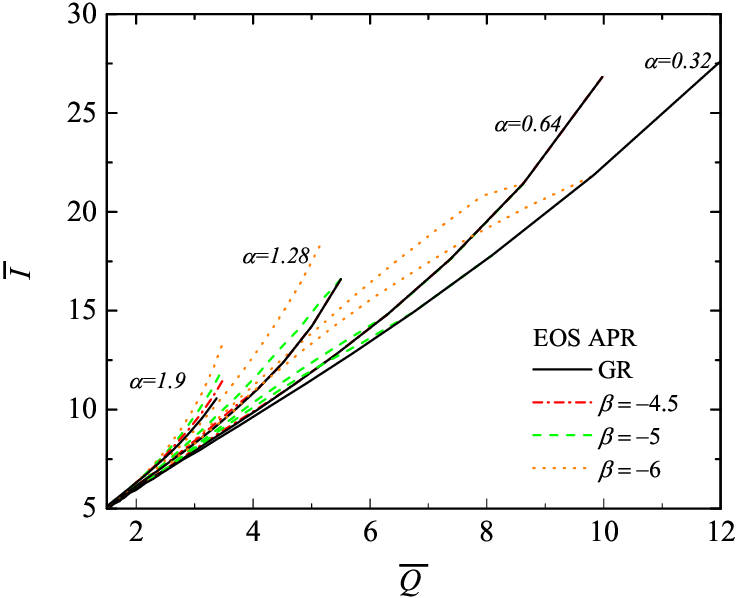}
\caption{Comparison between ${\bar I}-{\bar Q}$ relations for \gr and scalar-tensor theory for several values of $\beta_0$, where ${\bar I}=I/M^3$ and ${\bar Q}=QM/J^2$. The presented data are restricted to APR \EOS in order to provide better visibility, but the deviations for other \EOS{s} are small, typically below roughly $2\%$. Sequences for different values of the normalized rotational parameter $\alpha=M\Omega/2\pi$ are given, with $\Omega$ the angular velocity of the star. The dash-dotted lines with $\beta=-4.5$ appear for each value of $\alpha$, but they are limited to the low  ${\bar Q}$ region and deviate only slightly from GR. From \textcite{Doneva:2014faa}}.
\label{fig:NS_STT_ILoveQ}
\end{figure}

A class of universal relations that has attracted considerable attention is the I-Love-Q relations, which connect the normalized moment of inertia, the tidal Love number, and the quadrupole moment of the \ns{s}~\cite{Yagi:2013bca,Yagi:2013awa}. If any two quantities in the I-Love-Q trio are measured independently, one can constrain deviations from \gr by determining whether these observations agree with the \gr I-Love-Q relation within observational accuracy. \textcite{Silva:2020acr} applied this strategy to constrain dynamical Chern-Simons gravity~\cite{Jackiw:2003pm,Alexander:2009tp} using observational data from \LVK and \NICER. The application to scalarization models has not yet been done.

The I-Love-Q relations for scalarized \ns{s} were studied in slow~\cite{Pani:2014jra} and rapid rotations~\cite{Doneva:2014faa}, including the case of massive scalar fields~\cite{Hu:2021tyw,Doneva:2016xmf}. In Fig.~\ref{fig:NS_STT_ILoveQ} we show the relation between the normalized moment of inertia $\bar{I}$ and the quadrupole moment $\bar{Q}$ for sequences of scalarized \ns{s} with a fixed rotational parameter $\alpha=M\Omega/2\pi$.
We see that for small absolute values of $\beta$, for example, $\beta_0=-4.5$, the deviations from \gr are practically negligible for slow rotation (such as $\alpha=0.32$) while they are enhanced by rapid rotation. As we have seen, massive scalar fields allow for much smaller values $\beta_0$ while remaining consistent with observations. Hence, these scalarization models allow for larger differences between the scalarized \ns universal relations and the \gr ones~\cite{Doneva:2016xmf}.

The moment of inertia $I$ and the quadrupole moment $Q$ are among the leading-order multipole moments in the asymptotic expansion of the metric functions at infinity. There are infinitely many other higher-order multipole moments, however, and it is interesting to explore whether similar universal relations hold for them. This was first addressed in \gr~\cite{Pappas:2013naa,Yagi:2014bxa}, where such universal relations were derived for the higher multipole moments when proper normalization was applied. \textcite{Pappas:2014gca} and \textcite{Pappas:2018csu} extended these results to scalarized \ns{s}. This required the generalization of the multipole moment formalism of~\textcite{Geroch:1970cd,Hansen:1974zz} to scalar-tensor gravity.
\textcite{Pappas:2018csu} showed that future observations of \qpos of low-mass x-ray binaries can
in principle be used to measure different \ns properties and distinguish different gravity theories.

Another class of universal relations studied in the context of the \DEF model connects the \ns moment of inertia and compactness. This was endorsed by \textcite{Lattimer:2004nj} [see also \textcite{Breu:2016ufb}], who used this relation to argue that pulsar-timing
observations could lead to a measurement of the moment of inertia to within $10\%$.
These relations were generalized to scalarized \ns{s} with massless scalar-field potential by \textcite{Motahar:2017blm}, and later to massive self-interacting scalar fields by~\textcite{Popchev:2018fwu}. As with the I-Love-Q relations, only the massive scalar-field case leads to large deviations from \gr when observational constraints are taken into account.

\textcite{Ofengeim:2020zuc} proposed connecting the physical parameters of static \ns{s}, such as mass, radius, central energy density, pressure, and sound speed, at the maximum-mass point for a given \EOS. This is based on the observations that the nuclear matter \EOS (without phase transitions) can be well parametrized by only two parameters~\cite{Lindblom:2010bb}.  As a result, \textcite{Ofengeim:2020zuc} derived multiple constraints on the nuclear matter \EOS. \textcite{Danchev:2020zwn} generalized these relations to scalarized \ns{s}. They showed how sensitive these relations can be to the underlying theory of gravity. Thus, all of the \EOS restrictions derived in \gr should be taken with care since the possibility for \gr modifications is rarely taken into account when interpreting observational data.

\subsection{Extended scalar-tensor theories beyond the Damour--Esposito-Far\`ese model}
\label{sec:ns_extended}

Thus far we have discussed scalarization in the \DEF model or other \DEF-inspired models. However, as we saw in Sec.~\ref{models}, spontaneous scalarization is also possible for other modified theories of gravity. Here we discuss the theories where scalarized \ns solutions were obtained.

\paragraph{Scalar-tensor theories with disformal coupling}

\textcite{Minamitsuji:2016hkk} studied \ns scalarization in scalar-tensor theories with disformal coupling of the form of Eq.~\eqref{eq:disformal}. More specifically, instead of the standard conformal transformation between the Einstein-frame metric ($g_{\mu\nu}$) and the Jordan-frame metric (${\tilde g}_{\mu\nu}$)
\begin{equation}
   g_{\mu\nu}={\cal A}^2(\varphi){\tilde g}_{\mu\nu},
\end{equation}
where ${\mathcal A}(\varphi)$ is a function of the scalar field related to the Jordan-frame coupling between the scalar field and the Ricci scalar, we have a more general and complicated transformation that also involves scalar-field derivatives,
\begin{equation}
    g_{\mu\nu}=A^2(\varphi) \left[{\tilde g}_{\mu\nu} + \Lambda B^2(\varphi)\nabla_\mu \varphi \nabla_\nu \varphi \right],
\end{equation}
where $B(\varphi)$ is a function of the scalar field and $\Lambda$ is a parameter of dimensions of [length]$^{2}$. The motivation behind this modification of the conformal factor comes from \textcite{Bekenstein:1992pj} who aimed to find the most general coupling constructed from the metric $g_{\mu\nu}$ and the scalar field $\varphi$ that respected causality and the \WEP. More recently \textcite{Andreou:2019ikc} showed that such disformal coupling is actually equivalent to a kinetic coupling with the Ricci scalar and thus is contained in the minimal action \eqref{eq:ActionCaseI}; see Sec. \ref{sec:minimal_action} for a discussion.

\textcite{Minamitsuji:2016hkk} showed that, for negative values of the disformal coupling parameter $\Lambda$, scalarization can be suppressed, while for large positive values of $\Lambda$ the stellar structure equation becomes singular. Thus,
regular \ns solutions cannot be found in the latter case, which can be used to impose the upper limit $\Lambda \lesssim 100$~km$^{2}$. They also explored the universal relations between the moment of inertia and the compactness of \ns{s}, and they determined the range of parameters where these relations can deviate significantly from \gr.

\paragraph{Scalar-Gauss-Bonnet theory:}

Another well-studied modified gravity theory that allows for \ns scalarization is the scalar-Gauss-Bonnet gravity discussed in Sec. \ref{models}. Spacetime curvature itself, as well as matter, can be the source of the scalar field in this theory. Scalarized \ns solutions in scalar-Gauss-Bonnet gravity were studied for the first time by \textcite{Doneva:2017duq,Silva:2017uqg}. The solutions exhibit much different features than to the \DEF model while being qualitatively similar to \ns solutions in other classes of Gauss-Bonnet theories where scalarization is not possible~\cite{Pani:2011xm,Kleihaus:2016dui,Saffer:2019hqn}. For example, for a given central energy density the stellar mass is always smaller than in \gr. Moreover, there is normally only one bifurcation point at small central energy densities, and afterward the branches of solutions are terminated because of violation of the regularity conditions.

\begin{figure}
\includegraphics[width=0.95\columnwidth]{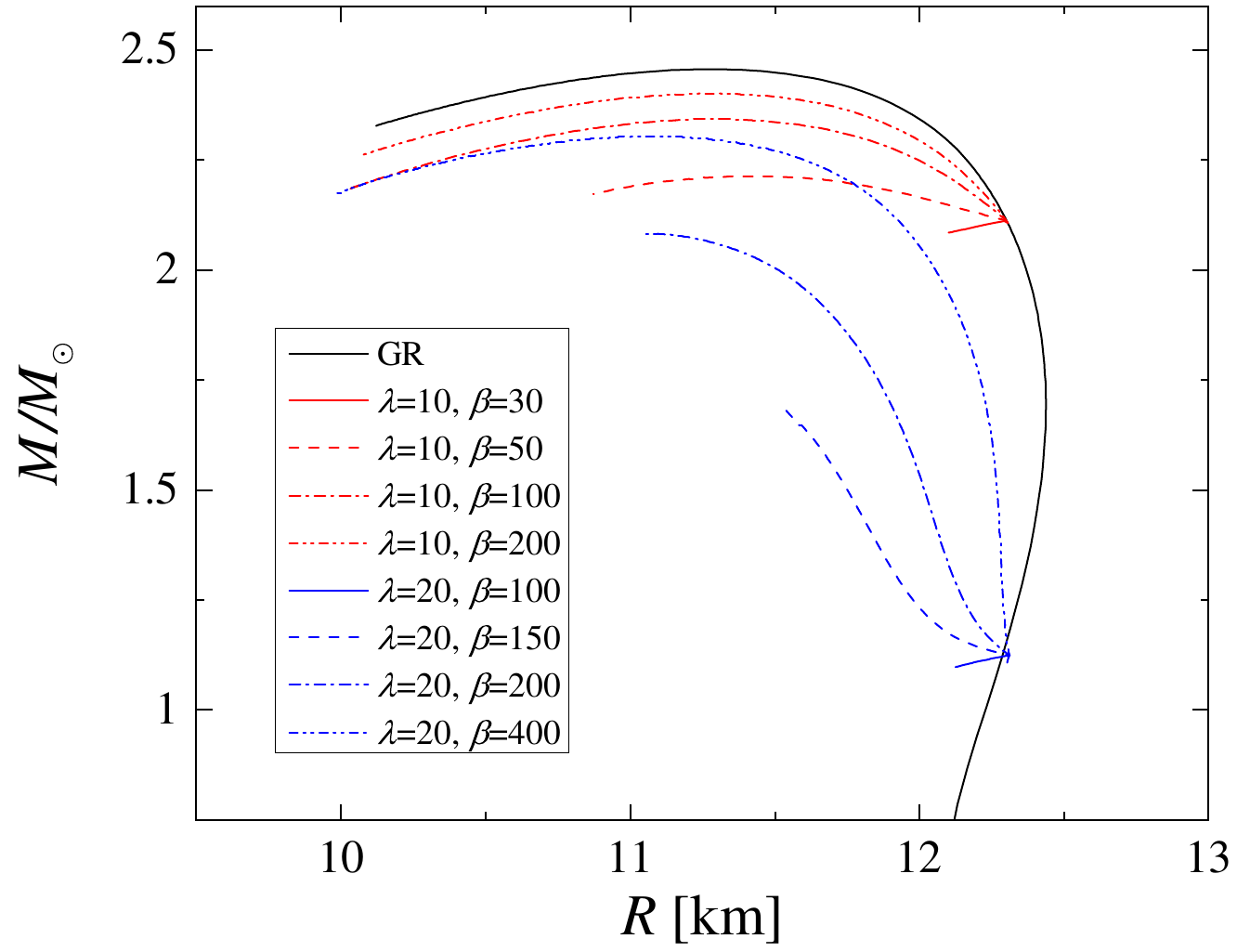}
\caption{Scalarized \ns{s} in scalar-Gauss-Bonnet gravity. We show the mass as a function of the radius for \gr (trivial branch with $\varphi=0$) as well as several scalarized branches with different values of the scalar-field coupling parameters $\lambda$ and $\beta$. From \textcite{Doneva:2017duq}.}
\label{fig:NS_MR_GB}
\end{figure}

In Fig.~\ref{fig:NS_MR_GB}, we show the branches of scalarized \ns solutions for a coupling function $f(\varphi) = - \lambda^2 (2 \beta)^{-1} [1 - \exp(-\beta \varphi^2/4)]$. Here $\lambda$ is a parameter with dimensions of [length]$^{-1}$ that controls the coupling strength in the action \eqref{actionsgb}. Note that the sign of $f(\varphi)$ is the opposite of that discussed in Sec.~\ref{sec:theory_sgb}, and thus cannot lead to \bh spontaneous scalarization, since in vacuum $\mu_{\eff}^2>0$, as defined in Eq.~\eqref{eq:mueffsqr_sgb}. However, owing the presence of matter, \ns{s} can scalarize. Scalarized \ns solutions were also found for the more ``conventional'' sign of $f(\varphi)$, namely, $f(\varphi) = \lambda^2 (2 \beta)^{-1} [1 - \exp(-\beta \varphi^2/4)]$ \cite{Doneva:2017duq}. As in the \DEF model, it was demonstrated that binary pulsars strongly constrain the coupling parameters of the theory, thereby leaving a small window for scalarization \cite{Danchev:2021tew}.

Little is currently known about the astrophysical implications of these stars. \textcite{Kuan:2021lol} considered the spherically symmetric core collapse of a noncompact star either to a protoneutron star or to a \bh in Gauss-Bonnet theory. They also proposed a realistic physical mechanism for the formation of isolated scalarized \bh{s} and \ns{s} The complexity of the problem is greatly increased with respect to \gr, though, and there are fundamental difficulties such as the loss of hyperbolicity of the evolution equations for certain ranges of parameters, effectively limiting the maximum possible scalar field to relatively low values; see also \textcite{East:2021bqk,East:2020hgw} and \textcite{Ripley:2020vpk} for the case of \bh{s}. Core-collapse simulations show that the remnant in scalar-Gauss-Bonnet theory can be rich, with (de)scalarization happening at the intermediate or final stages of the collapse, depending on the properties of the progenitor and the theory parameters. Since breathing modes are absent from this theory, the effect on \gw emission can be estimated only if one drops the assumption of spherical symmetry. This is something that has not yet been done in any modified gravity theories due to the complexity of the problem.

\paragraph{Ricci-Gauss-Bonnet model:} One can further modify the action in scalar-Gauss-Bonnet gravity to include additional terms. This is the case for the Ricci-Gauss-Bonnet model discussed in Sec.~\ref{sec:mixed_model}. This model has advantages such as the possibility of reconciling scalarization with cosmology. Here we focus on the \ns solutions within this theory that were considered by \textcite{Ventagli:2021ubn}. They conducted a thorough exploration of the theory parameter space in order to find the sectors where \ns solutions exist. Since their scalar charge is nonzero, in these sectors one can put severe constraints on the theory based on binary-pulsar observations, as in the \DEF model; see Sec. \ref{sec:ns_binary_pulsar}.

\begin{figure}[t]
	\includegraphics[width=\columnwidth]{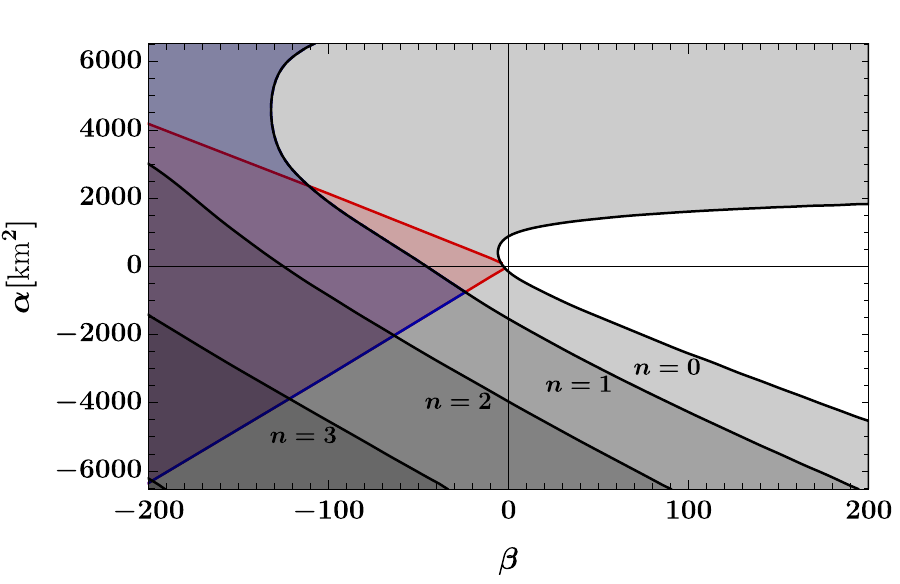}%
    \caption{Regions of existence of scalarized solutions in the $(\alpha,\beta)$ plane for the MPA1 \EOS with central energy density $\varepsilon_{\rm c}=6.3\times 10^{17}~\text{kg}/\text{m}^3$. In \gr, a star with this choice of $\varepsilon_{\rm c}$ and \EOS is light, with $M_\text{GR}=1.12~\msun$. The wedgelike red (wedgelike and upward-curving blue) region is the region where scalarized solutions with $0$ ($1$) scalar-field nodes exist. The gray contours obtained by~\textcite{Ventagli:2020rnx} are superimposed and represent the lines beyond which \gr solutions with the same density are unstable to scalar perturbations with $0, 1, 2,$ etc., nodes. We see that the region where there are scalarized solutions with $n$ nodes is included in the region where the \gr solutions are unstable to scalar perturbations with $n$ nodes but much smaller. The dashed boundary of the blue region corresponds to a breakdown of the integration inside the star. From \textcite{Ventagli:2021ubn}.}
	\label{fig:NS_MixedModel_MPA1}
\end{figure}

The free parameters in the theory, as evident from the action \eqref{action:RGmixed}, are $\beta$ and $\alpha$ which control the coupling strength to the Ricci scalar and the Gauss-Bonnet invariant, respectively. In Fig.~\ref{fig:NS_MixedModel_MPA1} the existence of \ns solutions is shown in a two-dimensional plot spanning the parameters $\alpha$ and $\beta$ in a broad range for a fixed central energy density and the MPA1 \EOS~\cite{Muther:1987xaa}. The white area Fig.~\ref{fig:NS_MixedModel_MPA1} corresponds to the region of the parameter space where the \gr solution is stable against scalar perturbations. Throughout the parameter space that is spanned by $\alpha$ and $\beta$, a new unstable mode appears every time that one crosses a black line. These unstable modes can be labeled by the number of scalar-field nodes (denoted by $n=0,1,2$ and $3$ in the figure).  Any point in the parameter space that lies within a gray region corresponds to a configuration where the \gr neutron star solution is unstable. The red (blue) area corresponds to the region where scalarized solutions with $n=0$ ($n=1$) nodes exist.

It is evident that scalarized solutions exist in only part of the parameter space. In the $\beta>0$ and small-$\alpha$ range (the white region in Fig.~\ref{fig:NS_MixedModel_MPA1}) the \gr neutron stars are stable. This is the region that is most noteworthy from the point of view of \bh scalarization and where we can reconcile scalarization
with cosmology \cite{Antoniou:2020nax,Antoniou:2021zoy}; see also Sec.~\ref{sec:bhs_stability_nr}. Thus, including a Ricci coupling to the scalar-Gauss-Bonnet action seems to allow for \bh scalarization while evading the binary-pulsar constraints.

Figure~\ref{fig:NS_MixedModel_MPA1} is for only one \EOS and a specific central energy density. Different \EOS{s} and central densities can lead to significant deformations of the existence and stability regions. The general point made by~\textcite{Ventagli:2021ubn}, however, is that the stability (white) region always survives.

\paragraph{Tensor-multiscalar theories:} \label{sec:TMST}
Another class of gravity theories where \ns scalarization was considered is the \TMST{s}~\cite{Damour:1992we}, whose basics were discussed in Sec.~\ref{models}. In this theory, the gravitational interaction is mediated  by the spacetime metric $g_{\mu\nu}$ and $N$ scalar fields $\varphi^{a}$, which take values in a coordinate patch of an $N$-dimensional Riemannian target manifold ${\cal E}_{N}$  with a positive-definite metric $\gamma_{ab}(\varphi)$ defined on it \cite{Damour:1992we,Horbatsch:2015bua}.

\begin{figure}
	\includegraphics[width=0.94\columnwidth]{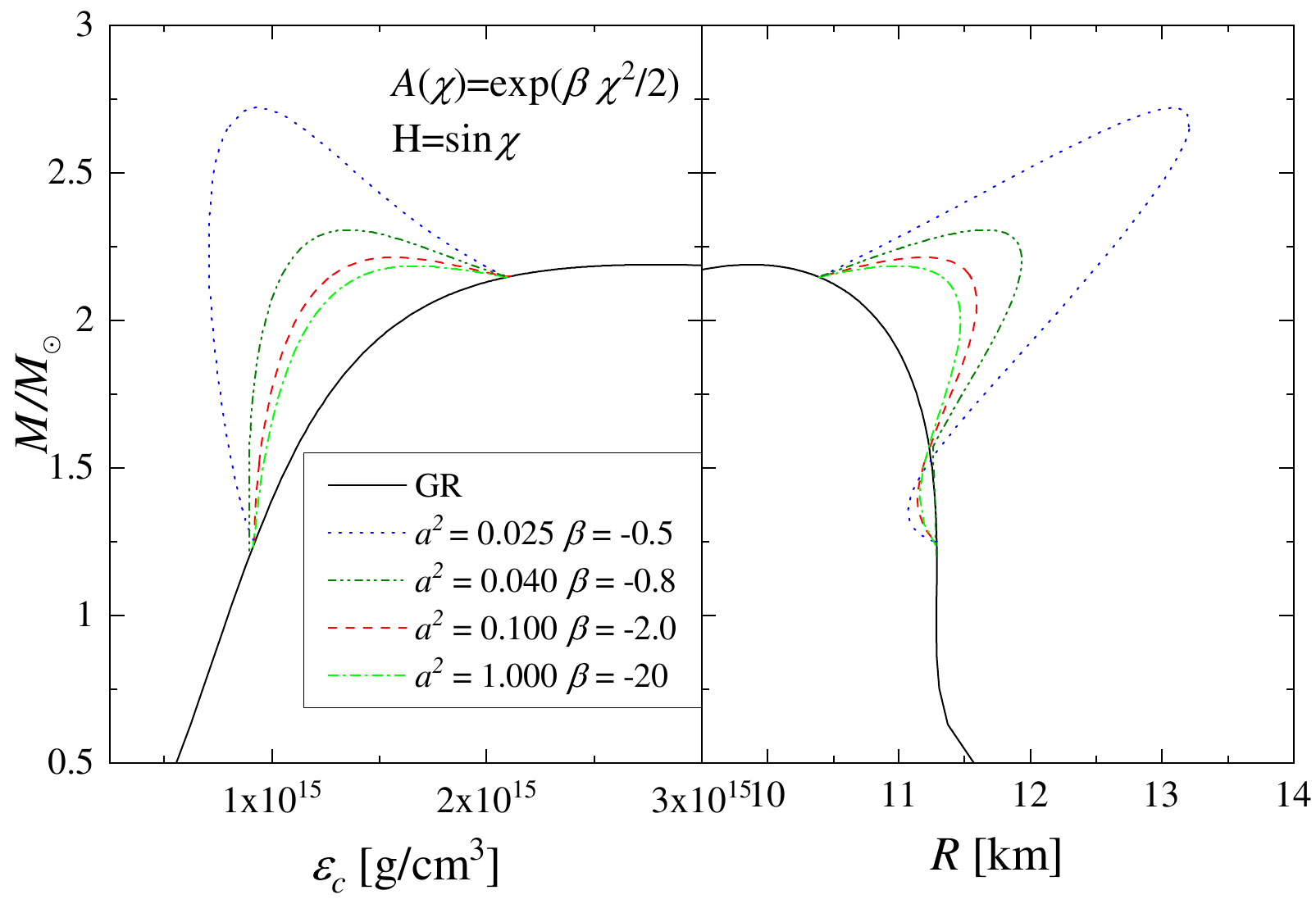}
\caption{Tensor-multiscalar theories: the mass as a function of the central energy density (left panel) and as a function of the stellar radius (right panel) for scalarized \ns{s} in tensor-multiscalar theories where the target-space metric represents a three-dimensional maximally symmetric space and we have a nontrivial map $\varphi:$~spacetime~$\rightarrow$~target~space.  Sequences of scalarized solutions for different combinations of parameters are given. Observe the appearance of nonuniqueness of the solutions with a nonzero scalar field for certain combinations of the parameters (for example, the small central energy-density region indicated by the dotted line). From \textcite{Doneva:2020afj}.}
	\label{fig:TMST_Scalarization_NS}
\end{figure}

The main features of \TMST are the inclusion of more than one scalar field and a structure called the target-space metric. In that sense, there are two directions to go in order to obtain scalarized \ns{s}. The first one is to consider a mixture of several more or less equivalent scalar fields; this was the approach followed by \textcite{Horbatsch:2015bua}. They, the authors examined the case of two scalar fields in the form of a complex scalar with a maximally symmetric target-space metric $\gamma_{ab}$.
\textcite{Doneva:2020afj} studied scalarization in tensor-multiscalar gravity in a more complicated setup, namely, when $\gamma_{ab}$ is a three-dimensional maximally symmetric space together with a nontrivial map $\varphi:$~spacetime~$\rightarrow$~target~space.
While the solutions given by \textcite{Horbatsch:2015bua} can be viewed as a generalization of the \DEF model to multiple scalar fields, the scalarized \ns{s} given by \textcite{Doneva:2020afj} have some distinct properties and they can be considered more as a limiting case of the topological \ns{s} discovered by \textcite{Doneva:2019ltb}. More specifically the scalar field is zero in the stellar center, while only its first derivative is zero in the \DEF model. The scalar charge for the scalarized stars in \TMST is zero as well. This automatically reconciles this theory with the binary-pulsar observations (due to the absence of scalar-dipole radiation) while still allowing for large deviations from \gr. Another interesting property is the fact that there is nonuniqueness of the scalarized solutions with respect to the central energy density, something that had never been observed before, at least for scalarized \ns{s}. Figure~\ref{fig:TMST_Scalarization_NS} illustrates this property. Interestingly, \textcite{Kuan:2021yih} showed that all of these solutions are stable until the maximum \ns, mass regardless of the nonuniqueness.

\paragraph{Other scalarization models:}
Apart from the previously discussed models of scalarized \ns{s}, other works have attempted to consider scalarization in more exotic theories of gravity, or even different types of scalarization.

For example,~\textcite{Azri:2020agc} considered \ns scalarization in what is called scalar-connection gravity, where gravity is mediated by a scalar field and the connection.
Other types of scalarization can happen in standard scalar-tensor theories if we allow for different scalar-field potentials. \textcite{Minamitsuji:2022qku} showed this when
considering the potential $V(\varphi)=m_\varphi^2 f_B^2 [1 + \cos(\varphi/f_B)]$ of a pseudo-Nambu-Goldstone boson. Here $(\mu_\varphi$, $f_B)$ are constants with dimensions of length and the scalar field has a negative mass $\mu^2 = - m_\varphi^2$. \textcite{Minamitsuji:2022qku} showed that in this case the scalar field sits at its vacuum expectation value far from the source, while inside a \ns a symmetry restoration can take place, resulting in a new type of scalarization. This theory also has an advantage in that it the avoids cosmological instabilities present in certain other scalarization models.

\section{Black-hole scalarization}
\label{sec:bh_scalarization}
In this section, we discuss the spontaneous scalarization of \bh{s} in three parts.
To begin, in Sec.~\ref{sec:bhs_vac} we consider models in which \bh{s} can scalarize in vacuum
due to couplings between the scalar field and curvature scalars.
Next, in Sec.~\ref{sec:bhs_nvac} we discuss models in which scalarization is induced by the presence of
extra fields (such as gauge or matter fields) in the \bh spacetime.
Finally, in Sec.~\ref{sec:bh_scalarization_others} we review a selection of other models
of \bh scalarization.

\subsection{Black-hole scalarization: Vacuum spacetimes}
\label{sec:bhs_vac}

As we saw in Sec.~\ref{sec:theory_bg}, some gravity theories admit the
same vacuum \bh solutions as \gr, yet can give rise to new branches of
solutions with scalar hair once certain conditions are met.
The prototypical example is described by scalar-Gauss-Bonnet theories, whose action in the absence of matter is given by Eq.~\eqref{actionsgb}, namely,
\begin{align*}
    S &= \frac{1}{16 \pi G} \int \dV \left[ R - \tfrac{1}{2} g^{\mu\nu}\partial_{\nu} \varphi \partial_{\mu} \varphi + f(\varphi) \mathscr{G} \right].
\end{align*}
As discussed in Sec.~\ref{sec:intro}, the first models exhibiting \bh
scalarization in vacuum were proposed by~\textcite{Doneva:2017bvd} and
\textcite{Silva:2017uqg}.
These models remain the most studied \bh scalarization models to date in
the literature. For this reason, they will be our main focus in this
section.\footnote{We remark that~\textcite{Doneva:2017bvd}
and~\textcite{Silva:2017uqg} worked with different scalar-field normalizations, with the former being
 $1/2$ times the latter. Here we use the latter; hence, the scalar charge $D$ reported in Fig.~\ref{fig:scalarized_bh_D_M} is twice what one would obtain with our normalization.
}

As explained in Sec.~\ref{sec:theory_sgb}, in theories described by the
action~\eqref{actionsgb}, a no-hair theorem guarantees that
the \bh solutions of \gr are unique to the theory, as long
as $(\dd^2 f / \dd \varphi^2)_{\varphi = \varphi_0} \mathscr{G} < 0$, for some constant $\varphi_0$~\cite{Silva:2017uqg}.
If this condition is violated, scalar-field perturbations can become
tachyonic unstable and the endpoint is expected to be a nonlinear, scalarized \bh~\cite{Ripley:2020vpk}.
Scalarized \bh{s} have been shown to form dynamically, as outcomes from
the core collapse of an initially unscalarized star~\cite{Kuan:2021lol}.

Here we first review our understanding of isolated scalarized \bh{s}
in scalar-Gauss-Bonnet theories, focusing on their properties and their stability.
We then review what is understood about when they are found in binaries.
Finally, we give an overview of vacuum \bh solutions in models that
generalize the action~\eqref{actionsgb}.

\subsubsection{Scalarized black holes}
\label{sec:bh_scalarization_sgb}

\begin{figure}[t]
\includegraphics[width=\columnwidth]{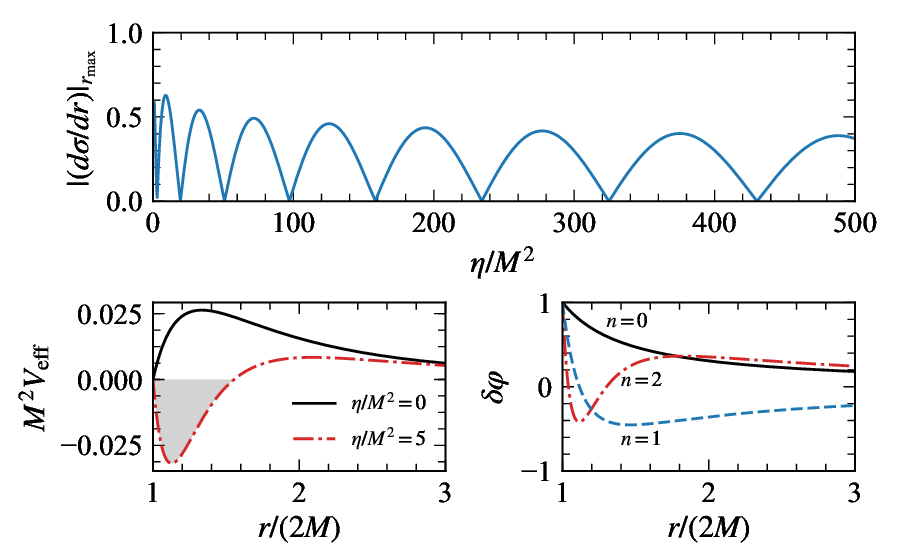}
\caption{Results of the numerical integration of the equation
of a monopole ($\ell = m = 0$) scalar perturbation $\delta \varphi$
on a Schwarzschild \bh background.
Top panel: asymptotic value of $|{\rm d} \sigma / {\rm d} r|$ (with $\sigma \equiv r \, \delta \varphi$)
evaluated at a point $r_{\rm max} \gg r_{\rm h}$,
far from the event horizon $r_{\rm h} = 2M$, as a function of $\eta / M^2$.
Cusps mark when bound state solutions form. They signal the transition
between stable Schwarzschild \bh{s} and scalarized solutions.
Bottom-left panel: effective potential $\Veff$  (see Sec.~\ref{sec:tachcurved}) for
$(\eta / M)^2 = 0$ and $5$. In the latter case, $\Veff$ develops
a negative region and can support bound states.
Bottom-right panel: radial profiles of $\delta \varphi$ for the first three
bound states corresponding to $\eta /M^2 = 2.902$, $19.50$, and
$50.93$. These profiles have $0$, $1$, and $2$ nodes, respectively.
The properties of these solutions are shared between the models
of~\textcite{Doneva:2017bvd,Silva:2017uqg}.
From \textcite{Silva:2017uqg}.
}
\label{fig:decoupl_bh_scalarization}
\end{figure}

Spontaneously scalarized \bh solutions were first found by \textcite{Doneva:2017bvd,Silva:2017uqg}.
What distinguishes these works is the choice of the coupling function
$f(\varphi)$ that couples the scalar field to the Gauss-Bonnet invariant
[cf.~Eq.~\eqref{actionsgb}],
\begin{subequations}
\label{eq:gaussian_and_quadratic_couplings}
\begin{align}
    f &= (\lambda^2/12) \, [1 - \varepsilon \exp(-3\varphi^2/2) ],
    \label{eq:gaussian_coupling}
    \\
    f &= (\eta /8)  \, \varepsilon \, \varphi^2,
    \label{eq:quadratic_coupling}
\end{align}
\end{subequations}
where $\lambda^2$ and $\eta$, positive by definition, are coupling constants with dimensions of [length]$^2$ and $\varepsilon = \pm 1$,
which should not be confused with the energy density of Sec.~\ref{sec:ns_scalarization}.
The two coupling functions agree in the small-$\varphi$ limit (i.e., when $\varphi \ll 1$)
and therefore result in the same prediction for the onset of scalarization of \gr \bh{s}.
This threshold can be found by searching for bound state solutions, i.e., time-independent
solutions of the linearized field equation for $\varphi$ around a \gr \bh, which are
regular at the event horizon $r_{\rm h}$ and that vanish at spatial infinity.
Hence, the determination of the scalarization threshold reduces to a boundary value problem.
In the simplest case of a Schwarzschild \bh of mass $M$, the dimensionless quantity $\eta / M^2$ plays the role of the eigenvalue and can be determined numerically with standard techniques such as the shooting method.
The smallest eigenvalue gives the threshold for the formation of the
``fundamental'' scalarized \bh solution (in the sense that the radial
profile of the scalar field has $n=0$ nodes).
The other eigenvalues give the thresholds for the formation of ``excited states,''
that is, solutions with one or more nodes.
In Fig.~\ref{fig:decoupl_bh_scalarization} we show the results of such a calculation
in the spacetime of a Schwarzschild \bh and $\varepsilon = 1$,
as done by~\textcite{Silva:2017uqg}.
We remark that no bound states can form for $\varepsilon = -1$, because the effective
mass of scalar-field perturbations is positive; therefore, the effective potential
is positive definite. See the discussion in Sec.~\ref{sec:tachcurved}.

\begin{figure}[t]
\includegraphics[width=0.80\columnwidth]{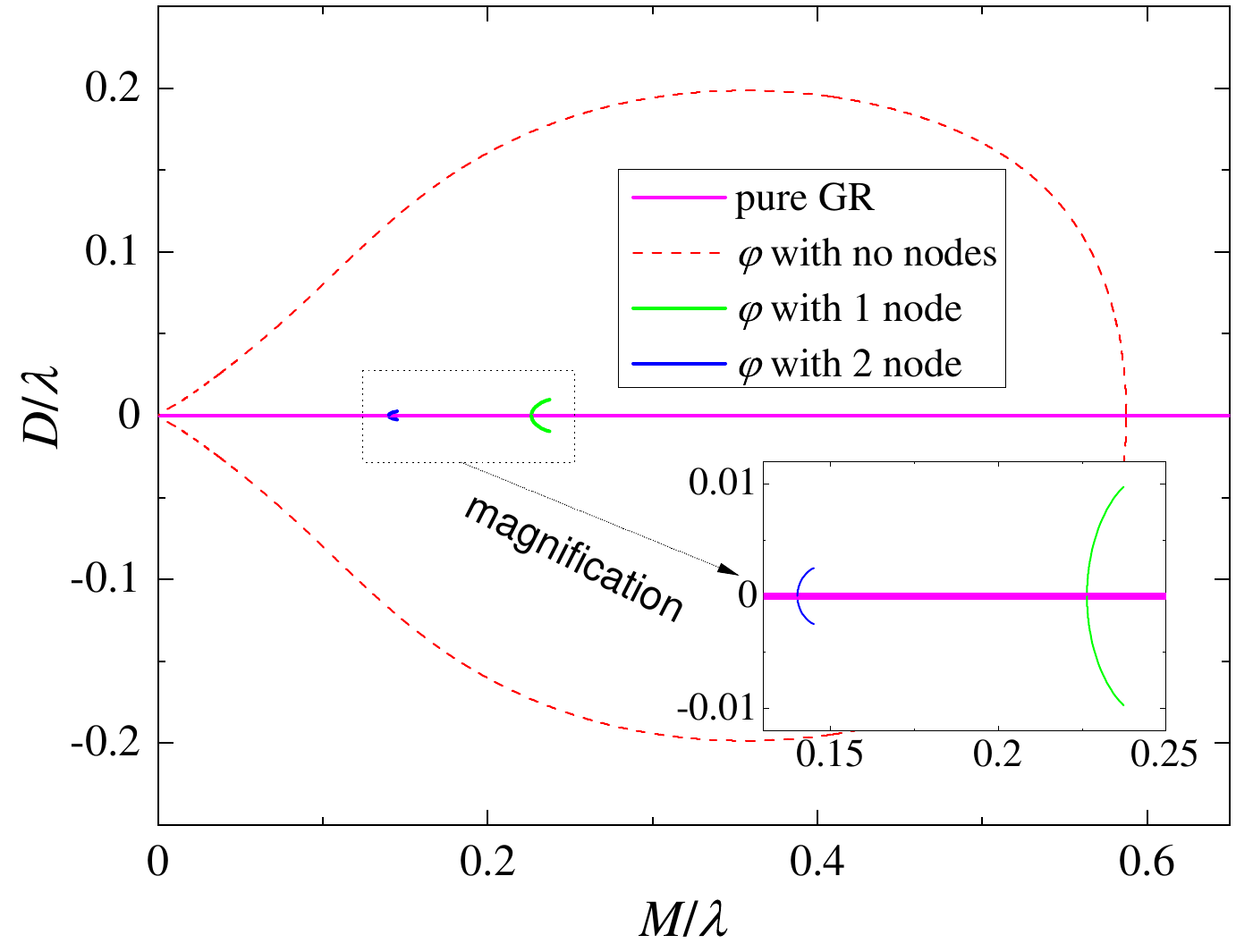}
\caption{
Scalar charge $D$ normalized to the coupling parameter $\lambda$ as a function of the normalized \bh mass $M/\lambda$ for sequences of scalarized \bh solutions.
The \gr solutions have zero scalar field and thus zero $D$ and lie along the $x$ axis, which is depicted by the horizontal magenta line.
The dashed line corresponds to solutions with nodeless scalar-field profiles, which was found to be the only stable branch~\cite{Blazquez-Salcedo:2018jnn}.
The curves with different colors correspond to scalarized solutions with scalar fields having different numbers of nodes.
Solutions with one and two nodes branch off the \gr solution around $M/\lambda \simeq 0.15$ and $M/\lambda \simeq 0.24$,
respectively.
The scalarized branches are symmetric with respect to the $y$ axis due to the reflection symmetry ($\varphi \to -\varphi$) in the theory.
Adapted from \textcite{Doneva:2017bvd}.
}
\label{fig:scalarized_bh_D_M}
\end{figure}

While both models in Eqs.~\eqref{eq:gaussian_and_quadratic_couplings} agree
in their prediction of the threshold for scalarization, they differ
significantly in their prediction of the properties of the nonlinear, scalarized solution, as discussed in Sec.~\ref{sec:scalarization_mechanism}.
In Fig.~\ref{fig:scalarized_bh_D_M} we show the branches of scalarized \bh{s},
in a parameter space spanned by the dimensionless scalar charge and \bh mass.
The solutions were obtained by solving the full system of field equations; see
\textcite{Doneva:2017bvd,Julie:2022huo} for detailed discussions.
The results are for the Gaussian coupling~\eqref{eq:gaussian_coupling} with $\varepsilon = 1$.
We show the fundamental, zero-node scalar-field solution (dashed red curves) and
also the excited scalarized \bh{s} (solid green and blue curves).
We see that the branches of excited \bh{s} can terminate at finite masses $M$.
This is due to a violation of an inequality at the \bh horizon that must be satisfied for the scalar field to be real valued and is typical of scalar-Gauss-Bonnet theories; see e.g.,~\textcite{Kanti:1995vq} and~\textcite{Antoniou:2017acq}.
The parameters in the coupling function~\eqref{eq:gaussian_coupling} are chosen in such a way that this disappearance of \bh solutions does not affect the fundamental branch.
For other couplings, such as Eq.~\eqref{eq:quadratic_coupling}, even this branch
can violate the regularity condition shortly after bifurcation; see \textcite{Doneva:2018rou,Silva:2018qhn}.
Hence, the domain of existence of scalarized solutions in the Gaussian model~\eqref{eq:gaussian_coupling} is larger
with respect to that of the ``quadratic'' model~\eqref{eq:quadratic_coupling}.

\begin{figure}[t]
\includegraphics[width=0.9\columnwidth]{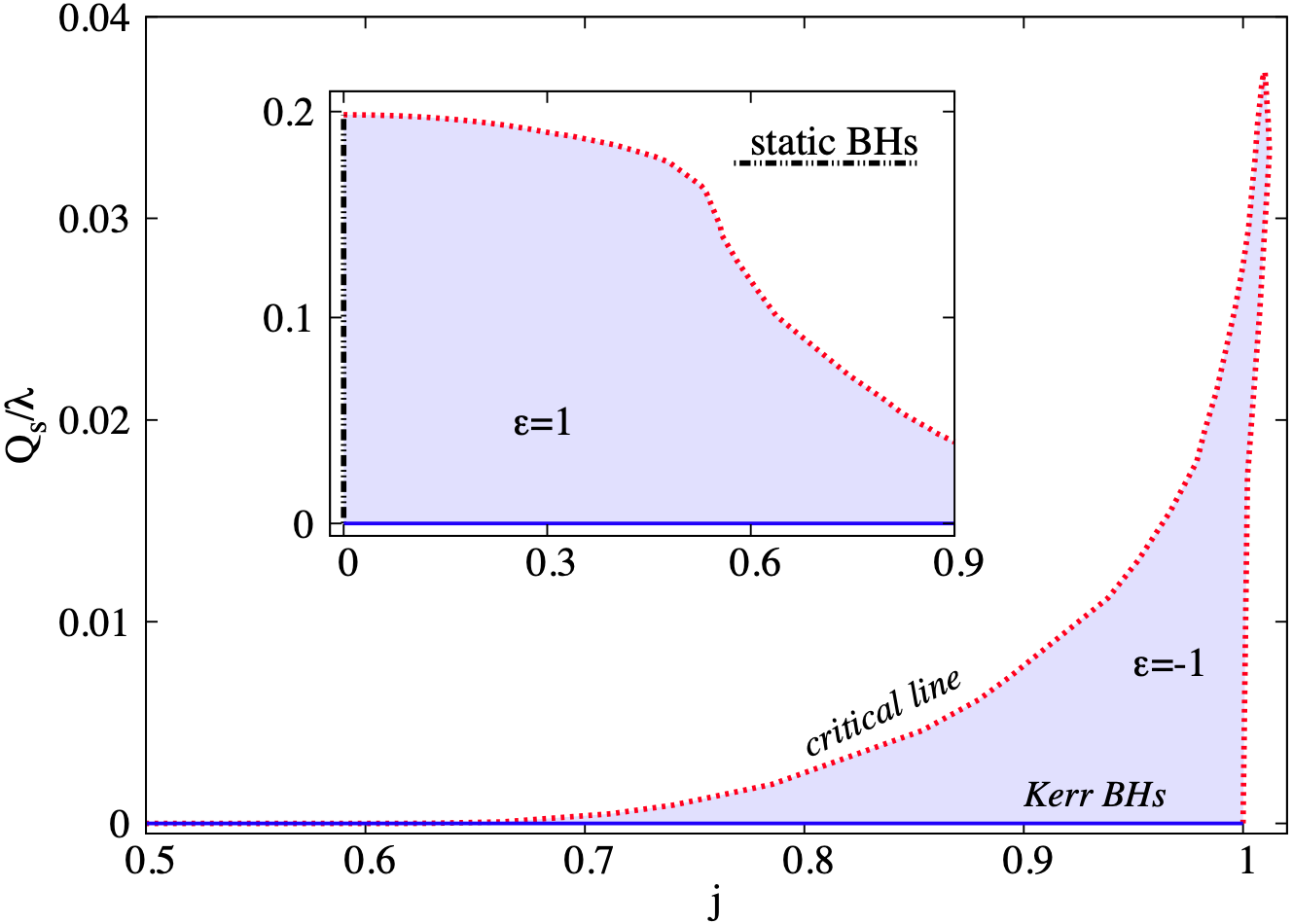}
\caption{Rotating scalarized \bh{s} in the Gaussian model~\eqref{eq:gaussian_coupling}.
The shaded regions represent the domain of existence of scalarized rotating BHs in the
plane spanned by the dimensionless scalar charge ($Q_{s} / \lambda$) and spin ($j = a / M$).
These regions are bound by the Kerr family of solutions $Q_{s}/\lambda = 0$ (solid blue line) and, in the nonrotating limit, by the Schwarzschild solution (black dot-dashed line),
and also when the regularity condition for the scalar field is violated (dotted curve).
The inset shows the case for positive coupling constant studied by~\textcite{Cunha:2019dwb}, while
the main panel shows nonlinear spin-induced scalarized \bh{s}.
From \textcite{Herdeiro:2020wei}.
}
\label{fig:rotating_scalarized_bhs}
\end{figure}

\begin{figure}[t]
\includegraphics[width=0.48\columnwidth]{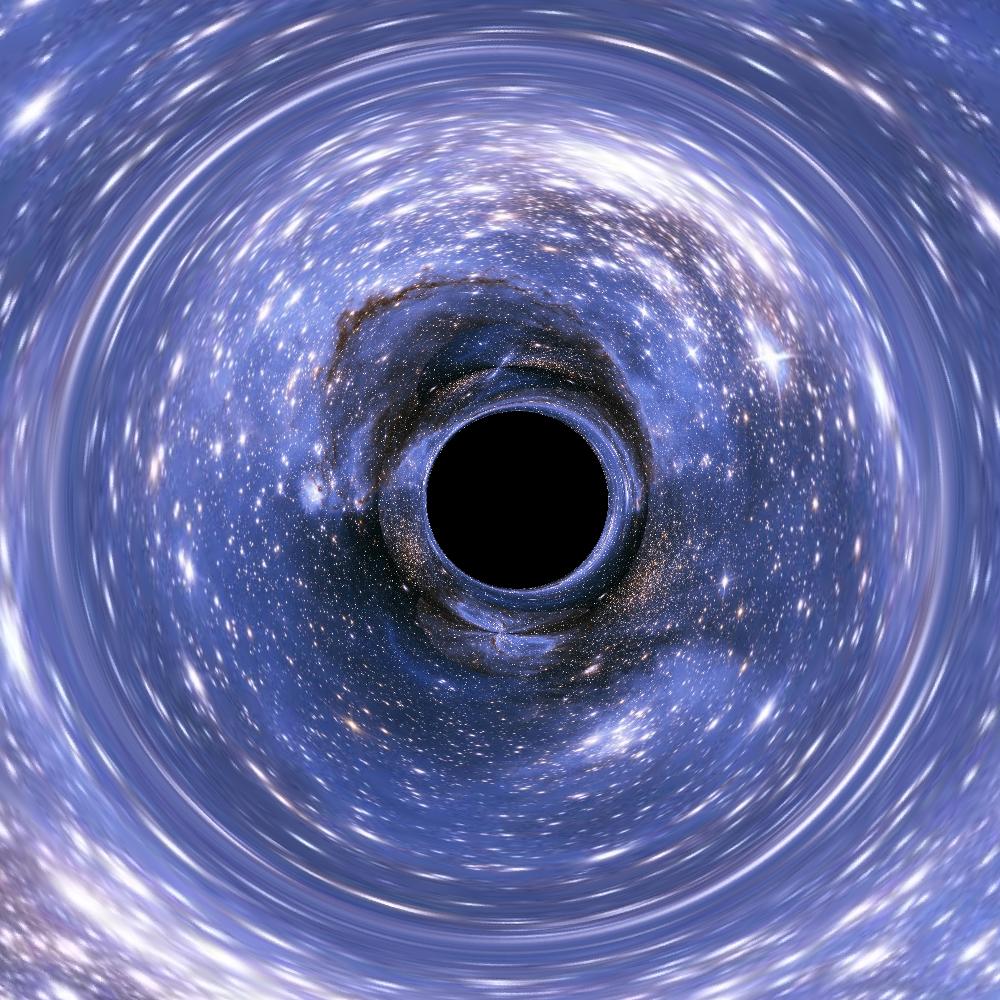}
\includegraphics[width=0.48\columnwidth]{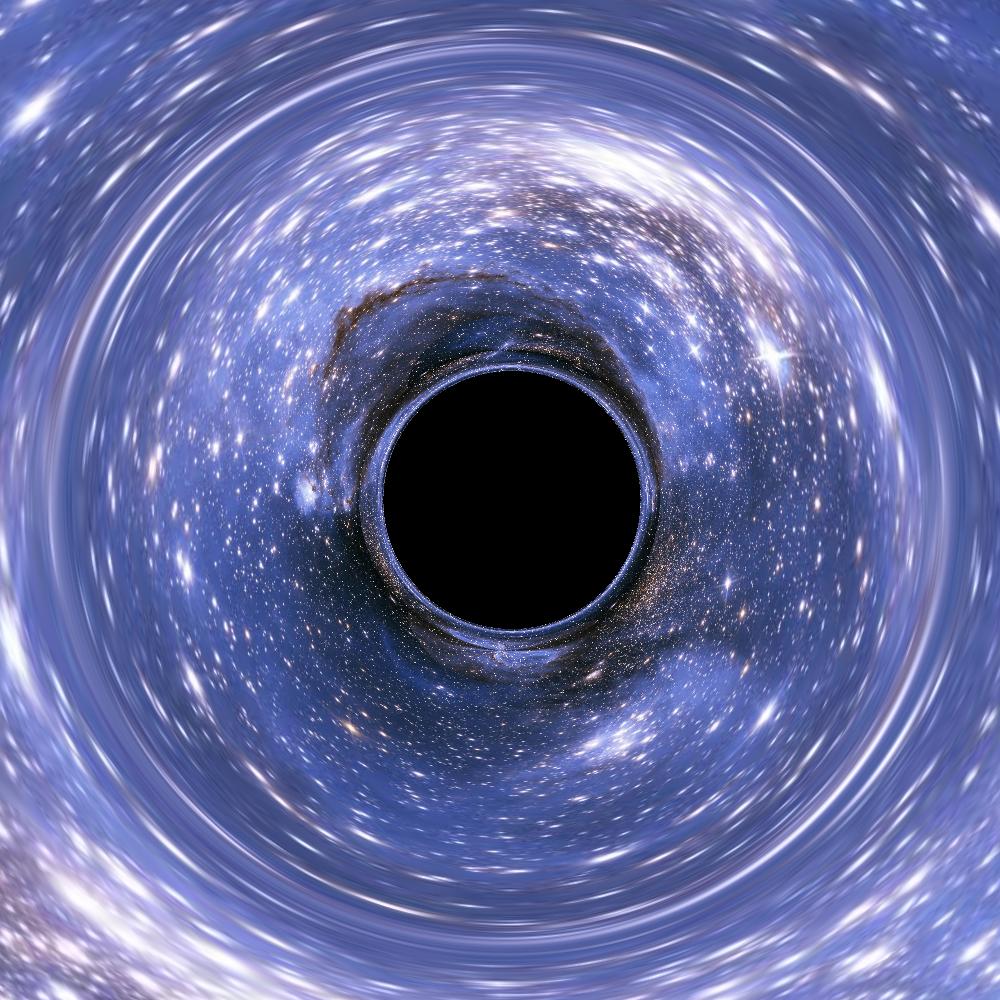}
\caption{Gravitational lensing and shadow produced by rotating \bh{s}
under similar observation conditions. Left panel: scalarized
\bh with mass $M / \lambda \approx 0.237$ and spin $j = 0.24$. Right panel:
a comparable Kerr \bh.
From \textcite{Cunha:2019dwb}.
}
\label{fig:scalarized_shadows}
\end{figure}

Asymptotically flat, spinning scalarized \bh{s} in the Gaussian models (with $\varepsilon = 1$) and quadratic theories were explored, respectively, by~\textcite{Cunha:2019dwb} and~\textcite{Collodel:2019kkx}.
In discussing these solutions, we consider a plane spanned by the dimensionless scalar charge $Q_s / \lambda$ and dimensionless spin parameter $j \equiv J / M^2$, where $J$ is the angular momentum of the \bh.
Figure~\ref{fig:rotating_scalarized_bhs} shows the domain of existence of scalarized rotating \bh
solutions. In particular, we focus on the inset, which corresponds to $\varepsilon = 1$ in Eq.~\eqref{eq:gaussian_coupling}.
We see that as the spin increases the existence domain of the solutions
(the shaded region) becomes smaller.
This result can be understood in terms
of the Gauss-Bonnet invariant not being positive definite for the Kerr metric~\cite{Cherubini:2002gen}.
More specifically, a Kerr \bh develops increasingly large regions where the Gauss-Bonnet
invariant is negative as the spin increases and hence suppresses scalarization (or triggers scalarization, when $\varepsilon = -1$, as discussed in Sec.~\ref{sec:bh_spin_induced}.)
As a consequence, the deviations relative to \gr predictions on physical observables
(for example, the location of the innermost stable circular orbit of massive particles or the \bh shadow) are smaller for rapidly rotating \bh{s}. See Fig.~\ref{fig:scalarized_shadows} for an example.
\textcite{Collodel:2019kkx} studied the quadratic coupling \eqref{eq:quadratic_coupling} and reached similar conclusions. They obtained radially excited solutions (i.e., scalar-field profiles with nodes in the radial direction) and also notes the existence of ``angularly'' excited rotating \bh{s} (i.e., scalar-field profiles with nodes in the angular directions).

\subsubsection{Stability of scalarized black holes
and implications for model building}
\label{sec:bhs_stability_nr}

Having established the existence of scalarized \bh{s}, the natural next task is to study their
stability.
A first indication comes from the study of the thermodynamical properties of such \bh{s}.
\textcite{Doneva:2017bvd} found that the scalarized solutions belonging to the fundamental branch in the Gaussian model are thermodynamically favored relative to the Kerr solution~\eqref{eq:gaussian_coupling},
in the sense that they have a smaller Wald entropy~\cite{Wald:1993nt}.
This suggests the stability of the former; i.e., in a dynamical process \bh{s} in this model
would evolve toward a state of smaller entropy.
In contrast, the quadratic coupling~\eqref{eq:quadratic_coupling} model always results in scalarized \bh{s} with higher entropy relative to their \gr counterparts.
These observations remain true in the case of rotating \bh{s}~\cite{Herdeiro:2020wei,Collodel:2019kkx}.

\begin{figure}[t]
\includegraphics[width=0.82\columnwidth]{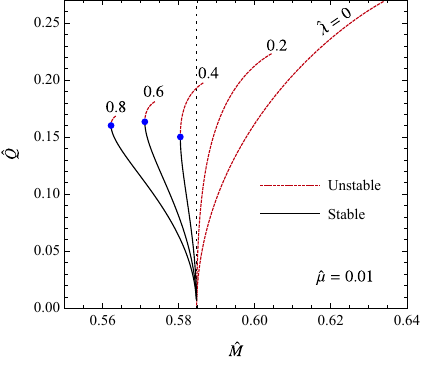}
\caption{
Scalar-charge-mass diagram for scalarized \bh{s} with a quadratic scalar-Gauss-Bonnet model
with a scalar-field potential $V(\varphi) = (\mu^2 / 2) \varphi^2 + (\lambda / 2) \varphi^4$.
Dimensionless quantities obtained by factors of $\eta^{1/2}$ are used: \bh charge $\hat{Q} = Q / \eta^{1/2}$ and mass $\hat{M} / \eta^{1/2}$, and scalar-field parameters $\hat{\mu} = \mu \eta^{1/2}$ and $\hat{\lambda} = \lambda \eta^{1/2}$.
The vertical line marks the threshold for scalarization $\hat{M}_{\rm th}$.
\bh{s} with $\hat{M} \geqslant \hat{M}_{\rm th}$ are radially unstable, while the Schwarzschild solution
is stable. For large $\hat{\lambda}$, we can obtain solutions with $\hat{M} < \hat{M}_{\rm th}$ and form two branches.
The ones that are radially unstable solutions are denoted with a dashed line (upper branch) and those stable to radial
perturbations are denoted with a solid line (lower branch).
The dot marks are the marginally stable solutions, which correspond to a minimum mass but maximally scalar charged
\bh{s} for fixed $\hat{\mu}$ and $\hat{\lambda}$.
From \textcite{Macedo:2019sem}.
}
\label{fig:eft_stabilization}
\end{figure}

\begin{figure}[t]
\includegraphics[width=\columnwidth]{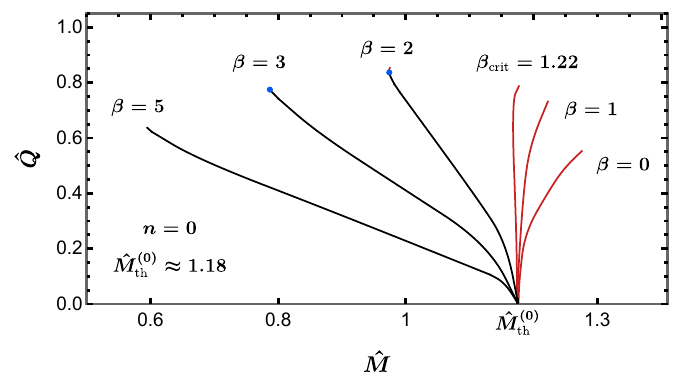}
\caption{
Like Fig.~\ref{fig:eft_stabilization}, but now in a ``mixed'' model where a
quadratic coupling between the scalar field and the Ricci tensor is added to the
quadratic scalar-Gauss-Bonnet interaction.
As in Fig.~\ref{fig:eft_stabilization}, we show the charge
$\hat{Q} = Q / \alpha^{1/2}$ and mass $\hat{M} / \alpha^{1/2}$ made dimensionless
by division by $\alpha^{1/2}$, for different values of scalar-Ricci coupling
$\beta$. See Eq.~\eqref{action:RGmixed} for the theory's action.
The left-bending black lines correspond to values of $\beta$ for which all scalarized \bh{s} have masses below the \gr instability
mass threshold, while the right-bending red lines mark values of $\beta$ that lead to scalarized \bh masses that are larger than the \gr mass
threshold. Past the turning points (marked with blue dots), the \bh{s} are expected to be unstable to radial perturbations.
From \textcite{Antoniou:2021zoy}.
}
\label{fig:ricci_stabilization}
\end{figure}

A complementary question is whether scalarized \bh{s} are stable to linear perturbations.
The first step in this direction was carried out by~\textcite{Blazquez-Salcedo:2018jnn}, who considered radial scalar-field and metric perturbations.
They showed that Schwarzschild \bh{s} became unstable at the scalarization threshold and that the stability of the fundamental (i.e., the zero-node) scalarized solution depended on which model in Eq.~\eqref{eq:gaussian_and_quadratic_couplings} was considered: in the quadratic case all solutions were unstable, whereas stable solutions existed in the Gaussian model.

This important difference between the two models can be traced back to the higher-than-quadratic terms in the scalar field in the theory's action.
More concretely \textcite{Silva:2018qhn} and \textcite{Minamitsuji:2018xde} showed that the inclusion of a quartic interaction,
say, $(\zeta/16) \, \varphi^4 \, {\mathscr G}$, is sufficient to yield
stable \bh solutions as long as $\zeta$ is negative and satisfies $\zeta / \eta < -0.8$.
These two conditions are satisfied by the Gaussian model. Indeed, expanding Eqs.~\eqref{eq:gaussian_coupling} to order ${\cal O}(\varphi^4)$, we find that $\zeta / \eta = - 3/2$,
which helps to explain the stability of \bh{s} in this model.

It was later shown by~\textcite{Macedo:2019sem} [see also~\textcite{Macedo:2020tbm}] that the original quadratic model can be made stable under radial perturbations by making the scalar field massive and self-interacting, $V(\varphi) \propto \mu^2 \varphi^2 + \lambda \varphi^4$,
as discussed in Fig.~\ref{fig:eft_stabilization}.
From an \EFT perspective, these terms are of lower order than the
quartic scalar-Gauss-Bonnet interaction and hence should be included.
A similar analysis was also performed for the Ricci-Gauss-Bonnet model by~\textcite{Antoniou:2021zoy} and is shown in Fig.~\ref{fig:ricci_stabilization}.
See also~\textcite{Doneva:2019vuh} for a study of scalarization with massive scalar fields.

The nonlinear stability of \bh{s} in this model was also studied in the time domain by~\textcite{Ripley:2020vpk}.
They found evidence for regions in mass-coupling parameter space in which the end state
of the radial instability of the Schwarzschild \bh is a stable scalarized \bh.
For larger couplings however, regions where the time-evolution equations change character from
hyperbolic to elliptic appear outside the \bh horizon. This signals a regime in which
the theory does not have a well-posed initial-value problem. See~\textcite{Hilditch:2013sba,Ripley:2022cdh} for discussions.
These results are in agreement with earlier findings by~\textcite{Blazquez-Salcedo:2018jnn}, but in the linear regime.
The radial stability of scalarized \bh{s} in the Ricci-Gauss-Bonnet model introduced in Sec.~\ref{sec:mixed_model} was analyzed by~\textcite{Antoniou:2022agj}.
They found that although the Ricci scalar does not affect the scalarization threshold, at
the nonlinear level it can stabilize the \bh{s} in the quadratic model.

The problem of nonradial perturbations of scalarized \bh solutions,
which is relevant in the context of testing such theories with ringdown \gw
signals, was addressed by \textcite{Blazquez-Salcedo:2020rhf,Blazquez-Salcedo:2020caw}.
However, the stability of rotating scalarized \bh{s} remains
an open problem, even in the slow rotating limit.

\subsubsection{Spin-induced scalarization}
\label{sec:bh_spin_induced}

One property of the Gauss-Bonnet invariant of the Schwarzschild metric is that it
is positive definite everywhere and, as a consequence, spontaneous scalarization
can occur only for positive values of the coupling constant.
Does this always have to be the case? In the case of the Kerr metric, it is known
that regions where $\mathscr{G} < 0$ outside the outer horizon are possible if
the \bh spins with $a / M \geqslant 0.5$~\cite{Cherubini:2002gen},
suggesting the possibility of a
\emph{spin-induced scalarization} when the coupling constant is negative, i.e.,
$\varepsilon = -1$ in Eqs.~\eqref{eq:gaussian_and_quadratic_couplings}.

\begin{figure}[t]
\includegraphics[width=\columnwidth]{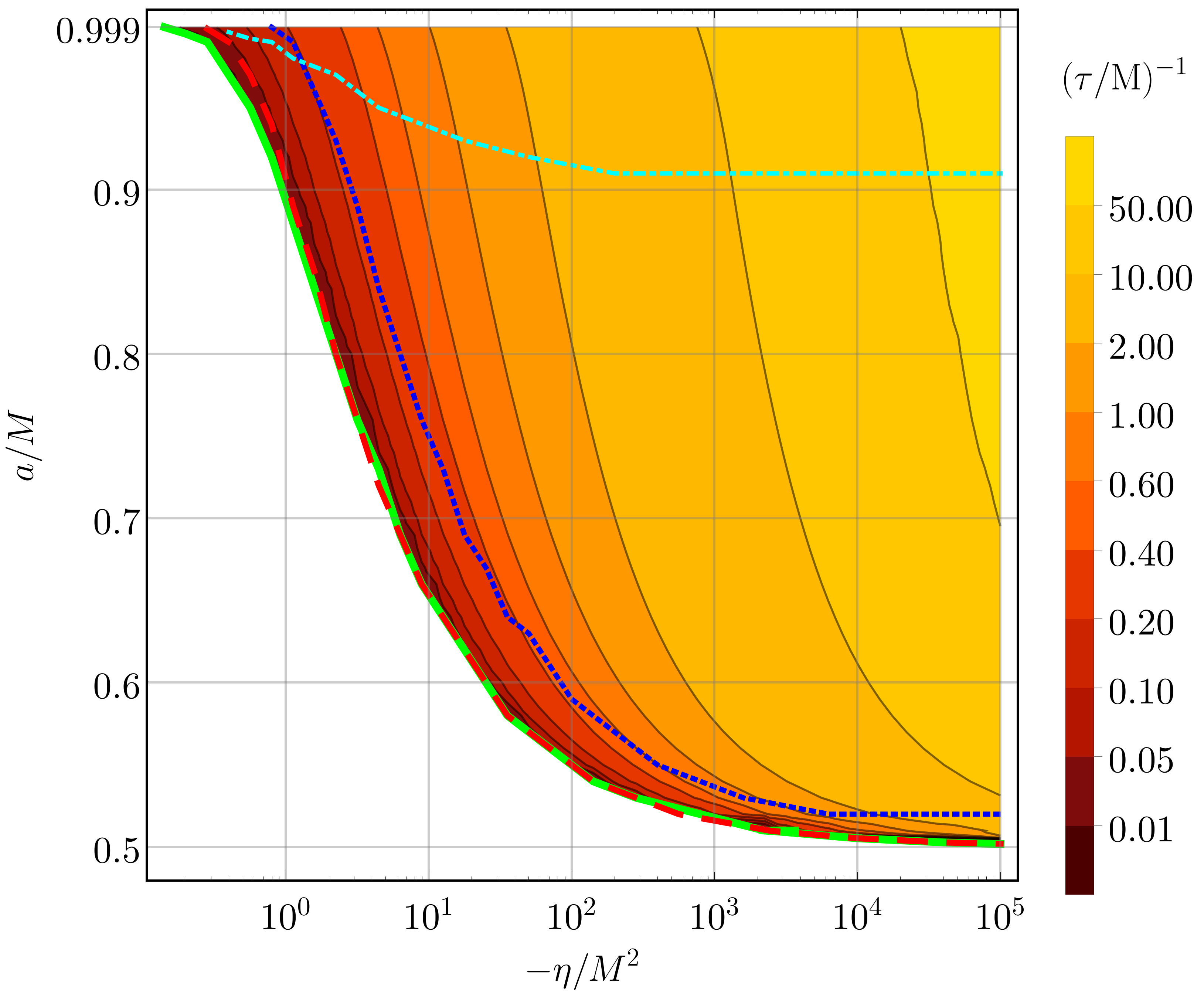}
\caption{Instability timescale $\tau$ for spin-induced scalarization
in the dimensionless spin $a/M$ and dimensionless coupling constant $-\eta / M^2$ plane.
The instability threshold for the total reconstructed field is
shown as a solid green line, while the threshold when the
$m = 0$ modes are excluded is shown as a blue dotted line.
The red dashed line corresponds to the instability threshold for
the $m = 0$ odd modes, while the dot-dashed cyan line marks
the instability threshold for the spherical mode $\ell = m = 0$.
From \textcite{Dima:2020yac}.
}
\label{fig:domain_spin_induced}
\end{figure}

Although conceptually simple, the problem is not straightforward to analyze because
(i) the analytical form of $\mathscr{G}$ forbids the separation of variables as done
in Sec.~\ref{sec:intro}, and thus the determination of the scalarization threshold will
in principle require more sophisticated numerical methods than the Schwarzschild case, and
(ii) the Kerr metric (with $a / M \geqslant 0.5$) has regions where $\mathscr{G}$ can be either negative or positive, where in the latter regions, for negative coupling constants,
the effective mass may trigger a superradiant instability~\cite{Brito:2015oca}.
To overcome these difficulties,~\textcite{Dima:2020yac} reduced the $(2+1)$-dimensional
nonseparable scalar-field equation to a $(1+1)$-dimensional system of equations that are
coupled through spherical harmonic multipole indices $l$ and $m$. They then performed time-domain
numerical integration of the scalar field borrowing methods that were previously developed to study superradiant instability~\cite{Dolan:2012yt}.
\textcite{Dima:2020yac} showed that the tachyonic instability is the dominant one and charted the
parameter space in which spin-induced scalarization would occur. This is shown in Fig.~\ref{fig:domain_spin_induced}.
\textcite{Hod:2020jjy} worked analytically in the infinitely large coupling limit ($\eta / M^2 \to \infty$) and
confirmed the expectation that $a / M \gtrsim 0.5$ is the minimal necessary spin value for which spin-induced scalarization occurs.
The corrections for large but finite coupling were obtained by~\textcite{Hod:2022hfm}.

Nonlinear spin-induced scalarized \bh solutions were obtained by~\textcite{Herdeiro:2020wei}
and by~\textcite{Berti:2020kgk}.
They confirmed the existence of scalarized \bh solutions in the parameter space region
in which linear theory predicts the spin-induced tachyonic instability.
In Fig.~\ref{fig:rotating_scalarized_bhs} we show
the domain of the spin-induced scalarized \bh{s} in the Gaussian model~\eqref{eq:gaussian_coupling}.
We see that scalarized solutions exist only for $j \gtrsim 0.5$, and that their
scalar charge increases with spin.
Moreover, there are solutions that can violate the Kerr bound $j \leq 1$ on \bh spins.

A approach complementary to that of~\textcite{Dima:2020yac} was used by~\textcite{Doneva:2020nbb,Doneva:2021dqn}. They evolved the scalar-field equation
in 2+1 dimensions, i.e., without doing a multipolar decomposition.
The influence of a nonvanishing scalar-field mass was explored by~\textcite{Doneva:2020kfv}.
A similar time-domain study that instead used the hyperboloidal foliation method~\cite{Zenginoglu:2007jw}
was performed by~\textcite{Zhang:2020pko} and a model with both the Pontryagin and Gauss-Bonnet invariants was studied by~\textcite{Myung:2020etf}.
\textcite{Annulli:2022ivr} studied the effect of strong magnetic fields
on spin-induced scalarization. They found that the magnetic field suppressed this effect; i.e., it pushed the scalarization threshold to larger values of the dimensionless spin
$j$ of the \bh.
The case of spin-induced scalarization has been much less studied in terms on its stability and on how additional nonlinear interactions affect the scalar charge.

\subsubsection{Scalarized black holes in binary systems}

Most work in \bh scalarization has focused on isolated \bh{s},
but to be able to confront these models against \gw observations
one has to turn to the two-body problem.
What phenomenology would we expect in \bh binaries?
To answer this question we first consider work that explored the
nonlinear regime of the late inspiral and merger of \bh binaries, a regime
that relies on \emph{numerical relativity}.
Next we consider work that focused on the inspiral alone, a regime that
can be modeled with \pn theory.

\begin{figure*}[t]
\includegraphics[width=\textwidth]{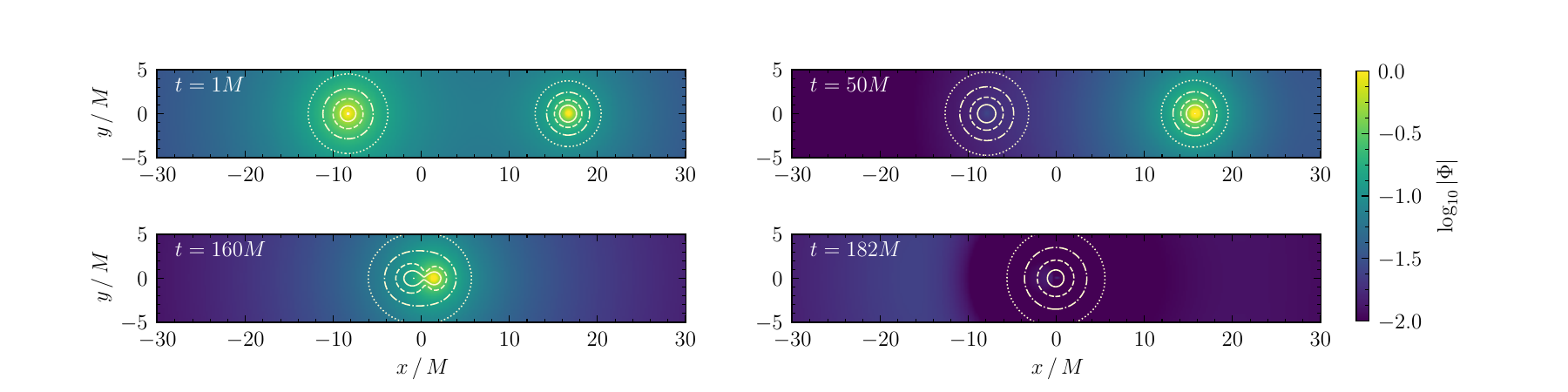}
\caption{Dynamical descalarization in binary \bh head-on collisions.
Scalar-field and Gauss-Bonnet dynamics on the $x$-$y$ plane of the \bh{s} with initial mass ratio $q = m_1 / m_2 = 1/2$.
We show the scalar-field amplitude $\log_{10}|\Phi|$ (color map) together with the isocurvature levels
of the Gauss-Bonnet invariant at the beginning of the evolution (top-left panel), during the \bh
approach (top-right panel), shortly before the collision (bottom-left panel) and shortly after
the merger (bottom-right panel).
These levels correspond to
$1 M^{-4}$ (solid line),
$10^{-1} M^{-4}$ (dashed line),
$10^{-2} M^{-4}$ (dot-dashed line) and
$10^{-3} M^{-4}$ (dotted line).
From \textcite{Silva:2020omi}.
}
\label{fig:descalarization_snapshot}
\end{figure*}

The first work in this area was done by~\textcite{Silva:2020omi}, who
studied the scalar-field dynamics (i.e., in the decoupling limit) in head-on
collisions and nonspinning, quasicircular inspirals of binary \bh spacetimes obtained
from numerical relativity simulations. This work adapted the methods developed by \textcite{Witek:2018dmd}
for shift-symmetric scalar-Gauss-Bonnet theory to theories which that allow for spontaneous scalarization.
In particular, because they were interested at phenomenology near the scalarization
threshold, \textcite{Silva:2020omi} adopted the quadratic
model~\eqref{eq:quadratic_coupling}, with a coupling strength $\eta$ such that
both (or one of them, depending on the mass ratio) binary components or the remnant \bh
can carry a scalar-field bound state through the simulation.
They showed that \bh binaries can either form a scalarized remnant or
dynamically descalarize by shedding off the initial scalar hair (i.e., the scalar
bound state configuration) depending on the values of the coupling constants and
the mass ratio between the two \bh{s}.
An example of dynamical descalarization is shown in Fig.~\ref{fig:descalarization_snapshot}.
Dynamical descalarization was also shown to occur in nonlinear scalarization
models (see~Sec.~\ref{sec:beyond_scalarization}) by \textcite{Doneva:2022byd}
that simulated head-on \bh collisions and worked in the decoupling limit.

A natural question that follows is: What would happen when $\varepsilon = -1$, i.e., the case in which
spin-induced scalarization occurs? This is relevant from an observational point of view because \bh remnants of binary \bh coalescences have typical dimensionless spins of $j \sim 0.7$. This value comfortably meets the criteria for spin-induced scalarization to happen.
This question was explored by \textcite{Elley:2022ept}, who found through a suite of numerical relativity simulations that:
(i) spin-induced dynamical descalarization can happen when the remnant \bh spin is smaller
than that of the initially spinning (and scalarized) binary components, and
(ii) spin-induced dynamical scalarization in an initially nonspinning \bh binary system
can result in a scalarized rotating \bh.

\begin{figure}[t]
\includegraphics[width=\columnwidth]{"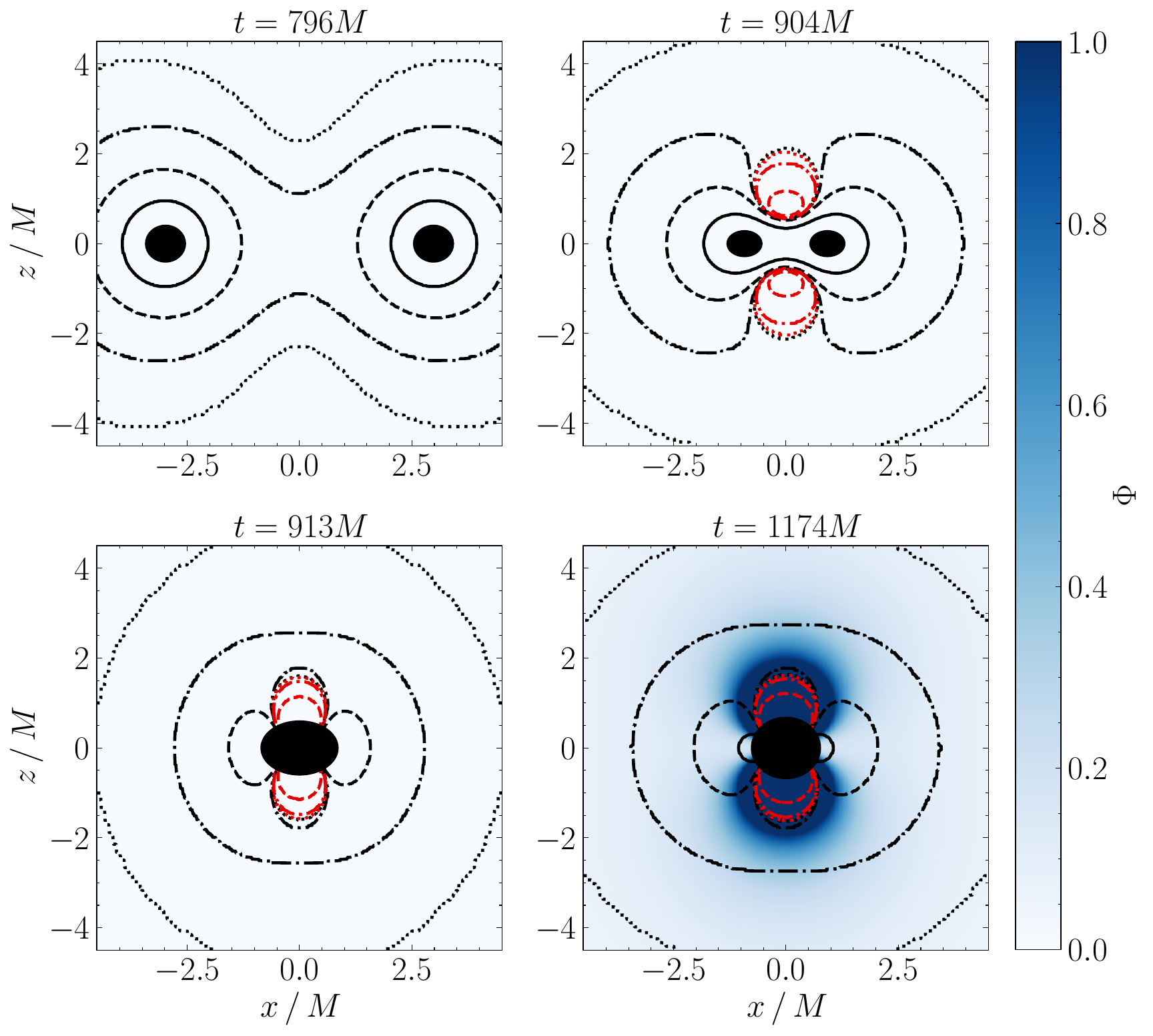"}
\caption{Snapshots of the scalar field, here labeled $\Phi$, and the Gauss-Bonnet invariant $\mathscr{G}$ in the $x$-$z$ plane llustrating the stealth scalarization.
The color map indicates the amplitude of the scalar field.
The curves represent isocurvature contours for the following values of $|\mathscr{G}|$:
$1 M^{-4}$ (solid line),
$10^{-1} M^{-4}$ (dashed line),
$10^{-2} M^{-4}$ (dot-dashed line), and
$10^{-3} M^{-4}$ (dotted line).
Black (red) lines correspond to positive (negative) values of $\mathscr{G}$.
We show the inspiral (top-left panel), half an orbit before merger (top-right panel),
formation of the first common apparent horizon (bottom-left panel), and at $200M$
after the merger.
From \textcite{Elley:2022ept}.
}
\label{fig:spin_induced_scalarization}
\end{figure}

In the latter case, if the value of the coupling constant is sufficiently large the scalar field may affect the late inspiral and ringdown of the binary. For values of the coupling constant near the scalarization threshold for the remnant \bh, the tachyonic instability can be delayed to some time $\approx 100M$ after the binary merger, and can thus hide the scalar field throughout the binary’s inspiral. This was termed stealth scalarization by \textcite{Elley:2022ept}; Fig.~\ref{fig:spin_induced_scalarization} displays an example.

A restriction of the decoupling limit is that it does not capture the
backreaction of the scalar field onto the spacetime metric, and thus does
not allow one to derive the modifications to the gravitational waves in
the previously mentioned binary setups.
These difficulties can be partially overcome by employing
a formulation of the theory's field equations in a generalized harmonic
gauge that was first developed by \textcite{Kovacs:2020pns,Kovacs:2020ywu};
see also~\textcite{Ripley:2022cdh}.
This formulation in principle allows one to evolve the complete scalar-field and metric dynamics
in scalar-Gauss-Bonnet gravity.
However, the first binary \bh coalescence simulations using this formalism  in scalar-Gauss-Bonnet
theories that allow for \bh scalarization~\cite{East:2020hgw,East:2021bqk} demonstrated that evolution ceases to be well posed for larger values of the coupling. How different gauge choices and different models perform in this respect has not been well explored.

\textcite{Franchini:2022ukz}, motivated to go beyond the small-coupling approximation, explored the \emph{fixing-the-equations} approach to perform numerical simulations,
an approach inspired by work in dissipative relativistic hydrodynamics~\cite{Cayuso:2017iqc}.
The main idea is to modify the evolution equations of the theory such that
the short wavelength modes (associated with strong coupling and the breakdown of well-posedness) with respect to some chosen scale are somewhat controlled, while longer wavelength modes become tractable.
\textcite{Franchini:2022ukz} were able to use the formalism to study scalar-field collapse in spherical symmetry in the scalarization model of~\textcite{Silva:2018qhn}.
This allowed them to evolve the system past situations where the original equations fail due to the appearance of regions in spacetime
where the evolution changed character from hyperbolic to elliptic, which prevented evolution of the system in time~\cite{Ripley:2020vpk}.
Moreover, the ``fixed-equation''  system evolves toward the static \bh{s} predicted by the original system of equations, thus indicating
that it may provide a suitable ``completion'' of the original theory.
Whether this approach remains valid in a situation with less symmetry, for example the case of \bh binaries,
remains to be explored.

Progress has also been made in modeling compact binary dynamics in the regime of wider separation and lower velocities, i.e., in the domain of validity of \pn theory.
In particular, \textcite{Shiralilou:2020gah,Shiralilou:2021mfl} extended to a general $f(\varphi)$ coupling the calculations of~\textcite{Yagi:2011xp}, which applied for the shift-symmetric scalar-Gauss-Bonnet theory, i.e., $f(\varphi) \sim \varphi$.
These works calculated the leading-order corrections due to curvature nonlinearities in the \gw and scalar waveforms,
finding that corrections due to the Gauss-Bonnet term appear at 1\pn order in \gw{s}.
They also obtained the \gw polarization and phasing.
In addition,~\textcite{Julie:2019sab} argued that during the adiabatic inspiral of two \bh{s} the Wald entropy
of each \bh is constant. This allows a precise definition of the sensitivities of \bh{s} in scalar-Gauss-Bonnet gravity in terms of the variation of the \ADM mass as a function of the ambient scalar field that the \bh is embedded in at fixed Wald entropy.
This parallels the notion that the baryonic mass of \ns{s} is constant during
the inspiral~\cite{Damour:1996ke}. They also derived the two-body Lagrangian at 1\pn order.
\textcite{Julie:2022huo} used this result in conjunction with the numerical calculation of sensitivities of spontaneously scalarized \bh{s} to study binaries. They found that in principle \bh{s} can evolve toward a situation in which the inner singularity approaches the event horizon of the \bh before the merger.
This suggest the possibility that, depending on the model of scalar-Gauss-Bonnet gravity and the binary parameters, the field equations might simply not be able to predict a full binary evolution ranging from inspiral to merger.

\subsection{Black-hole scalarization in the presence of matter}
\label{sec:bhs_nvac}

\textcite{Stefanov:2007eq} suggested the earliest model of \bh scalarization, which
consisted of the \DEF model coupled to nonlinear electromagnetism~\cite{Stefanov:2007bn,Stefanov:2007qw}.
They introduced a coupling between the scalar field and the Born-Infeld Lagrangian and found
that \bh solutions with scalar hair branch off from the general-relativistic sequence of solutions. These \bh{s} violate the no-scalar-hair theorems of \textcite{Mayo:1996mv,Sen:1998ftn}, which are valid for
charged, nonrotating \bh{s} in scalar-tensor theory.
The bifurcation point occurs where the \gr sequence of solutions becomes unstable to scalar perturbations~\cite{Doneva:2010ke}.

A variety of scalarization models have been studied in Einstein-Maxwell-scalar theory following~\textcite{Herdeiro:2018wub},
who introduced a coupling between the scalar field and the Lorentz invariant ${\cal F}^2 = F_{\mu\nu}F^{\mu\nu}$ of the form $\exp(- \alpha \varphi^2) {\cal F}^2$,
where $\alpha$ is a dimensionless constant.
In this model \bh{s} can scalarize, with ${\cal F}^2$ playing the role
of the trace of the energy-momentum tensor in the \ns scalarization in \DEF-like models, or the Gauss-Bonnet invariant in \bh scalarization models.

The simpler nature of the model allowed~\textcite{Herdeiro:2018wub} to perform numerical relativity simulations to
study the scalarization process in the time domain by expanding upon already available code bases~\cite{Sanchis-Gual:2015lje,Sanchis-Gual:2016tcm}.
This allowed~\textcite{Herdeiro:2018wub} to show that the end point of the instability
of the Reissner-Nordstr\"om \bh was indeed a \bh
with scalar hair.
Additional studies of this model were done by~\textcite{Fernandes:2019rez,Astefanesei:2019pfq,Herdeiro:2020htm,Xiong:2022ozw,Luo:2022roz,Niu:2022zlf}.
Within this model,~\textcite{Myung:2018vug} studied the instability of the Reissner-Nordstr\"om solution, while~\textcite{Myung:2018iyq,Myung:2019oua,Blazquez-Salcedo:2020jee} analyzed the stability of the scalarized solutions.
\textcite{Fernandes:2019kmh,Herdeiro:2020iyi} explored other variations on this theme, including
scalar-axion couplings, i.e., $\propto h(\varphi) F^{\mu\nu} \tilde{F}_{\mu\nu}$, where
$\tilde{F}_{\mu\nu} = (1/2) \epsilon\indices{_{\mu\nu}^{\alpha\beta}} F_{\alpha\beta}$.
In addition, \textcite{Ikeda:2019okp} considered the coupling between the double dual of the Riemann tensor and $F_{\mu\nu}$.

\textcite{Zhang:2021btn} studied an Einstein-Maxwell-scalar model, with a coupling
between scalar and Maxwell fields of the form $\exp(\beta \varphi^4) {\cal F}^2$. In this model, the Reissner-Nordstr\"om solution is linear stable against scalar perturbations, but it can become unstable under large, nonlinear perturbations.~\textcite{Zhang:2021btn} observed a critical phenomena separating  unscalarized and scalarized BHs. This establishes an interesting connection between nonlinear scalarization with the result of critical phenomena in gravitational collapse, which was first observed by~\textcite{Choptuik:1992jv}. Similar investigations were carried out in Einstein-Maxwell-scalar the Reissner--Nordstr\"om--anti-de Sitter \bh{s}~\cite{Zhang:2022cmu} and in the Ricci-Gauss-Bonnet model~\cite{Liu:2022fxy,Liu:2022eri}.

\textcite{Cardoso:2013fwa,Cardoso:2013opa} explored a relevant scenario from an astrophysical perspective: matter in the vicinity of a \bh could trigger scalarization and in turn endow the \bh with scalar hair. They showed that this ``matter-induced'' scalarization (see Sec.~\ref{sec:matter_induced_other_field}) is in principle possible for the \DEF model.
The viability of this proposal in realistic matter configurations, such as an accretion disk or a dark matter halo, has not yet been explored. Matter-induced scalarization seems unlikely to happen in light of the severe constraints on the \DEF model. However, this might not be the case for other scalarization models, and the topic deserves further investigation.

After it was understood that in scalar-Gauss-Bonnet theories the Schwarzschild solution can scalarize, it was natural to ask if the same can happen to its charged counterpart, i.e., the Reissner-Nordstr{\"o}m solution.
\textcite{Doneva:2018rou} analyzed this and found the existence of
two bifurcation points, one at larger masses where the scalarized solutions
bifurcated from the Reissner-Nordstr{\"o}m one, and one at smaller masses where the scalar
charge of the solution decreases again to zero and the branch merges again with the
\gr one.
Scalarized charged \bh{s} in the fundamental mode are also thermodynamically favored over the
Reissner-Nordstr{\"o}m solution.
\textcite{Brihaye:2019kvj} showed that for the near-extremal Reissner-Nordstr{\"o}m case, scalarization can happen for either sign of the scalar-to-Gauss-Bonnet coupling constant~\eqref{eq:quadratic_coupling}.
Other works varying the choice of the coupling function $f$ between the scalar field (or the axion field) and
the Maxwell invariant and studying the scalarization of dyonic \bh{s} include \textcite{Fernandes:2019rez},~\textcite{ Fernandes:2019kmh},~\textcite{Blazquez-Salcedo:2020nhs} and~\textcite{Blazquez-Salcedo:2020jee}.

\textcite{Herdeiro:2019yjy} put forward their motivation for studying these models.
They noted that quantum effects can break the scale invariance and vanishing energy-momentum trace properties of electro-vacuum \gr \bh{s}.
As a consequence, even the simplest nonminimally scalar-field coupling $\xi \varphi^2 R$ can result in \bh scalarization when these quantum effects are taken into account.
Within these models, they discussed the scalarization of the Reissner-N\"ordstrom solution
and a noncommutative generalization of the Schwarzschild solution.
They found that the resulting scalarized \bh{s} are in general entropically favored over the \gr solutions.

\subsection{Variations of the curvature-induced scalarization model}
\label{sec:bh_scalarization_others}

In light of the nontrivial effect of rotation on scalarization, it is
natural to consider what happens when the Gauss-Bonnet invariant is replaced by the
Pontryagin invariant $\pnt$ as the ``curvature source'' to which the scalar field is couple.
The Pontryagin density is known to vanish in spherically symmetric spacetimes
(such as that of a Schwarzschild \bh) and becomes nonzero in nonspherically symmetric spacetimes (such as that of a Kerr \bh).
In this sense, all scalarized BHs in theories that replace the
Gauss-Bonnet invariant with the Pontryagin density are necessarily ``spin
induced''; a characteristic shared with \bh solutions in dynamical Chern-Simons
gravity [see e.g., \textcite{Yunes:2009hc,Konno:2009kg},
and \textcite{Alexander:2009tp} for a review] which features a linear coupling
between the scalar field and the Pontryagin invariant.
The coupling between a scalar field and the Pontryagin density leads to equations that are of third order in derivatives [see \textcite{Motohashi:2011ds,Delsate:2014hba}], which is not the case for a coupling with the Gauss-Bonnet invariant. Thus, although both of these couplings can be seen as part of an EFT, they lead to distinct challenges when it comes to nonlinear evolution.
\textcite{Myung:2019wvb} studied the combined effect of Gauss-Bonnet and Pontryagin densities.
In the test field limit, the scalar-field dynamics with an effective mass proportional
to $\phi^2 \, \pnt$ was also studied in a Kerr background \cite{Gao:2018acg,Doneva:2021dcc}
and in the Schwarzschild-Newman-Unti-Tamburino background \cite{Brihaye:2018bgc}.

Another extension of the original Gauss-Bonnet spontaneous scalarization
model involves the consideration of $n>1$ scalar fields $\varphi^{a}$, whose interaction
is determined by their ``target space'' $\gamma_{ab}(\varphi^{c})$, an
$n$-dimensional Riemannian manifold~\cite{Damour:1992we,Horbatsch:2015bua}.
These are similar to the tensor-multiscalar \DEF models for \ns scalarization presented in Sec.~\ref{sec:ns_extended}.
In these models, the scalar-field dynamics is determined by the quantity $\gamma_{ab}(\varphi) g^{\rho\sigma}\partial_{\rho}\varphi^{a}\partial_{\sigma}\varphi^{b}$.
\textcite{Doneva:2020qww} numerically obtained a scalarized \bh for the case $n=3$ and maximally symmetric target-space geometries.
An important feature of these solutions is that the scalar fields decay asymptotically as $1 / r^2$ (i.e., the \bh{s} do not have a monopole scalar charge). This shows that \bh{s} in these theories will not emit dipole scalar radiation when they are placed in binaries.

\section{Generalizations of scalarization to other fields and instabilities }
\label{sec:other}
We identified the underlying reason for spontaneous scalarization in its various forms to be a tachyonic instability. However, the scalar nature of the field did not play a special role in the mechanism. This suggests that other fields, such as vectors might also spontaneously develop nontrivial configurations around compact objects when they exhibit suitable couplings to curvature.
In this section, we investigate this idea, which leads to the so-called spontaneous tensorization theories.

A second type of generalization of spontaneous scalarization considers new types of instabilities, as opposed to new types of fields. For instance, instead of replacing the scalar field with, say, a vector, we replace the tachyonic instability with, say, a ghost instability. We later see that a surprising key result of spontaneous tensorization is that these two paths are intimately connected. Namely, even if we intend only to have a theory of tachyons living on nonscalar fields, ghost and gradient instabilities necessarily arise in almost all models.

\subsection{Spontaneous vectorization}
\label{sec:vectorization}
What happens if we replace the scalar field of spontaneous scalarization with a vector? In scalarization, we started with the most general context in Sec.~\ref{sec:minimal_action}, considering all the allowed couplings to the metric. Here we follow the reverse path, starting with individual models of vector-tensor theories and considering the more encompassing theory later. This exposition is preferable because the study of specific models that are straightforward generalizations of known scalarization models reveals some pathologies and provides guidance for further model building.

The idea and the first specific model of vectorization were introduced in analogy with a massive version of the \DEF model in Eq.~\eqref{eq:action_st}.
Consider the vector-tensor theory action~\cite{BeltranJimenez:2012kby,Ramazanoglu:2017xbl}
\begin{align}\label{action_vt}
S &= \frac{1}{16\pi G_{*}} \int \dV \left[R - F^{\mu\nu} F_{\mu\nu} - 2 \mu_X^2 X^2 \right] \nonumber \\
&\quad + S_{\mm}\left[\Psi_{\mm};\, {\cal A}^{2}(X^2) g_{\mu \nu} \right],
\end{align}
where $F_{\mu\nu} = \nabla_\mu X_\nu -\nabla_\nu X_\mu$ and $X^2 =g^{\mu\nu}X_\mu X_\nu = X_\mu X^\mu$. The vector field $X_\mu$ has the canonical kinetic term in this frame, and the matter degrees of freedom couple to the metric $\tilde{g}_{\mu\nu}={\cal A}^{2}(X^2) g_{\mu \nu}$, which is conformally scaled with respect to $g_{\mu\nu}$. The vector field equation is
\begin{align} \label{eom_vt1}
\nabla_\rho F^{\rho \mu}& = [-8\pi G_{*}{\cal A}^4 \Lambda \tilde{T} + \mu_X^2 ] \, X^\mu,
\end{align}
where $\Lambda = \dd \ln {\cal A} / \dd x$, $T_{\mu \nu}$ is the energy-momentum tensor in the Einstein frame, and $\tilde{T}$ is the trace of the stress-energy tensor in the Jordan frame, i.e., with respect to the metric $\tilde{g}_{\mu\nu}$.

Equation~\eqref{eom_vt1} is that of a massive vector (Proca) field where the expression inside the square brackets acts as an effective mass $\mu_{\rm eff}^2$. In parallel with the \DEF model, when $\cal A$ has an appropriate form, for instance, ${\cal A}= \exp(\beta_0 X^2/2)$ with sufficiently negative $\beta_0$, $\mu_{\rm eff}^2$ becomes negative. Furthermore, for dense enough objects such as \ns{s} this occurs for order-of-unity values of $\beta_0$. The expectation is that the vector field will grow from arbitrarily small perturbations around $X_\mu = 0$ due to this tachyonic behavior, which can be called spontaneous vectorization, in exact analogy with spontaneous scalarization. However, we later see that there are many subtle points in this narrative.

Even though we presented spontaneous vectorization in the Einstein frame with a nonminimal coupling to matter, it can be converted to a theory of minimal matter coupling and vector-curvature couplings as in spontaneous scalarization~\cite{Ramazanoglu:2019tyi}. Even more directly, there are theories of spontaneous vectorization that are purely conceived through curvature coupling, with the first example being the Hellings-Nordtvedt theory~\cite{Hellings:1973zz} studied by~\textcite{Annulli:2019fzq}
\begin{align}\label{action_vec_hl}
    S &= \frac{1}{16 \pi G} \int \dV
    \left[ R - F^{\mu\nu} F_{\mu\nu} -\Omega X^2 R \right.
    \nonumber \\
    &\quad \left. - \, \eta X^\mu X^\nu R_{\mu\nu} \right] + S_{\mm} [\Psi_{\mm};\, g_{\mu\nu}],
\end{align}
where $\Omega$ and $\eta$ are coupling constants. Scalarization through coupling to the Gauss-Bonnet term also has a vector analog as in the action~\cite{Ramazanoglu:2019gbz}
\begin{align}\label{action_vec_gb}
    S &= \frac{1}{16 \pi G} \int d^4x \sqrt{-g} \left[ R - F^{\mu\nu} F_{\mu\nu} + f(X^2) \mathscr{G} \right]
    \nonumber \\
    &\quad + S_{\mm} [\Psi_{\mm};\, g_{\mu\nu}],
\end{align}
which has a similar structure as Eq.~\eqref{actionsgb}. The action~\eqref{action_vec_gb} leads to \bh vectorization as well, unlike the other models we have seen thus far.

The models of vectorization that we have examined are merely specific examples, and any theory of scalarization can be generalized to vector fields in principle. In this regard, \textcite{Brihaye:2020oxh} studied the coupling of a vector field to a scalar, and \textcite{Oliveira:2020dru} and \textcite{Brihaye:2021qvc} studied the coupling between two vector fields to obtain vectorized compact objects. \textcite{Kase:2020yhw} considered generalized Proca theories~\cite{Heisenberg:2014rta,Tasinato:2014eka} with various couplings of a massive vector field and showed that spontaneous vectorization occurred in those models as well. \textcite{Ramazanoglu:2018tig} investigated models where the effective mass of the vector field was generated by the Higgs mechanism, thus preserving the gauge symmetry. On a separate front, conformal scaling of the metric in the matter action~\eqref{action_vt} can be generalized to disformal transformations, and spontaneous vectorization still occurs~\cite{Ramazanoglu:2019jrr,Minamitsuji:2021rtw}. In terms of approximate solutions, \textcite{Garcia-Saenz:2022wsl} recently calculated the Schwarzschild \qnm{s} of nonminimally coupled vector fields. There have also been efforts to study all these phenomena in a unified manner in compact binaries, using more generic \EFT tools~\cite{Khalil:2019wyy}.

To summarize, coupling of the vector field to any nonvanishing term in the action can be considered for spontaneous vectorization~\cite{Ramazanoglu:2019gbz}, and most options have been considered in at least a preliminary sense. The case of all possible couplings, an analog of the minimal action of scalarization in Sec.~\ref{sec:minimal_action}, was studied by~\textcite{Garcia-Saenz:2021uyv}. They also revealed some with the fundamental problems of vectorization, as we later discuss more broadly.

Vectorization models strongly resemble scalarization models in terms of their actions, but this can be misleading. One major difference is in the number of additional degrees of freedom. Scalar fields contribute a single new degree of freedom irrespective of whether they are part of spontaneous scalarization provided that the field equations are of second order in derivatives. For vectorization, this is not the case. The action~\eqref{eom_vt1} and all of the vector-tensor theory actions that we have discussed break the gauge freedom found in massless vector fields.\footnote{Even though we have an intrinsic mass $\mu_X$ for the vector field in our discussion, vectorization can occur without this term, as is the case for scalarization.} As a result, the vectors of vectorization models carry 3 degrees of freedom instead of the 2 found in electromagnetism. This is not an immediate reason of concern, since minimally coupled  massive vectors, known as Proca fields~\cite{Proca:1936fbw}, also break the gauge symmetry, and they still provide a classical field theory with well-behaved dynamics. However, as we later see, the extra degree of freedom is far more problematic in vectorization.

Although the vectorization models are designed to have a linear tachyonic instability in analogy with scalarization, the presence of the third degree of freedom, whose dynamics is not obvious, casts doubt on the success of this design. Namely, whether vectorization can indeed be described as a tachyonic instability of the vector modes that is then quenched nonlinearly is not apparent. This concern is amplified by the fact that vectorized compact objects in the models studied thus far seem to have some striking differences with respect to their scalarized counterparts.

As an example, consider vectorized \bh{s} in the theory described by the action~\eqref{action_vec_gb}. Entropy can be a measure of stability for \bh{s}, where higher entropy solutions are favored in terms of stability. Entropies of spherically symmetric vectorized \bh{s} were numerically computed by \textcite{Barton:2021wfj}, and \gr \bh{s} were shown to be entropically favored over vectorized ones in all cases. The verdict of stability is not definite without time evolution, but this is a strong indication for the instability of vectorized \bh{s}. In contrast, scalarized \bh{s} in scalar-Gauss-Bonnet theories, such as Eq.~\eqref{actionsgb} can be entropically favored over those in \gr for appropriate coupling functions~\cite{Doneva:2017bvd}, and their stability is also indicated through other methods (see Sec.~\ref{sec:bhs_stability_nr}), which makes the stability of vectorized objects even more suspect.

\begin{figure}
\includegraphics[width=\columnwidth]{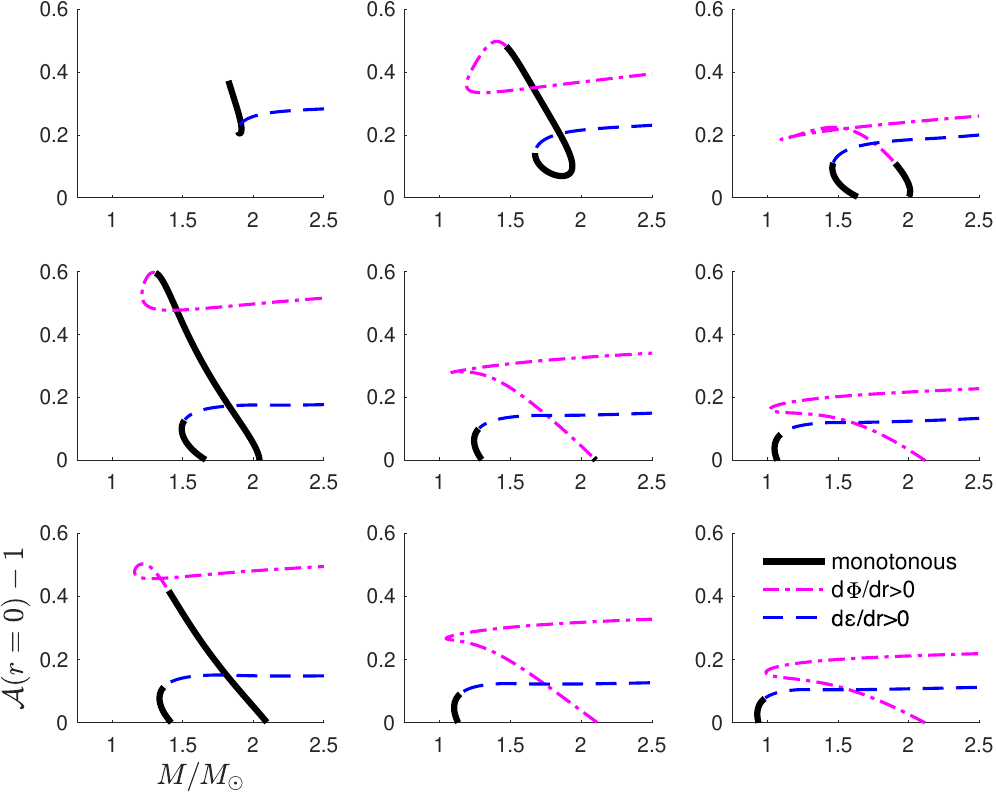}
\caption{
The strength of vectorization measured by ${\cal A}(r=0)-1$ as a function the \ns mass $M$ for various values of $\beta_0$ and $\mu_X$ in Eq.~\eqref{eom_vt1}. Top row: $\mu_{X}= 1.6 \times 10^{-11}$~eV. Middle row: $\mu_{X}= 8.0 \times 10^{-12}$~eV. Bottom row: $\mu_{X}= 4.8 \times 10^{-12}$~eV. Left column: $\beta_0 =-4$. Middle column: $\beta_0=-5$. Right column: $\beta_0=-6$. The dashed and dot-dashed lines, respectively, indicate solutions where the energy density $\varepsilon$ and $\Phi=-n^\mu X_\mu$ (with $n^\mu$ the normal vector field to spatial hypersurfaces) do not monotonically decrease with radius.
From \textcite{Ramazanoglu:2017xbl}.
}
\label{vec_parameter_space}
\end{figure}

Scalarized and vectorized \ns{s} also show major differences. For example, the dependence of the strength of vectorization\footnote{There is no analog of the scalar charge in Eq.~\eqref{eq:scalar_charge} for massive scalarization or vectorization theories due to different asymptotic behavior at $r \to \infty$. However, ${\cal A}-1$ is a monotonic function of the norm of the vector field in this case and is known to be a good indicator of the deviation of vectorized star structures from those of GR~\cite{Ramazanoglu:2017xbl}. Thus, we use it as a measure of vectorization strength.} on the \ns mass has no discernible pattern, as can be seen in Fig.~\ref{vec_parameter_space}, whereas scalarization occurs in a finite \ns mass interval and vanishes at the boundary mass values; see Fig.~\ref{fig:DEF_NSmodels_Nonrot}. It is also known that while scalarized \ns{s} are continuously connected to those of \gr as the theory parameters are smoothly changed (for example, in the limit $\beta \to 0$ in Fig.~\ref{fig:DEF_NSmodels_Nonrot}), this is not the case in vectorization~\cite{Ramazanoglu:2017xbl,Minamitsuji:2020pak}. The latter is known to be an indication of instability in certain scalarization theories; see~\textcite{Mendes:2016fby}. Thus, qualitative differences between scalarized and vectorized objects are also apparent for \ns and, moreover, point to the instability of the vectorized ones.

These observations suggest that the vectorized compact object solutions obtained thus far may not be stable. Hence, they are not astrophysically relevant, at least in the case of spherically symmetric spacetimes. A close look at the instability mechanism of vectorization reveals the underlying reasons for this in Sec.~\ref{sec:vectorization_ghosts}.

\subsection{Vectorization and ghosts}
\label{sec:vectorization_ghosts}
Despite the fact that the vectorization phenomena that we have presented are directly inspired by scalarization, we have seen noteworthy differences between the two mechanisms. Even though it was puzzling for a time, this is now strongly suspected to be related to the fact that vectorization models suffer from ghost and/or gradient instabilities in addition to tachyonic ones, as demonstrated by~\textcite{Garcia-Saenz:2021uyv} and~\textcite{Silva:2021jya}.
Here we first follow the latter and later outline the methodology of the former.\footnote{\textcite{Esposito-Farese:2009wbc} is an earlier similar study that specifically investigated cosmological scenarios.}

We begin by reviewing the basic aspects of different types of instabilities~\cite{Demirboga:2021nrc}.
Recalling our discussion of the tachyon in Secs.~\ref{tachyonquenched} and~\ref{sec:tachcurved}, we consider the linearized scalar-field equation~\eqref{phi4lincurv} in $1+1$ dimensions,
\begin{align}
\label{eq:instability_eom}
g^{tt} \partial_t^2 \delta\varphi
+ g^{xx} \partial_x^2 \delta\varphi
=\mu^2 \varphi,
\end{align}
where we assume for simplicity that the metric is diagonal with constant components. The common case without instabilities is when $g^{tt}<0$, $g^{xx}>0$, and $\mu^2>0$. What happens when each of these three terms changes its sign while the other two stay the same?

For a plane wave mode $\delta\varphi(t,x) \propto \exp[\ii\left(\omega t - kx\right)]$, the dispersion relation is given by
\begin{align}
\omega(k) = \sqrt{ (\mu^2+g^{xx} k^2) / (-g^{tt}) }.
\end{align}
We saw that for $\mu^2<0$ the mode behaves as a tachyon, where $\omega(k)$ becomes imaginary for sufficiently small $|k|$. This leads to exponential growth as discussed, and the fastest growing mode behaves as $\sim \exp[\sqrt{\mu^2/g^{tt}} \, t]$; i.e., the growth rate has an upper limit.

A \emph{ghost} occurs when $g^{tt}>0$. There is also exponential growth in this case; however, the growth rate diverges with increasing wave number as $\sim \exp[\sqrt{g^{xx}/g^{tt}}k t]$ and hence has no upper bound.
The case $g^{xx}<0$ also gives the same asymptotic growth rate as the ghost and is called a gradient instability.\footnote{For the simple equation~\eqref{eq:instability_eom}, a simultaneous change of sign of two of the coefficients is equivalent to changing the sign of the third one; i.e., $g^{tt}>0$, $g^{xx}<0$, and $\mu^2>0$ is also a tachyon. Similarly, all three coefficients changing signs gives a stable theory. This is due to the fact that the equation as a whole or the action that leads to it can be multiplied by an overall factor of $-1$ without changing the physics. However, the scalar, or any investigated field that carries instabilities, is always coupled to the metric and other terms in the action, which means that the overall sign is meaningful. Hence, we use the aforementioned classification here and say that a theory with $g^{tt}>0$, $g^{xx}<0$, and $\mu^2>0$ carries both ghost and gradient instabilities according to the literature~\cite{Demirboga:2021nrc}.} In summary, ghost and gradient instabilities as we presented them here have qualitative differences from tachyons, a with main one being the arbitrarily fast growth rates. This means that a well-defined time evolution can be problematic in such theories, and it is more difficult to devise nonlinear quenching mechanisms.

The spontaneous vectorization model of Eq.~\eqref{action_vt} modifies the ``mass term'' in the field equation for the vector field. Hence, the first impression might be that it leads to a tachyonic instability as in scalarization. However, this approach is naive, and a thorough analysis reveals that theories of vectorization necessarily carry ghost instabilities in addition to tachyonic ones. Following \textcite{Silva:2021jya}, one way to understand this is to use the Stueckelberg trick~\cite{Ruegg:2003ps} where we introduce a scalar field $\psi$ into the action~\eqref{action_vt} using the substitution
\begin{align}
X_{\alpha} \to X_{\alpha} + \mu_X^{-1} \nabla_{\alpha} \psi.
\label{eq:trick}
\end{align}
This leads to the scalar equation of motion
\begin{align} \label{eq:box_psi_eff_metric}
\overline{\Box}\psi = -\bar{g}^{\mu\nu} \,
[\, \mu_X \overline{\nabla}_\mu X_\nu +\tfrac{1}{2}
( \overline{\nabla}_\mu \log \hat{z} )
( \overline{\nabla}_\nu \psi+\mu_X X_\nu )
\,],
\nonumber \\
\end{align}
where the covariant derivative $\overline{\nabla}$ is compatible with the new metric
\begin{equation}\label{eq:geff}
\bar{g}^{\mu\nu}= \hat{z} \, g^{\mu\nu}, \qquad
\hat{z} \equiv \frac{\mu^2_{\rm eff}}{\mu_X^2} = 1-\frac{8\pi G_{*} {\cal A}^4 \Lambda \tilde{T}}{\mu_X^2}.
\end{equation}
The signature of $\bar{g}_{\alpha\beta}$ changes with the sign of $\hat{z}$, which controls the change of sign of the effective mass of $X_\mu$ in Eq.~\eqref{eom_vt1}. In other words, when we try to instigate a tachyon on $X_\mu$ by having negative $\hat{z}$ in parts of the spacetime, $\psi$ carries ghost and gradient instabilities; cf. Eq.~\eqref{eq:instability_eom}. This is a degree of freedom whose dynamics is governed by the effective metric $\bar{g}_{\mu\nu}$, not the spacetime metric $g_{\mu\nu}$.

It is curious to see that the Stueckelberg trick reveals the ghost, even though it is not apparent in the vector field equation~\eqref{eom_vt1}. However, a more careful analysis can clarify this issue~\cite{Silva:2021jya}. Note first that the vector field equation~\eqref{eom_vt1} is not manifestly hyperbolic; i.e., its principal part, the term with the highest derivatives, is not the wave operator, since
\begin{equation}
\nabla_\mu F^{\mu\nu} = \nabla_\mu \nabla^\mu X^\nu -\nabla_\mu \nabla^\nu X^\mu.
\end{equation}
The second term on the right-hand side of Eq.~\eqref{eom_vt1} vanishes in Proca theory (${\cal A}=1$) making the principal part the wave operator. This is not the case in general due to the so-called (generalized) Lorenz condition
\begin{equation} \label{lorenz_vt}
\nabla_\mu \left(\hat{z}  X^\mu \right) = 0,
\end{equation}
which is obtained by acting on both sides of Eq.~\eqref{eom_vt1} with $\nabla_\mu$ and recalling the antisymmetry of $F^{\rho\mu}$. We can use Eq.~\eqref{lorenz_vt} to obtain a manifestly hyperbolic form of the linearized field equations on a fixed background of vanishing vector fields as
\begin{align}
\nabla_\mu \nabla^\mu X_\nu
+ ( \nabla_\mu \ln \hat{z} ) \nabla_\nu X^\mu
= \mathcal{M}_{\nu\mu} X^\mu ,
\label{eq:vec_component}
\end{align}
where we defined the mass-squared tensor
\begin{align}
\mathcal{M}_{\nu\mu} = \hat{z} \mu_X^2 g_{\nu\mu}  + R_{\nu\mu}
- \nabla_\nu \nabla_\mu \ln |\hat{z}| .
\label{eq:mass-tensor}
\end{align}
In Eq.~\eqref{eq:vec_component} and \eqref{eq:mass-tensor} the metric and $\hat{z}$ have their fixed background \gr values. This is a generalized massive wave equation.

Since $\hat{z}=1$ in the absence of matter and it becomes negative in a continuous manner inside matter for an instability to exist at all, $\hat{z}$ necessarily vanishes at some points of a \ns spacetime. The $\nabla_\nu \nabla_\mu \ln \hat{z}$ term introduced by the vectorization contains powers of $\hat{z}^{-1}$, which means that the effective mass-squared tensor diverges at such points. Alternatively, if we move the $\hat{z}^{-1}$ terms to the other side of the equation, the principle part becomes $\hat{z}\Box X_\nu$; i.e., it changes its sign with $\hat{z}$ the same way as the $\psi$ field in Eq.~\eqref{eq:box_psi_eff_metric}. Hence, if we properly rewrite the vector field equation into a manifestly hyperbolic form, the ghost is apparent.

Equations~\eqref{eq:vec_component} and~\eqref{eq:mass-tensor} show that calling $\mu^2_\eff = \hat{z} \mu_X^2$ the effective mass squared of the theory is misleading, unlike in the case of scalarization. Even though $\mu^2_\eff$ replaces the mass term in the vector field equation of a minimally coupled Proca field, it is not actually the effective mass of physically propagating degrees of freedom. The physical mass is determined by Eq.~\eqref{eq:mass-tensor}, which diverges at points where $\hat{z}=0$.

We have presented the appearance of the ghost in two separate ways, the Stueckelberg mechanism of Eq.~\eqref{eq:box_psi_eff_metric} and the direct vector field formulation of Eq.~\eqref{eq:vec_component}.  The degrees of freedom in the two pictures are related to each other in a nontrivial manner through Eq.~\eqref{eq:trick}. Hence the presence of the ghost manifests differently. Nevertheless, the criterion for the appearance of the ghost $\hat{z}=0$ is the same in both cases. The Stueckelberg picture has an advantage in that it can straightforwardly identify the ghost as a scalar degree of freedom $\psi$ at decoupling, which is well known in the case of the Proca field.

One might at first think that the ghost and gradient instabilities can be regularized by nonlinearities. After all, the tachyon of scalarization is also an instability, but it is eventually quenched by nonlinear effects as explained in Sec.~\ref{sec:scalarization_mechanism}. In such a scenario, one would expect the vectorized objects to be free of instabilities even though the \gr solutions are unstable. However, current evidence points away from this. Even though detailed mathematical analyses of the partial differential equations have not been performed to ensure nonlinear instability, \textcite{Demirboga:2021nrc} showed that spherically symmetric vectorized \ns{s} in the theory~\eqref{action_vt}, which could be potential stable endpoints of vectorization, are also unstable to ghosts and gradient instabilities at the perturbative level.

Even if stable solutions can be found in other vectorization models, there is a more fundamental problem due to the modification to the principal part of the vector equations in equation~\eqref{eq:box_psi_eff_metric} or~\eqref{eq:vec_component}. The coefficient of the wave operator vanishes at certain points in spacetime or, equivalently, the physical effective mass diverges if we move these coefficients to the other side of the equation. In  either case, this leads to divergent time derivatives for arbitrarily small perturbative vector field values, which renders the initial assumptions of a perturbative approach invalid. Overall, it is not clear whether one can even define a well-posed time evolution, let alone one that can somehow suppress the exponential growth of the vector field~\cite{Silva:2021jya}. Further elucidation of these issues requires a more rigorous mathematical investigation, which has not yet been attempted.

Thus far we have investigated the ghost and gradient instabilities of the specific vectorization theory in the action~\eqref{action_vt}, but similar results are known to exist for all theories of vectorization~\cite{Garcia-Saenz:2021uyv,Silva:2021jya} due to similar mechanisms. To summarize, replacing the scalar of scalarization with a vector field results in theories  where the central issue is not the nature of the field the instability lives on, but rather the type of instabilities that arise. We cannot construct vectorization models that exhibit only the tachyonic instability commonly associated with the scalarization mechanism, which can be benign and tamed by nonlinear effects. Rather, vectorization exhibits far more threatening instabilities.

We have closely followed \textcite{Silva:2021jya}, but \textcite{Garcia-Saenz:2021uyv} reached similar conclusions via an alternative approach. They considered the most general action that contained a vector field and the metric~\cite{Heisenberg:2014rta,Tasinato:2014eka} and truncated its action at the quadratic order around \gr to obtain the linearized field equations. The main qualitative difference from the scalar minimal action of Sec.~\ref{sec:minimal_action} is that one can demonstrate that if any of the physical degrees of freedom have a tachyonic instability, then there are also degrees of freedom with ghost or gradient instabilities, which are manifested by a change of sign of coefficients in the dispersion relationship. The conclusion is the same: vectorization based on tachyonic instabilities necessarily has ghost or gradient instabilities as well.

Recent work has shown that the issues with vectorization also happen in simpler theories, with an example being minimally coupled self-interacting vector field theories. For instance, for $\mathcal{A}(x)=1$ the action~\eqref{action_vt} becomes the minimally coupled Proca theory, which is well posed. However, generalizations of the potential beyond a mass term, that is, $2 \mu_X^2 x\to 4V(x)$ for some generic function $V$, results in ill-posedness~\cite{Mou:2022hqb,Clough:2022ygm,Coates:2022qia}. The underlying reason for this is the fact that there is an analog of the generalized Lorenz condition~\eqref{lorenz_vt} in this case as well, which ultimately leads to the same result: there is a degree of freedom in the theory whose dynamics is governed by an effective metric, an analog of $\bar{g}_{\mu\nu}$ in Eq.~\eqref{eq:geff}. This new metric depends on the vector field itself and can lose its Lorentzian nature depending on how $X_\mu$ evolves in time~\cite{Coates:2022qia}.

\textcite{Esposito-Farese:2009wbc} provided some of the earliest examples of the ill-posedness of self-interacting vector field theories that we have mentioned in a cosmological context,  and it was recently demonstrated that initially healthy configurations of self-interacting vectors naturally evolve to a point where hyperbolic evolution becomes impossible~\cite{Mou:2022hqb,Clough:2022ygm,Coates:2022qia,Coates:2022nif}. This means that nonlinear extensions of the Proca theory are more delicate than their scalar counterparts in terms of providing physical theories. It seems that the underlying reason for the ``failure'' of vectorization is a much more general phenomenon in theories with conditions such as Eq.~\eqref{lorenz_vt}, at least for the models conceived thus far. In this sense, exploration of vectorization has been leading to a deeper understanding of vector fields in general, which has implications beyond theories of gravitation.

We close this section with a pertinent comment on ghost instabilities. We encountered them as an unappealing artifact in vectorization theories, but one could consider making them the driver of a spontaneous growth mechanism. This option was indeed considered prior to their discovery in vectorization models. The simplest example is another analog of the DEF model where the conformal scaling function ${\cal A}$ depends on derivative terms~\cite{Ramazanoglu:2017yun}
\begin{align}\label{eq:action_scalar_ ghost}
S&= \frac{1}{16\pi G_{*}} \int d^4x \sqrt{-g} \left[R -
2g^{\mu\nu}\partial_{\mu}\varphi \partial_{\nu}\varphi -
2\mu^2 \varphi^2\right] \nonumber \\
&\quad + S_{\mm}[\Psi_{\mm}; {\cal A}^{2}(K)g_{\mu\nu}],
\end{align}
with $K=-g^{\mu\nu}\partial_\mu\varphi\partial_\nu\varphi/2$. The scalar equation of motion is
\begin{align}\label{eq:ghost_EOM}
\nabla_\mu \left(\hat{z} \nabla^\mu \varphi\right) = \mu^2 \varphi,
\end{align}
where $\hat{z} \equiv 1+4\pi G_* \tilde{T}{\cal A}^4 \left(\dd \ln{\cal A}/\dd K\right)$. Unlike for the tachyonic instability where the mass term is modified, the derivative terms are modified in this case.

The nature of Eq.~\eqref{eq:ghost_EOM} can be better understood when we consider the linearized equation for perturbative values of the scalar around a \gr background and highlight the principal part as
\begin{align}\label{eq:ghost_EOM2}
\hat{z} \, \Box \delta\varphi + \dots = 0.
\end{align}
For an appropriate choice such as ${\cal A} = \exp(\beta_0 K)$ with large enough $\beta_0>0$, the sign of $\hat{z}$ can become negative in the presence of matter. In all spontaneous scalarization theories considered thus far, the source of the instability was the change of sign of the effective mass term, which resulted in a tachyon. In Eq.~\eqref{eq:ghost_EOM2} the overall sign of the wave operator changes; hence, we have both ghost and gradient instabilities, recalling Eq.~\eqref{eq:instability_eom} and the subsequent discussion. Spontaneous scalarization that might arise from this new mechanism is called ghost-based spontaneous scalarization~\cite{Ramazanoglu:2017yun}. It has similar problems as vectorization in that it is hard to tame the ghost and other instabilities; hence, we do not end up with a theory with sensible dynamics.

\subsection{Spontaneous tensorization}
The basic mechanism of spontaneous scalarization can be generalized beyond vector fields, resulting in \emph{spontaneous tensorization}. However, all such known theories also feature ghosts and suffer from similar problems with vectorization, aside from a single possible example of spinor-tensor theory.

The simplest mathematical generalization of scalarization beyond vectors occurs for $p$-form fields, i.e., totally antisymmetric tensors $X_{\mu_1 \dots \mu_p}$ of rank $(0,p)$. This is due to the fact that a vector field $X_\mu$ is a $1$-form field, and the actions that we encountered in spontaneous vectorization can be readily generalized to this class of fields as~\cite{Ramazanoglu:2019jfy}
\begin{align}\label{action_pform}
S &= \frac{1}{16\pi} \int \dV \left[R - F_{\mu_1\dots \mu_{p+1}}F^{\mu_1\dots \mu_{p+1}} \right] \nonumber \\
&\quad + S_{\mm} \left[\Psi_{\mm};\, {\cal A}^{2}(X^2) g_{\mu \nu} \right],
\end{align}
where $F_{\mu_1\dots \mu_{p+1}} = (p+1) \nabla_{[\mu_1} X_{\mu_2\dots \mu_{p+1}]}$, with the square brackets denoting antisymmetrization and $X^2=X_{\mu_1\dots \mu_{p}} X^{\mu_1\dots \mu_{p}}$. It is then straightforward to show that the analog of $\mu^2_\eff$ becomes negative inside \ns{s} for an appropriate choice of ${\cal A}(x)$. However, these theories carry ghost instabilities similar to those of vectors when we try to establish spontaneous growth from tachyonic instabilities since they possess an analog of the Lorenz condition~\eqref{lorenz_vt}~\cite{Silva:2021jya}.

Spontaneous tensorization of a symmetric rank $(0,2)$ tensor field $f_{\mu\nu}$, which is spin $2$, is a natural avenue to investigate after spin-$0$ scalars and spin-$1$ vectors. Ghosts appear in this case once more since the metric in our gravity theories $g_{\mu\nu}$ is also a spin-$2$ field. Note that spontaneous tensorization theories include terms where the field that tensorizes, in this case $f_{\mu\nu}$, is coupled to the metric $g_{\mu\nu}$. However, interacting spin-$2$ fields are known to generically lead to ghost degrees of freedom, rendering most such theories unphysical~\cite{Boulware:1972yco,deRham:2014zqa} and have led to ghost-free theories being discovered more recently in the form of massive gravity and bigravity~\cite{deRham:2010kj, Hassan:2011zd, deRham:2014zqa}. Matter can couple to one of the spin-$2$ fields in the novel ghost-free theories, but coupling to both metrics generically invokes a ghost again~\cite{Yamashita:2014fga}. For example, trying to mimic the \DEF model by having a conformal scaling function that depends on $f_{\mu\nu}$, i.e.,~changing the matter action as $g_{\mu \nu} \to {\cal A}(f_{\mu\nu},g_{\mu\nu})\ g_{\mu \nu}$, leads to ghosts. Overall, there is no known form of bigravity theory that features an analog of spontaneous scalarization.

Spinor fields present the most interesting case of generalizing scalarization, perhaps aside from vectors.\footnote{Although a spinor is not a tensor in the technical sense, we classify spinorization here as an example of tensorization.}
The classical Lagrangian for a Dirac spinor is
\begin{align}\label{eq:spinor_action}
\mathcal{L}_\psi = \tfrac{1}{2} \left[ \psib \gamma^\mu (\nabla_\mu \psi)
- \nabla_\mu \gammab^5 \gamma^\mu \psi \right] -\mu \psib \psi,
\end{align}
where the conventions for gamma matrices and the effect of covariant derivatives on spinors are as given by \textcite{Ramazanoglu:2018hwk}.
This action provides the usual dispersion relation $\omega^2 = k_ik^i + \mu^2$ for a plane wave of the form $\psi \propto \exp[\ii (\omega t-k_i x^i)]$. Since the mass term appears linearly, not quadratically, changing its sign results in the same dispersion relation. However, a tachyonic spinor action is still possible in the following form:
\begin{align}\label{eq:spinor_action_tachyon}
\mathcal{L}_\psi^{5} = \tfrac{1}{2} \left[ \psib \gammab^5 \gamma^\mu (\nabla_\mu \psi)
- (\nabla_\mu \psib) \gammab^5 \gamma^\mu \psi \right] -\mu \psib \psi
,
\end{align}
with the field equation
\begin{align}\label{tachtonic_spin_feqn}
 \gamma^\mu \nabla_\mu \psi -   \gammab^5\mu \psi= 0.
\end{align}
In Eq.~\eqref{tachtonic_spin_feqn} $\omega^2 = k_ik^i - \mu^2$; that is, the dispersion relation is tachyonic~\cite{Chodos:1984cy,Jentschura:2011ga}.

\textcite{Ramazanoglu:2018hwk} used Eq.~\eqref{eq:spinor_action_tachyon} to obtain the first \emph{spontaneous spinorization} theory where spinor fields are unstable to growth around \gr backgrounds of \ns{s}. Even though the meaning of a ghost as opposed to a tachyon is a subtle issue for spinors, the equation of motion in this theory has divergent coefficients as in the vector field case~\eqref{eq:vec_component}. Hence, this form of spinorization suffers from similar problems to vectorization.

\textcite{Minamitsuji:2020hpl} proposed an alternative model of spinorization given by the action
\begin{align}\label{action_spinor2}
S &= \frac{1}{16\pi G_*} \int \dV
\left\{ R +\tfrac{1}{2} \left[\psib \gammab^5 \gamma^\mu (\nabla_\mu \psi) \right. \right. \nonumber \\
&\quad \left.\left. - \, (\nabla_\mu \psib) \gammab^5 \gamma^\mu \psi \right]  \right\}
+ S_{\rm m} \left[\Psi_{\mm},\, {\cal A}^{2}(\psib\psi) g_{\mu \nu} \right].
\end{align}
This model has an unusual feature in that the derivative part of the spinor action is not the canonical one in Eq.~\eqref{eq:spinor_action}, but rather the tachyonic one in Eq.~\eqref{eq:spinor_action_tachyon}. The resulting tachyonic equation of motion is
\begin{align}\label{dirac_eqn_spinorization2}
 \gamma^\mu \nabla_\mu \psi - \gammab^5 ( 4\pi {\cal A}^4 \Lambda \tilde{T} ) \psi= 0,
\end{align}
where $\Lambda = \dd \ln{\cal A}/\dd(\psib\psi)$. The interpretation is straightforward: we have a tachyonic field with an effective mass $\mu_{\rm eff}=4\pi {\cal A}^4 \Lambda \tilde{T}$; cf.~Eq.~\eqref{tachtonic_spin_feqn}. The most important aspect of this theory is the fact that the effective mass term and all coefficients of the field equation are regular everywhere; hence, the dynamics does not have any signs of the ill-posedness that we have found in all generalizations of scalarization thus far. In other words, the action~\eqref{dirac_eqn_spinorization2} is the only known analog of spontaneous scalarization for nonscalar fields that does not suffer from ghost or gradient instabilities.

Our results on spontaneous tensorization suggest a no-go theorem for generalizing scalarization beyond scalar fields, at least if we want to avoid ghosts. It is curious that the only exception to this trend is the relatively exotic case of spinorization, which invites studies of the deeper reasons that make spontaneous scalarization difficult to replicate for other types of fields.

\section{Open problems and future perspectives}
\label{sec:conclusion}
We chose to review the literature by starting from the theoretical
underpinnings of the scalarization mechanism, moving on to discussing \ns{s}
and \bh{s} separately, and then returning to model building in order to discuss
generalization of the mechanism to other fields. Most but not all open problems
have already been mentioned in one or more of the previous sections and also
discussed to some extent.
Nevertheless, we revisit them in this section and cover any additional areas
that require further development, opting for a concise summary of future perspectives.

\subsection{Scalarization and cosmology}

As discussed in Sec.~\ref{sec:mixed_model}, one of the key challenges for
scalarization is understanding whether it is compatible with cosmology. Recall
that the main premise of scalarization is that there is a constant value of the
scalar field $\varphi_0$ for which spacetimes of \gr become admissible
solutions to the field equations, and it is these solutions that describe
stationary objects that we expect (from an observational perspective) to carry
no scalar charge.
For this to be true, cosmic evolution needs to comply and drive $\varphi$ to
$\varphi_0$ in the late Universe; otherwise, the entire Universe will in effect
be ``scalarized.'' It was pointed out early on by~\textcite{Damour:1992kf} and
more recently by~\textcite{Anderson:2016aoi, Franchini:2019npi} that reaching
this preferred value for the scalar fields in the late Universe is not generic
for simple models of scalarization and instead requires severe fine-tuning of
initial conditions.

This could well be an artifact of not having the complete theory, and ideally
one would expect the need for fine-tuned initial conditions to disappear by
addressing further corrections to the model. A first step in this direction was
made by~\textcite{Antoniou:2020nax} [who were inspired
by~\textcite{Damour:1992kf}], who showed that the mixed model of scalarization
discussed in Sec.~\ref{sec:mixed_model} has \gr with a constant scalar as a
late-time cosmic attractor for the right sign of the coupling between the
scalar and the Ricci scalar. This demonstrated that in principle scalarization
models can be made compatible with late-time cosmology and provided a recipe
for doing so: include in the action terms that will dominate in late cosmology
(for example, Ricci coupling) over the terms that control the onset of
scalarization (for example, Gauss-Bonnet invariant) and hence impose the
desired cosmological behavior.
However, the specific model discussed as an example by
\textcite{Antoniou:2020nax} is by no means unique, and there have been many
other attempts to address similar
concerns~\cite{Chen:2015zmx,Minamitsuji:2022qku,Erices:2022bws}. Perhaps more
importantly, understanding the efficiency of the attractor behavior in
reproducing the behavior of \gr (with a cosmological constant) quantitatively,
down to evolution of perturbations, structure formations, etc., certainly
merits further investigation and could lead to constraints on realistic models
of scalarization.

A second point of friction between scalarization and cosmology relates to the
early Universe. Broadly speaking, scalarization is controlled by the coupling
between the scalar-field and curvature invariants, so it is inevitable that
these couplings will become increasingly important as one runs cosmic evolution
backward and move to higher and higher curvatures. In particular, they will
tend to dominate when the size of the Universe becomes comparable to the size
of the compact objects that we want scalarization to affect today
\cite{Antoniou:2020nax}.
It was pointed out
by~\textcite{Anson:2019uto,Anson:2019ebp} that quantum fluctuations could then
seed a scalarizationlike instability during inflation, and that it would be
hard to prevent this linear instability by adding corrections to the action. It
should be stressed, however, that one does not need the scalar field to
necessarily remain constant or have its current value through the evolution of
the Universe. Hence, one might not need to prevent such an instability, but
might instead just quench it nonlinearly, exactly as it happens in
scalarization itself. The scalar field could then smoothly evolve away from the
vacuum that leads to this instability before it becomes relevant as one moves
backward in cosmic time. The key question here is whether there is an \EFT
applicable to the early Universe that contains the scalarization models as late
Universe limits but that can also host an inflationary scenario compatible with
observations.

As a final remark in relation to cosmology, we emphasize that we have discussed
here only the cosmological implications of known scalarization models and only
under the assumption that the scalar field is cosmologically subdominant at
late times. There are a plethora of generalized scalar-tensor theories that
have been used in the context of inflation or dark energy and that exhibit
couplings between the scalar and curvature invariants similar to those employed
for scalarization models. In most cases, however, the nature of the couplings
differs significantly from that of scalarization models. Reviewing such
attempts goes well beyond the scope of this work. Exploring whether a scalar
field that exhibits scalarization around compact objects could also play a role
in inflation or account for dark energy would certainly be interesting, but
putting together such a model (with a single scalar) seems particularly
challenging. Note that in this context evading constraints on the coupling
between a scalar and the Gauss-Bonnet invariant coming from the speed of \gw{s}
requires the scalar to be cosmologically subdominant \cite{Franchini:2019npi}.

\subsection{Dynamical evolution}

The study of the dynamical nonlinear regime of scalarization requires
nonperturbative time-evolution schemes, whose availability varies among
models. Finding formulations of gravity theories that are amenable to numerical
time evolution, i.e.,~numerical relativity, is nontrivial even in
\gr~\cite{Pretorius:2005gq,Campanelli:2005dd,Baker:2005vv}, and the issue can
become even more complicated for alternative gravity theories.

Time-domain, nonlinear evolutions of DEF-like models have been made since
the early work by~\textcite{Novak:1997hw,Novak:1998rk,Novak:1999jg}, and binary inspirals have been performed
for nearly a decade~\cite{Barausse:2012da}.
This was possible thanks to the relative simplicity of the field equations, which allows one
to establish their well-posedness~\cite{Salgado:2005hx,Salgado:2008xh}.
However, detailed and fully nonlinear numerical results in relation to
\gw{s}, a primary tool to test scalarization, are still lacking, aside from the study by~\textcite{Shibata:2013pra}.

The picture is much different for \bh scalarization, where the coupling to the Gauss-Bonnet term can result in a complicated causal structure of the spacetime in dynamical situations. It was demonstrated at both linear~\cite{Blazquez-Salcedo:2018jnn,Blazquez-Salcedo:2020rhf,Blazquez-Salcedo:2020caw} and nonlinear~\cite{East:2021bqk,East:2020hgw,Ripley:2020vpk} levels that part, but not all, of the parameter space of spontaneous scalarization where scalarized \bh{s} can be found loses hyperbolicity when the theory is evolved in time in a certain set of gauges, which means that no predictions can be obtained.
The well-posedness of the initial-value problem in broader classes of
modified gravity theories, including scalar-Gauss-Bonnet gravity, was studied
by \textcite{Papallo:2017ddx,Kovacs:2019jqj,Kovacs:2020ywu,Kovacs:2020pns}.
At the moment there is no consensus on whether this loss of hyperbolicity can be cured with a better gauge choice or if it is intrinsic to the evolution equations, which calls for further work on the issue. Furthermore, there are no detailed studies of binary evolution and merger even for the part of the parameter space for which hyperbolic time evolution has been shown to exist.
The dynamics of more general models where various coupling terms are present as in Eq.~\eqref{eq:ActionCaseI} are not available either. The lack of results on all of these fronts presents important future research directions.

It can be the case that an \EFT, obtained as a certain limit or truncation of a more fundamental theory, is ill posed while the latter is or is expected to be well posed. A typical example is relativistic hydrodynamics once viscosity has been taken into account. Indeed, it has been suggested to introduce techniques used in hydrodynamics to gravity in order to render the problematic time evolution of some theories hyperbolic by suitably modifying the equations while keeping the end point of evolution the same~\cite{Cayuso:2017iqc}.
This approach has seen increasing use in alternatives to \gr, including the specific case of scalar-Gauss-Bonnet theories under certain symmetries~\cite{Franchini:2022ukz}, and it is another avenue to explore the dynamics of scalarization in cases where current methods are inadequate. The applicability and power of such methods to general cases, for instance, fully nonlinear 3+1 evolution of a highly dynamical system, are yet to be confirmed.

\subsection{Model building in and beyond scalarization}

We have covered several models of spontaneous scalarization and their phenomenology for \ns{s} and \bh{s}. What they have in common is that scalarization is triggered by a tachyonic instability, whose threshold is controlled by fewer than a handful of terms in the action \cite{Andreou:2019ikc}. Yet, the properties of the final scalarized configuration depend on the nonlinearities and coupling terms that contribute beyond the linear level. The specific choices of these nonlinear interactions have been shown to crucially affect the stability  \cite{Blazquez-Salcedo:2018jnn,Silva:2018qhn,Antoniou:2022agj} and the scalar charge \cite{Silva:2018qhn,Macedo:2019sem,Antoniou:2021zoy,Doneva:2018rou} of scalarized solutions, as well as the cosmological behavior of the corresponding theory \cite{Antoniou:2020nax,Anson:2019uto,Erices:2022bws}. Thus, such choices affect the viability and observability of the models. As a result, although certain models of scalarization, such as the original DEF model, have been studied fairly exhaustively, the exploration of the broader class of theories that exhibit scalarization, and their phenomenology, has only started.

Furthermore, one could consider introducing new classes of scalarization based on instabilities that are not tachyonic in nature. As discussed in Sec.~\ref{sec:other}, success in this direction has thus far been limited, but it is nevertheless an interesting prospect. Alternatively, one can abandon the idea of a linear instability quenched by nonlinearities altogether, as discussed in Sec.~\ref{sec:beyond_scalarization}. \textcite{Doneva:2021tvn} demostrated that a theory featuring a coupling to the Gauss-Bonnet invariant with a leading-order expansion with respect to the scalar field that is proportional to $\varphi^4$ can still have scalarized \bh{s} below a certain mass threshold, although the Kerr solution is linearly stable with respect to massless scalar perturbations~\cite{Blazquez-Salcedo:2022omw}. This observation remains true in more general theories, such as Gauss-Bonnet gravity with multiple scalar fields \cite{Staykov:2022uwq}. This strongly suggests the existence of a class of theories in which scalarization is a purely nonlinear phenomenon.

Scalarization models that have been considered thus far typically assume that the scalar field does not couple to matter directly in a suitable choice of variables, usually called the Jordan frame, which suffices for the theory to satisfy the \WEP. However, as pointed out by \textcite{Coates:2016ktu}, this assumption might not be necessary because scalarization itself forces the scalar field into a trivial configuration away from a specific compact object. Therefore, scalarization models can be indistinguishable from \gr in regard to the \WEP for all current observations. It was further argued by \textcite{Coates:2016ktu} and \textcite{Franchini:2017zzx} that particular couplings to matter that do not disrupt scalarization could introduce a Higgs-like mechanism in gravity: the scalar field changes value only near compact objects, and its coupling to matter changes the properties of the standard model in these regions. This is a largely unexplored possibility.

As discussed in Sec.~\ref{sec:other}, another important open question is whether one can generalize the scalarization mechanism to other fields. Doing so in the context of a theory that is free of pathologies has thus far proven to be challenging, except perhaps in the case of spinors~\cite{Minamitsuji:2020hpl}. This is in part because controlling the dynamics of the extra degree of freedom is notoriously difficult in general in gravity theories with additional vector or tensor fields~\cite{deRham:2014zqa}, and vectorization or tensorization models are no exception.

We stress that the work on scalarization and its extensions has focused on isolated compact objects. In fact, theoretical considerations for putting together scalarization models have also been heavily influence by studies of isolated \bh{s} and \ns{s}. However, as previously discussed,
scalarization could happen dynamically in a binary \cite{Barausse:2012da,Palenzuela:2013hsa,Shibata:2013pra,Taniguchi:2014fqa,Elley:2022ept,Silva:2020omi,Doneva:2022byd}. Apart from finding ways to model this effect~\cite{Khalil:2022sii} and potentially constraining it with \gw observations,
it would be particularly interesting to further explore which properties of a binary control dynamical scalarization, and how this differs among the various scalarization models.

\subsection{Observational prospects}

Perhaps the most attractive feature of scalarization is that the tachyonic instability typically leads to large scalar-field amplitudes before it is quenched, which results in nonperturbative deviations from \gr in high curvature regions.
At the same time, the theory mimics \gr closely for weak gravity, easily satisfying existing tests.
This suggests that new fundamental scalar fields that exhibit this behavior could be discovered with strong-gravity observations.

Current observations put stringent bounds on scalar radiation from pulsars that have already almost completely ruled out the original (massless) DEF model~\cite{Zhao:2022vig}. Electromagnetic observations of compact objects in binaries or surrounded by some form of matter also have the potential to extend such results to more general models of scalarization.
For example, x-ray data from \ns surfaces have recently been considered as a possible way to detect scalarized objects \cite{Silva:2018yxz,Silva:2019leq}.
The masses, radii, and moment of inertia of \ns{s} can be inferred through such observations~\cite{Bogdanov:2019ixe,Bogdanov:2019qjb,Bogdanov:2021yip}. Future observation
could be used to distinguish scalarized \ns{s} from nonscalarized ones by, for instance, using
\EOS-independent relations. The astrophysical signatures of most extensions of the \DEF model, especially those that allow \bh scalarization, have not yet been studied in detail, because
they are recent.

The most interesting current development in strong gravity is the advent of \gw astronomy, which is an arena for exploring spontaneous scalarization. The effects of scalarization on \gw{s} of coalescing \ns{s} and/or \bh{s} have been investigated to some extent, mostly in regard to the \DEF model \cite{Sennett:2016rwa,Khalil:2019wyy,Khalil:2022sii,Niu:2021nic} or scalar-Gauss-Bonnet gravity~\cite{Wong:2022wni}.
The most pressing issue on this front is the development of accurate waveform models that
ideally would cover the inspiral, merger, and ringdown stages of coalescing binaries.
Developing such waveform models in \gr has been (and remains) a formidable task that combines tools from post-Newtonian theory, numerical relativity, black-hole perturbation theory, and gravitational self-force. Progress on all these fronts is necessary for one to develop accurate scalarized waveform models.
Stars that undergo core collapse are another interesting source of \gw{s}. Albeit expected to
be rare, such events may produce large bursts of scalar radiation~\cite{Sperhake:2017itk,Rosca-Mead:2019seq,Kuan:2022oxs} and may contribute to the stochastic \gw background, which could in principle be probed by pulsar-timing arrays.
In addition, some models of scalarization (but not exclusively) predict the existence of extra polarizations in \gw{s}, which provides another way to test them.

Last, future spaceborne \gw detectors such as LISA will bring another channel for testing modified gravity. Especially interesting with respect to scalarization is the case of extreme mass-ratio inspirals when the secondary object is scalarized~\cite{Maselli:2020zgv,Maselli:2021men}.
Data from future \gw detectors such as LISA or third generation ground-based detectors such as the Einstein Telescope and the Cosmic Explorer, combined with more precise electromagnetic observations such as those from improved x-ray observatories, will be crucial for detecting or ruling out scalar fields that exhibit scalarization.

\begin{acknowledgments}
D.D.D. acknowledges financial support via the Emmy Noether Research Group, which was
funded by the German Research Foundation (DFG) under Grant No.~DO 1771/1-1.
S.S.Y. thanks the University of T\"ubingen for the financial support
and acknowledges partial support from Bulgarian NSF Grant No.~KP-06-H28/7.
T.P.S. acknowledges partial support from STFC Consolidated Grants No.~ST/T000732/1 and No.~ST/V005596/1.
F.M.R acknowledges support from 2020 Bilim Akademisi GEB\.IP award and T\"UB\.ITAK Project No.~122F097.
H.O.S. acknowledges funding from Deutsche Forschungsgemeinschaft (DFG) Project No.~386119226.
We also acknowledge the networking support of COST Actions No.~CA16104 and No.~CA15117.
\end{acknowledgments}

\bibliographystyle{apsrmp4-2}
\bibliography{references}

\begin{thebibliography}{394}%
\makeatletter
\providecommand \@ifxundefined [1]{%
 \@ifx{#1\undefined}
}%
\providecommand \@ifnum [1]{%
 \ifnum #1\expandafter \@firstoftwo
 \else \expandafter \@secondoftwo
 \fi
}%
\providecommand \@ifx [1]{%
 \ifx #1\expandafter \@firstoftwo
 \else \expandafter \@secondoftwo
 \fi
}%
\providecommand \natexlab [1]{#1}%
\providecommand \enquote  [1]{``#1''}%
\providecommand \bibnamefont  [1]{#1}%
\providecommand \bibfnamefont [1]{#1}%
\providecommand \citenamefont [1]{#1}%
\providecommand \href@noop [0]{\@secondoftwo}%
\providecommand \href [0]{\begingroup \@sanitize@url \@href}%
\providecommand \@href[1]{\@@startlink{#1}\@@href}%
\providecommand \@@href[1]{\endgroup#1\@@endlink}%
\providecommand \@sanitize@url [0]{\catcode `\\12\catcode `\$12\catcode
  `\&12\catcode `\#12\catcode `\^12\catcode `\_12\catcode `\%12\relax}%
\providecommand \@@startlink[1]{}%
\providecommand \@@endlink[0]{}%
\providecommand \url  [0]{\begingroup\@sanitize@url \@url }%
\providecommand \@url [1]{\endgroup\@href {#1}{\urlprefix }}%
\providecommand \urlprefix  [0]{URL }%
\providecommand \Eprint [0]{\href }%
\providecommand \doibase [0]{https://doi.org/}%
\providecommand \selectlanguage [0]{\@gobble}%
\providecommand \bibinfo  [0]{\@secondoftwo}%
\providecommand \bibfield  [0]{\@secondoftwo}%
\providecommand \translation [1]{[#1]}%
\providecommand \BibitemOpen [0]{}%
\providecommand \bibitemStop [0]{}%
\providecommand \bibitemNoStop [0]{.\EOS\space}%
\providecommand \EOS [0]{\spacefactor3000\relax}%
\providecommand \BibitemShut  [1]{\csname bibitem#1\endcsname}%
\let\auto@bib@innerbib\@empty
\bibitem [{\citenamefont {Abbott}\ \emph {et~al.}(2019)\citenamefont {Abbott}
  \emph {et~al.}}]{LIGOScientific:2018mvr}%
  \BibitemOpen
  \bibfield  {author} {\bibinfo {author} {\bibnamefont {Abbott}, \bibfnamefont
  {B.~P.}},  \emph {et~al.} (\bibinfo {collaboration} {LIGO Scientific,
  Virgo})} (\bibinfo {year} {2019}),\ \href
  {https://doi.org/10.1103/PhysRevX.9.031040} {\bibfield  {journal} {\bibinfo
  {journal} {Phys. Rev. X}\ }\textbf {\bibinfo {volume} {9}}~(\bibinfo {number}
  {3}),\ \bibinfo {pages} {031040}},\ \Eprint
  {https://arxiv.org/abs/1811.12907} {arXiv:1811.12907 [astro-ph.HE]}
  \BibitemShut {NoStop}%
\bibitem [{\citenamefont {Abbott}\ \emph
  {et~al.}(2021{\natexlab{a}})\citenamefont {Abbott} \emph
  {et~al.}}]{LIGOScientific:2020ibl}%
  \BibitemOpen
  \bibfield  {author} {\bibinfo {author} {\bibnamefont {Abbott}, \bibfnamefont
  {R.}},  \emph {et~al.} (\bibinfo {collaboration} {LIGO Scientific, Virgo})}
  (\bibinfo {year} {2021}{\natexlab{a}}),\ \href
  {https://doi.org/10.1103/PhysRevX.11.021053} {\bibfield  {journal} {\bibinfo
  {journal} {Phys. Rev. X}\ }\textbf {\bibinfo {volume} {11}},\ \bibinfo
  {pages} {021053}},\ \Eprint {https://arxiv.org/abs/2010.14527}
  {arXiv:2010.14527 [gr-qc]} \BibitemShut {NoStop}%
\bibitem [{\citenamefont {Abbott}\ \emph
  {et~al.}(2021{\natexlab{b}})\citenamefont {Abbott} \emph
  {et~al.}}]{LIGOScientific:2021djp}%
  \BibitemOpen
  \bibfield  {author} {\bibinfo {author} {\bibnamefont {Abbott}, \bibfnamefont
  {R.}},  \emph {et~al.} (\bibinfo {collaboration} {LIGO Scientific, VIRGO,
  KAGRA})} (\bibinfo {year} {2021}{\natexlab{b}}),\ \href@noop {} {\enquote
  {\bibinfo {title} {{GWTC-3: Compact Binary Coalescences Observed by LIGO and
  Virgo During the Second Part of the Third Observing Run}},}\ }\Eprint
  {https://arxiv.org/abs/2111.03606} {arXiv:2111.03606 [gr-qc]} \BibitemShut
  {NoStop}%
\bibitem [{\citenamefont {Abuter}\ \emph {et~al.}(2018)\citenamefont {Abuter}
  \emph {et~al.}}]{GRAVITY:2018ofz}%
  \BibitemOpen
  \bibfield  {author} {\bibinfo {author} {\bibnamefont {Abuter}, \bibfnamefont
  {R.}},  \emph {et~al.} (\bibinfo {collaboration} {GRAVITY})} (\bibinfo {year}
  {2018}),\ \href {https://doi.org/10.1051/0004-6361/201833718} {\bibfield
  {journal} {\bibinfo  {journal} {Astron. Astrophys.}\ }\textbf {\bibinfo
  {volume} {615}},\ \bibinfo {pages} {L15}},\ \Eprint
  {https://arxiv.org/abs/1807.09409} {arXiv:1807.09409 [astro-ph.GA]}
  \BibitemShut {NoStop}%
\bibitem [{\citenamefont {Abuter}\ \emph {et~al.}(2020)\citenamefont {Abuter}
  \emph {et~al.}}]{GRAVITY:2020gka}%
  \BibitemOpen
  \bibfield  {author} {\bibinfo {author} {\bibnamefont {Abuter}, \bibfnamefont
  {R.}},  \emph {et~al.} (\bibinfo {collaboration} {GRAVITY})} (\bibinfo {year}
  {2020}),\ \href {https://doi.org/10.1051/0004-6361/202037813} {\bibfield
  {journal} {\bibinfo  {journal} {Astron. Astrophys.}\ }\textbf {\bibinfo
  {volume} {636}},\ \bibinfo {pages} {L5}},\ \Eprint
  {https://arxiv.org/abs/2004.07187} {arXiv:2004.07187 [astro-ph.GA]}
  \BibitemShut {NoStop}%
\bibitem [{\citenamefont {Akiyama}\ \emph {et~al.}(2019)\citenamefont {Akiyama}
  \emph {et~al.}}]{EventHorizonTelescope:2019dse}%
  \BibitemOpen
  \bibfield  {author} {\bibinfo {author} {\bibnamefont {Akiyama}, \bibfnamefont
  {K.}},  \emph {et~al.} (\bibinfo {collaboration} {Event Horizon Telescope})}
  (\bibinfo {year} {2019}),\ \href {https://doi.org/10.3847/2041-8213/ab0ec7}
  {\bibfield  {journal} {\bibinfo  {journal} {Astrophys. J. Lett.}\ }\textbf
  {\bibinfo {volume} {875}},\ \bibinfo {pages} {L1}},\ \Eprint
  {https://arxiv.org/abs/1906.11238} {arXiv:1906.11238 [astro-ph.GA]}
  \BibitemShut {NoStop}%
\bibitem [{\citenamefont {Akmal}\ \emph {et~al.}(1998)\citenamefont {Akmal},
  \citenamefont {Pandharipande},\ and\ \citenamefont
  {Ravenhall}}]{Akmal:1998cf}%
  \BibitemOpen
  \bibfield  {author} {\bibinfo {author} {\bibnamefont {Akmal}, \bibfnamefont
  {A.}}, \bibinfo {author} {\bibfnamefont {V.~R.}\ \bibnamefont
  {Pandharipande}}, and\ \bibinfo {author} {\bibfnamefont {D.~G.}\ \bibnamefont
  {Ravenhall}}} (\bibinfo {year} {1998}),\ \href
  {https://doi.org/10.1103/PhysRevC.58.1804} {\bibfield  {journal} {\bibinfo
  {journal} {Phys. Rev. C}\ }\textbf {\bibinfo {volume} {58}},\ \bibinfo
  {pages} {1804}},\ \Eprint {https://arxiv.org/abs/nucl-th/9804027}
  {arXiv:nucl-th/9804027} \BibitemShut {NoStop}%
\bibitem [{\citenamefont {Alcubierre}\ \emph {et~al.}(2010)\citenamefont
  {Alcubierre}, \citenamefont {Degollado}, \citenamefont {Nunez}, \citenamefont
  {Ruiz},\ and\ \citenamefont {Salgado}}]{Alcubierre:2010ea}%
  \BibitemOpen
  \bibfield  {author} {\bibinfo {author} {\bibnamefont {Alcubierre},
  \bibfnamefont {M.}}, \bibinfo {author} {\bibfnamefont {J.~C.}\ \bibnamefont
  {Degollado}}, \bibinfo {author} {\bibfnamefont {D.}~\bibnamefont {Nunez}},
  \bibinfo {author} {\bibfnamefont {M.}~\bibnamefont {Ruiz}}, and\ \bibinfo
  {author} {\bibfnamefont {M.}~\bibnamefont {Salgado}}} (\bibinfo {year}
  {2010}),\ \href {https://doi.org/10.1103/PhysRevD.81.124018} {\bibfield
  {journal} {\bibinfo  {journal} {Phys. Rev. D}\ }\textbf {\bibinfo {volume}
  {81}},\ \bibinfo {pages} {124018}},\ \Eprint
  {https://arxiv.org/abs/1003.4767} {arXiv:1003.4767 [gr-qc]} \BibitemShut
  {NoStop}%
\bibitem [{\citenamefont {Alexander}\ and\ \citenamefont
  {Yunes}(2009)}]{Alexander:2009tp}%
  \BibitemOpen
  \bibfield  {author} {\bibinfo {author} {\bibnamefont {Alexander},
  \bibfnamefont {S.}}, and\ \bibinfo {author} {\bibfnamefont {N.}~\bibnamefont
  {Yunes}}} (\bibinfo {year} {2009}),\ \href
  {https://doi.org/10.1016/j.physrep.2009.07.002} {\bibfield  {journal}
  {\bibinfo  {journal} {Phys. Rept.}\ }\textbf {\bibinfo {volume} {480}},\
  \bibinfo {pages} {1}},\ \Eprint {https://arxiv.org/abs/0907.2562}
  {arXiv:0907.2562 [hep-th]} \BibitemShut {NoStop}%
\bibitem [{\citenamefont {Alsing}\ \emph {et~al.}(2012)\citenamefont {Alsing},
  \citenamefont {Berti}, \citenamefont {Will},\ and\ \citenamefont
  {Zaglauer}}]{Alsing:2011er}%
  \BibitemOpen
  \bibfield  {author} {\bibinfo {author} {\bibnamefont {Alsing}, \bibfnamefont
  {J.}}, \bibinfo {author} {\bibfnamefont {E.}~\bibnamefont {Berti}}, \bibinfo
  {author} {\bibfnamefont {C.~M.}\ \bibnamefont {Will}}, and\ \bibinfo {author}
  {\bibfnamefont {H.}~\bibnamefont {Zaglauer}}} (\bibinfo {year} {2012}),\
  \href {https://doi.org/10.1103/PhysRevD.85.064041} {\bibfield  {journal}
  {\bibinfo  {journal} {Phys. Rev. D}\ }\textbf {\bibinfo {volume} {85}},\
  \bibinfo {pages} {064041}},\ \Eprint {https://arxiv.org/abs/1112.4903}
  {arXiv:1112.4903 [gr-qc]} \BibitemShut {NoStop}%
\bibitem [{\citenamefont {Altaha~Motahar}\ \emph {et~al.}(2019)\citenamefont
  {Altaha~Motahar}, \citenamefont {Bl\'azquez-Salcedo}, \citenamefont {Doneva},
  \citenamefont {Kunz},\ and\ \citenamefont
  {Yazadjiev}}]{AltahaMotahar:2019ekm}%
  \BibitemOpen
  \bibfield  {author} {\bibinfo {author} {\bibnamefont {Altaha~Motahar},
  \bibfnamefont {Z.}}, \bibinfo {author} {\bibfnamefont {J.~L.}\ \bibnamefont
  {Bl\'azquez-Salcedo}}, \bibinfo {author} {\bibfnamefont {D.~D.}\ \bibnamefont
  {Doneva}}, \bibinfo {author} {\bibfnamefont {J.}~\bibnamefont {Kunz}}, and\
  \bibinfo {author} {\bibfnamefont {S.~S.}\ \bibnamefont {Yazadjiev}}}
  (\bibinfo {year} {2019}),\ \href {https://doi.org/10.1103/PhysRevD.99.104006}
  {\bibfield  {journal} {\bibinfo  {journal} {Phys. Rev. D}\ }\textbf {\bibinfo
  {volume} {99}}~(\bibinfo {number} {10}),\ \bibinfo {pages} {104006}},\
  \Eprint {https://arxiv.org/abs/1902.01277} {arXiv:1902.01277 [gr-qc]}
  \BibitemShut {NoStop}%
\bibitem [{\citenamefont {Altaha~Motahar}\ \emph {et~al.}(2017)\citenamefont
  {Altaha~Motahar}, \citenamefont {Bl\'azquez-Salcedo}, \citenamefont
  {Kleihaus},\ and\ \citenamefont {Kunz}}]{Motahar:2017blm}%
  \BibitemOpen
  \bibfield  {author} {\bibinfo {author} {\bibnamefont {Altaha~Motahar},
  \bibfnamefont {Z.}}, \bibinfo {author} {\bibfnamefont {J.~L.}\ \bibnamefont
  {Bl\'azquez-Salcedo}}, \bibinfo {author} {\bibfnamefont {B.}~\bibnamefont
  {Kleihaus}}, and\ \bibinfo {author} {\bibfnamefont {J.}~\bibnamefont {Kunz}}}
  (\bibinfo {year} {2017}),\ \href {https://doi.org/10.1103/PhysRevD.96.064046}
  {\bibfield  {journal} {\bibinfo  {journal} {Phys. Rev. D}\ }\textbf {\bibinfo
  {volume} {96}}~(\bibinfo {number} {6}),\ \bibinfo {pages} {064046}},\ \Eprint
  {https://arxiv.org/abs/1707.05280} {arXiv:1707.05280 [gr-qc]} \BibitemShut
  {NoStop}%
\bibitem [{\citenamefont {Altaha~Motahar}\ \emph {et~al.}(2018)\citenamefont
  {Altaha~Motahar}, \citenamefont {Bl\'azquez-Salcedo}, \citenamefont
  {Kleihaus},\ and\ \citenamefont {Kunz}}]{AltahaMotahar:2018djk}%
  \BibitemOpen
  \bibfield  {author} {\bibinfo {author} {\bibnamefont {Altaha~Motahar},
  \bibfnamefont {Z.}}, \bibinfo {author} {\bibfnamefont {J.~L.}\ \bibnamefont
  {Bl\'azquez-Salcedo}}, \bibinfo {author} {\bibfnamefont {B.}~\bibnamefont
  {Kleihaus}}, and\ \bibinfo {author} {\bibfnamefont {J.}~\bibnamefont {Kunz}}}
  (\bibinfo {year} {2018}),\ \href {https://doi.org/10.1103/PhysRevD.98.044032}
  {\bibfield  {journal} {\bibinfo  {journal} {Phys. Rev. D}\ }\textbf {\bibinfo
  {volume} {98}}~(\bibinfo {number} {4}),\ \bibinfo {pages} {044032}},\ \Eprint
  {https://arxiv.org/abs/1807.02598} {arXiv:1807.02598 [gr-qc]} \BibitemShut
  {NoStop}%
\bibitem [{\citenamefont {Anderson}\ and\ \citenamefont
  {Yunes}(2019)}]{Anderson:2019hio}%
  \BibitemOpen
  \bibfield  {author} {\bibinfo {author} {\bibnamefont {Anderson},
  \bibfnamefont {D.}}, and\ \bibinfo {author} {\bibfnamefont {N.}~\bibnamefont
  {Yunes}}} (\bibinfo {year} {2019}),\ \href
  {https://doi.org/10.1088/1361-6382/ab2eda} {\bibfield  {journal} {\bibinfo
  {journal} {Class. Quant. Grav.}\ }\textbf {\bibinfo {volume} {36}}~(\bibinfo
  {number} {16}),\ \bibinfo {pages} {165003}},\ \Eprint
  {https://arxiv.org/abs/1901.00937} {arXiv:1901.00937 [gr-qc]} \BibitemShut
  {NoStop}%
\bibitem [{\citenamefont {Anderson}\ \emph {et~al.}(2016)\citenamefont
  {Anderson}, \citenamefont {Yunes},\ and\ \citenamefont
  {Barausse}}]{Anderson:2016aoi}%
  \BibitemOpen
  \bibfield  {author} {\bibinfo {author} {\bibnamefont {Anderson},
  \bibfnamefont {D.}}, \bibinfo {author} {\bibfnamefont {N.}~\bibnamefont
  {Yunes}}, and\ \bibinfo {author} {\bibfnamefont {E.}~\bibnamefont
  {Barausse}}} (\bibinfo {year} {2016}),\ \href
  {https://doi.org/10.1103/PhysRevD.94.104064} {\bibfield  {journal} {\bibinfo
  {journal} {Phys. Rev. D}\ }\textbf {\bibinfo {volume} {94}}~(\bibinfo
  {number} {10}),\ \bibinfo {pages} {104064}},\ \Eprint
  {https://arxiv.org/abs/1607.08888} {arXiv:1607.08888 [gr-qc]} \BibitemShut
  {NoStop}%
\bibitem [{\citenamefont {Andersson}\ and\ \citenamefont
  {Kokkotas}(1998)}]{Andersson:1997rn}%
  \BibitemOpen
  \bibfield  {author} {\bibinfo {author} {\bibnamefont {Andersson},
  \bibfnamefont {N.}}, and\ \bibinfo {author} {\bibfnamefont {K.~D.}\
  \bibnamefont {Kokkotas}}} (\bibinfo {year} {1998}),\ \href
  {https://doi.org/10.1046/j.1365-8711.1998.01840.x} {\bibfield  {journal}
  {\bibinfo  {journal} {Mon. Not. Roy. Astron. Soc.}\ }\textbf {\bibinfo
  {volume} {299}},\ \bibinfo {pages} {1059}},\ \Eprint
  {https://arxiv.org/abs/gr-qc/9711088} {arXiv:gr-qc/9711088} \BibitemShut
  {NoStop}%
\bibitem [{\citenamefont {Andreou}\ \emph {et~al.}(2019)\citenamefont
  {Andreou}, \citenamefont {Franchini}, \citenamefont {Ventagli},\ and\
  \citenamefont {Sotiriou}}]{Andreou:2019ikc}%
  \BibitemOpen
  \bibfield  {author} {\bibinfo {author} {\bibnamefont {Andreou}, \bibfnamefont
  {N.}}, \bibinfo {author} {\bibfnamefont {N.}~\bibnamefont {Franchini}},
  \bibinfo {author} {\bibfnamefont {G.}~\bibnamefont {Ventagli}}, and\ \bibinfo
  {author} {\bibfnamefont {T.~P.}\ \bibnamefont {Sotiriou}}} (\bibinfo {year}
  {2019}),\ \href {https://doi.org/10.1103/PhysRevD.99.124022} {\bibfield
  {journal} {\bibinfo  {journal} {Phys. Rev. D}\ }\textbf {\bibinfo {volume}
  {99}}~(\bibinfo {number} {12}),\ \bibinfo {pages} {124022}},\ \bibinfo {note}
  {[Erratum: Phys.Rev.D 101, 109903 (2020)]},\ \Eprint
  {https://arxiv.org/abs/1904.06365} {arXiv:1904.06365 [gr-qc]} \BibitemShut
  {NoStop}%
\bibitem [{\citenamefont {Annulli}\ \emph {et~al.}(2019)\citenamefont
  {Annulli}, \citenamefont {Cardoso},\ and\ \citenamefont
  {Gualtieri}}]{Annulli:2019fzq}%
  \BibitemOpen
  \bibfield  {author} {\bibinfo {author} {\bibnamefont {Annulli}, \bibfnamefont
  {L.}}, \bibinfo {author} {\bibfnamefont {V.}~\bibnamefont {Cardoso}}, and\
  \bibinfo {author} {\bibfnamefont {L.}~\bibnamefont {Gualtieri}}} (\bibinfo
  {year} {2019}),\ \href {https://doi.org/10.1103/PhysRevD.99.044038}
  {\bibfield  {journal} {\bibinfo  {journal} {Phys. Rev. D}\ }\textbf {\bibinfo
  {volume} {99}}~(\bibinfo {number} {4}),\ \bibinfo {pages} {044038}},\ \Eprint
  {https://arxiv.org/abs/1901.02461} {arXiv:1901.02461 [gr-qc]} \BibitemShut
  {NoStop}%
\bibitem [{\citenamefont {Annulli}\ \emph {et~al.}(2022)\citenamefont
  {Annulli}, \citenamefont {Herdeiro},\ and\ \citenamefont
  {Radu}}]{Annulli:2022ivr}%
  \BibitemOpen
  \bibfield  {author} {\bibinfo {author} {\bibnamefont {Annulli}, \bibfnamefont
  {L.}}, \bibinfo {author} {\bibfnamefont {C.~A.~R.}\ \bibnamefont {Herdeiro}},
  and\ \bibinfo {author} {\bibfnamefont {E.}~\bibnamefont {Radu}}} (\bibinfo
  {year} {2022}),\ \href {https://doi.org/10.1016/j.physletb.2022.137227}
  {\bibfield  {journal} {\bibinfo  {journal} {Phys. Lett. B}\ }\textbf
  {\bibinfo {volume} {832}},\ \bibinfo {pages} {137227}},\ \Eprint
  {https://arxiv.org/abs/2203.13267} {arXiv:2203.13267 [gr-qc]} \BibitemShut
  {NoStop}%
\bibitem [{\citenamefont {Anson}\ \emph
  {et~al.}(2019{\natexlab{a}})\citenamefont {Anson}, \citenamefont {Babichev},
  \citenamefont {Charmousis},\ and\ \citenamefont {Ramazanov}}]{Anson:2019uto}%
  \BibitemOpen
  \bibfield  {author} {\bibinfo {author} {\bibnamefont {Anson}, \bibfnamefont
  {T.}}, \bibinfo {author} {\bibfnamefont {E.}~\bibnamefont {Babichev}},
  \bibinfo {author} {\bibfnamefont {C.}~\bibnamefont {Charmousis}}, and\
  \bibinfo {author} {\bibfnamefont {S.}~\bibnamefont {Ramazanov}}} (\bibinfo
  {year} {2019}{\natexlab{a}}),\ \href
  {https://doi.org/10.1088/1475-7516/2019/06/023} {\bibfield  {journal}
  {\bibinfo  {journal} {JCAP}\ }\textbf {\bibinfo {volume} {06}},\ \bibinfo
  {pages} {023}},\ \Eprint {https://arxiv.org/abs/1903.02399} {arXiv:1903.02399
  [gr-qc]} \BibitemShut {NoStop}%
\bibitem [{\citenamefont {Anson}\ \emph
  {et~al.}(2019{\natexlab{b}})\citenamefont {Anson}, \citenamefont {Babichev},\
  and\ \citenamefont {Ramazanov}}]{Anson:2019ebp}%
  \BibitemOpen
  \bibfield  {author} {\bibinfo {author} {\bibnamefont {Anson}, \bibfnamefont
  {T.}}, \bibinfo {author} {\bibfnamefont {E.}~\bibnamefont {Babichev}}, and\
  \bibinfo {author} {\bibfnamefont {S.}~\bibnamefont {Ramazanov}}} (\bibinfo
  {year} {2019}{\natexlab{b}}),\ \href
  {https://doi.org/10.1103/PhysRevD.100.104051} {\bibfield  {journal} {\bibinfo
   {journal} {Phys. Rev. D}\ }\textbf {\bibinfo {volume} {100}}~(\bibinfo
  {number} {10}),\ \bibinfo {pages} {104051}},\ \Eprint
  {https://arxiv.org/abs/1905.10393} {arXiv:1905.10393 [gr-qc]} \BibitemShut
  {NoStop}%
\bibitem [{\citenamefont {Antoniadis}\ \emph {et~al.}(2013)\citenamefont
  {Antoniadis} \emph {et~al.}}]{Antoniadis:2013pzd}%
  \BibitemOpen
  \bibfield  {author} {\bibinfo {author} {\bibnamefont {Antoniadis},
  \bibfnamefont {J.}},  \emph {et~al.}} (\bibinfo {year} {2013}),\ \href
  {https://doi.org/10.1126/science.1233232} {\bibfield  {journal} {\bibinfo
  {journal} {Science}\ }\textbf {\bibinfo {volume} {340}},\ \bibinfo {pages}
  {6131}},\ \Eprint {https://arxiv.org/abs/1304.6875} {arXiv:1304.6875
  [astro-ph.HE]} \BibitemShut {NoStop}%
\bibitem [{\citenamefont {Antoniou}\ \emph {et~al.}(2018)\citenamefont
  {Antoniou}, \citenamefont {Bakopoulos},\ and\ \citenamefont
  {Kanti}}]{Antoniou:2017acq}%
  \BibitemOpen
  \bibfield  {author} {\bibinfo {author} {\bibnamefont {Antoniou},
  \bibfnamefont {G.}}, \bibinfo {author} {\bibfnamefont {A.}~\bibnamefont
  {Bakopoulos}}, and\ \bibinfo {author} {\bibfnamefont {P.}~\bibnamefont
  {Kanti}}} (\bibinfo {year} {2018}),\ \href
  {https://doi.org/10.1103/PhysRevLett.120.131102} {\bibfield  {journal}
  {\bibinfo  {journal} {Phys. Rev. Lett.}\ }\textbf {\bibinfo {volume}
  {120}}~(\bibinfo {number} {13}),\ \bibinfo {pages} {131102}},\ \Eprint
  {https://arxiv.org/abs/1711.03390} {arXiv:1711.03390 [hep-th]} \BibitemShut
  {NoStop}%
\bibitem [{\citenamefont {Antoniou}\ \emph
  {et~al.}(2021{\natexlab{a}})\citenamefont {Antoniou}, \citenamefont
  {Bordin},\ and\ \citenamefont {Sotiriou}}]{Antoniou:2020nax}%
  \BibitemOpen
  \bibfield  {author} {\bibinfo {author} {\bibnamefont {Antoniou},
  \bibfnamefont {G.}}, \bibinfo {author} {\bibfnamefont {L.}~\bibnamefont
  {Bordin}}, and\ \bibinfo {author} {\bibfnamefont {T.~P.}\ \bibnamefont
  {Sotiriou}}} (\bibinfo {year} {2021}{\natexlab{a}}),\ \href
  {https://doi.org/10.1103/PhysRevD.103.024012} {\bibfield  {journal} {\bibinfo
   {journal} {Phys. Rev. D}\ }\textbf {\bibinfo {volume} {103}}~(\bibinfo
  {number} {2}),\ \bibinfo {pages} {024012}},\ \Eprint
  {https://arxiv.org/abs/2004.14985} {arXiv:2004.14985 [gr-qc]} \BibitemShut
  {NoStop}%
\bibitem [{\citenamefont {Antoniou}\ \emph
  {et~al.}(2021{\natexlab{b}})\citenamefont {Antoniou}, \citenamefont
  {Leh\'ebel}, \citenamefont {Ventagli},\ and\ \citenamefont
  {Sotiriou}}]{Antoniou:2021zoy}%
  \BibitemOpen
  \bibfield  {author} {\bibinfo {author} {\bibnamefont {Antoniou},
  \bibfnamefont {G.}}, \bibinfo {author} {\bibfnamefont {A.}~\bibnamefont
  {Leh\'ebel}}, \bibinfo {author} {\bibfnamefont {G.}~\bibnamefont {Ventagli}},
  and\ \bibinfo {author} {\bibfnamefont {T.~P.}\ \bibnamefont {Sotiriou}}}
  (\bibinfo {year} {2021}{\natexlab{b}}),\ \href
  {https://doi.org/10.1103/PhysRevD.104.044002} {\bibfield  {journal} {\bibinfo
   {journal} {Phys. Rev. D}\ }\textbf {\bibinfo {volume} {104}}~(\bibinfo
  {number} {4}),\ \bibinfo {pages} {044002}},\ \Eprint
  {https://arxiv.org/abs/2105.04479} {arXiv:2105.04479 [gr-qc]} \BibitemShut
  {NoStop}%
\bibitem [{\citenamefont {Antoniou}\ \emph {et~al.}(2022)\citenamefont
  {Antoniou}, \citenamefont {Macedo}, \citenamefont {McManus},\ and\
  \citenamefont {Sotiriou}}]{Antoniou:2022agj}%
  \BibitemOpen
  \bibfield  {author} {\bibinfo {author} {\bibnamefont {Antoniou},
  \bibfnamefont {G.}}, \bibinfo {author} {\bibfnamefont {C.~F.~B.}\
  \bibnamefont {Macedo}}, \bibinfo {author} {\bibfnamefont {R.}~\bibnamefont
  {McManus}}, and\ \bibinfo {author} {\bibfnamefont {T.~P.}\ \bibnamefont
  {Sotiriou}}} (\bibinfo {year} {2022}),\ \href
  {https://doi.org/10.1103/PhysRevD.106.024029} {\bibfield  {journal} {\bibinfo
   {journal} {Phys. Rev. D}\ }\textbf {\bibinfo {volume} {106}}~(\bibinfo
  {number} {2}),\ \bibinfo {pages} {024029}},\ \Eprint
  {https://arxiv.org/abs/2204.01684} {arXiv:2204.01684 [gr-qc]} \BibitemShut
  {NoStop}%
\bibitem [{\citenamefont {Arun}\ \emph {et~al.}(2022)\citenamefont {Arun} \emph
  {et~al.}}]{LISA:2022kgy}%
  \BibitemOpen
  \bibfield  {author} {\bibinfo {author} {\bibnamefont {Arun}, \bibfnamefont
  {K.~G.}},  \emph {et~al.} (\bibinfo {collaboration} {LISA})} (\bibinfo {year}
  {2022}),\ \href@noop {} {\enquote {\bibinfo {title} {{New Horizons for
  Fundamental Physics with LISA}},}\ }\Eprint
  {https://arxiv.org/abs/2205.01597} {arXiv:2205.01597 [gr-qc]} \BibitemShut
  {NoStop}%
\bibitem [{\citenamefont {{Arzoumanian}}\ \emph {et~al.}(2014)\citenamefont
  {{Arzoumanian}} \emph {et~al.}}]{Arzoumanian2014:SPIE}%
  \BibitemOpen
  \bibfield  {author} {\bibinfo {author} {\bibnamefont {{Arzoumanian}},
  \bibfnamefont {Z.}},  \emph {et~al.}} (\bibinfo {year} {2014}),\ in\ \href
  {https://doi.org/10.1117/12.2056811} {\emph {\bibinfo {booktitle} {Space
  Telescopes and Instrumentation 2014: Ultraviolet to Gamma Ray}}},\ \bibinfo
  {series} {Proc.~SPIE}, Vol.\ \bibinfo {volume} {9144},\ p.\ \bibinfo {pages}
  {914420}\BibitemShut {NoStop}%
\bibitem [{\citenamefont {Astefanesei}\ \emph {et~al.}(2019)\citenamefont
  {Astefanesei}, \citenamefont {Herdeiro}, \citenamefont {Pombo},\ and\
  \citenamefont {Radu}}]{Astefanesei:2019pfq}%
  \BibitemOpen
  \bibfield  {author} {\bibinfo {author} {\bibnamefont {Astefanesei},
  \bibfnamefont {D.}}, \bibinfo {author} {\bibfnamefont {C.}~\bibnamefont
  {Herdeiro}}, \bibinfo {author} {\bibfnamefont {A.}~\bibnamefont {Pombo}},
  and\ \bibinfo {author} {\bibfnamefont {E.}~\bibnamefont {Radu}}} (\bibinfo
  {year} {2019}),\ \href {https://doi.org/10.1007/JHEP10(2019)078} {\bibfield
  {journal} {\bibinfo  {journal} {JHEP}\ }\textbf {\bibinfo {volume} {10}},\
  \bibinfo {pages} {078}},\ \Eprint {https://arxiv.org/abs/1905.08304}
  {arXiv:1905.08304 [hep-th]} \BibitemShut {NoStop}%
\bibitem [{\citenamefont {Azri}\ and\ \citenamefont
  {Nasri}(2021)}]{Azri:2020agc}%
  \BibitemOpen
  \bibfield  {author} {\bibinfo {author} {\bibnamefont {Azri}, \bibfnamefont
  {H.}}, and\ \bibinfo {author} {\bibfnamefont {S.}~\bibnamefont {Nasri}}}
  (\bibinfo {year} {2021}),\ \href
  {https://doi.org/10.1103/PhysRevD.103.024035} {\bibfield  {journal} {\bibinfo
   {journal} {Phys. Rev. D}\ }\textbf {\bibinfo {volume} {103}}~(\bibinfo
  {number} {2}),\ \bibinfo {pages} {024035}},\ \Eprint
  {https://arxiv.org/abs/2012.04694} {arXiv:2012.04694 [gr-qc]} \BibitemShut
  {NoStop}%
\bibitem [{\citenamefont {Baker}\ \emph {et~al.}(2006)\citenamefont {Baker},
  \citenamefont {Centrella}, \citenamefont {Choi}, \citenamefont {Koppitz},\
  and\ \citenamefont {van Meter}}]{Baker:2005vv}%
  \BibitemOpen
  \bibfield  {author} {\bibinfo {author} {\bibnamefont {Baker}, \bibfnamefont
  {J.~G.}}, \bibinfo {author} {\bibfnamefont {J.}~\bibnamefont {Centrella}},
  \bibinfo {author} {\bibfnamefont {D.-I.}\ \bibnamefont {Choi}}, \bibinfo
  {author} {\bibfnamefont {M.}~\bibnamefont {Koppitz}}, and\ \bibinfo {author}
  {\bibfnamefont {J.}~\bibnamefont {van Meter}}} (\bibinfo {year} {2006}),\
  \href {https://doi.org/10.1103/PhysRevLett.96.111102} {\bibfield  {journal}
  {\bibinfo  {journal} {Phys. Rev. Lett.}\ }\textbf {\bibinfo {volume} {96}},\
  \bibinfo {pages} {111102}},\ \Eprint {https://arxiv.org/abs/gr-qc/0511103}
  {arXiv:gr-qc/0511103} \BibitemShut {NoStop}%
\bibitem [{\citenamefont {Barack}\ \emph {et~al.}(2019)\citenamefont {Barack}
  \emph {et~al.}}]{Barack:2018yly}%
  \BibitemOpen
  \bibfield  {author} {\bibinfo {author} {\bibnamefont {Barack}, \bibfnamefont
  {L.}},  \emph {et~al.}} (\bibinfo {year} {2019}),\ \href
  {https://doi.org/10.1088/1361-6382/ab0587} {\bibfield  {journal} {\bibinfo
  {journal} {Class. Quant. Grav.}\ }\textbf {\bibinfo {volume} {36}}~(\bibinfo
  {number} {14}),\ \bibinfo {pages} {143001}},\ \Eprint
  {https://arxiv.org/abs/1806.05195} {arXiv:1806.05195 [gr-qc]} \BibitemShut
  {NoStop}%
\bibitem [{\citenamefont {Barausse}\ \emph {et~al.}(2013)\citenamefont
  {Barausse}, \citenamefont {Palenzuela}, \citenamefont {Ponce},\ and\
  \citenamefont {Lehner}}]{Barausse:2012da}%
  \BibitemOpen
  \bibfield  {author} {\bibinfo {author} {\bibnamefont {Barausse},
  \bibfnamefont {E.}}, \bibinfo {author} {\bibfnamefont {C.}~\bibnamefont
  {Palenzuela}}, \bibinfo {author} {\bibfnamefont {M.}~\bibnamefont {Ponce}},
  and\ \bibinfo {author} {\bibfnamefont {L.}~\bibnamefont {Lehner}}} (\bibinfo
  {year} {2013}),\ \href {https://doi.org/10.1103/PhysRevD.87.081506}
  {\bibfield  {journal} {\bibinfo  {journal} {Phys. Rev. D}\ }\textbf {\bibinfo
  {volume} {87}},\ \bibinfo {pages} {081506}},\ \Eprint
  {https://arxiv.org/abs/1212.5053} {arXiv:1212.5053 [gr-qc]} \BibitemShut
  {NoStop}%
\bibitem [{\citenamefont {Barausse}\ \emph {et~al.}(2020)\citenamefont
  {Barausse} \emph {et~al.}}]{Barausse:2020rsu}%
  \BibitemOpen
  \bibfield  {author} {\bibinfo {author} {\bibnamefont {Barausse},
  \bibfnamefont {E.}},  \emph {et~al.}} (\bibinfo {year} {2020}),\ \href
  {https://doi.org/10.1007/s10714-020-02691-1} {\bibfield  {journal} {\bibinfo
  {journal} {Gen. Rel. Grav.}\ }\textbf {\bibinfo {volume} {52}}~(\bibinfo
  {number} {8}),\ \bibinfo {pages} {81}},\ \Eprint
  {https://arxiv.org/abs/2001.09793} {arXiv:2001.09793 [gr-qc]} \BibitemShut
  {NoStop}%
\bibitem [{\citenamefont {Barton}\ \emph {et~al.}(2021)\citenamefont {Barton},
  \citenamefont {Hartmann}, \citenamefont {Kleihaus},\ and\ \citenamefont
  {Kunz}}]{Barton:2021wfj}%
  \BibitemOpen
  \bibfield  {author} {\bibinfo {author} {\bibnamefont {Barton}, \bibfnamefont
  {S.}}, \bibinfo {author} {\bibfnamefont {B.}~\bibnamefont {Hartmann}},
  \bibinfo {author} {\bibfnamefont {B.}~\bibnamefont {Kleihaus}}, and\ \bibinfo
  {author} {\bibfnamefont {J.}~\bibnamefont {Kunz}}} (\bibinfo {year} {2021}),\
  \href {https://doi.org/10.1016/j.physletb.2021.136336} {\bibfield  {journal}
  {\bibinfo  {journal} {Phys. Lett. B}\ }\textbf {\bibinfo {volume} {817}},\
  \bibinfo {pages} {136336}},\ \Eprint {https://arxiv.org/abs/2103.01651}
  {arXiv:2103.01651 [gr-qc]} \BibitemShut {NoStop}%
\bibitem [{\citenamefont {Bauswein}\ \emph {et~al.}(2013)\citenamefont
  {Bauswein}, \citenamefont {Baumgarte},\ and\ \citenamefont
  {Janka}}]{Bauswein:2013jpa}%
  \BibitemOpen
  \bibfield  {author} {\bibinfo {author} {\bibnamefont {Bauswein},
  \bibfnamefont {A.}}, \bibinfo {author} {\bibfnamefont {T.~W.}\ \bibnamefont
  {Baumgarte}}, and\ \bibinfo {author} {\bibfnamefont {H.~T.}\ \bibnamefont
  {Janka}}} (\bibinfo {year} {2013}),\ \href
  {https://doi.org/10.1103/PhysRevLett.111.131101} {\bibfield  {journal}
  {\bibinfo  {journal} {Phys. Rev. Lett.}\ }\textbf {\bibinfo {volume}
  {111}}~(\bibinfo {number} {13}),\ \bibinfo {pages} {131101}},\ \Eprint
  {https://arxiv.org/abs/1307.5191} {arXiv:1307.5191 [astro-ph.SR]}
  \BibitemShut {NoStop}%
\bibitem [{\citenamefont {Bauswein}\ and\ \citenamefont
  {Janka}(2012)}]{Bauswein:2011tp}%
  \BibitemOpen
  \bibfield  {author} {\bibinfo {author} {\bibnamefont {Bauswein},
  \bibfnamefont {A.}}, and\ \bibinfo {author} {\bibfnamefont {H.~T.}\
  \bibnamefont {Janka}}} (\bibinfo {year} {2012}),\ \href
  {https://doi.org/10.1103/PhysRevLett.108.011101} {\bibfield  {journal}
  {\bibinfo  {journal} {Phys. Rev. Lett.}\ }\textbf {\bibinfo {volume} {108}},\
  \bibinfo {pages} {011101}},\ \Eprint {https://arxiv.org/abs/1106.1616}
  {arXiv:1106.1616 [astro-ph.SR]} \BibitemShut {NoStop}%
\bibitem [{\citenamefont {Bauswein}\ \emph {et~al.}(2012)\citenamefont
  {Bauswein}, \citenamefont {Janka}, \citenamefont {Hebeler},\ and\
  \citenamefont {Schwenk}}]{Bauswein:2012ya}%
  \BibitemOpen
  \bibfield  {author} {\bibinfo {author} {\bibnamefont {Bauswein},
  \bibfnamefont {A.}}, \bibinfo {author} {\bibfnamefont {H.~T.}\ \bibnamefont
  {Janka}}, \bibinfo {author} {\bibfnamefont {K.}~\bibnamefont {Hebeler}}, and\
  \bibinfo {author} {\bibfnamefont {A.}~\bibnamefont {Schwenk}}} (\bibinfo
  {year} {2012}),\ \href {https://doi.org/10.1103/PhysRevD.86.063001}
  {\bibfield  {journal} {\bibinfo  {journal} {Phys. Rev. D}\ }\textbf {\bibinfo
  {volume} {86}},\ \bibinfo {pages} {063001}},\ \Eprint
  {https://arxiv.org/abs/1204.1888} {arXiv:1204.1888 [astro-ph.SR]}
  \BibitemShut {NoStop}%
\bibitem [{\citenamefont {Bauswein}\ and\ \citenamefont
  {Stergioulas}(2017)}]{Bauswein:2017aur}%
  \BibitemOpen
  \bibfield  {author} {\bibinfo {author} {\bibnamefont {Bauswein},
  \bibfnamefont {A.}}, and\ \bibinfo {author} {\bibfnamefont {N.}~\bibnamefont
  {Stergioulas}}} (\bibinfo {year} {2017}),\ \href
  {https://doi.org/10.1093/mnras/stx1983} {\bibfield  {journal} {\bibinfo
  {journal} {Mon. Not. Roy. Astron. Soc.}\ }\textbf {\bibinfo {volume}
  {471}}~(\bibinfo {number} {4}),\ \bibinfo {pages} {4956}},\ \Eprint
  {https://arxiv.org/abs/1702.02567} {arXiv:1702.02567 [astro-ph.HE]}
  \BibitemShut {NoStop}%
\bibitem [{\citenamefont {Bauswein}\ \emph {et~al.}(2014)\citenamefont
  {Bauswein}, \citenamefont {Stergioulas},\ and\ \citenamefont
  {Janka}}]{Bauswein:2014qla}%
  \BibitemOpen
  \bibfield  {author} {\bibinfo {author} {\bibnamefont {Bauswein},
  \bibfnamefont {A.}}, \bibinfo {author} {\bibfnamefont {N.}~\bibnamefont
  {Stergioulas}}, and\ \bibinfo {author} {\bibfnamefont {H.~T.}\ \bibnamefont
  {Janka}}} (\bibinfo {year} {2014}),\ \href
  {https://doi.org/10.1103/PhysRevD.90.023002} {\bibfield  {journal} {\bibinfo
  {journal} {Phys. Rev. D}\ }\textbf {\bibinfo {volume} {90}}~(\bibinfo
  {number} {2}),\ \bibinfo {pages} {023002}},\ \Eprint
  {https://arxiv.org/abs/1403.5301} {arXiv:1403.5301 [astro-ph.SR]}
  \BibitemShut {NoStop}%
\bibitem [{\citenamefont {Baym}\ \emph {et~al.}(2018)\citenamefont {Baym},
  \citenamefont {Hatsuda}, \citenamefont {Kojo}, \citenamefont {Powell},
  \citenamefont {Song},\ and\ \citenamefont {Takatsuka}}]{Baym:2017whm}%
  \BibitemOpen
  \bibfield  {author} {\bibinfo {author} {\bibnamefont {Baym}, \bibfnamefont
  {G.}}, \bibinfo {author} {\bibfnamefont {T.}~\bibnamefont {Hatsuda}},
  \bibinfo {author} {\bibfnamefont {T.}~\bibnamefont {Kojo}}, \bibinfo {author}
  {\bibfnamefont {P.~D.}\ \bibnamefont {Powell}}, \bibinfo {author}
  {\bibfnamefont {Y.}~\bibnamefont {Song}}, and\ \bibinfo {author}
  {\bibfnamefont {T.}~\bibnamefont {Takatsuka}}} (\bibinfo {year} {2018}),\
  \href {https://doi.org/10.1088/1361-6633/aaae14} {\bibfield  {journal}
  {\bibinfo  {journal} {Rept. Prog. Phys.}\ }\textbf {\bibinfo {volume}
  {81}}~(\bibinfo {number} {5}),\ \bibinfo {pages} {056902}},\ \Eprint
  {https://arxiv.org/abs/1707.04966} {arXiv:1707.04966 [astro-ph.HE]}
  \BibitemShut {NoStop}%
\bibitem [{\citenamefont {Bekenstein}(1993)}]{Bekenstein:1992pj}%
  \BibitemOpen
  \bibfield  {author} {\bibinfo {author} {\bibnamefont {Bekenstein},
  \bibfnamefont {J.~D.}}} (\bibinfo {year} {1993}),\ \href
  {https://doi.org/10.1103/PhysRevD.48.3641} {\bibfield  {journal} {\bibinfo
  {journal} {Phys. Rev. D}\ }\textbf {\bibinfo {volume} {48}},\ \bibinfo
  {pages} {3641}},\ \Eprint {https://arxiv.org/abs/gr-qc/9211017}
  {arXiv:gr-qc/9211017} \BibitemShut {NoStop}%
\bibitem [{\citenamefont {Beltran~Jimenez}\ \emph {et~al.}(2013)\citenamefont
  {Beltran~Jimenez}, \citenamefont {Delvas~Froes},\ and\ \citenamefont
  {Mota}}]{BeltranJimenez:2012kby}%
  \BibitemOpen
  \bibfield  {author} {\bibinfo {author} {\bibnamefont {Beltran~Jimenez},
  \bibfnamefont {J.}}, \bibinfo {author} {\bibfnamefont {A.~L.}\ \bibnamefont
  {Delvas~Froes}}, and\ \bibinfo {author} {\bibfnamefont {D.~F.}\ \bibnamefont
  {Mota}}} (\bibinfo {year} {2013}),\ \href
  {https://doi.org/10.1016/j.physletb.2013.07.032} {\bibfield  {journal}
  {\bibinfo  {journal} {Phys. Lett. B}\ }\textbf {\bibinfo {volume} {725}},\
  \bibinfo {pages} {212}},\ \Eprint {https://arxiv.org/abs/1212.1923}
  {arXiv:1212.1923 [astro-ph.CO]} \BibitemShut {NoStop}%
\bibitem [{\citenamefont {Berti}\ \emph {et~al.}(2021)\citenamefont {Berti},
  \citenamefont {Collodel}, \citenamefont {Kleihaus},\ and\ \citenamefont
  {Kunz}}]{Berti:2020kgk}%
  \BibitemOpen
  \bibfield  {author} {\bibinfo {author} {\bibnamefont {Berti}, \bibfnamefont
  {E.}}, \bibinfo {author} {\bibfnamefont {L.~G.}\ \bibnamefont {Collodel}},
  \bibinfo {author} {\bibfnamefont {B.}~\bibnamefont {Kleihaus}}, and\ \bibinfo
  {author} {\bibfnamefont {J.}~\bibnamefont {Kunz}}} (\bibinfo {year} {2021}),\
  \href {https://doi.org/10.1103/PhysRevLett.126.011104} {\bibfield  {journal}
  {\bibinfo  {journal} {Phys. Rev. Lett.}\ }\textbf {\bibinfo {volume}
  {126}}~(\bibinfo {number} {1}),\ \bibinfo {pages} {011104}},\ \Eprint
  {https://arxiv.org/abs/2009.03905} {arXiv:2009.03905 [gr-qc]} \BibitemShut
  {NoStop}%
\bibitem [{\citenamefont {Berti}\ \emph {et~al.}(2015)\citenamefont {Berti}
  \emph {et~al.}}]{Berti:2015itd}%
  \BibitemOpen
  \bibfield  {author} {\bibinfo {author} {\bibnamefont {Berti}, \bibfnamefont
  {E.}},  \emph {et~al.}} (\bibinfo {year} {2015}),\ \href
  {https://doi.org/10.1088/0264-9381/32/24/243001} {\bibfield  {journal}
  {\bibinfo  {journal} {Class. Quant. Grav.}\ }\textbf {\bibinfo {volume}
  {32}},\ \bibinfo {pages} {243001}},\ \Eprint
  {https://arxiv.org/abs/1501.07274} {arXiv:1501.07274 [gr-qc]} \BibitemShut
  {NoStop}%
\bibitem [{\citenamefont {Bettoni}\ and\ \citenamefont
  {Liberati}(2013)}]{Bettoni:2013diz}%
  \BibitemOpen
  \bibfield  {author} {\bibinfo {author} {\bibnamefont {Bettoni}, \bibfnamefont
  {D.}}, and\ \bibinfo {author} {\bibfnamefont {S.}~\bibnamefont {Liberati}}}
  (\bibinfo {year} {2013}),\ \href {https://doi.org/10.1103/PhysRevD.88.084020}
  {\bibfield  {journal} {\bibinfo  {journal} {Phys. Rev. D}\ }\textbf {\bibinfo
  {volume} {88}},\ \bibinfo {pages} {084020}},\ \Eprint
  {https://arxiv.org/abs/1306.6724} {arXiv:1306.6724 [gr-qc]} \BibitemShut
  {NoStop}%
\bibitem [{\citenamefont {Bl\'azquez-Salcedo}\ \emph
  {et~al.}(2020{\natexlab{a}})\citenamefont {Bl\'azquez-Salcedo}, \citenamefont
  {Doneva}, \citenamefont {Kahlen}, \citenamefont {Kunz}, \citenamefont
  {Nedkova},\ and\ \citenamefont {Yazadjiev}}]{Blazquez-Salcedo:2020rhf}%
  \BibitemOpen
  \bibfield  {author} {\bibinfo {author} {\bibnamefont {Bl\'azquez-Salcedo},
  \bibfnamefont {J.~L.}}, \bibinfo {author} {\bibfnamefont {D.~D.}\
  \bibnamefont {Doneva}}, \bibinfo {author} {\bibfnamefont {S.}~\bibnamefont
  {Kahlen}}, \bibinfo {author} {\bibfnamefont {J.}~\bibnamefont {Kunz}},
  \bibinfo {author} {\bibfnamefont {P.}~\bibnamefont {Nedkova}}, and\ \bibinfo
  {author} {\bibfnamefont {S.~S.}\ \bibnamefont {Yazadjiev}}} (\bibinfo {year}
  {2020}{\natexlab{a}}),\ \href {https://doi.org/10.1103/PhysRevD.101.104006}
  {\bibfield  {journal} {\bibinfo  {journal} {Phys. Rev. D}\ }\textbf {\bibinfo
  {volume} {101}}~(\bibinfo {number} {10}),\ \bibinfo {pages} {104006}},\
  \Eprint {https://arxiv.org/abs/2003.02862} {arXiv:2003.02862 [gr-qc]}
  \BibitemShut {NoStop}%
\bibitem [{\citenamefont {Bl\'azquez-Salcedo}\ \emph
  {et~al.}(2020{\natexlab{b}})\citenamefont {Bl\'azquez-Salcedo}, \citenamefont
  {Doneva}, \citenamefont {Kahlen}, \citenamefont {Kunz}, \citenamefont
  {Nedkova},\ and\ \citenamefont {Yazadjiev}}]{Blazquez-Salcedo:2020caw}%
  \BibitemOpen
  \bibfield  {author} {\bibinfo {author} {\bibnamefont {Bl\'azquez-Salcedo},
  \bibfnamefont {J.~L.}}, \bibinfo {author} {\bibfnamefont {D.~D.}\
  \bibnamefont {Doneva}}, \bibinfo {author} {\bibfnamefont {S.}~\bibnamefont
  {Kahlen}}, \bibinfo {author} {\bibfnamefont {J.}~\bibnamefont {Kunz}},
  \bibinfo {author} {\bibfnamefont {P.}~\bibnamefont {Nedkova}}, and\ \bibinfo
  {author} {\bibfnamefont {S.~S.}\ \bibnamefont {Yazadjiev}}} (\bibinfo {year}
  {2020}{\natexlab{b}}),\ \href {https://doi.org/10.1103/PhysRevD.102.024086}
  {\bibfield  {journal} {\bibinfo  {journal} {Phys. Rev. D}\ }\textbf {\bibinfo
  {volume} {102}}~(\bibinfo {number} {2}),\ \bibinfo {pages} {024086}},\
  \Eprint {https://arxiv.org/abs/2006.06006} {arXiv:2006.06006 [gr-qc]}
  \BibitemShut {NoStop}%
\bibitem [{\citenamefont {Bl\'azquez-Salcedo}\ \emph
  {et~al.}(2018)\citenamefont {Bl\'azquez-Salcedo}, \citenamefont {Doneva},
  \citenamefont {Kunz},\ and\ \citenamefont
  {Yazadjiev}}]{Blazquez-Salcedo:2018jnn}%
  \BibitemOpen
  \bibfield  {author} {\bibinfo {author} {\bibnamefont {Bl\'azquez-Salcedo},
  \bibfnamefont {J.~L.}}, \bibinfo {author} {\bibfnamefont {D.~D.}\
  \bibnamefont {Doneva}}, \bibinfo {author} {\bibfnamefont {J.}~\bibnamefont
  {Kunz}}, and\ \bibinfo {author} {\bibfnamefont {S.~S.}\ \bibnamefont
  {Yazadjiev}}} (\bibinfo {year} {2018}),\ \href
  {https://doi.org/10.1103/PhysRevD.98.084011} {\bibfield  {journal} {\bibinfo
  {journal} {Phys. Rev. D}\ }\textbf {\bibinfo {volume} {98}}~(\bibinfo
  {number} {8}),\ \bibinfo {pages} {084011}},\ \Eprint
  {https://arxiv.org/abs/1805.05755} {arXiv:1805.05755 [gr-qc]} \BibitemShut
  {NoStop}%
\bibitem [{\citenamefont {Bl\'azquez-Salcedo}\ \emph
  {et~al.}(2022)\citenamefont {Bl\'azquez-Salcedo}, \citenamefont {Doneva},
  \citenamefont {Kunz},\ and\ \citenamefont
  {Yazadjiev}}]{Blazquez-Salcedo:2022omw}%
  \BibitemOpen
  \bibfield  {author} {\bibinfo {author} {\bibnamefont {Bl\'azquez-Salcedo},
  \bibfnamefont {J.~L.}}, \bibinfo {author} {\bibfnamefont {D.~D.}\
  \bibnamefont {Doneva}}, \bibinfo {author} {\bibfnamefont {J.}~\bibnamefont
  {Kunz}}, and\ \bibinfo {author} {\bibfnamefont {S.~S.}\ \bibnamefont
  {Yazadjiev}}} (\bibinfo {year} {2022}),\ \href
  {https://doi.org/10.1103/PhysRevD.105.124005} {\bibfield  {journal} {\bibinfo
   {journal} {Phys. Rev. D}\ }\textbf {\bibinfo {volume} {105}}~(\bibinfo
  {number} {12}),\ \bibinfo {pages} {124005}},\ \Eprint
  {https://arxiv.org/abs/2203.00709} {arXiv:2203.00709 [gr-qc]} \BibitemShut
  {NoStop}%
\bibitem [{\citenamefont {Bl\'azquez-Salcedo}\ \emph
  {et~al.}(2021)\citenamefont {Bl\'azquez-Salcedo}, \citenamefont {Herdeiro},
  \citenamefont {Kahlen}, \citenamefont {Kunz}, \citenamefont {Pombo},\ and\
  \citenamefont {Radu}}]{Blazquez-Salcedo:2020jee}%
  \BibitemOpen
  \bibfield  {author} {\bibinfo {author} {\bibnamefont {Bl\'azquez-Salcedo},
  \bibfnamefont {J.~L.}}, \bibinfo {author} {\bibfnamefont {C.~A.~R.}\
  \bibnamefont {Herdeiro}}, \bibinfo {author} {\bibfnamefont {S.}~\bibnamefont
  {Kahlen}}, \bibinfo {author} {\bibfnamefont {J.}~\bibnamefont {Kunz}},
  \bibinfo {author} {\bibfnamefont {A.~M.}\ \bibnamefont {Pombo}}, and\
  \bibinfo {author} {\bibfnamefont {E.}~\bibnamefont {Radu}}} (\bibinfo {year}
  {2021}),\ \href {https://doi.org/10.1140/epjc/s10052-021-08952-w} {\bibfield
  {journal} {\bibinfo  {journal} {Eur. Phys. J. C}\ }\textbf {\bibinfo {volume}
  {81}}~(\bibinfo {number} {2}),\ \bibinfo {pages} {155}},\ \Eprint
  {https://arxiv.org/abs/2008.11744} {arXiv:2008.11744 [gr-qc]} \BibitemShut
  {NoStop}%
\bibitem [{\citenamefont {Bl\'azquez-Salcedo}\ \emph
  {et~al.}(2020{\natexlab{c}})\citenamefont {Bl\'azquez-Salcedo}, \citenamefont
  {Herdeiro}, \citenamefont {Kunz}, \citenamefont {Pombo},\ and\ \citenamefont
  {Radu}}]{Blazquez-Salcedo:2020nhs}%
  \BibitemOpen
  \bibfield  {author} {\bibinfo {author} {\bibnamefont {Bl\'azquez-Salcedo},
  \bibfnamefont {J.~L.}}, \bibinfo {author} {\bibfnamefont {C.~A.~R.}\
  \bibnamefont {Herdeiro}}, \bibinfo {author} {\bibfnamefont {J.}~\bibnamefont
  {Kunz}}, \bibinfo {author} {\bibfnamefont {A.~M.}\ \bibnamefont {Pombo}},
  and\ \bibinfo {author} {\bibfnamefont {E.}~\bibnamefont {Radu}}} (\bibinfo
  {year} {2020}{\natexlab{c}}),\ \href
  {https://doi.org/10.1016/j.physletb.2020.135493} {\bibfield  {journal}
  {\bibinfo  {journal} {Phys. Lett. B}\ }\textbf {\bibinfo {volume} {806}},\
  \bibinfo {pages} {135493}},\ \Eprint {https://arxiv.org/abs/2002.00963}
  {arXiv:2002.00963 [gr-qc]} \BibitemShut {NoStop}%
\bibitem [{\citenamefont {Bl\'azquez-Salcedo}\ \emph
  {et~al.}(2020{\natexlab{d}})\citenamefont {Bl\'azquez-Salcedo}, \citenamefont
  {Scen~Khoo},\ and\ \citenamefont {Kunz}}]{Blazquez-Salcedo:2020ibb}%
  \BibitemOpen
  \bibfield  {author} {\bibinfo {author} {\bibnamefont {Bl\'azquez-Salcedo},
  \bibfnamefont {J.~L.}}, \bibinfo {author} {\bibfnamefont {F.}~\bibnamefont
  {Scen~Khoo}}, and\ \bibinfo {author} {\bibfnamefont {J.}~\bibnamefont
  {Kunz}}} (\bibinfo {year} {2020}{\natexlab{d}}),\ \href
  {https://doi.org/10.1209/0295-5075/130/50002} {\bibfield  {journal} {\bibinfo
   {journal} {EPL}\ }\textbf {\bibinfo {volume} {130}}~(\bibinfo {number}
  {5}),\ \bibinfo {pages} {50002}},\ \Eprint {https://arxiv.org/abs/2001.09117}
  {arXiv:2001.09117 [gr-qc]} \BibitemShut {NoStop}%
\bibitem [{\citenamefont {Bogdanov}\ \emph {et~al.}(2016)\citenamefont
  {Bogdanov}, \citenamefont {Heinke}, \citenamefont {\"Ozel},\ and\
  \citenamefont {G\"uver}}]{Bogdanov:2016nle}%
  \BibitemOpen
  \bibfield  {author} {\bibinfo {author} {\bibnamefont {Bogdanov},
  \bibfnamefont {S.}}, \bibinfo {author} {\bibfnamefont {C.~O.}\ \bibnamefont
  {Heinke}}, \bibinfo {author} {\bibfnamefont {F.}~\bibnamefont {\"Ozel}}, and\
  \bibinfo {author} {\bibfnamefont {T.}~\bibnamefont {G\"uver}}} (\bibinfo
  {year} {2016}),\ \href {https://doi.org/10.3847/0004-637X/831/2/184}
  {\bibfield  {journal} {\bibinfo  {journal} {Astrophys. J.}\ }\textbf
  {\bibinfo {volume} {831}}~(\bibinfo {number} {2}),\ \bibinfo {pages} {184}},\
  \Eprint {https://arxiv.org/abs/1603.01630} {arXiv:1603.01630 [astro-ph.HE]}
  \BibitemShut {NoStop}%
\bibitem [{\citenamefont {Bogdanov}\ \emph
  {et~al.}(2019{\natexlab{a}})\citenamefont {Bogdanov} \emph
  {et~al.}}]{Bogdanov:2019ixe}%
  \BibitemOpen
  \bibfield  {author} {\bibinfo {author} {\bibnamefont {Bogdanov},
  \bibfnamefont {S.}},  \emph {et~al.}} (\bibinfo {year}
  {2019}{\natexlab{a}}),\ \href {https://doi.org/10.3847/2041-8213/ab53eb}
  {\bibfield  {journal} {\bibinfo  {journal} {Astrophys. J. Lett.}\ }\textbf
  {\bibinfo {volume} {887}}~(\bibinfo {number} {1}),\ \bibinfo {pages} {L25}},\
  \Eprint {https://arxiv.org/abs/1912.05706} {arXiv:1912.05706 [astro-ph.HE]}
  \BibitemShut {NoStop}%
\bibitem [{\citenamefont {Bogdanov}\ \emph
  {et~al.}(2019{\natexlab{b}})\citenamefont {Bogdanov} \emph
  {et~al.}}]{Bogdanov:2019qjb}%
  \BibitemOpen
  \bibfield  {author} {\bibinfo {author} {\bibnamefont {Bogdanov},
  \bibfnamefont {S.}},  \emph {et~al.}} (\bibinfo {year}
  {2019}{\natexlab{b}}),\ \href {https://doi.org/10.3847/2041-8213/ab5968}
  {\bibfield  {journal} {\bibinfo  {journal} {Astrophys. J. Lett.}\ }\textbf
  {\bibinfo {volume} {887}}~(\bibinfo {number} {1}),\ \bibinfo {pages} {L26}},\
  \Eprint {https://arxiv.org/abs/1912.05707} {arXiv:1912.05707 [astro-ph.HE]}
  \BibitemShut {NoStop}%
\bibitem [{\citenamefont {Bogdanov}\ \emph {et~al.}(2021)\citenamefont
  {Bogdanov} \emph {et~al.}}]{Bogdanov:2021yip}%
  \BibitemOpen
  \bibfield  {author} {\bibinfo {author} {\bibnamefont {Bogdanov},
  \bibfnamefont {S.}},  \emph {et~al.}} (\bibinfo {year} {2021}),\ \href
  {https://doi.org/10.3847/2041-8213/abfb79} {\bibfield  {journal} {\bibinfo
  {journal} {Astrophys. J. Lett.}\ }\textbf {\bibinfo {volume} {914}}~(\bibinfo
  {number} {1}),\ \bibinfo {pages} {L15}},\ \Eprint
  {https://arxiv.org/abs/2104.06928} {arXiv:2104.06928 [astro-ph.HE]}
  \BibitemShut {NoStop}%
\bibitem [{\citenamefont {Boulware}\ and\ \citenamefont
  {Deser}(1972)}]{Boulware:1972yco}%
  \BibitemOpen
  \bibfield  {author} {\bibinfo {author} {\bibnamefont {Boulware},
  \bibfnamefont {D.~G.}}, and\ \bibinfo {author} {\bibfnamefont
  {S.}~\bibnamefont {Deser}}} (\bibinfo {year} {1972}),\ \href
  {https://doi.org/10.1103/PhysRevD.6.3368} {\bibfield  {journal} {\bibinfo
  {journal} {Phys. Rev. D}\ }\textbf {\bibinfo {volume} {6}},\ \bibinfo {pages}
  {3368}}\BibitemShut {NoStop}%
\bibitem [{\citenamefont {Breu}\ and\ \citenamefont
  {Rezzolla}(2016)}]{Breu:2016ufb}%
  \BibitemOpen
  \bibfield  {author} {\bibinfo {author} {\bibnamefont {Breu}, \bibfnamefont
  {C.}}, and\ \bibinfo {author} {\bibfnamefont {L.}~\bibnamefont {Rezzolla}}}
  (\bibinfo {year} {2016}),\ \href {https://doi.org/10.1093/mnras/stw575}
  {\bibfield  {journal} {\bibinfo  {journal} {Mon. Not. Roy. Astron. Soc.}\
  }\textbf {\bibinfo {volume} {459}}~(\bibinfo {number} {1}),\ \bibinfo {pages}
  {646}},\ \Eprint {https://arxiv.org/abs/1601.06083} {arXiv:1601.06083
  [gr-qc]} \BibitemShut {NoStop}%
\bibitem [{\citenamefont {Brihaye}\ and\ \citenamefont
  {Hartmann}(2019)}]{Brihaye:2019kvj}%
  \BibitemOpen
  \bibfield  {author} {\bibinfo {author} {\bibnamefont {Brihaye}, \bibfnamefont
  {Y.}}, and\ \bibinfo {author} {\bibfnamefont {B.}~\bibnamefont {Hartmann}}}
  (\bibinfo {year} {2019}),\ \href
  {https://doi.org/10.1016/j.physletb.2019.03.043} {\bibfield  {journal}
  {\bibinfo  {journal} {Phys. Lett. B}\ }\textbf {\bibinfo {volume} {792}},\
  \bibinfo {pages} {244}},\ \Eprint {https://arxiv.org/abs/1902.05760}
  {arXiv:1902.05760 [gr-qc]} \BibitemShut {NoStop}%
\bibitem [{\citenamefont {Brihaye}\ \emph {et~al.}(2022)\citenamefont
  {Brihaye}, \citenamefont {Hartmann}, \citenamefont {Kleihaus},\ and\
  \citenamefont {Kunz}}]{Brihaye:2021qvc}%
  \BibitemOpen
  \bibfield  {author} {\bibinfo {author} {\bibnamefont {Brihaye}, \bibfnamefont
  {Y.}}, \bibinfo {author} {\bibfnamefont {B.}~\bibnamefont {Hartmann}},
  \bibinfo {author} {\bibfnamefont {B.}~\bibnamefont {Kleihaus}}, and\ \bibinfo
  {author} {\bibfnamefont {J.}~\bibnamefont {Kunz}}} (\bibinfo {year} {2022}),\
  \href {https://doi.org/10.1103/PhysRevD.105.044050} {\bibfield  {journal}
  {\bibinfo  {journal} {Phys. Rev. D}\ }\textbf {\bibinfo {volume}
  {105}}~(\bibinfo {number} {4}),\ \bibinfo {pages} {044050}},\ \Eprint
  {https://arxiv.org/abs/2109.12345} {arXiv:2109.12345 [gr-qc]} \BibitemShut
  {NoStop}%
\bibitem [{\citenamefont {Brihaye}\ \emph {et~al.}(2019)\citenamefont
  {Brihaye}, \citenamefont {Herdeiro},\ and\ \citenamefont
  {Radu}}]{Brihaye:2018bgc}%
  \BibitemOpen
  \bibfield  {author} {\bibinfo {author} {\bibnamefont {Brihaye}, \bibfnamefont
  {Y.}}, \bibinfo {author} {\bibfnamefont {C.}~\bibnamefont {Herdeiro}}, and\
  \bibinfo {author} {\bibfnamefont {E.}~\bibnamefont {Radu}}} (\bibinfo {year}
  {2019}),\ \href {https://doi.org/10.1016/j.physletb.2018.11.022} {\bibfield
  {journal} {\bibinfo  {journal} {Phys. Lett. B}\ }\textbf {\bibinfo {volume}
  {788}},\ \bibinfo {pages} {295}},\ \Eprint {https://arxiv.org/abs/1810.09560}
  {arXiv:1810.09560 [gr-qc]} \BibitemShut {NoStop}%
\bibitem [{\citenamefont {Brihaye}\ and\ \citenamefont
  {Verbin}(2020)}]{Brihaye:2020oxh}%
  \BibitemOpen
  \bibfield  {author} {\bibinfo {author} {\bibnamefont {Brihaye}, \bibfnamefont
  {Y.}}, and\ \bibinfo {author} {\bibfnamefont {Y.}~\bibnamefont {Verbin}}}
  (\bibinfo {year} {2020}),\ \href
  {https://doi.org/10.1103/PhysRevD.102.124021} {\bibfield  {journal} {\bibinfo
   {journal} {Phys. Rev. D}\ }\textbf {\bibinfo {volume} {102}},\ \bibinfo
  {pages} {124021}},\ \Eprint {https://arxiv.org/abs/2004.01681}
  {arXiv:2004.01681 [gr-qc]} \BibitemShut {NoStop}%
\bibitem [{\citenamefont {Brito}\ \emph {et~al.}(2015)\citenamefont {Brito},
  \citenamefont {Cardoso},\ and\ \citenamefont {Pani}}]{Brito:2015oca}%
  \BibitemOpen
  \bibfield  {author} {\bibinfo {author} {\bibnamefont {Brito}, \bibfnamefont
  {R.}}, \bibinfo {author} {\bibfnamefont {V.}~\bibnamefont {Cardoso}}, and\
  \bibinfo {author} {\bibfnamefont {P.}~\bibnamefont {Pani}}} (\bibinfo {year}
  {2015}),\ \href {https://doi.org/10.1007/978-3-319-19000-6} {\bibfield
  {journal} {\bibinfo  {journal} {Lect. Notes Phys.}\ }\textbf {\bibinfo
  {volume} {906}},\ \bibinfo {pages} {pp.1}},\ \Eprint
  {https://arxiv.org/abs/1501.06570} {arXiv:1501.06570 [gr-qc]} \BibitemShut
  {NoStop}%
\bibitem [{\citenamefont {Bucciantini}\ and\ \citenamefont
  {Soldateschi}(2020)}]{Bucciantini:2020owm}%
  \BibitemOpen
  \bibfield  {author} {\bibinfo {author} {\bibnamefont {Bucciantini},
  \bibfnamefont {N.}}, and\ \bibinfo {author} {\bibfnamefont {J.}~\bibnamefont
  {Soldateschi}}} (\bibinfo {year} {2020}),\ \href
  {https://doi.org/10.1093/mnrasl/slaa059} {\bibfield  {journal} {\bibinfo
  {journal} {Mon. Not. Roy. Astron. Soc.}\ }\textbf {\bibinfo {volume}
  {495}}~(\bibinfo {number} {1}),\ \bibinfo {pages} {L56}},\ \Eprint
  {https://arxiv.org/abs/2004.00322} {arXiv:2004.00322 [astro-ph.HE]}
  \BibitemShut {NoStop}%
\bibitem [{\citenamefont {Campanelli}\ \emph {et~al.}(2006)\citenamefont
  {Campanelli}, \citenamefont {Lousto}, \citenamefont {Marronetti},\ and\
  \citenamefont {Zlochower}}]{Campanelli:2005dd}%
  \BibitemOpen
  \bibfield  {author} {\bibinfo {author} {\bibnamefont {Campanelli},
  \bibfnamefont {M.}}, \bibinfo {author} {\bibfnamefont {C.~O.}\ \bibnamefont
  {Lousto}}, \bibinfo {author} {\bibfnamefont {P.}~\bibnamefont {Marronetti}},
  and\ \bibinfo {author} {\bibfnamefont {Y.}~\bibnamefont {Zlochower}}}
  (\bibinfo {year} {2006}),\ \href
  {https://doi.org/10.1103/PhysRevLett.96.111101} {\bibfield  {journal}
  {\bibinfo  {journal} {Phys. Rev. Lett.}\ }\textbf {\bibinfo {volume} {96}},\
  \bibinfo {pages} {111101}},\ \Eprint {https://arxiv.org/abs/gr-qc/0511048}
  {arXiv:gr-qc/0511048} \BibitemShut {NoStop}%
\bibitem [{\citenamefont {Cardoso}\ \emph
  {et~al.}(2013{\natexlab{a}})\citenamefont {Cardoso}, \citenamefont {Carucci},
  \citenamefont {Pani},\ and\ \citenamefont {Sotiriou}}]{Cardoso:2013fwa}%
  \BibitemOpen
  \bibfield  {author} {\bibinfo {author} {\bibnamefont {Cardoso}, \bibfnamefont
  {V.}}, \bibinfo {author} {\bibfnamefont {I.~P.}\ \bibnamefont {Carucci}},
  \bibinfo {author} {\bibfnamefont {P.}~\bibnamefont {Pani}}, and\ \bibinfo
  {author} {\bibfnamefont {T.~P.}\ \bibnamefont {Sotiriou}}} (\bibinfo {year}
  {2013}{\natexlab{a}}),\ \href
  {https://doi.org/10.1103/PhysRevLett.111.111101} {\bibfield  {journal}
  {\bibinfo  {journal} {Phys. Rev. Lett.}\ }\textbf {\bibinfo {volume} {111}},\
  \bibinfo {pages} {111101}},\ \Eprint {https://arxiv.org/abs/1308.6587}
  {arXiv:1308.6587 [gr-qc]} \BibitemShut {NoStop}%
\bibitem [{\citenamefont {Cardoso}\ \emph
  {et~al.}(2013{\natexlab{b}})\citenamefont {Cardoso}, \citenamefont {Carucci},
  \citenamefont {Pani},\ and\ \citenamefont {Sotiriou}}]{Cardoso:2013opa}%
  \BibitemOpen
  \bibfield  {author} {\bibinfo {author} {\bibnamefont {Cardoso}, \bibfnamefont
  {V.}}, \bibinfo {author} {\bibfnamefont {I.~P.}\ \bibnamefont {Carucci}},
  \bibinfo {author} {\bibfnamefont {P.}~\bibnamefont {Pani}}, and\ \bibinfo
  {author} {\bibfnamefont {T.~P.}\ \bibnamefont {Sotiriou}}} (\bibinfo {year}
  {2013}{\natexlab{b}}),\ \href {https://doi.org/10.1103/PhysRevD.88.044056}
  {\bibfield  {journal} {\bibinfo  {journal} {Phys. Rev. D}\ }\textbf {\bibinfo
  {volume} {88}},\ \bibinfo {pages} {044056}},\ \Eprint
  {https://arxiv.org/abs/1305.6936} {arXiv:1305.6936 [gr-qc]} \BibitemShut
  {NoStop}%
\bibitem [{\citenamefont {Cardoso}\ \emph {et~al.}(2020)\citenamefont
  {Cardoso}, \citenamefont {Foschi},\ and\ \citenamefont
  {Zilhao}}]{Cardoso:2020cwo}%
  \BibitemOpen
  \bibfield  {author} {\bibinfo {author} {\bibnamefont {Cardoso}, \bibfnamefont
  {V.}}, \bibinfo {author} {\bibfnamefont {A.}~\bibnamefont {Foschi}}, and\
  \bibinfo {author} {\bibfnamefont {M.}~\bibnamefont {Zilhao}}} (\bibinfo
  {year} {2020}),\ \href {https://doi.org/10.1103/PhysRevLett.124.221104}
  {\bibfield  {journal} {\bibinfo  {journal} {Phys. Rev. Lett.}\ }\textbf
  {\bibinfo {volume} {124}}~(\bibinfo {number} {22}),\ \bibinfo {pages}
  {221104}},\ \Eprint {https://arxiv.org/abs/2005.12284} {arXiv:2005.12284
  [gr-qc]} \BibitemShut {NoStop}%
\bibitem [{\citenamefont {Cayuso}\ \emph {et~al.}(2017)\citenamefont {Cayuso},
  \citenamefont {Ortiz},\ and\ \citenamefont {Lehner}}]{Cayuso:2017iqc}%
  \BibitemOpen
  \bibfield  {author} {\bibinfo {author} {\bibnamefont {Cayuso}, \bibfnamefont
  {J.}}, \bibinfo {author} {\bibfnamefont {N.}~\bibnamefont {Ortiz}}, and\
  \bibinfo {author} {\bibfnamefont {L.}~\bibnamefont {Lehner}}} (\bibinfo
  {year} {2017}),\ \href {https://doi.org/10.1103/PhysRevD.96.084043}
  {\bibfield  {journal} {\bibinfo  {journal} {Phys. Rev. D}\ }\textbf {\bibinfo
  {volume} {96}}~(\bibinfo {number} {8}),\ \bibinfo {pages} {084043}},\ \Eprint
  {https://arxiv.org/abs/1706.07421} {arXiv:1706.07421 [gr-qc]} \BibitemShut
  {NoStop}%
\bibitem [{\citenamefont {Chandrasekhar}(1970)}]{Chandrasekhar:1992pr}%
  \BibitemOpen
  \bibfield  {author} {\bibinfo {author} {\bibnamefont {Chandrasekhar},
  \bibfnamefont {S.}}} (\bibinfo {year} {1970}),\ \href
  {https://doi.org/10.1103/PhysRevLett.24.611} {\bibfield  {journal} {\bibinfo
  {journal} {Phys. Rev. Lett.}\ }\textbf {\bibinfo {volume} {24}},\ \bibinfo
  {pages} {611}}\BibitemShut {NoStop}%
\bibitem [{\citenamefont {Chen}\ \emph {et~al.}(2015)\citenamefont {Chen},
  \citenamefont {Suyama},\ and\ \citenamefont {Yokoyama}}]{Chen:2015zmx}%
  \BibitemOpen
  \bibfield  {author} {\bibinfo {author} {\bibnamefont {Chen}, \bibfnamefont
  {P.}}, \bibinfo {author} {\bibfnamefont {T.}~\bibnamefont {Suyama}}, and\
  \bibinfo {author} {\bibfnamefont {J.}~\bibnamefont {Yokoyama}}} (\bibinfo
  {year} {2015}),\ \href {https://doi.org/10.1103/PhysRevD.92.124016}
  {\bibfield  {journal} {\bibinfo  {journal} {Phys. Rev. D}\ }\textbf {\bibinfo
  {volume} {92}},\ \bibinfo {pages} {124016}},\ \Eprint
  {https://arxiv.org/abs/1508.01384} {arXiv:1508.01384 [gr-qc]} \BibitemShut
  {NoStop}%
\bibitem [{\citenamefont {Cheong}\ and\ \citenamefont
  {Li}(2019)}]{Cheong:2018gzn}%
  \BibitemOpen
  \bibfield  {author} {\bibinfo {author} {\bibnamefont {Cheong}, \bibfnamefont
  {P.~C.-K.}}, and\ \bibinfo {author} {\bibfnamefont {T.~G.~F.}\ \bibnamefont
  {Li}}} (\bibinfo {year} {2019}),\ \href
  {https://doi.org/10.1103/PhysRevD.100.024027} {\bibfield  {journal} {\bibinfo
   {journal} {Phys. Rev. D}\ }\textbf {\bibinfo {volume} {100}}~(\bibinfo
  {number} {2}),\ \bibinfo {pages} {024027}},\ \Eprint
  {https://arxiv.org/abs/1812.04835} {arXiv:1812.04835 [gr-qc]} \BibitemShut
  {NoStop}%
\bibitem [{\citenamefont {Cherubini}\ \emph {et~al.}(2002)\citenamefont
  {Cherubini}, \citenamefont {Bini}, \citenamefont {Capozziello},\ and\
  \citenamefont {Ruffini}}]{Cherubini:2002gen}%
  \BibitemOpen
  \bibfield  {author} {\bibinfo {author} {\bibnamefont {Cherubini},
  \bibfnamefont {C.}}, \bibinfo {author} {\bibfnamefont {D.}~\bibnamefont
  {Bini}}, \bibinfo {author} {\bibfnamefont {S.}~\bibnamefont {Capozziello}},
  and\ \bibinfo {author} {\bibfnamefont {R.}~\bibnamefont {Ruffini}}} (\bibinfo
  {year} {2002}),\ \href {https://doi.org/10.1142/S0218271802002037} {\bibfield
   {journal} {\bibinfo  {journal} {Int. J. Mod. Phys. D}\ }\textbf {\bibinfo
  {volume} {11}},\ \bibinfo {pages} {827}},\ \Eprint
  {https://arxiv.org/abs/gr-qc/0302095} {arXiv:gr-qc/0302095} \BibitemShut
  {NoStop}%
\bibitem [{\citenamefont {Chiba}(2022)}]{Chiba:2021rqa}%
  \BibitemOpen
  \bibfield  {author} {\bibinfo {author} {\bibnamefont {Chiba}, \bibfnamefont
  {T.}}} (\bibinfo {year} {2022}),\ \href
  {https://doi.org/10.1093/ptep/ptab138} {\bibfield  {journal} {\bibinfo
  {journal} {PTEP}\ }\textbf {\bibinfo {volume} {2022}},\ \bibinfo {pages}
  {013E01}},\ \Eprint {https://arxiv.org/abs/2104.11362} {arXiv:2104.11362
  [gr-qc]} \BibitemShut {NoStop}%
\bibitem [{\citenamefont {Chodos}\ \emph {et~al.}(1985)\citenamefont {Chodos},
  \citenamefont {Hauser},\ and\ \citenamefont {Kostelecky}}]{Chodos:1984cy}%
  \BibitemOpen
  \bibfield  {author} {\bibinfo {author} {\bibnamefont {Chodos}, \bibfnamefont
  {A.}}, \bibinfo {author} {\bibfnamefont {A.~I.}\ \bibnamefont {Hauser}}, and\
  \bibinfo {author} {\bibfnamefont {V.~A.}\ \bibnamefont {Kostelecky}}}
  (\bibinfo {year} {1985}),\ \href
  {https://doi.org/10.1016/0370-2693(85)90460-5} {\bibfield  {journal}
  {\bibinfo  {journal} {Phys. Lett.}\ }\textbf {\bibinfo {volume} {150B}},\
  \bibinfo {pages} {431}}\BibitemShut {NoStop}%
\bibitem [{\citenamefont {Choptuik}(1993)}]{Choptuik:1992jv}%
  \BibitemOpen
  \bibfield  {author} {\bibinfo {author} {\bibnamefont {Choptuik},
  \bibfnamefont {M.~W.}}} (\bibinfo {year} {1993}),\ \href
  {https://doi.org/10.1103/PhysRevLett.70.9} {\bibfield  {journal} {\bibinfo
  {journal} {Phys. Rev. Lett.}\ }\textbf {\bibinfo {volume} {70}},\ \bibinfo
  {pages} {9}}\BibitemShut {NoStop}%
\bibitem [{\citenamefont {Clark}\ \emph {et~al.}(2014)\citenamefont {Clark},
  \citenamefont {Bauswein}, \citenamefont {Cadonati}, \citenamefont {Janka},
  \citenamefont {Pankow},\ and\ \citenamefont {Stergioulas}}]{Clark:2014wua}%
  \BibitemOpen
  \bibfield  {author} {\bibinfo {author} {\bibnamefont {Clark}, \bibfnamefont
  {J.}}, \bibinfo {author} {\bibfnamefont {A.}~\bibnamefont {Bauswein}},
  \bibinfo {author} {\bibfnamefont {L.}~\bibnamefont {Cadonati}}, \bibinfo
  {author} {\bibfnamefont {H.~T.}\ \bibnamefont {Janka}}, \bibinfo {author}
  {\bibfnamefont {C.}~\bibnamefont {Pankow}}, and\ \bibinfo {author}
  {\bibfnamefont {N.}~\bibnamefont {Stergioulas}}} (\bibinfo {year} {2014}),\
  \href {https://doi.org/10.1103/PhysRevD.90.062004} {\bibfield  {journal}
  {\bibinfo  {journal} {Phys. Rev. D}\ }\textbf {\bibinfo {volume}
  {90}}~(\bibinfo {number} {6}),\ \bibinfo {pages} {062004}},\ \Eprint
  {https://arxiv.org/abs/1406.5444} {arXiv:1406.5444 [astro-ph.HE]}
  \BibitemShut {NoStop}%
\bibitem [{\citenamefont {Clifton}\ \emph {et~al.}(2012)\citenamefont
  {Clifton}, \citenamefont {Ferreira}, \citenamefont {Padilla},\ and\
  \citenamefont {Skordis}}]{Clifton:2011jh}%
  \BibitemOpen
  \bibfield  {author} {\bibinfo {author} {\bibnamefont {Clifton}, \bibfnamefont
  {T.}}, \bibinfo {author} {\bibfnamefont {P.~G.}\ \bibnamefont {Ferreira}},
  \bibinfo {author} {\bibfnamefont {A.}~\bibnamefont {Padilla}}, and\ \bibinfo
  {author} {\bibfnamefont {C.}~\bibnamefont {Skordis}}} (\bibinfo {year}
  {2012}),\ \href {https://doi.org/10.1016/j.physrep.2012.01.001} {\bibfield
  {journal} {\bibinfo  {journal} {Phys. Rept.}\ }\textbf {\bibinfo {volume}
  {513}},\ \bibinfo {pages} {1}},\ \Eprint {https://arxiv.org/abs/1106.2476}
  {arXiv:1106.2476 [astro-ph.CO]} \BibitemShut {NoStop}%
\bibitem [{\citenamefont {Clough}\ \emph {et~al.}(2022)\citenamefont {Clough},
  \citenamefont {Helfer}, \citenamefont {Witek},\ and\ \citenamefont
  {Berti}}]{Clough:2022ygm}%
  \BibitemOpen
  \bibfield  {author} {\bibinfo {author} {\bibnamefont {Clough}, \bibfnamefont
  {K.}}, \bibinfo {author} {\bibfnamefont {T.}~\bibnamefont {Helfer}}, \bibinfo
  {author} {\bibfnamefont {H.}~\bibnamefont {Witek}}, and\ \bibinfo {author}
  {\bibfnamefont {E.}~\bibnamefont {Berti}}} (\bibinfo {year} {2022}),\ \href
  {https://doi.org/10.1103/PhysRevLett.129.151102} {\bibfield  {journal}
  {\bibinfo  {journal} {Phys. Rev. Lett.}\ }\textbf {\bibinfo {volume}
  {129}}~(\bibinfo {number} {15}),\ \bibinfo {pages} {151102}},\ \Eprint
  {https://arxiv.org/abs/2204.10868} {arXiv:2204.10868 [gr-qc]} \BibitemShut
  {NoStop}%
\bibitem [{\citenamefont {Coates}\ \emph {et~al.}(2017)\citenamefont {Coates},
  \citenamefont {Horbartsch},\ and\ \citenamefont {Sotiriou}}]{Coates:2016ktu}%
  \BibitemOpen
  \bibfield  {author} {\bibinfo {author} {\bibnamefont {Coates}, \bibfnamefont
  {A.}}, \bibinfo {author} {\bibfnamefont {M.~W.}\ \bibnamefont {Horbartsch}},
  and\ \bibinfo {author} {\bibfnamefont {T.~P.}\ \bibnamefont {Sotiriou}}}
  (\bibinfo {year} {2017}),\ \href {https://doi.org/10.1103/PhysRevD.95.084003}
  {\bibfield  {journal} {\bibinfo  {journal} {Phys. Rev. D}\ }\textbf {\bibinfo
  {volume} {95}}~(\bibinfo {number} {8}),\ \bibinfo {pages} {084003}},\ \Eprint
  {https://arxiv.org/abs/1606.03981} {arXiv:1606.03981 [gr-qc]} \BibitemShut
  {NoStop}%
\bibitem [{\citenamefont {Coates}\ and\ \citenamefont
  {Ramazano\u{g}lu}(2022)}]{Coates:2022qia}%
  \BibitemOpen
  \bibfield  {author} {\bibinfo {author} {\bibnamefont {Coates}, \bibfnamefont
  {A.}}, and\ \bibinfo {author} {\bibfnamefont {F.~M.}\ \bibnamefont
  {Ramazano\u{g}lu}}} (\bibinfo {year} {2022}),\ \href
  {https://doi.org/10.1103/PhysRevLett.129.151103} {\bibfield  {journal}
  {\bibinfo  {journal} {Phys. Rev. Lett.}\ }\textbf {\bibinfo {volume}
  {129}}~(\bibinfo {number} {15}),\ \bibinfo {pages} {151103}},\ \Eprint
  {https://arxiv.org/abs/2205.07784} {arXiv:2205.07784 [gr-qc]} \BibitemShut
  {NoStop}%
\bibitem [{\citenamefont {Coates}\ and\ \citenamefont
  {Ramazano\u{g}lu}(2023)}]{Coates:2022nif}%
  \BibitemOpen
  \bibfield  {author} {\bibinfo {author} {\bibnamefont {Coates}, \bibfnamefont
  {A.}}, and\ \bibinfo {author} {\bibfnamefont {F.~M.}\ \bibnamefont
  {Ramazano\u{g}lu}}} (\bibinfo {year} {2023}),\ \href
  {https://doi.org/10.1103/PhysRevLett.130.021401} {\bibfield  {journal}
  {\bibinfo  {journal} {Phys. Rev. Lett.}\ }\textbf {\bibinfo {volume}
  {130}}~(\bibinfo {number} {2}),\ \bibinfo {pages} {021401}},\ \Eprint
  {https://arxiv.org/abs/2211.08027} {arXiv:2211.08027 [gr-qc]} \BibitemShut
  {NoStop}%
\bibitem [{\citenamefont {Collodel}\ \emph {et~al.}(2020)\citenamefont
  {Collodel}, \citenamefont {Kleihaus}, \citenamefont {Kunz},\ and\
  \citenamefont {Berti}}]{Collodel:2019kkx}%
  \BibitemOpen
  \bibfield  {author} {\bibinfo {author} {\bibnamefont {Collodel},
  \bibfnamefont {L.~G.}}, \bibinfo {author} {\bibfnamefont {B.}~\bibnamefont
  {Kleihaus}}, \bibinfo {author} {\bibfnamefont {J.}~\bibnamefont {Kunz}}, and\
  \bibinfo {author} {\bibfnamefont {E.}~\bibnamefont {Berti}}} (\bibinfo {year}
  {2020}),\ \href {https://doi.org/10.1088/1361-6382/ab74f9} {\bibfield
  {journal} {\bibinfo  {journal} {Class. Quant. Grav.}\ }\textbf {\bibinfo
  {volume} {37}}~(\bibinfo {number} {7}),\ \bibinfo {pages} {075018}},\ \Eprint
  {https://arxiv.org/abs/1912.05382} {arXiv:1912.05382 [gr-qc]} \BibitemShut
  {NoStop}%
\bibitem [{\citenamefont {{Cowling}}(1941)}]{Cowling:1941MNRAS}%
  \BibitemOpen
  \bibfield  {author} {\bibinfo {author} {\bibnamefont {{Cowling}},
  \bibfnamefont {T.~G.}}} (\bibinfo {year} {1941}),\ \href
  {https://doi.org/10.1093/mnras/101.8.367} {\bibfield  {journal} {\bibinfo
  {journal} {{Mon. Not. Roy. Astron. Soc.}}\ }\textbf {\bibinfo {volume}
  {101}},\ \bibinfo {pages} {367}}\BibitemShut {NoStop}%
\bibitem [{\citenamefont {Cunha}\ \emph {et~al.}(2019)\citenamefont {Cunha},
  \citenamefont {Herdeiro},\ and\ \citenamefont {Radu}}]{Cunha:2019dwb}%
  \BibitemOpen
  \bibfield  {author} {\bibinfo {author} {\bibnamefont {Cunha}, \bibfnamefont
  {P.~V.~P.}}, \bibinfo {author} {\bibfnamefont {C.~A.~R.}\ \bibnamefont
  {Herdeiro}}, and\ \bibinfo {author} {\bibfnamefont {E.}~\bibnamefont {Radu}}}
  (\bibinfo {year} {2019}),\ \href
  {https://doi.org/10.1103/PhysRevLett.123.011101} {\bibfield  {journal}
  {\bibinfo  {journal} {Phys. Rev. Lett.}\ }\textbf {\bibinfo {volume}
  {123}}~(\bibinfo {number} {1}),\ \bibinfo {pages} {011101}},\ \Eprint
  {https://arxiv.org/abs/1904.09997} {arXiv:1904.09997 [gr-qc]} \BibitemShut
  {NoStop}%
\bibitem [{\citenamefont {Damour}(2015)}]{Damour:2014tpa}%
  \BibitemOpen
  \bibfield  {author} {\bibinfo {author} {\bibnamefont {Damour}, \bibfnamefont
  {T.}}} (\bibinfo {year} {2015}),\ \href
  {https://doi.org/10.1088/0264-9381/32/12/124009} {\bibfield  {journal}
  {\bibinfo  {journal} {Class. Quant. Grav.}\ }\textbf {\bibinfo {volume}
  {32}}~(\bibinfo {number} {12}),\ \bibinfo {pages} {124009}},\ \Eprint
  {https://arxiv.org/abs/1411.3930} {arXiv:1411.3930 [gr-qc]} \BibitemShut
  {NoStop}%
\bibitem [{\citenamefont {Damour}\ and\ \citenamefont
  {Esposito-Far\`ese}(1992)}]{Damour:1992we}%
  \BibitemOpen
  \bibfield  {author} {\bibinfo {author} {\bibnamefont {Damour}, \bibfnamefont
  {T.}}, and\ \bibinfo {author} {\bibfnamefont {G.}~\bibnamefont
  {Esposito-Far\`ese}}} (\bibinfo {year} {1992}),\ \href
  {https://doi.org/10.1088/0264-9381/9/9/015} {\bibfield  {journal} {\bibinfo
  {journal} {Class. Quant. Grav.}\ }\textbf {\bibinfo {volume} {9}},\ \bibinfo
  {pages} {2093}}\BibitemShut {NoStop}%
\bibitem [{\citenamefont {Damour}\ and\ \citenamefont
  {Esposito-Far\`ese}(1993)}]{Damour:1993hw}%
  \BibitemOpen
  \bibfield  {author} {\bibinfo {author} {\bibnamefont {Damour}, \bibfnamefont
  {T.}}, and\ \bibinfo {author} {\bibfnamefont {G.}~\bibnamefont
  {Esposito-Far\`ese}}} (\bibinfo {year} {1993}),\ \href
  {https://doi.org/10.1103/PhysRevLett.70.2220} {\bibfield  {journal} {\bibinfo
   {journal} {Phys. Rev. Lett.}\ }\textbf {\bibinfo {volume} {70}},\ \bibinfo
  {pages} {2220}}\BibitemShut {NoStop}%
\bibitem [{\citenamefont {Damour}\ and\ \citenamefont
  {Esposito-Far\`ese}(1996{\natexlab{a}})}]{Damour:1996ke}%
  \BibitemOpen
  \bibfield  {author} {\bibinfo {author} {\bibnamefont {Damour}, \bibfnamefont
  {T.}}, and\ \bibinfo {author} {\bibfnamefont {G.}~\bibnamefont
  {Esposito-Far\`ese}}} (\bibinfo {year} {1996}{\natexlab{a}}),\ \href
  {https://doi.org/10.1103/PhysRevD.54.1474} {\bibfield  {journal} {\bibinfo
  {journal} {Phys. Rev. D}\ }\textbf {\bibinfo {volume} {54}},\ \bibinfo
  {pages} {1474}},\ \Eprint {https://arxiv.org/abs/gr-qc/9602056}
  {arXiv:gr-qc/9602056} \BibitemShut {NoStop}%
\bibitem [{\citenamefont {Damour}\ and\ \citenamefont
  {Esposito-Far\`ese}(1996{\natexlab{b}})}]{Damour:1995kt}%
  \BibitemOpen
  \bibfield  {author} {\bibinfo {author} {\bibnamefont {Damour}, \bibfnamefont
  {T.}}, and\ \bibinfo {author} {\bibfnamefont {G.}~\bibnamefont
  {Esposito-Far\`ese}}} (\bibinfo {year} {1996}{\natexlab{b}}),\ \href
  {https://doi.org/10.1103/PhysRevD.53.5541} {\bibfield  {journal} {\bibinfo
  {journal} {Phys. Rev. D}\ }\textbf {\bibinfo {volume} {53}},\ \bibinfo
  {pages} {5541}},\ \Eprint {https://arxiv.org/abs/gr-qc/9506063}
  {arXiv:gr-qc/9506063} \BibitemShut {NoStop}%
\bibitem [{\citenamefont {Damour}\ and\ \citenamefont
  {Esposito-Far\`ese}(1998)}]{Damour:1998jk}%
  \BibitemOpen
  \bibfield  {author} {\bibinfo {author} {\bibnamefont {Damour}, \bibfnamefont
  {T.}}, and\ \bibinfo {author} {\bibfnamefont {G.}~\bibnamefont
  {Esposito-Far\`ese}}} (\bibinfo {year} {1998}),\ \href
  {https://doi.org/10.1103/PhysRevD.58.042001} {\bibfield  {journal} {\bibinfo
  {journal} {Phys. Rev. D}\ }\textbf {\bibinfo {volume} {58}},\ \bibinfo
  {pages} {042001}},\ \Eprint {https://arxiv.org/abs/gr-qc/9803031}
  {arXiv:gr-qc/9803031} \BibitemShut {NoStop}%
\bibitem [{\citenamefont {Damour}\ and\ \citenamefont
  {Nordtvedt}(1993)}]{Damour:1992kf}%
  \BibitemOpen
  \bibfield  {author} {\bibinfo {author} {\bibnamefont {Damour}, \bibfnamefont
  {T.}}, and\ \bibinfo {author} {\bibfnamefont {K.}~\bibnamefont {Nordtvedt}}}
  (\bibinfo {year} {1993}),\ \href
  {https://doi.org/10.1103/PhysRevLett.70.2217} {\bibfield  {journal} {\bibinfo
   {journal} {Phys. Rev. Lett.}\ }\textbf {\bibinfo {volume} {70}},\ \bibinfo
  {pages} {2217}}\BibitemShut {NoStop}%
\bibitem [{\citenamefont {Damour}\ and\ \citenamefont
  {Taylor}(1992)}]{Damour:1991rd}%
  \BibitemOpen
  \bibfield  {author} {\bibinfo {author} {\bibnamefont {Damour}, \bibfnamefont
  {T.}}, and\ \bibinfo {author} {\bibfnamefont {J.~H.}\ \bibnamefont {Taylor}}}
  (\bibinfo {year} {1992}),\ \href {https://doi.org/10.1103/PhysRevD.45.1840}
  {\bibfield  {journal} {\bibinfo  {journal} {Phys. Rev. D}\ }\textbf {\bibinfo
  {volume} {45}},\ \bibinfo {pages} {1840}}\BibitemShut {NoStop}%
\bibitem [{\citenamefont {Danchev}\ and\ \citenamefont
  {Doneva}(2021)}]{Danchev:2020zwn}%
  \BibitemOpen
  \bibfield  {author} {\bibinfo {author} {\bibnamefont {Danchev}, \bibfnamefont
  {V.~I.}}, and\ \bibinfo {author} {\bibfnamefont {D.~D.}\ \bibnamefont
  {Doneva}}} (\bibinfo {year} {2021}),\ \href
  {https://doi.org/10.1103/PhysRevD.103.024049} {\bibfield  {journal} {\bibinfo
   {journal} {Phys. Rev. D}\ }\textbf {\bibinfo {volume} {103}}~(\bibinfo
  {number} {2}),\ \bibinfo {pages} {024049}},\ \Eprint
  {https://arxiv.org/abs/2010.07392} {arXiv:2010.07392 [gr-qc]} \BibitemShut
  {NoStop}%
\bibitem [{\citenamefont {Danchev}\ \emph {et~al.}(2022)\citenamefont
  {Danchev}, \citenamefont {Doneva},\ and\ \citenamefont
  {Yazadjiev}}]{Danchev:2021tew}%
  \BibitemOpen
  \bibfield  {author} {\bibinfo {author} {\bibnamefont {Danchev}, \bibfnamefont
  {V.~I.}}, \bibinfo {author} {\bibfnamefont {D.~D.}\ \bibnamefont {Doneva}},
  and\ \bibinfo {author} {\bibfnamefont {S.~S.}\ \bibnamefont {Yazadjiev}}}
  (\bibinfo {year} {2022}),\ \href
  {https://doi.org/10.1103/PhysRevD.106.124001} {\bibfield  {journal} {\bibinfo
   {journal} {Phys. Rev. D}\ }\textbf {\bibinfo {volume} {106}}~(\bibinfo
  {number} {12}),\ \bibinfo {pages} {124001}},\ \Eprint
  {https://arxiv.org/abs/2112.03869} {arXiv:2112.03869 [gr-qc]} \BibitemShut
  {NoStop}%
\bibitem [{\citenamefont {DeDeo}\ and\ \citenamefont
  {Psaltis}(2003)}]{DeDeo:2003ju}%
  \BibitemOpen
  \bibfield  {author} {\bibinfo {author} {\bibnamefont {DeDeo}, \bibfnamefont
  {S.}}, and\ \bibinfo {author} {\bibfnamefont {D.}~\bibnamefont {Psaltis}}}
  (\bibinfo {year} {2003}),\ \href
  {https://doi.org/10.1103/PhysRevLett.90.141101} {\bibfield  {journal}
  {\bibinfo  {journal} {Phys. Rev. Lett.}\ }\textbf {\bibinfo {volume} {90}},\
  \bibinfo {pages} {141101}},\ \Eprint {https://arxiv.org/abs/astro-ph/0302095}
  {arXiv:astro-ph/0302095} \BibitemShut {NoStop}%
\bibitem [{\citenamefont {DeDeo}\ and\ \citenamefont
  {Psaltis}(2004)}]{DeDeo:2004kk}%
  \BibitemOpen
  \bibfield  {author} {\bibinfo {author} {\bibnamefont {DeDeo}, \bibfnamefont
  {S.}}, and\ \bibinfo {author} {\bibfnamefont {D.}~\bibnamefont {Psaltis}}}
  (\bibinfo {year} {2004}),\ \href@noop {} {\enquote {\bibinfo {title}
  {{Testing strong-field gravity with quasiperiodic oscillations}},}\ }\Eprint
  {https://arxiv.org/abs/astro-ph/0405067} {arXiv:astro-ph/0405067}
  \BibitemShut {NoStop}%
\bibitem [{\citenamefont {Deffayet}\ \emph {et~al.}(2009)\citenamefont
  {Deffayet}, \citenamefont {Deser},\ and\ \citenamefont
  {Esposito-Far{\`e}se}}]{Deffayet:2009mn}%
  \BibitemOpen
  \bibfield  {author} {\bibinfo {author} {\bibnamefont {Deffayet},
  \bibfnamefont {C.}}, \bibinfo {author} {\bibfnamefont {S.}~\bibnamefont
  {Deser}}, and\ \bibinfo {author} {\bibfnamefont {G.}~\bibnamefont
  {Esposito-Far{\`e}se}}} (\bibinfo {year} {2009}),\ \href
  {https://doi.org/10.1103/PhysRevD.80.064015} {\bibfield  {journal} {\bibinfo
  {journal} {Phys. Rev. D}\ }\textbf {\bibinfo {volume} {80}},\ \bibinfo
  {pages} {064015}},\ \Eprint {https://arxiv.org/abs/0906.1967}
  {arXiv:0906.1967 [gr-qc]} \BibitemShut {NoStop}%
\bibitem [{\citenamefont {Degollado}\ \emph {et~al.}(2020)\citenamefont
  {Degollado}, \citenamefont {Salgado},\ and\ \citenamefont
  {Alcubierre}}]{Degollado:2020lsa}%
  \BibitemOpen
  \bibfield  {author} {\bibinfo {author} {\bibnamefont {Degollado},
  \bibfnamefont {J.~C.}}, \bibinfo {author} {\bibfnamefont {M.}~\bibnamefont
  {Salgado}}, and\ \bibinfo {author} {\bibfnamefont {M.}~\bibnamefont
  {Alcubierre}}} (\bibinfo {year} {2020}),\ \href
  {https://doi.org/10.1016/j.physletb.2020.135666} {\bibfield  {journal}
  {\bibinfo  {journal} {Phys. Lett. B}\ }\textbf {\bibinfo {volume} {808}},\
  \bibinfo {pages} {135666}},\ \Eprint {https://arxiv.org/abs/2008.10683}
  {arXiv:2008.10683 [gr-qc]} \BibitemShut {NoStop}%
\bibitem [{\citenamefont {Delsate}\ \emph {et~al.}(2015)\citenamefont
  {Delsate}, \citenamefont {Hilditch},\ and\ \citenamefont
  {Witek}}]{Delsate:2014hba}%
  \BibitemOpen
  \bibfield  {author} {\bibinfo {author} {\bibnamefont {Delsate}, \bibfnamefont
  {T.}}, \bibinfo {author} {\bibfnamefont {D.}~\bibnamefont {Hilditch}}, and\
  \bibinfo {author} {\bibfnamefont {H.}~\bibnamefont {Witek}}} (\bibinfo {year}
  {2015}),\ \href {https://doi.org/10.1103/PhysRevD.91.024027} {\bibfield
  {journal} {\bibinfo  {journal} {Phys. Rev. D}\ }\textbf {\bibinfo {volume}
  {91}}~(\bibinfo {number} {2}),\ \bibinfo {pages} {024027}},\ \Eprint
  {https://arxiv.org/abs/1407.6727} {arXiv:1407.6727 [gr-qc]} \BibitemShut
  {NoStop}%
\bibitem [{\citenamefont {Demirbo\u{g}a}\ \emph {et~al.}(2022)\citenamefont
  {Demirbo\u{g}a}, \citenamefont {Coates},\ and\ \citenamefont
  {Ramazano\u{g}lu}}]{Demirboga:2021nrc}%
  \BibitemOpen
  \bibfield  {author} {\bibinfo {author} {\bibnamefont {Demirbo\u{g}a},
  \bibfnamefont {E.~S.}}, \bibinfo {author} {\bibfnamefont {A.}~\bibnamefont
  {Coates}}, and\ \bibinfo {author} {\bibfnamefont {F.~M.}\ \bibnamefont
  {Ramazano\u{g}lu}}} (\bibinfo {year} {2022}),\ \href
  {https://doi.org/10.1103/PhysRevD.105.024057} {\bibfield  {journal} {\bibinfo
   {journal} {Phys. Rev. D}\ }\textbf {\bibinfo {volume} {105}}~(\bibinfo
  {number} {2}),\ \bibinfo {pages} {024057}},\ \Eprint
  {https://arxiv.org/abs/2112.04269} {arXiv:2112.04269 [gr-qc]} \BibitemShut
  {NoStop}%
\bibitem [{\citenamefont {{Diaz Alonso}}\ and\ \citenamefont {{Ibanez
  Cabanell}}(1985)}]{1985ApJ...291..308D}%
  \BibitemOpen
  \bibfield  {author} {\bibinfo {author} {\bibnamefont {{Diaz Alonso}},
  \bibfnamefont {J.}}, and\ \bibinfo {author} {\bibfnamefont {J.~M.}\
  \bibnamefont {{Ibanez Cabanell}}}} (\bibinfo {year} {1985}),\ \href
  {https://doi.org/10.1086/163070} {\bibfield  {journal} {\bibinfo  {journal}
  {\apj}\ }\textbf {\bibinfo {volume} {291}},\ \bibinfo {pages}
  {308}}\BibitemShut {NoStop}%
\bibitem [{\citenamefont {Dima}\ \emph {et~al.}(2020)\citenamefont {Dima},
  \citenamefont {Barausse}, \citenamefont {Franchini},\ and\ \citenamefont
  {Sotiriou}}]{Dima:2020yac}%
  \BibitemOpen
  \bibfield  {author} {\bibinfo {author} {\bibnamefont {Dima}, \bibfnamefont
  {A.}}, \bibinfo {author} {\bibfnamefont {E.}~\bibnamefont {Barausse}},
  \bibinfo {author} {\bibfnamefont {N.}~\bibnamefont {Franchini}}, and\
  \bibinfo {author} {\bibfnamefont {T.~P.}\ \bibnamefont {Sotiriou}}} (\bibinfo
  {year} {2020}),\ \href {https://doi.org/10.1103/PhysRevLett.125.231101}
  {\bibfield  {journal} {\bibinfo  {journal} {Phys. Rev. Lett.}\ }\textbf
  {\bibinfo {volume} {125}}~(\bibinfo {number} {23}),\ \bibinfo {pages}
  {231101}},\ \Eprint {https://arxiv.org/abs/2006.03095} {arXiv:2006.03095
  [gr-qc]} \BibitemShut {NoStop}%
\bibitem [{\citenamefont {Do}\ \emph {et~al.}(2019)\citenamefont {Do} \emph
  {et~al.}}]{Do:2019txf}%
  \BibitemOpen
  \bibfield  {author} {\bibinfo {author} {\bibnamefont {Do}, \bibfnamefont
  {T.}},  \emph {et~al.}} (\bibinfo {year} {2019}),\ \href
  {https://doi.org/10.1126/science.aav8137} {\bibfield  {journal} {\bibinfo
  {journal} {Science}\ }\textbf {\bibinfo {volume} {365}}~(\bibinfo {number}
  {6454}),\ \bibinfo {pages} {664}},\ \Eprint
  {https://arxiv.org/abs/1907.10731} {arXiv:1907.10731 [astro-ph.GA]}
  \BibitemShut {NoStop}%
\bibitem [{\citenamefont {Dolan}(2013)}]{Dolan:2012yt}%
  \BibitemOpen
  \bibfield  {author} {\bibinfo {author} {\bibnamefont {Dolan}, \bibfnamefont
  {S.~R.}}} (\bibinfo {year} {2013}),\ \href
  {https://doi.org/10.1103/PhysRevD.87.124026} {\bibfield  {journal} {\bibinfo
  {journal} {Phys. Rev. D}\ }\textbf {\bibinfo {volume} {87}}~(\bibinfo
  {number} {12}),\ \bibinfo {pages} {124026}},\ \Eprint
  {https://arxiv.org/abs/1212.1477} {arXiv:1212.1477 [gr-qc]} \BibitemShut
  {NoStop}%
\bibitem [{\citenamefont {Doneva}\ \emph
  {et~al.}(2020{\natexlab{a}})\citenamefont {Doneva}, \citenamefont {Collodel},
  \citenamefont {Kr\"uger},\ and\ \citenamefont {Yazadjiev}}]{Doneva:2020nbb}%
  \BibitemOpen
  \bibfield  {author} {\bibinfo {author} {\bibnamefont {Doneva}, \bibfnamefont
  {D.~D.}}, \bibinfo {author} {\bibfnamefont {L.~G.}\ \bibnamefont {Collodel}},
  \bibinfo {author} {\bibfnamefont {C.~J.}\ \bibnamefont {Kr\"uger}}, and\
  \bibinfo {author} {\bibfnamefont {S.~S.}\ \bibnamefont {Yazadjiev}}}
  (\bibinfo {year} {2020}{\natexlab{a}}),\ \href
  {https://doi.org/10.1103/PhysRevD.102.104027} {\bibfield  {journal} {\bibinfo
   {journal} {Phys. Rev. D}\ }\textbf {\bibinfo {volume} {102}}~(\bibinfo
  {number} {10}),\ \bibinfo {pages} {104027}},\ \Eprint
  {https://arxiv.org/abs/2008.07391} {arXiv:2008.07391 [gr-qc]} \BibitemShut
  {NoStop}%
\bibitem [{\citenamefont {Doneva}\ \emph
  {et~al.}(2020{\natexlab{b}})\citenamefont {Doneva}, \citenamefont {Collodel},
  \citenamefont {Kr\"uger},\ and\ \citenamefont {Yazadjiev}}]{Doneva:2020kfv}%
  \BibitemOpen
  \bibfield  {author} {\bibinfo {author} {\bibnamefont {Doneva}, \bibfnamefont
  {D.~D.}}, \bibinfo {author} {\bibfnamefont {L.~G.}\ \bibnamefont {Collodel}},
  \bibinfo {author} {\bibfnamefont {C.~J.}\ \bibnamefont {Kr\"uger}}, and\
  \bibinfo {author} {\bibfnamefont {S.~S.}\ \bibnamefont {Yazadjiev}}}
  (\bibinfo {year} {2020}{\natexlab{b}}),\ \href
  {https://doi.org/10.1140/epjc/s10052-020-08765-3} {\bibfield  {journal}
  {\bibinfo  {journal} {Eur. Phys. J. C}\ }\textbf {\bibinfo {volume}
  {80}}~(\bibinfo {number} {12}),\ \bibinfo {pages} {1205}},\ \Eprint
  {https://arxiv.org/abs/2009.03774} {arXiv:2009.03774 [gr-qc]} \BibitemShut
  {NoStop}%
\bibitem [{\citenamefont {Doneva}\ \emph
  {et~al.}(2018{\natexlab{a}})\citenamefont {Doneva}, \citenamefont
  {Kiorpelidi}, \citenamefont {Nedkova}, \citenamefont {Papantonopoulos},\ and\
  \citenamefont {Yazadjiev}}]{Doneva:2018rou}%
  \BibitemOpen
  \bibfield  {author} {\bibinfo {author} {\bibnamefont {Doneva}, \bibfnamefont
  {D.~D.}}, \bibinfo {author} {\bibfnamefont {S.}~\bibnamefont {Kiorpelidi}},
  \bibinfo {author} {\bibfnamefont {P.~G.}\ \bibnamefont {Nedkova}}, \bibinfo
  {author} {\bibfnamefont {E.}~\bibnamefont {Papantonopoulos}}, and\ \bibinfo
  {author} {\bibfnamefont {S.~S.}\ \bibnamefont {Yazadjiev}}} (\bibinfo {year}
  {2018}{\natexlab{a}}),\ \href {https://doi.org/10.1103/PhysRevD.98.104056}
  {\bibfield  {journal} {\bibinfo  {journal} {Phys. Rev. D}\ }\textbf {\bibinfo
  {volume} {98}}~(\bibinfo {number} {10}),\ \bibinfo {pages} {104056}},\
  \Eprint {https://arxiv.org/abs/1809.00844} {arXiv:1809.00844 [gr-qc]}
  \BibitemShut {NoStop}%
\bibitem [{\citenamefont {Doneva}\ and\ \citenamefont
  {Pappas}(2018)}]{Doneva:2017jop}%
  \BibitemOpen
  \bibfield  {author} {\bibinfo {author} {\bibnamefont {Doneva}, \bibfnamefont
  {D.~D.}}, and\ \bibinfo {author} {\bibfnamefont {G.}~\bibnamefont {Pappas}}}
  (\bibinfo {year} {2018}),\ \href
  {https://doi.org/10.1007/978-3-319-97616-7_13} {\bibfield  {journal}
  {\bibinfo  {journal} {Astrophys. Space Sci. Libr.}\ }\textbf {\bibinfo
  {volume} {457}},\ \bibinfo {pages} {737}},\ \Eprint
  {https://arxiv.org/abs/1709.08046} {arXiv:1709.08046 [gr-qc]} \BibitemShut
  {NoStop}%
\bibitem [{\citenamefont {Doneva}\ \emph {et~al.}(2019)\citenamefont {Doneva},
  \citenamefont {Staykov},\ and\ \citenamefont {Yazadjiev}}]{Doneva:2019vuh}%
  \BibitemOpen
  \bibfield  {author} {\bibinfo {author} {\bibnamefont {Doneva}, \bibfnamefont
  {D.~D.}}, \bibinfo {author} {\bibfnamefont {K.~V.}\ \bibnamefont {Staykov}},
  and\ \bibinfo {author} {\bibfnamefont {S.~S.}\ \bibnamefont {Yazadjiev}}}
  (\bibinfo {year} {2019}),\ \href {https://doi.org/10.1103/PhysRevD.99.104045}
  {\bibfield  {journal} {\bibinfo  {journal} {Phys. Rev. D}\ }\textbf {\bibinfo
  {volume} {99}}~(\bibinfo {number} {10}),\ \bibinfo {pages} {104045}},\
  \Eprint {https://arxiv.org/abs/1903.08119} {arXiv:1903.08119 [gr-qc]}
  \BibitemShut {NoStop}%
\bibitem [{\citenamefont {Doneva}\ \emph
  {et~al.}(2020{\natexlab{c}})\citenamefont {Doneva}, \citenamefont {Staykov},
  \citenamefont {Yazadjiev},\ and\ \citenamefont {Zheleva}}]{Doneva:2020qww}%
  \BibitemOpen
  \bibfield  {author} {\bibinfo {author} {\bibnamefont {Doneva}, \bibfnamefont
  {D.~D.}}, \bibinfo {author} {\bibfnamefont {K.~V.}\ \bibnamefont {Staykov}},
  \bibinfo {author} {\bibfnamefont {S.~S.}\ \bibnamefont {Yazadjiev}}, and\
  \bibinfo {author} {\bibfnamefont {R.~Z.}\ \bibnamefont {Zheleva}}} (\bibinfo
  {year} {2020}{\natexlab{c}}),\ \href
  {https://doi.org/10.1103/PhysRevD.102.064042} {\bibfield  {journal} {\bibinfo
   {journal} {Phys. Rev. D}\ }\textbf {\bibinfo {volume} {102}}~(\bibinfo
  {number} {6}),\ \bibinfo {pages} {064042}},\ \Eprint
  {https://arxiv.org/abs/2006.11515} {arXiv:2006.11515 [gr-qc]} \BibitemShut
  {NoStop}%
\bibitem [{\citenamefont {Doneva}\ \emph {et~al.}(2022)\citenamefont {Doneva},
  \citenamefont {Va\~n\'o Vi\~nuales},\ and\ \citenamefont
  {Yazadjiev}}]{Doneva:2022byd}%
  \BibitemOpen
  \bibfield  {author} {\bibinfo {author} {\bibnamefont {Doneva}, \bibfnamefont
  {D.~D.}}, \bibinfo {author} {\bibfnamefont {A.}~\bibnamefont {Va\~n\'o
  Vi\~nuales}}, and\ \bibinfo {author} {\bibfnamefont {S.~S.}\ \bibnamefont
  {Yazadjiev}}} (\bibinfo {year} {2022}),\ \href
  {https://doi.org/10.1103/PhysRevD.106.L061502} {\bibfield  {journal}
  {\bibinfo  {journal} {Phys. Rev. D}\ }\textbf {\bibinfo {volume}
  {106}}~(\bibinfo {number} {6}),\ \bibinfo {pages} {L061502}},\ \Eprint
  {https://arxiv.org/abs/2204.05333} {arXiv:2204.05333 [gr-qc]} \BibitemShut
  {NoStop}%
\bibitem [{\citenamefont {Doneva}\ and\ \citenamefont
  {Yazadjiev}(2016)}]{Doneva:2016xmf}%
  \BibitemOpen
  \bibfield  {author} {\bibinfo {author} {\bibnamefont {Doneva}, \bibfnamefont
  {D.~D.}}, and\ \bibinfo {author} {\bibfnamefont {S.~S.}\ \bibnamefont
  {Yazadjiev}}} (\bibinfo {year} {2016}),\ \href
  {https://doi.org/10.1088/1475-7516/2016/11/019} {\bibfield  {journal}
  {\bibinfo  {journal} {JCAP}\ }\textbf {\bibinfo {volume} {11}},\ \bibinfo
  {pages} {019}},\ \Eprint {https://arxiv.org/abs/1607.03299} {arXiv:1607.03299
  [gr-qc]} \BibitemShut {NoStop}%
\bibitem [{\citenamefont {Doneva}\ and\ \citenamefont
  {Yazadjiev}(2018{\natexlab{a}})}]{Doneva:2017duq}%
  \BibitemOpen
  \bibfield  {author} {\bibinfo {author} {\bibnamefont {Doneva}, \bibfnamefont
  {D.~D.}}, and\ \bibinfo {author} {\bibfnamefont {S.~S.}\ \bibnamefont
  {Yazadjiev}}} (\bibinfo {year} {2018}{\natexlab{a}}),\ \href
  {https://doi.org/10.1088/1475-7516/2018/04/011} {\bibfield  {journal}
  {\bibinfo  {journal} {JCAP}\ }\textbf {\bibinfo {volume} {04}},\ \bibinfo
  {pages} {011}},\ \Eprint {https://arxiv.org/abs/1712.03715} {arXiv:1712.03715
  [gr-qc]} \BibitemShut {NoStop}%
\bibitem [{\citenamefont {Doneva}\ and\ \citenamefont
  {Yazadjiev}(2018{\natexlab{b}})}]{Doneva:2017bvd}%
  \BibitemOpen
  \bibfield  {author} {\bibinfo {author} {\bibnamefont {Doneva}, \bibfnamefont
  {D.~D.}}, and\ \bibinfo {author} {\bibfnamefont {S.~S.}\ \bibnamefont
  {Yazadjiev}}} (\bibinfo {year} {2018}{\natexlab{b}}),\ \href
  {https://doi.org/10.1103/PhysRevLett.120.131103} {\bibfield  {journal}
  {\bibinfo  {journal} {Phys. Rev. Lett.}\ }\textbf {\bibinfo {volume}
  {120}}~(\bibinfo {number} {13}),\ \bibinfo {pages} {131103}},\ \Eprint
  {https://arxiv.org/abs/1711.01187} {arXiv:1711.01187 [gr-qc]} \BibitemShut
  {NoStop}%
\bibitem [{\citenamefont {Doneva}\ and\ \citenamefont
  {Yazadjiev}(2020{\natexlab{a}})}]{Doneva:2020afj}%
  \BibitemOpen
  \bibfield  {author} {\bibinfo {author} {\bibnamefont {Doneva}, \bibfnamefont
  {D.~D.}}, and\ \bibinfo {author} {\bibfnamefont {S.~S.}\ \bibnamefont
  {Yazadjiev}}} (\bibinfo {year} {2020}{\natexlab{a}}),\ \href
  {https://doi.org/10.1103/PhysRevD.101.104010} {\bibfield  {journal} {\bibinfo
   {journal} {Phys. Rev. D}\ }\textbf {\bibinfo {volume} {101}}~(\bibinfo
  {number} {10}),\ \bibinfo {pages} {104010}},\ \Eprint
  {https://arxiv.org/abs/2004.03956} {arXiv:2004.03956 [gr-qc]} \BibitemShut
  {NoStop}%
\bibitem [{\citenamefont {Doneva}\ and\ \citenamefont
  {Yazadjiev}(2020{\natexlab{b}})}]{Doneva:2019ltb}%
  \BibitemOpen
  \bibfield  {author} {\bibinfo {author} {\bibnamefont {Doneva}, \bibfnamefont
  {D.~D.}}, and\ \bibinfo {author} {\bibfnamefont {S.~S.}\ \bibnamefont
  {Yazadjiev}}} (\bibinfo {year} {2020}{\natexlab{b}}),\ \href
  {https://doi.org/10.1103/PhysRevD.101.064072} {\bibfield  {journal} {\bibinfo
   {journal} {Phys. Rev. D}\ }\textbf {\bibinfo {volume} {101}}~(\bibinfo
  {number} {6}),\ \bibinfo {pages} {064072}},\ \Eprint
  {https://arxiv.org/abs/1911.06908} {arXiv:1911.06908 [gr-qc]} \BibitemShut
  {NoStop}%
\bibitem [{\citenamefont {Doneva}\ and\ \citenamefont
  {Yazadjiev}(2021{\natexlab{a}})}]{Doneva:2021dqn}%
  \BibitemOpen
  \bibfield  {author} {\bibinfo {author} {\bibnamefont {Doneva}, \bibfnamefont
  {D.~D.}}, and\ \bibinfo {author} {\bibfnamefont {S.~S.}\ \bibnamefont
  {Yazadjiev}}} (\bibinfo {year} {2021}{\natexlab{a}}),\ \href
  {https://doi.org/10.1103/PhysRevD.103.064024} {\bibfield  {journal} {\bibinfo
   {journal} {Phys. Rev. D}\ }\textbf {\bibinfo {volume} {103}}~(\bibinfo
  {number} {6}),\ \bibinfo {pages} {064024}},\ \Eprint
  {https://arxiv.org/abs/2101.03514} {arXiv:2101.03514 [gr-qc]} \BibitemShut
  {NoStop}%
\bibitem [{\citenamefont {Doneva}\ and\ \citenamefont
  {Yazadjiev}(2021{\natexlab{b}})}]{Doneva:2021dcc}%
  \BibitemOpen
  \bibfield  {author} {\bibinfo {author} {\bibnamefont {Doneva}, \bibfnamefont
  {D.~D.}}, and\ \bibinfo {author} {\bibfnamefont {S.~S.}\ \bibnamefont
  {Yazadjiev}}} (\bibinfo {year} {2021}{\natexlab{b}}),\ \href
  {https://doi.org/10.1103/PhysRevD.103.083007} {\bibfield  {journal} {\bibinfo
   {journal} {Phys. Rev. D}\ }\textbf {\bibinfo {volume} {103}}~(\bibinfo
  {number} {8}),\ \bibinfo {pages} {083007}},\ \Eprint
  {https://arxiv.org/abs/2102.03940} {arXiv:2102.03940 [gr-qc]} \BibitemShut
  {NoStop}%
\bibitem [{\citenamefont {Doneva}\ and\ \citenamefont
  {Yazadjiev}(2022)}]{Doneva:2021tvn}%
  \BibitemOpen
  \bibfield  {author} {\bibinfo {author} {\bibnamefont {Doneva}, \bibfnamefont
  {D.~D.}}, and\ \bibinfo {author} {\bibfnamefont {S.~S.}\ \bibnamefont
  {Yazadjiev}}} (\bibinfo {year} {2022}),\ \href
  {https://doi.org/10.1103/PhysRevD.105.L041502} {\bibfield  {journal}
  {\bibinfo  {journal} {Phys. Rev. D}\ }\textbf {\bibinfo {volume}
  {105}}~(\bibinfo {number} {4}),\ \bibinfo {pages} {L041502}},\ \Eprint
  {https://arxiv.org/abs/2107.01738} {arXiv:2107.01738 [gr-qc]} \BibitemShut
  {NoStop}%
\bibitem [{\citenamefont {Doneva}\ \emph {et~al.}(2010)\citenamefont {Doneva},
  \citenamefont {Yazadjiev}, \citenamefont {Kokkotas},\ and\ \citenamefont
  {Stefanov}}]{Doneva:2010ke}%
  \BibitemOpen
  \bibfield  {author} {\bibinfo {author} {\bibnamefont {Doneva}, \bibfnamefont
  {D.~D.}}, \bibinfo {author} {\bibfnamefont {S.~S.}\ \bibnamefont
  {Yazadjiev}}, \bibinfo {author} {\bibfnamefont {K.~D.}\ \bibnamefont
  {Kokkotas}}, and\ \bibinfo {author} {\bibfnamefont {I.~Z.}\ \bibnamefont
  {Stefanov}}} (\bibinfo {year} {2010}),\ \href
  {https://doi.org/10.1103/PhysRevD.82.064030} {\bibfield  {journal} {\bibinfo
  {journal} {Phys. Rev. D}\ }\textbf {\bibinfo {volume} {82}},\ \bibinfo
  {pages} {064030}},\ \Eprint {https://arxiv.org/abs/1007.1767}
  {arXiv:1007.1767 [gr-qc]} \BibitemShut {NoStop}%
\bibitem [{\citenamefont {Doneva}\ \emph
  {et~al.}(2014{\natexlab{a}})\citenamefont {Doneva}, \citenamefont
  {Yazadjiev}, \citenamefont {Staykov},\ and\ \citenamefont
  {Kokkotas}}]{Doneva:2014faa}%
  \BibitemOpen
  \bibfield  {author} {\bibinfo {author} {\bibnamefont {Doneva}, \bibfnamefont
  {D.~D.}}, \bibinfo {author} {\bibfnamefont {S.~S.}\ \bibnamefont
  {Yazadjiev}}, \bibinfo {author} {\bibfnamefont {K.~V.}\ \bibnamefont
  {Staykov}}, and\ \bibinfo {author} {\bibfnamefont {K.~D.}\ \bibnamefont
  {Kokkotas}}} (\bibinfo {year} {2014}{\natexlab{a}}),\ \href
  {https://doi.org/10.1103/PhysRevD.90.104021} {\bibfield  {journal} {\bibinfo
  {journal} {Phys. Rev. D}\ }\textbf {\bibinfo {volume} {90}}~(\bibinfo
  {number} {10}),\ \bibinfo {pages} {104021}},\ \Eprint
  {https://arxiv.org/abs/1408.1641} {arXiv:1408.1641 [gr-qc]} \BibitemShut
  {NoStop}%
\bibitem [{\citenamefont {Doneva}\ \emph {et~al.}(2013)\citenamefont {Doneva},
  \citenamefont {Yazadjiev}, \citenamefont {Stergioulas},\ and\ \citenamefont
  {Kokkotas}}]{Doneva:2013qva}%
  \BibitemOpen
  \bibfield  {author} {\bibinfo {author} {\bibnamefont {Doneva}, \bibfnamefont
  {D.~D.}}, \bibinfo {author} {\bibfnamefont {S.~S.}\ \bibnamefont
  {Yazadjiev}}, \bibinfo {author} {\bibfnamefont {N.}~\bibnamefont
  {Stergioulas}}, and\ \bibinfo {author} {\bibfnamefont {K.~D.}\ \bibnamefont
  {Kokkotas}}} (\bibinfo {year} {2013}),\ \href
  {https://doi.org/10.1103/PhysRevD.88.084060} {\bibfield  {journal} {\bibinfo
  {journal} {Phys. Rev. D}\ }\textbf {\bibinfo {volume} {88}}~(\bibinfo
  {number} {8}),\ \bibinfo {pages} {084060}},\ \Eprint
  {https://arxiv.org/abs/1309.0605} {arXiv:1309.0605 [gr-qc]} \BibitemShut
  {NoStop}%
\bibitem [{\citenamefont {Doneva}\ \emph
  {et~al.}(2018{\natexlab{b}})\citenamefont {Doneva}, \citenamefont
  {Yazadjiev}, \citenamefont {Stergioulas},\ and\ \citenamefont
  {Kokkotas}}]{Doneva:2018ouu}%
  \BibitemOpen
  \bibfield  {author} {\bibinfo {author} {\bibnamefont {Doneva}, \bibfnamefont
  {D.~D.}}, \bibinfo {author} {\bibfnamefont {S.~S.}\ \bibnamefont
  {Yazadjiev}}, \bibinfo {author} {\bibfnamefont {N.}~\bibnamefont
  {Stergioulas}}, and\ \bibinfo {author} {\bibfnamefont {K.~D.}\ \bibnamefont
  {Kokkotas}}} (\bibinfo {year} {2018}{\natexlab{b}}),\ \href
  {https://doi.org/10.1103/PhysRevD.98.104039} {\bibfield  {journal} {\bibinfo
  {journal} {Phys. Rev. D}\ }\textbf {\bibinfo {volume} {98}}~(\bibinfo
  {number} {10}),\ \bibinfo {pages} {104039}},\ \Eprint
  {https://arxiv.org/abs/1807.05449} {arXiv:1807.05449 [gr-qc]} \BibitemShut
  {NoStop}%
\bibitem [{\citenamefont {Doneva}\ \emph
  {et~al.}(2014{\natexlab{b}})\citenamefont {Doneva}, \citenamefont
  {Yazadjiev}, \citenamefont {Stergioulas}, \citenamefont {Kokkotas},\ and\
  \citenamefont {Athanasiadis}}]{Doneva:2014uma}%
  \BibitemOpen
  \bibfield  {author} {\bibinfo {author} {\bibnamefont {Doneva}, \bibfnamefont
  {D.~D.}}, \bibinfo {author} {\bibfnamefont {S.~S.}\ \bibnamefont
  {Yazadjiev}}, \bibinfo {author} {\bibfnamefont {N.}~\bibnamefont
  {Stergioulas}}, \bibinfo {author} {\bibfnamefont {K.~D.}\ \bibnamefont
  {Kokkotas}}, and\ \bibinfo {author} {\bibfnamefont {T.~M.}\ \bibnamefont
  {Athanasiadis}}} (\bibinfo {year} {2014}{\natexlab{b}}),\ \href
  {https://doi.org/10.1103/PhysRevD.90.044004} {\bibfield  {journal} {\bibinfo
  {journal} {Phys. Rev. D}\ }\textbf {\bibinfo {volume} {90}}~(\bibinfo
  {number} {4}),\ \bibinfo {pages} {044004}},\ \Eprint
  {https://arxiv.org/abs/1405.6976} {arXiv:1405.6976 [astro-ph.HE]}
  \BibitemShut {NoStop}%
\bibitem [{\citenamefont {Douchin}\ and\ \citenamefont
  {Haensel}(2001)}]{Douchin:2001sv}%
  \BibitemOpen
  \bibfield  {author} {\bibinfo {author} {\bibnamefont {Douchin}, \bibfnamefont
  {F.}}, and\ \bibinfo {author} {\bibfnamefont {P.}~\bibnamefont {Haensel}}}
  (\bibinfo {year} {2001}),\ \href {https://doi.org/10.1051/0004-6361:20011402}
  {\bibfield  {journal} {\bibinfo  {journal} {Astron. Astrophys.}\ }\textbf
  {\bibinfo {volume} {380}},\ \bibinfo {pages} {151}},\ \Eprint
  {https://arxiv.org/abs/astro-ph/0111092} {arXiv:astro-ph/0111092}
  \BibitemShut {NoStop}%
\bibitem [{\citenamefont {East}\ and\ \citenamefont
  {Ripley}(2021{\natexlab{a}})}]{East:2021bqk}%
  \BibitemOpen
  \bibfield  {author} {\bibinfo {author} {\bibnamefont {East}, \bibfnamefont
  {W.~E.}}, and\ \bibinfo {author} {\bibfnamefont {J.~L.}\ \bibnamefont
  {Ripley}}} (\bibinfo {year} {2021}{\natexlab{a}}),\ \href
  {https://doi.org/10.1103/PhysRevLett.127.101102} {\bibfield  {journal}
  {\bibinfo  {journal} {Phys. Rev. Lett.}\ }\textbf {\bibinfo {volume}
  {127}}~(\bibinfo {number} {10}),\ \bibinfo {pages} {101102}},\ \Eprint
  {https://arxiv.org/abs/2105.08571} {arXiv:2105.08571 [gr-qc]} \BibitemShut
  {NoStop}%
\bibitem [{\citenamefont {East}\ and\ \citenamefont
  {Ripley}(2021{\natexlab{b}})}]{East:2020hgw}%
  \BibitemOpen
  \bibfield  {author} {\bibinfo {author} {\bibnamefont {East}, \bibfnamefont
  {W.~E.}}, and\ \bibinfo {author} {\bibfnamefont {J.~L.}\ \bibnamefont
  {Ripley}}} (\bibinfo {year} {2021}{\natexlab{b}}),\ \href
  {https://doi.org/10.1103/PhysRevD.103.044040} {\bibfield  {journal} {\bibinfo
   {journal} {Phys. Rev. D}\ }\textbf {\bibinfo {volume} {103}}~(\bibinfo
  {number} {4}),\ \bibinfo {pages} {044040}},\ \Eprint
  {https://arxiv.org/abs/2011.03547} {arXiv:2011.03547 [gr-qc]} \BibitemShut
  {NoStop}%
\bibitem [{\citenamefont {Elley}\ \emph {et~al.}(2022)\citenamefont {Elley},
  \citenamefont {Silva}, \citenamefont {Witek},\ and\ \citenamefont
  {Yunes}}]{Elley:2022ept}%
  \BibitemOpen
  \bibfield  {author} {\bibinfo {author} {\bibnamefont {Elley}, \bibfnamefont
  {M.}}, \bibinfo {author} {\bibfnamefont {H.~O.}\ \bibnamefont {Silva}},
  \bibinfo {author} {\bibfnamefont {H.}~\bibnamefont {Witek}}, and\ \bibinfo
  {author} {\bibfnamefont {N.}~\bibnamefont {Yunes}}} (\bibinfo {year}
  {2022}),\ \href {https://doi.org/10.1103/PhysRevD.106.044018} {\bibfield
  {journal} {\bibinfo  {journal} {Phys. Rev. D}\ }\textbf {\bibinfo {volume}
  {106}}~(\bibinfo {number} {4}),\ \bibinfo {pages} {044018}},\ \Eprint
  {https://arxiv.org/abs/2205.06240} {arXiv:2205.06240 [gr-qc]} \BibitemShut
  {NoStop}%
\bibitem [{\citenamefont {Erices}\ \emph {et~al.}(2022)\citenamefont {Erices},
  \citenamefont {Riquelme},\ and\ \citenamefont {Zalaquett}}]{Erices:2022bws}%
  \BibitemOpen
  \bibfield  {author} {\bibinfo {author} {\bibnamefont {Erices}, \bibfnamefont
  {C.}}, \bibinfo {author} {\bibfnamefont {S.}~\bibnamefont {Riquelme}}, and\
  \bibinfo {author} {\bibfnamefont {N.}~\bibnamefont {Zalaquett}}} (\bibinfo
  {year} {2022}),\ \href {https://doi.org/10.1103/PhysRevD.106.044046}
  {\bibfield  {journal} {\bibinfo  {journal} {Phys. Rev. D}\ }\textbf {\bibinfo
  {volume} {106}}~(\bibinfo {number} {4}),\ \bibinfo {pages} {044046}},\
  \Eprint {https://arxiv.org/abs/2203.06030} {arXiv:2203.06030 [gr-qc]}
  \BibitemShut {NoStop}%
\bibitem [{\citenamefont {Esposito-Far{\`e}se}(2004)}]{EspositoFarese:2004cc}%
  \BibitemOpen
  \bibfield  {author} {\bibinfo {author} {\bibnamefont {Esposito-Far{\`e}se},
  \bibfnamefont {G.}}} (\bibinfo {year} {2004}),\ \href
  {https://doi.org/10.1063/1.1835173} {\bibfield  {journal} {\bibinfo
  {journal} {AIP Conf. Proc.}\ }\textbf {\bibinfo {volume} {736}}~(\bibinfo
  {number} {1}),\ \bibinfo {pages} {35}},\ \Eprint
  {https://arxiv.org/abs/gr-qc/0409081} {arXiv:gr-qc/0409081} \BibitemShut
  {NoStop}%
\bibitem [{\citenamefont {Esposito-Far{\`e}se}\ \emph
  {et~al.}(2010)\citenamefont {Esposito-Far{\`e}se}, \citenamefont {Pitrou},\
  and\ \citenamefont {Uzan}}]{Esposito-Farese:2009wbc}%
  \BibitemOpen
  \bibfield  {author} {\bibinfo {author} {\bibnamefont {Esposito-Far{\`e}se},
  \bibfnamefont {G.}}, \bibinfo {author} {\bibfnamefont {C.}~\bibnamefont
  {Pitrou}}, and\ \bibinfo {author} {\bibfnamefont {J.-P.}\ \bibnamefont
  {Uzan}}} (\bibinfo {year} {2010}),\ \href
  {https://doi.org/10.1103/PhysRevD.81.063519} {\bibfield  {journal} {\bibinfo
  {journal} {Phys. Rev. D}\ }\textbf {\bibinfo {volume} {81}},\ \bibinfo
  {pages} {063519}},\ \Eprint {https://arxiv.org/abs/0912.0481}
  {arXiv:0912.0481 [gr-qc]} \BibitemShut {NoStop}%
\bibitem [{\citenamefont {Fernandes}\ \emph
  {et~al.}(2019{\natexlab{a}})\citenamefont {Fernandes}, \citenamefont
  {Herdeiro}, \citenamefont {Pombo}, \citenamefont {Radu},\ and\ \citenamefont
  {Sanchis-Gual}}]{Fernandes:2019kmh}%
  \BibitemOpen
  \bibfield  {author} {\bibinfo {author} {\bibnamefont {Fernandes},
  \bibfnamefont {P.~G.~S.}}, \bibinfo {author} {\bibfnamefont {C.~A.~R.}\
  \bibnamefont {Herdeiro}}, \bibinfo {author} {\bibfnamefont {A.~M.}\
  \bibnamefont {Pombo}}, \bibinfo {author} {\bibfnamefont {E.}~\bibnamefont
  {Radu}}, and\ \bibinfo {author} {\bibfnamefont {N.}~\bibnamefont
  {Sanchis-Gual}}} (\bibinfo {year} {2019}{\natexlab{a}}),\ \href
  {https://doi.org/10.1103/PhysRevD.100.084045} {\bibfield  {journal} {\bibinfo
   {journal} {Phys. Rev. D}\ }\textbf {\bibinfo {volume} {100}}~(\bibinfo
  {number} {8}),\ \bibinfo {pages} {084045}},\ \Eprint
  {https://arxiv.org/abs/1908.00037} {arXiv:1908.00037 [gr-qc]} \BibitemShut
  {NoStop}%
\bibitem [{\citenamefont {Fernandes}\ \emph
  {et~al.}(2019{\natexlab{b}})\citenamefont {Fernandes}, \citenamefont
  {Herdeiro}, \citenamefont {Pombo}, \citenamefont {Radu},\ and\ \citenamefont
  {Sanchis-Gual}}]{Fernandes:2019rez}%
  \BibitemOpen
  \bibfield  {author} {\bibinfo {author} {\bibnamefont {Fernandes},
  \bibfnamefont {P.~G.~S.}}, \bibinfo {author} {\bibfnamefont {C.~A.~R.}\
  \bibnamefont {Herdeiro}}, \bibinfo {author} {\bibfnamefont {A.~M.}\
  \bibnamefont {Pombo}}, \bibinfo {author} {\bibfnamefont {E.}~\bibnamefont
  {Radu}}, and\ \bibinfo {author} {\bibfnamefont {N.}~\bibnamefont
  {Sanchis-Gual}}} (\bibinfo {year} {2019}{\natexlab{b}}),\ \href
  {https://doi.org/10.1088/1361-6382/ab23a1} {\bibfield  {journal} {\bibinfo
  {journal} {Class. Quant. Grav.}\ }\textbf {\bibinfo {volume} {36}}~(\bibinfo
  {number} {13}),\ \bibinfo {pages} {134002}},\ \bibinfo {note} {[Erratum:
  Class.Quant.Grav. 37, 049501 (2020)]},\ \Eprint
  {https://arxiv.org/abs/1902.05079} {arXiv:1902.05079 [gr-qc]} \BibitemShut
  {NoStop}%
\bibitem [{\citenamefont {Franchini}\ \emph {et~al.}(2022)\citenamefont
  {Franchini}, \citenamefont {Bezares}, \citenamefont {Barausse},\ and\
  \citenamefont {Lehner}}]{Franchini:2022ukz}%
  \BibitemOpen
  \bibfield  {author} {\bibinfo {author} {\bibnamefont {Franchini},
  \bibfnamefont {N.}}, \bibinfo {author} {\bibfnamefont {M.}~\bibnamefont
  {Bezares}}, \bibinfo {author} {\bibfnamefont {E.}~\bibnamefont {Barausse}},
  and\ \bibinfo {author} {\bibfnamefont {L.}~\bibnamefont {Lehner}}} (\bibinfo
  {year} {2022}),\ \href {https://doi.org/10.1103/PhysRevD.106.064061}
  {\bibfield  {journal} {\bibinfo  {journal} {Phys. Rev. D}\ }\textbf {\bibinfo
  {volume} {106}}~(\bibinfo {number} {6}),\ \bibinfo {pages} {064061}},\
  \Eprint {https://arxiv.org/abs/2206.00014} {arXiv:2206.00014 [gr-qc]}
  \BibitemShut {NoStop}%
\bibitem [{\citenamefont {Franchini}\ \emph {et~al.}(2018)\citenamefont
  {Franchini}, \citenamefont {Coates},\ and\ \citenamefont
  {Sotiriou}}]{Franchini:2017zzx}%
  \BibitemOpen
  \bibfield  {author} {\bibinfo {author} {\bibnamefont {Franchini},
  \bibfnamefont {N.}}, \bibinfo {author} {\bibfnamefont {A.}~\bibnamefont
  {Coates}}, and\ \bibinfo {author} {\bibfnamefont {T.~P.}\ \bibnamefont
  {Sotiriou}}} (\bibinfo {year} {2018}),\ \href
  {https://doi.org/10.1103/PhysRevD.97.064013} {\bibfield  {journal} {\bibinfo
  {journal} {Phys. Rev. D}\ }\textbf {\bibinfo {volume} {97}}~(\bibinfo
  {number} {6}),\ \bibinfo {pages} {064013}},\ \Eprint
  {https://arxiv.org/abs/1708.02113} {arXiv:1708.02113 [gr-qc]} \BibitemShut
  {NoStop}%
\bibitem [{\citenamefont {Franchini}\ and\ \citenamefont
  {Sotiriou}(2020)}]{Franchini:2019npi}%
  \BibitemOpen
  \bibfield  {author} {\bibinfo {author} {\bibnamefont {Franchini},
  \bibfnamefont {N.}}, and\ \bibinfo {author} {\bibfnamefont {T.~P.}\
  \bibnamefont {Sotiriou}}} (\bibinfo {year} {2020}),\ \href
  {https://doi.org/10.1103/PhysRevD.101.064068} {\bibfield  {journal} {\bibinfo
   {journal} {Phys. Rev. D}\ }\textbf {\bibinfo {volume} {101}}~(\bibinfo
  {number} {6}),\ \bibinfo {pages} {064068}},\ \Eprint
  {https://arxiv.org/abs/1903.05427} {arXiv:1903.05427 [gr-qc]} \BibitemShut
  {NoStop}%
\bibitem [{\citenamefont {Freire}(2022)}]{FreireWebSite}%
  \BibitemOpen
  \bibfield  {author} {\bibinfo {author} {\bibnamefont {Freire}, \bibfnamefont
  {P.~C.~C.}}} (\bibinfo {year} {2022}),\ \href@noop {} {\enquote {\bibinfo
  {title} {{Pulsar mass measurements and tests of general relativity}},}\
  }\bibinfo {howpublished}
  {\url{https://www3.mpifr-bonn.mpg.de/staff/pfreire/NS_masses.html}},\
  \bibinfo {note} {accessed: 2021-12-28}\BibitemShut {NoStop}%
\bibitem [{\citenamefont {Freire}\ \emph {et~al.}(2012)\citenamefont {Freire},
  \citenamefont {Wex}, \citenamefont {Esposito-Far\`ese}, \citenamefont
  {Verbiest}, \citenamefont {Bailes}, \citenamefont {Jacoby}, \citenamefont
  {Kramer}, \citenamefont {Stairs}, \citenamefont {Antoniadis},\ and\
  \citenamefont {Janssen}}]{Freire:2012mg}%
  \BibitemOpen
  \bibfield  {author} {\bibinfo {author} {\bibnamefont {Freire}, \bibfnamefont
  {P.~C.~C.}}, \bibinfo {author} {\bibfnamefont {N.}~\bibnamefont {Wex}},
  \bibinfo {author} {\bibfnamefont {G.}~\bibnamefont {Esposito-Far\`ese}},
  \bibinfo {author} {\bibfnamefont {J.~P.~W.}\ \bibnamefont {Verbiest}},
  \bibinfo {author} {\bibfnamefont {M.}~\bibnamefont {Bailes}}, \bibinfo
  {author} {\bibfnamefont {B.~A.}\ \bibnamefont {Jacoby}}, \bibinfo {author}
  {\bibfnamefont {M.}~\bibnamefont {Kramer}}, \bibinfo {author} {\bibfnamefont
  {I.~H.}\ \bibnamefont {Stairs}}, \bibinfo {author} {\bibfnamefont
  {J.}~\bibnamefont {Antoniadis}}, and\ \bibinfo {author} {\bibfnamefont
  {G.~H.}\ \bibnamefont {Janssen}}} (\bibinfo {year} {2012}),\ \href
  {https://doi.org/10.1111/j.1365-2966.2012.21253.x} {\bibfield  {journal}
  {\bibinfo  {journal} {Mon. Not. Roy. Astron. Soc.}\ }\textbf {\bibinfo
  {volume} {423}},\ \bibinfo {pages} {3328}},\ \Eprint
  {https://arxiv.org/abs/1205.1450} {arXiv:1205.1450 [astro-ph.GA]}
  \BibitemShut {NoStop}%
\bibitem [{\citenamefont {Friedman}\ and\ \citenamefont
  {Schutz}(1978)}]{Friedman:1978hf}%
  \BibitemOpen
  \bibfield  {author} {\bibinfo {author} {\bibnamefont {Friedman},
  \bibfnamefont {J.~L.}}, and\ \bibinfo {author} {\bibfnamefont {B.~F.}\
  \bibnamefont {Schutz}}} (\bibinfo {year} {1978}),\ \href
  {https://doi.org/10.1086/156143} {\bibfield  {journal} {\bibinfo  {journal}
  {Astrophys. J.}\ }\textbf {\bibinfo {volume} {222}},\ \bibinfo {pages}
  {281}}\BibitemShut {NoStop}%
\bibitem [{\citenamefont {Gao}\ \emph {et~al.}(2019)\citenamefont {Gao},
  \citenamefont {Huang},\ and\ \citenamefont {Liu}}]{Gao:2018acg}%
  \BibitemOpen
  \bibfield  {author} {\bibinfo {author} {\bibnamefont {Gao}, \bibfnamefont
  {Y.-X.}}, \bibinfo {author} {\bibfnamefont {Y.}~\bibnamefont {Huang}}, and\
  \bibinfo {author} {\bibfnamefont {D.-J.}\ \bibnamefont {Liu}}} (\bibinfo
  {year} {2019}),\ \href {https://doi.org/10.1103/PhysRevD.99.044020}
  {\bibfield  {journal} {\bibinfo  {journal} {Phys. Rev. D}\ }\textbf {\bibinfo
  {volume} {99}}~(\bibinfo {number} {4}),\ \bibinfo {pages} {044020}},\ \Eprint
  {https://arxiv.org/abs/1808.01433} {arXiv:1808.01433 [gr-qc]} \BibitemShut
  {NoStop}%
\bibitem [{\citenamefont {Garcia-Saenz}\ \emph {et~al.}(2021)\citenamefont
  {Garcia-Saenz}, \citenamefont {Held},\ and\ \citenamefont
  {Zhang}}]{Garcia-Saenz:2021uyv}%
  \BibitemOpen
  \bibfield  {author} {\bibinfo {author} {\bibnamefont {Garcia-Saenz},
  \bibfnamefont {S.}}, \bibinfo {author} {\bibfnamefont {A.}~\bibnamefont
  {Held}}, and\ \bibinfo {author} {\bibfnamefont {J.}~\bibnamefont {Zhang}}}
  (\bibinfo {year} {2021}),\ \href
  {https://doi.org/10.1103/PhysRevLett.127.131104} {\bibfield  {journal}
  {\bibinfo  {journal} {Phys. Rev. Lett.}\ }\textbf {\bibinfo {volume}
  {127}}~(\bibinfo {number} {13}),\ \bibinfo {pages} {131104}},\ \Eprint
  {https://arxiv.org/abs/2104.08049} {arXiv:2104.08049 [gr-qc]} \BibitemShut
  {NoStop}%
\bibitem [{\citenamefont {Garcia-Saenz}\ \emph {et~al.}(2022)\citenamefont
  {Garcia-Saenz}, \citenamefont {Held},\ and\ \citenamefont
  {Zhang}}]{Garcia-Saenz:2022wsl}%
  \BibitemOpen
  \bibfield  {author} {\bibinfo {author} {\bibnamefont {Garcia-Saenz},
  \bibfnamefont {S.}}, \bibinfo {author} {\bibfnamefont {A.}~\bibnamefont
  {Held}}, and\ \bibinfo {author} {\bibfnamefont {J.}~\bibnamefont {Zhang}}}
  (\bibinfo {year} {2022}),\ \href {https://doi.org/10.1007/JHEP05(2022)139}
  {\bibfield  {journal} {\bibinfo  {journal} {JHEP}\ }\textbf {\bibinfo
  {volume} {05}},\ \bibinfo {pages} {139}},\ \Eprint
  {https://arxiv.org/abs/2202.07131} {arXiv:2202.07131 [gr-qc]} \BibitemShut
  {NoStop}%
\bibitem [{\citenamefont {{Gendreau}}\ and\ \citenamefont
  {{Arzoumanian}}(2017)}]{Gendreau2017:NatAst}%
  \BibitemOpen
  \bibfield  {author} {\bibinfo {author} {\bibnamefont {{Gendreau}},
  \bibfnamefont {K.}}, and\ \bibinfo {author} {\bibfnamefont {Z.}~\bibnamefont
  {{Arzoumanian}}}} (\bibinfo {year} {2017}),\ \href
  {https://doi.org/10.1038/s41550-017-0301-3} {\bibfield  {journal} {\bibinfo
  {journal} {Nature Astronomy}\ }\textbf {\bibinfo {volume} {1}},\ \bibinfo
  {pages} {895}}\BibitemShut {NoStop}%
\bibitem [{\citenamefont {{Gendreau}}\ \emph {et~al.}(2012)\citenamefont
  {{Gendreau}}, \citenamefont {{Arzoumanian}},\ and\ \citenamefont
  {{Okajima}}}]{Gendreau2012:SPIE}%
  \BibitemOpen
  \bibfield  {author} {\bibinfo {author} {\bibnamefont {{Gendreau}},
  \bibfnamefont {K.~C.}}, \bibinfo {author} {\bibfnamefont {Z.}~\bibnamefont
  {{Arzoumanian}}}, and\ \bibinfo {author} {\bibfnamefont {T.}~\bibnamefont
  {{Okajima}}}} (\bibinfo {year} {2012}),\ in\ \href
  {https://doi.org/10.1117/12.926396} {\emph {\bibinfo {booktitle} {Space
  Telescopes and Instrumentation 2012: Ultraviolet to Gamma Ray}}},\ \bibinfo
  {series} {Proc.~SPIE}, Vol.\ \bibinfo {volume} {8443},\ p.\ \bibinfo {pages}
  {844313}\BibitemShut {NoStop}%
\bibitem [{\citenamefont {Geng}\ \emph {et~al.}(2020)\citenamefont {Geng},
  \citenamefont {Kuan},\ and\ \citenamefont {Luo}}]{Geng:2020slq}%
  \BibitemOpen
  \bibfield  {author} {\bibinfo {author} {\bibnamefont {Geng}, \bibfnamefont
  {C.-Q.}}, \bibinfo {author} {\bibfnamefont {H.-J.}\ \bibnamefont {Kuan}},
  and\ \bibinfo {author} {\bibfnamefont {L.-W.}\ \bibnamefont {Luo}}} (\bibinfo
  {year} {2020}),\ \href {https://doi.org/10.1140/epjc/s10052-020-8359-y}
  {\bibfield  {journal} {\bibinfo  {journal} {Eur. Phys. J. C}\ }\textbf
  {\bibinfo {volume} {80}}~(\bibinfo {number} {8}),\ \bibinfo {pages} {780}},\
  \Eprint {https://arxiv.org/abs/2005.11629} {arXiv:2005.11629 [gr-qc]}
  \BibitemShut {NoStop}%
\bibitem [{\citenamefont {Geroch}(1970)}]{Geroch:1970cd}%
  \BibitemOpen
  \bibfield  {author} {\bibinfo {author} {\bibnamefont {Geroch}, \bibfnamefont
  {R.~P.}}} (\bibinfo {year} {1970}),\ \href
  {https://doi.org/10.1063/1.1665427} {\bibfield  {journal} {\bibinfo
  {journal} {J. Math. Phys.}\ }\textbf {\bibinfo {volume} {11}},\ \bibinfo
  {pages} {2580}}\BibitemShut {NoStop}%
\bibitem [{\citenamefont {Gerosa}\ \emph {et~al.}(2016)\citenamefont {Gerosa},
  \citenamefont {Sperhake},\ and\ \citenamefont {Ott}}]{Gerosa:2016fri}%
  \BibitemOpen
  \bibfield  {author} {\bibinfo {author} {\bibnamefont {Gerosa}, \bibfnamefont
  {D.}}, \bibinfo {author} {\bibfnamefont {U.}~\bibnamefont {Sperhake}}, and\
  \bibinfo {author} {\bibfnamefont {C.~D.}\ \bibnamefont {Ott}}} (\bibinfo
  {year} {2016}),\ \href {https://doi.org/10.1088/0264-9381/33/13/135002}
  {\bibfield  {journal} {\bibinfo  {journal} {Class. Quant. Grav.}\ }\textbf
  {\bibinfo {volume} {33}}~(\bibinfo {number} {13}),\ \bibinfo {pages}
  {135002}},\ \Eprint {https://arxiv.org/abs/1602.06952} {arXiv:1602.06952
  [gr-qc]} \BibitemShut {NoStop}%
\bibitem [{\citenamefont {Gourgoulhon}\ \emph {et~al.}(2001)\citenamefont
  {Gourgoulhon}, \citenamefont {Grandclement}, \citenamefont {Taniguchi},
  \citenamefont {Marck},\ and\ \citenamefont {Bonazzola}}]{Gourgoulhon:2000nn}%
  \BibitemOpen
  \bibfield  {author} {\bibinfo {author} {\bibnamefont {Gourgoulhon},
  \bibfnamefont {E.}}, \bibinfo {author} {\bibfnamefont {P.}~\bibnamefont
  {Grandclement}}, \bibinfo {author} {\bibfnamefont {K.}~\bibnamefont
  {Taniguchi}}, \bibinfo {author} {\bibfnamefont {J.-A.}\ \bibnamefont
  {Marck}}, and\ \bibinfo {author} {\bibfnamefont {S.}~\bibnamefont
  {Bonazzola}}} (\bibinfo {year} {2001}),\ \href
  {https://doi.org/10.1103/PhysRevD.63.064029} {\bibfield  {journal} {\bibinfo
  {journal} {Phys. Rev. D}\ }\textbf {\bibinfo {volume} {63}},\ \bibinfo
  {pages} {064029}},\ \Eprint {https://arxiv.org/abs/gr-qc/0007028}
  {arXiv:gr-qc/0007028} \BibitemShut {NoStop}%
\bibitem [{\citenamefont {Guo}\ \emph {et~al.}(2021)\citenamefont {Guo},
  \citenamefont {Zhao},\ and\ \citenamefont {Shao}}]{Guo:2021leu}%
  \BibitemOpen
  \bibfield  {author} {\bibinfo {author} {\bibnamefont {Guo}, \bibfnamefont
  {M.}}, \bibinfo {author} {\bibfnamefont {J.}~\bibnamefont {Zhao}}, and\
  \bibinfo {author} {\bibfnamefont {L.}~\bibnamefont {Shao}}} (\bibinfo {year}
  {2021}),\ \href {https://doi.org/10.1103/PhysRevD.104.104065} {\bibfield
  {journal} {\bibinfo  {journal} {Phys. Rev. D}\ }\textbf {\bibinfo {volume}
  {104}}~(\bibinfo {number} {10}),\ \bibinfo {pages} {104065}},\ \Eprint
  {https://arxiv.org/abs/2106.01622} {arXiv:2106.01622 [gr-qc]} \BibitemShut
  {NoStop}%
\bibitem [{\citenamefont {Hansen}(1974)}]{Hansen:1974zz}%
  \BibitemOpen
  \bibfield  {author} {\bibinfo {author} {\bibnamefont {Hansen}, \bibfnamefont
  {R.~O.}}} (\bibinfo {year} {1974}),\ \href
  {https://doi.org/10.1063/1.1666501} {\bibfield  {journal} {\bibinfo
  {journal} {J. Math. Phys.}\ }\textbf {\bibinfo {volume} {15}},\ \bibinfo
  {pages} {46}}\BibitemShut {NoStop}%
\bibitem [{\citenamefont {Harada}(1997)}]{Harada:1997mr}%
  \BibitemOpen
  \bibfield  {author} {\bibinfo {author} {\bibnamefont {Harada}, \bibfnamefont
  {T.}}} (\bibinfo {year} {1997}),\ \href {https://doi.org/10.1143/PTP.98.359}
  {\bibfield  {journal} {\bibinfo  {journal} {Prog. Theor. Phys.}\ }\textbf
  {\bibinfo {volume} {98}},\ \bibinfo {pages} {359}},\ \Eprint
  {https://arxiv.org/abs/gr-qc/9706014} {arXiv:gr-qc/9706014} \BibitemShut
  {NoStop}%
\bibitem [{\citenamefont {Harada}(1998)}]{Harada:1998ge}%
  \BibitemOpen
  \bibfield  {author} {\bibinfo {author} {\bibnamefont {Harada}, \bibfnamefont
  {T.}}} (\bibinfo {year} {1998}),\ \href
  {https://doi.org/10.1103/PhysRevD.57.4802} {\bibfield  {journal} {\bibinfo
  {journal} {Phys. Rev. D}\ }\textbf {\bibinfo {volume} {57}},\ \bibinfo
  {pages} {4802}},\ \Eprint {https://arxiv.org/abs/gr-qc/9801049}
  {arXiv:gr-qc/9801049} \BibitemShut {NoStop}%
\bibitem [{\citenamefont {Hartle}(1967)}]{Hartle:1967he}%
  \BibitemOpen
  \bibfield  {author} {\bibinfo {author} {\bibnamefont {Hartle}, \bibfnamefont
  {J.~B.}}} (\bibinfo {year} {1967}),\ \href {https://doi.org/10.1086/149400}
  {\bibfield  {journal} {\bibinfo  {journal} {Astrophys. J.}\ }\textbf
  {\bibinfo {volume} {150}},\ \bibinfo {pages} {1005}}\BibitemShut {NoStop}%
\bibitem [{\citenamefont {Hartle}(1978)}]{Hartle:1978201}%
  \BibitemOpen
  \bibfield  {author} {\bibinfo {author} {\bibnamefont {Hartle}, \bibfnamefont
  {J.~B.}}} (\bibinfo {year} {1978}),\ \href
  {https://doi.org/https://doi.org/10.1016/0370-1573(78)90140-0} {\bibfield
  {journal} {\bibinfo  {journal} {Physics Reports}\ }\textbf {\bibinfo {volume}
  {46}}~(\bibinfo {number} {6}),\ \bibinfo {pages} {201}}\BibitemShut {NoStop}%
\bibitem [{\citenamefont {Hartle}\ and\ \citenamefont
  {Thorne}(1968)}]{Hartle:1968si}%
  \BibitemOpen
  \bibfield  {author} {\bibinfo {author} {\bibnamefont {Hartle}, \bibfnamefont
  {J.~B.}}, and\ \bibinfo {author} {\bibfnamefont {K.~S.}\ \bibnamefont
  {Thorne}}} (\bibinfo {year} {1968}),\ \href {https://doi.org/10.1086/149707}
  {\bibfield  {journal} {\bibinfo  {journal} {Astrophys. J.}\ }\textbf
  {\bibinfo {volume} {153}},\ \bibinfo {pages} {807}}\BibitemShut {NoStop}%
\bibitem [{\citenamefont {Hassan}\ and\ \citenamefont
  {Rosen}(2012)}]{Hassan:2011zd}%
  \BibitemOpen
  \bibfield  {author} {\bibinfo {author} {\bibnamefont {Hassan}, \bibfnamefont
  {S.~F.}}, and\ \bibinfo {author} {\bibfnamefont {R.~A.}\ \bibnamefont
  {Rosen}}} (\bibinfo {year} {2012}),\ \href
  {https://doi.org/10.1007/JHEP02(2012)126} {\bibfield  {journal} {\bibinfo
  {journal} {JHEP}\ }\textbf {\bibinfo {volume} {02}},\ \bibinfo {pages}
  {126}},\ \Eprint {https://arxiv.org/abs/1109.3515} {arXiv:1109.3515 [hep-th]}
  \BibitemShut {NoStop}%
\bibitem [{\citenamefont {Hawking}(1972)}]{Hawking:1972qk}%
  \BibitemOpen
  \bibfield  {author} {\bibinfo {author} {\bibnamefont {Hawking}, \bibfnamefont
  {S.~W.}}} (\bibinfo {year} {1972}),\ \href
  {https://doi.org/10.1007/BF01877518} {\bibfield  {journal} {\bibinfo
  {journal} {Commun. Math. Phys.}\ }\textbf {\bibinfo {volume} {25}},\ \bibinfo
  {pages} {167}}\BibitemShut {NoStop}%
\bibitem [{\citenamefont {Heisenberg}(2014)}]{Heisenberg:2014rta}%
  \BibitemOpen
  \bibfield  {author} {\bibinfo {author} {\bibnamefont {Heisenberg},
  \bibfnamefont {L.}}} (\bibinfo {year} {2014}),\ \href
  {https://doi.org/10.1088/1475-7516/2014/05/015} {\bibfield  {journal}
  {\bibinfo  {journal} {JCAP}\ }\textbf {\bibinfo {volume} {05}},\ \bibinfo
  {pages} {015}},\ \Eprint {https://arxiv.org/abs/1402.7026} {arXiv:1402.7026
  [hep-th]} \BibitemShut {NoStop}%
\bibitem [{\citenamefont {Hellings}\ and\ \citenamefont
  {Nordtvedt}(1973)}]{Hellings:1973zz}%
  \BibitemOpen
  \bibfield  {author} {\bibinfo {author} {\bibnamefont {Hellings},
  \bibfnamefont {R.~W.}}, and\ \bibinfo {author} {\bibfnamefont
  {K.}~\bibnamefont {Nordtvedt}}} (\bibinfo {year} {1973}),\ \href
  {https://doi.org/10.1103/PhysRevD.7.3593} {\bibfield  {journal} {\bibinfo
  {journal} {Phys. Rev. D}\ }\textbf {\bibinfo {volume} {7}},\ \bibinfo {pages}
  {3593}}\BibitemShut {NoStop}%
\bibitem [{\citenamefont {Herdeiro}\ \emph
  {et~al.}(2021{\natexlab{a}})\citenamefont {Herdeiro}, \citenamefont {Ikeda},
  \citenamefont {Minamitsuji}, \citenamefont {Nakamura},\ and\ \citenamefont
  {Radu}}]{Herdeiro:2020htm}%
  \BibitemOpen
  \bibfield  {author} {\bibinfo {author} {\bibnamefont {Herdeiro},
  \bibfnamefont {C.~A.~R.}}, \bibinfo {author} {\bibfnamefont {T.}~\bibnamefont
  {Ikeda}}, \bibinfo {author} {\bibfnamefont {M.}~\bibnamefont {Minamitsuji}},
  \bibinfo {author} {\bibfnamefont {T.}~\bibnamefont {Nakamura}}, and\ \bibinfo
  {author} {\bibfnamefont {E.}~\bibnamefont {Radu}}} (\bibinfo {year}
  {2021}{\natexlab{a}}),\ \href {https://doi.org/10.1103/PhysRevD.103.044019}
  {\bibfield  {journal} {\bibinfo  {journal} {Phys. Rev. D}\ }\textbf {\bibinfo
  {volume} {103}}~(\bibinfo {number} {4}),\ \bibinfo {pages} {044019}},\
  \Eprint {https://arxiv.org/abs/2009.06971} {arXiv:2009.06971 [gr-qc]}
  \BibitemShut {NoStop}%
\bibitem [{\citenamefont {Herdeiro}\ and\ \citenamefont
  {Oliveira}(2020)}]{Herdeiro:2020iyi}%
  \BibitemOpen
  \bibfield  {author} {\bibinfo {author} {\bibnamefont {Herdeiro},
  \bibfnamefont {C.~A.~R.}}, and\ \bibinfo {author} {\bibfnamefont {J.~a.
  M.~S.}\ \bibnamefont {Oliveira}}} (\bibinfo {year} {2020}),\ \href
  {https://doi.org/10.1007/JHEP07(2020)130} {\bibfield  {journal} {\bibinfo
  {journal} {JHEP}\ }\textbf {\bibinfo {volume} {07}},\ \bibinfo {pages}
  {130}},\ \Eprint {https://arxiv.org/abs/2005.05354} {arXiv:2005.05354
  [gr-qc]} \BibitemShut {NoStop}%
\bibitem [{\citenamefont {Herdeiro}\ and\ \citenamefont
  {Radu}(2015)}]{Herdeiro:2015waa}%
  \BibitemOpen
  \bibfield  {author} {\bibinfo {author} {\bibnamefont {Herdeiro},
  \bibfnamefont {C.~A.~R.}}, and\ \bibinfo {author} {\bibfnamefont
  {E.}~\bibnamefont {Radu}}} (\bibinfo {year} {2015}),\ \href
  {https://doi.org/10.1142/S0218271815420146} {\bibfield  {journal} {\bibinfo
  {journal} {Int. J. Mod. Phys. D}\ }\textbf {\bibinfo {volume} {24}}~(\bibinfo
  {number} {09}),\ \bibinfo {pages} {1542014}},\ \Eprint
  {https://arxiv.org/abs/1504.08209} {arXiv:1504.08209 [gr-qc]} \BibitemShut
  {NoStop}%
\bibitem [{\citenamefont {Herdeiro}\ and\ \citenamefont
  {Radu}(2019)}]{Herdeiro:2019yjy}%
  \BibitemOpen
  \bibfield  {author} {\bibinfo {author} {\bibnamefont {Herdeiro},
  \bibfnamefont {C.~A.~R.}}, and\ \bibinfo {author} {\bibfnamefont
  {E.}~\bibnamefont {Radu}}} (\bibinfo {year} {2019}),\ \href
  {https://doi.org/10.1103/PhysRevD.99.084039} {\bibfield  {journal} {\bibinfo
  {journal} {Phys. Rev. D}\ }\textbf {\bibinfo {volume} {99}}~(\bibinfo
  {number} {8}),\ \bibinfo {pages} {084039}},\ \Eprint
  {https://arxiv.org/abs/1901.02953} {arXiv:1901.02953 [gr-qc]} \BibitemShut
  {NoStop}%
\bibitem [{\citenamefont {Herdeiro}\ \emph {et~al.}(2018)\citenamefont
  {Herdeiro}, \citenamefont {Radu}, \citenamefont {Sanchis-Gual},\ and\
  \citenamefont {Font}}]{Herdeiro:2018wub}%
  \BibitemOpen
  \bibfield  {author} {\bibinfo {author} {\bibnamefont {Herdeiro},
  \bibfnamefont {C.~A.~R.}}, \bibinfo {author} {\bibfnamefont {E.}~\bibnamefont
  {Radu}}, \bibinfo {author} {\bibfnamefont {N.}~\bibnamefont {Sanchis-Gual}},
  and\ \bibinfo {author} {\bibfnamefont {J.~A.}\ \bibnamefont {Font}}}
  (\bibinfo {year} {2018}),\ \href
  {https://doi.org/10.1103/PhysRevLett.121.101102} {\bibfield  {journal}
  {\bibinfo  {journal} {Phys. Rev. Lett.}\ }\textbf {\bibinfo {volume}
  {121}}~(\bibinfo {number} {10}),\ \bibinfo {pages} {101102}},\ \Eprint
  {https://arxiv.org/abs/1806.05190} {arXiv:1806.05190 [gr-qc]} \BibitemShut
  {NoStop}%
\bibitem [{\citenamefont {Herdeiro}\ \emph
  {et~al.}(2021{\natexlab{b}})\citenamefont {Herdeiro}, \citenamefont {Radu},
  \citenamefont {Silva}, \citenamefont {Sotiriou},\ and\ \citenamefont
  {Yunes}}]{Herdeiro:2020wei}%
  \BibitemOpen
  \bibfield  {author} {\bibinfo {author} {\bibnamefont {Herdeiro},
  \bibfnamefont {C.~A.~R.}}, \bibinfo {author} {\bibfnamefont {E.}~\bibnamefont
  {Radu}}, \bibinfo {author} {\bibfnamefont {H.~O.}\ \bibnamefont {Silva}},
  \bibinfo {author} {\bibfnamefont {T.~P.}\ \bibnamefont {Sotiriou}}, and\
  \bibinfo {author} {\bibfnamefont {N.}~\bibnamefont {Yunes}}} (\bibinfo {year}
  {2021}{\natexlab{b}}),\ \href
  {https://doi.org/10.1103/PhysRevLett.126.011103} {\bibfield  {journal}
  {\bibinfo  {journal} {Phys. Rev. Lett.}\ }\textbf {\bibinfo {volume}
  {126}}~(\bibinfo {number} {1}),\ \bibinfo {pages} {011103}},\ \Eprint
  {https://arxiv.org/abs/2009.03904} {arXiv:2009.03904 [gr-qc]} \BibitemShut
  {NoStop}%
\bibitem [{\citenamefont {Herrera}\ and\ \citenamefont
  {Santos}(1997)}]{Herrera:1997plx}%
  \BibitemOpen
  \bibfield  {author} {\bibinfo {author} {\bibnamefont {Herrera}, \bibfnamefont
  {L.}}, and\ \bibinfo {author} {\bibfnamefont {N.~O.}\ \bibnamefont {Santos}}}
  (\bibinfo {year} {1997}),\ \href
  {https://doi.org/10.1016/S0370-1573(96)00042-7} {\bibfield  {journal}
  {\bibinfo  {journal} {Phys. Rept.}\ }\textbf {\bibinfo {volume} {286}},\
  \bibinfo {pages} {53}}\BibitemShut {NoStop}%
\bibitem [{\citenamefont {Hilditch}(2013)}]{Hilditch:2013sba}%
  \BibitemOpen
  \bibfield  {author} {\bibinfo {author} {\bibnamefont {Hilditch},
  \bibfnamefont {D.}}} (\bibinfo {year} {2013}),\ \href
  {https://doi.org/10.1142/S0217751X13400150} {\bibfield  {journal} {\bibinfo
  {journal} {Int. J. Mod. Phys. A}\ }\textbf {\bibinfo {volume} {28}},\
  \bibinfo {pages} {1340015}},\ \Eprint {https://arxiv.org/abs/1309.2012}
  {arXiv:1309.2012 [gr-qc]} \BibitemShut {NoStop}%
\bibitem [{\citenamefont {Hod}(2020)}]{Hod:2020jjy}%
  \BibitemOpen
  \bibfield  {author} {\bibinfo {author} {\bibnamefont {Hod}, \bibfnamefont
  {S.}}} (\bibinfo {year} {2020}),\ \href
  {https://doi.org/10.1103/PhysRevD.102.084060} {\bibfield  {journal} {\bibinfo
   {journal} {Phys. Rev. D}\ }\textbf {\bibinfo {volume} {102}}~(\bibinfo
  {number} {8}),\ \bibinfo {pages} {084060}},\ \Eprint
  {https://arxiv.org/abs/2006.09399} {arXiv:2006.09399 [gr-qc]} \BibitemShut
  {NoStop}%
\bibitem [{\citenamefont {Hod}(2022)}]{Hod:2022hfm}%
  \BibitemOpen
  \bibfield  {author} {\bibinfo {author} {\bibnamefont {Hod}, \bibfnamefont
  {S.}}} (\bibinfo {year} {2022}),\ \href
  {https://doi.org/10.1103/PhysRevD.105.024074} {\bibfield  {journal} {\bibinfo
   {journal} {Phys. Rev. D}\ }\textbf {\bibinfo {volume} {105}}~(\bibinfo
  {number} {2}),\ \bibinfo {pages} {024074}}\BibitemShut {NoStop}%
\bibitem [{\citenamefont {Horbatsch}\ \emph {et~al.}(2015)\citenamefont
  {Horbatsch}, \citenamefont {Silva}, \citenamefont {Gerosa}, \citenamefont
  {Pani}, \citenamefont {Berti}, \citenamefont {Gualtieri},\ and\ \citenamefont
  {Sperhake}}]{Horbatsch:2015bua}%
  \BibitemOpen
  \bibfield  {author} {\bibinfo {author} {\bibnamefont {Horbatsch},
  \bibfnamefont {M.}}, \bibinfo {author} {\bibfnamefont {H.~O.}\ \bibnamefont
  {Silva}}, \bibinfo {author} {\bibfnamefont {D.}~\bibnamefont {Gerosa}},
  \bibinfo {author} {\bibfnamefont {P.}~\bibnamefont {Pani}}, \bibinfo {author}
  {\bibfnamefont {E.}~\bibnamefont {Berti}}, \bibinfo {author} {\bibfnamefont
  {L.}~\bibnamefont {Gualtieri}}, and\ \bibinfo {author} {\bibfnamefont
  {U.}~\bibnamefont {Sperhake}}} (\bibinfo {year} {2015}),\ \href
  {https://doi.org/10.1088/0264-9381/32/20/204001} {\bibfield  {journal}
  {\bibinfo  {journal} {Class. Quant. Grav.}\ }\textbf {\bibinfo {volume}
  {32}}~(\bibinfo {number} {20}),\ \bibinfo {pages} {204001}},\ \Eprint
  {https://arxiv.org/abs/1505.07462} {arXiv:1505.07462 [gr-qc]} \BibitemShut
  {NoStop}%
\bibitem [{\citenamefont {Horbatsch}\ and\ \citenamefont
  {Burgess}(2011)}]{Horbatsch:2010hj}%
  \BibitemOpen
  \bibfield  {author} {\bibinfo {author} {\bibnamefont {Horbatsch},
  \bibfnamefont {M.~W.}}, and\ \bibinfo {author} {\bibfnamefont {C.~P.}\
  \bibnamefont {Burgess}}} (\bibinfo {year} {2011}),\ \href
  {https://doi.org/10.1088/1475-7516/2011/08/027} {\bibfield  {journal}
  {\bibinfo  {journal} {JCAP}\ }\textbf {\bibinfo {volume} {08}},\ \bibinfo
  {pages} {027}},\ \Eprint {https://arxiv.org/abs/1006.4411} {arXiv:1006.4411
  [gr-qc]} \BibitemShut {NoStop}%
\bibitem [{\citenamefont {Horbatsch}\ and\ \citenamefont
  {Burgess}(2012)}]{Horbatsch:2011nh}%
  \BibitemOpen
  \bibfield  {author} {\bibinfo {author} {\bibnamefont {Horbatsch},
  \bibfnamefont {M.~W.}}, and\ \bibinfo {author} {\bibfnamefont {C.~P.}\
  \bibnamefont {Burgess}}} (\bibinfo {year} {2012}),\ \href
  {https://doi.org/10.1088/0264-9381/29/24/245004} {\bibfield  {journal}
  {\bibinfo  {journal} {Class. Quant. Grav.}\ }\textbf {\bibinfo {volume}
  {29}},\ \bibinfo {pages} {245004}},\ \Eprint
  {https://arxiv.org/abs/1107.3585} {arXiv:1107.3585 [gr-qc]} \BibitemShut
  {NoStop}%
\bibitem [{\citenamefont {Horndeski}(1974)}]{Horndeski:1974wa}%
  \BibitemOpen
  \bibfield  {author} {\bibinfo {author} {\bibnamefont {Horndeski},
  \bibfnamefont {G.~W.}}} (\bibinfo {year} {1974}),\ \href
  {https://doi.org/10.1007/BF01807638} {\bibfield  {journal} {\bibinfo
  {journal} {Int. J. Theor. Phys.}\ }\textbf {\bibinfo {volume} {10}},\
  \bibinfo {pages} {363}}\BibitemShut {NoStop}%
\bibitem [{\citenamefont {Hotokezaka}\ \emph {et~al.}(2013)\citenamefont
  {Hotokezaka}, \citenamefont {Kiuchi}, \citenamefont {Kyutoku}, \citenamefont
  {Muranushi}, \citenamefont {Sekiguchi}, \citenamefont {Shibata},\ and\
  \citenamefont {Taniguchi}}]{Hotokezaka:2013iia}%
  \BibitemOpen
  \bibfield  {author} {\bibinfo {author} {\bibnamefont {Hotokezaka},
  \bibfnamefont {K.}}, \bibinfo {author} {\bibfnamefont {K.}~\bibnamefont
  {Kiuchi}}, \bibinfo {author} {\bibfnamefont {K.}~\bibnamefont {Kyutoku}},
  \bibinfo {author} {\bibfnamefont {T.}~\bibnamefont {Muranushi}}, \bibinfo
  {author} {\bibfnamefont {Y.-i.}\ \bibnamefont {Sekiguchi}}, \bibinfo {author}
  {\bibfnamefont {M.}~\bibnamefont {Shibata}}, and\ \bibinfo {author}
  {\bibfnamefont {K.}~\bibnamefont {Taniguchi}}} (\bibinfo {year} {2013}),\
  \href {https://doi.org/10.1103/PhysRevD.88.044026} {\bibfield  {journal}
  {\bibinfo  {journal} {Phys. Rev. D}\ }\textbf {\bibinfo {volume} {88}},\
  \bibinfo {pages} {044026}},\ \Eprint {https://arxiv.org/abs/1307.5888}
  {arXiv:1307.5888 [astro-ph.HE]} \BibitemShut {NoStop}%
\bibitem [{\citenamefont {Hu}\ \emph {et~al.}(2021)\citenamefont {Hu},
  \citenamefont {Gao}, \citenamefont {Xu},\ and\ \citenamefont
  {Shao}}]{Hu:2021tyw}%
  \BibitemOpen
  \bibfield  {author} {\bibinfo {author} {\bibnamefont {Hu}, \bibfnamefont
  {Z.}}, \bibinfo {author} {\bibfnamefont {Y.}~\bibnamefont {Gao}}, \bibinfo
  {author} {\bibfnamefont {R.}~\bibnamefont {Xu}}, and\ \bibinfo {author}
  {\bibfnamefont {L.}~\bibnamefont {Shao}}} (\bibinfo {year} {2021}),\ \href
  {https://doi.org/10.1103/PhysRevD.104.104014} {\bibfield  {journal} {\bibinfo
   {journal} {Phys. Rev. D}\ }\textbf {\bibinfo {volume} {104}}~(\bibinfo
  {number} {10}),\ \bibinfo {pages} {104014}},\ \Eprint
  {https://arxiv.org/abs/2109.13453} {arXiv:2109.13453 [gr-qc]} \BibitemShut
  {NoStop}%
\bibitem [{\citenamefont {Hulse}\ and\ \citenamefont
  {Taylor}(1975)}]{Hulse:1974eb}%
  \BibitemOpen
  \bibfield  {author} {\bibinfo {author} {\bibnamefont {Hulse}, \bibfnamefont
  {R.~A.}}, and\ \bibinfo {author} {\bibfnamefont {J.~H.}\ \bibnamefont
  {Taylor}}} (\bibinfo {year} {1975}),\ \href {https://doi.org/10.1086/181708}
  {\bibfield  {journal} {\bibinfo  {journal} {Astrophys. J. Lett.}\ }\textbf
  {\bibinfo {volume} {195}},\ \bibinfo {pages} {L51}}\BibitemShut {NoStop}%
\bibitem [{\citenamefont {Ikeda}\ \emph {et~al.}(2019)\citenamefont {Ikeda},
  \citenamefont {Nakamura},\ and\ \citenamefont {Minamitsuji}}]{Ikeda:2019okp}%
  \BibitemOpen
  \bibfield  {author} {\bibinfo {author} {\bibnamefont {Ikeda}, \bibfnamefont
  {T.}}, \bibinfo {author} {\bibfnamefont {T.}~\bibnamefont {Nakamura}}, and\
  \bibinfo {author} {\bibfnamefont {M.}~\bibnamefont {Minamitsuji}}} (\bibinfo
  {year} {2019}),\ \href {https://doi.org/10.1103/PhysRevD.100.104014}
  {\bibfield  {journal} {\bibinfo  {journal} {Phys. Rev. D}\ }\textbf {\bibinfo
  {volume} {100}}~(\bibinfo {number} {10}),\ \bibinfo {pages} {104014}},\
  \Eprint {https://arxiv.org/abs/1908.09394} {arXiv:1908.09394 [gr-qc]}
  \BibitemShut {NoStop}%
\bibitem [{\citenamefont {Israel}\ \emph {et~al.}(2005)\citenamefont {Israel},
  \citenamefont {Belloni}, \citenamefont {Stella}, \citenamefont {Rephaeli},
  \citenamefont {Gruber}, \citenamefont {Casella}, \citenamefont {Dall'Osso},
  \citenamefont {Rea}, \citenamefont {Persic},\ and\ \citenamefont
  {Rothschild}}]{Israel:2005av}%
  \BibitemOpen
  \bibfield  {author} {\bibinfo {author} {\bibnamefont {Israel}, \bibfnamefont
  {G.}}, \bibinfo {author} {\bibfnamefont {T.}~\bibnamefont {Belloni}},
  \bibinfo {author} {\bibfnamefont {L.}~\bibnamefont {Stella}}, \bibinfo
  {author} {\bibfnamefont {Y.}~\bibnamefont {Rephaeli}}, \bibinfo {author}
  {\bibfnamefont {D.}~\bibnamefont {Gruber}}, \bibinfo {author} {\bibfnamefont
  {P.~G.}\ \bibnamefont {Casella}}, \bibinfo {author} {\bibfnamefont
  {S.}~\bibnamefont {Dall'Osso}}, \bibinfo {author} {\bibfnamefont
  {N.}~\bibnamefont {Rea}}, \bibinfo {author} {\bibfnamefont {M.}~\bibnamefont
  {Persic}}, and\ \bibinfo {author} {\bibfnamefont {R.}~\bibnamefont
  {Rothschild}}} (\bibinfo {year} {2005}),\ \href
  {https://doi.org/10.1086/432615} {\bibfield  {journal} {\bibinfo  {journal}
  {Astrophys. J. Lett.}\ }\textbf {\bibinfo {volume} {628}},\ \bibinfo {pages}
  {L53}},\ \Eprint {https://arxiv.org/abs/astro-ph/0505255}
  {arXiv:astro-ph/0505255} \BibitemShut {NoStop}%
\bibitem [{\citenamefont {Jackiw}\ and\ \citenamefont
  {Pi}(2003)}]{Jackiw:2003pm}%
  \BibitemOpen
  \bibfield  {author} {\bibinfo {author} {\bibnamefont {Jackiw}, \bibfnamefont
  {R.}}, and\ \bibinfo {author} {\bibfnamefont {S.~Y.}\ \bibnamefont {Pi}}}
  (\bibinfo {year} {2003}),\ \href {https://doi.org/10.1103/PhysRevD.68.104012}
  {\bibfield  {journal} {\bibinfo  {journal} {Phys. Rev. D}\ }\textbf {\bibinfo
  {volume} {68}},\ \bibinfo {pages} {104012}},\ \Eprint
  {https://arxiv.org/abs/gr-qc/0308071} {arXiv:gr-qc/0308071} \BibitemShut
  {NoStop}%
\bibitem [{\citenamefont {Jentschura}\ and\ \citenamefont
  {Wundt}(2012)}]{Jentschura:2011ga}%
  \BibitemOpen
  \bibfield  {author} {\bibinfo {author} {\bibnamefont {Jentschura},
  \bibfnamefont {U.~D.}}, and\ \bibinfo {author} {\bibfnamefont {B.~J.}\
  \bibnamefont {Wundt}}} (\bibinfo {year} {2012}),\ \href
  {https://doi.org/10.1088/1751-8113/45/44/444017} {\bibfield  {journal}
  {\bibinfo  {journal} {J. Phys.}\ }\textbf {\bibinfo {volume} {A45}}~(\bibinfo
  {number} {44}),\ \bibinfo {pages} {444017}},\ \Eprint
  {https://arxiv.org/abs/1110.4171} {arXiv:1110.4171 [hep-ph]} \BibitemShut
  {NoStop}%
\bibitem [{\citenamefont {Juli\'e}\ and\ \citenamefont
  {Berti}(2019)}]{Julie:2019sab}%
  \BibitemOpen
  \bibfield  {author} {\bibinfo {author} {\bibnamefont {Juli\'e}, \bibfnamefont
  {F.-L.}}, and\ \bibinfo {author} {\bibfnamefont {E.}~\bibnamefont {Berti}}}
  (\bibinfo {year} {2019}),\ \href
  {https://doi.org/10.1103/PhysRevD.100.104061} {\bibfield  {journal} {\bibinfo
   {journal} {Phys. Rev. D}\ }\textbf {\bibinfo {volume} {100}}~(\bibinfo
  {number} {10}),\ \bibinfo {pages} {104061}},\ \Eprint
  {https://arxiv.org/abs/1909.05258} {arXiv:1909.05258 [gr-qc]} \BibitemShut
  {NoStop}%
\bibitem [{\citenamefont {Juli\'e}\ \emph {et~al.}(2022)\citenamefont
  {Juli\'e}, \citenamefont {Silva}, \citenamefont {Berti},\ and\ \citenamefont
  {Yunes}}]{Julie:2022huo}%
  \BibitemOpen
  \bibfield  {author} {\bibinfo {author} {\bibnamefont {Juli\'e}, \bibfnamefont
  {F.-L.}}, \bibinfo {author} {\bibfnamefont {H.~O.}\ \bibnamefont {Silva}},
  \bibinfo {author} {\bibfnamefont {E.}~\bibnamefont {Berti}}, and\ \bibinfo
  {author} {\bibfnamefont {N.}~\bibnamefont {Yunes}}} (\bibinfo {year}
  {2022}),\ \href {https://doi.org/10.1103/PhysRevD.105.124031} {\bibfield
  {journal} {\bibinfo  {journal} {Phys. Rev. D}\ }\textbf {\bibinfo {volume}
  {105}}~(\bibinfo {number} {12}),\ \bibinfo {pages} {124031}},\ \Eprint
  {https://arxiv.org/abs/2202.01329} {arXiv:2202.01329 [gr-qc]} \BibitemShut
  {NoStop}%
\bibitem [{\citenamefont {Kalogera}\ \emph {et~al.}(2021)\citenamefont
  {Kalogera} \emph {et~al.}}]{Kalogera:2021bya}%
  \BibitemOpen
  \bibfield  {author} {\bibinfo {author} {\bibnamefont {Kalogera},
  \bibfnamefont {V.}},  \emph {et~al.}} (\bibinfo {year} {2021}),\ \href@noop
  {} {\enquote {\bibinfo {title} {{The Next Generation Global Gravitational
  Wave Observatory: The Science Book}},}\ }\Eprint
  {https://arxiv.org/abs/2111.06990} {arXiv:2111.06990 [gr-qc]} \BibitemShut
  {NoStop}%
\bibitem [{\citenamefont {Kanti}\ \emph {et~al.}(1996)\citenamefont {Kanti},
  \citenamefont {Mavromatos}, \citenamefont {Rizos}, \citenamefont {Tamvakis},\
  and\ \citenamefont {Winstanley}}]{Kanti:1995vq}%
  \BibitemOpen
  \bibfield  {author} {\bibinfo {author} {\bibnamefont {Kanti}, \bibfnamefont
  {P.}}, \bibinfo {author} {\bibfnamefont {N.~E.}\ \bibnamefont {Mavromatos}},
  \bibinfo {author} {\bibfnamefont {J.}~\bibnamefont {Rizos}}, \bibinfo
  {author} {\bibfnamefont {K.}~\bibnamefont {Tamvakis}}, and\ \bibinfo {author}
  {\bibfnamefont {E.}~\bibnamefont {Winstanley}}} (\bibinfo {year} {1996}),\
  \href {https://doi.org/10.1103/PhysRevD.54.5049} {\bibfield  {journal}
  {\bibinfo  {journal} {Phys. Rev. D}\ }\textbf {\bibinfo {volume} {54}},\
  \bibinfo {pages} {5049}},\ \Eprint {https://arxiv.org/abs/hep-th/9511071}
  {arXiv:hep-th/9511071} \BibitemShut {NoStop}%
\bibitem [{\citenamefont {Kase}\ \emph {et~al.}(2020)\citenamefont {Kase},
  \citenamefont {Minamitsuji},\ and\ \citenamefont {Tsujikawa}}]{Kase:2020yhw}%
  \BibitemOpen
  \bibfield  {author} {\bibinfo {author} {\bibnamefont {Kase}, \bibfnamefont
  {R.}}, \bibinfo {author} {\bibfnamefont {M.}~\bibnamefont {Minamitsuji}},
  and\ \bibinfo {author} {\bibfnamefont {S.}~\bibnamefont {Tsujikawa}}}
  (\bibinfo {year} {2020}),\ \href
  {https://doi.org/10.1103/PhysRevD.102.024067} {\bibfield  {journal} {\bibinfo
   {journal} {Phys. Rev. D}\ }\textbf {\bibinfo {volume} {102}}~(\bibinfo
  {number} {2}),\ \bibinfo {pages} {024067}},\ \Eprint
  {https://arxiv.org/abs/2001.10701} {arXiv:2001.10701 [gr-qc]} \BibitemShut
  {NoStop}%
\bibitem [{\citenamefont {Khalil}\ \emph {et~al.}(2022)\citenamefont {Khalil},
  \citenamefont {Mendes}, \citenamefont {Ortiz},\ and\ \citenamefont
  {Steinhoff}}]{Khalil:2022sii}%
  \BibitemOpen
  \bibfield  {author} {\bibinfo {author} {\bibnamefont {Khalil}, \bibfnamefont
  {M.}}, \bibinfo {author} {\bibfnamefont {R.~F.~P.}\ \bibnamefont {Mendes}},
  \bibinfo {author} {\bibfnamefont {N.}~\bibnamefont {Ortiz}}, and\ \bibinfo
  {author} {\bibfnamefont {J.}~\bibnamefont {Steinhoff}}} (\bibinfo {year}
  {2022}),\ \href {https://doi.org/10.1103/PhysRevD.106.104016} {\bibfield
  {journal} {\bibinfo  {journal} {Phys. Rev. D}\ }\textbf {\bibinfo {volume}
  {106}}~(\bibinfo {number} {10}),\ \bibinfo {pages} {104016}},\ \Eprint
  {https://arxiv.org/abs/2206.13233} {arXiv:2206.13233 [gr-qc]} \BibitemShut
  {NoStop}%
\bibitem [{\citenamefont {Khalil}\ \emph {et~al.}(2019)\citenamefont {Khalil},
  \citenamefont {Sennett}, \citenamefont {Steinhoff},\ and\ \citenamefont
  {Buonanno}}]{Khalil:2019wyy}%
  \BibitemOpen
  \bibfield  {author} {\bibinfo {author} {\bibnamefont {Khalil}, \bibfnamefont
  {M.}}, \bibinfo {author} {\bibfnamefont {N.}~\bibnamefont {Sennett}},
  \bibinfo {author} {\bibfnamefont {J.}~\bibnamefont {Steinhoff}}, and\
  \bibinfo {author} {\bibfnamefont {A.}~\bibnamefont {Buonanno}}} (\bibinfo
  {year} {2019}),\ \href {https://doi.org/10.1103/PhysRevD.100.124013}
  {\bibfield  {journal} {\bibinfo  {journal} {Phys. Rev. D}\ }\textbf {\bibinfo
  {volume} {100}}~(\bibinfo {number} {12}),\ \bibinfo {pages} {124013}},\
  \Eprint {https://arxiv.org/abs/1906.08161} {arXiv:1906.08161 [gr-qc]}
  \BibitemShut {NoStop}%
\bibitem [{\citenamefont {Kleihaus}\ \emph {et~al.}(2016)\citenamefont
  {Kleihaus}, \citenamefont {Kunz}, \citenamefont {Mojica},\ and\ \citenamefont
  {Zagermann}}]{Kleihaus:2016dui}%
  \BibitemOpen
  \bibfield  {author} {\bibinfo {author} {\bibnamefont {Kleihaus},
  \bibfnamefont {B.}}, \bibinfo {author} {\bibfnamefont {J.}~\bibnamefont
  {Kunz}}, \bibinfo {author} {\bibfnamefont {S.}~\bibnamefont {Mojica}}, and\
  \bibinfo {author} {\bibfnamefont {M.}~\bibnamefont {Zagermann}}} (\bibinfo
  {year} {2016}),\ \href {https://doi.org/10.1103/PhysRevD.93.064077}
  {\bibfield  {journal} {\bibinfo  {journal} {Phys. Rev. D}\ }\textbf {\bibinfo
  {volume} {93}}~(\bibinfo {number} {6}),\ \bibinfo {pages} {064077}},\ \Eprint
  {https://arxiv.org/abs/1601.05583} {arXiv:1601.05583 [gr-qc]} \BibitemShut
  {NoStop}%
\bibitem [{\citenamefont {Kobayashi}(2019)}]{Kobayashi:2019hrl}%
  \BibitemOpen
  \bibfield  {author} {\bibinfo {author} {\bibnamefont {Kobayashi},
  \bibfnamefont {T.}}} (\bibinfo {year} {2019}),\ \href
  {https://doi.org/10.1088/1361-6633/ab2429} {\bibfield  {journal} {\bibinfo
  {journal} {Rept. Prog. Phys.}\ }\textbf {\bibinfo {volume} {82}}~(\bibinfo
  {number} {8}),\ \bibinfo {pages} {086901}},\ \Eprint
  {https://arxiv.org/abs/1901.07183} {arXiv:1901.07183 [gr-qc]} \BibitemShut
  {NoStop}%
\bibitem [{\citenamefont {Kobayashi}\ \emph {et~al.}(2011)\citenamefont
  {Kobayashi}, \citenamefont {Yamaguchi},\ and\ \citenamefont
  {Yokoyama}}]{Kobayashi:2011nu}%
  \BibitemOpen
  \bibfield  {author} {\bibinfo {author} {\bibnamefont {Kobayashi},
  \bibfnamefont {T.}}, \bibinfo {author} {\bibfnamefont {M.}~\bibnamefont
  {Yamaguchi}}, and\ \bibinfo {author} {\bibfnamefont {J.}~\bibnamefont
  {Yokoyama}}} (\bibinfo {year} {2011}),\ \href
  {https://doi.org/10.1143/PTP.126.511} {\bibfield  {journal} {\bibinfo
  {journal} {Prog. Theor. Phys.}\ }\textbf {\bibinfo {volume} {126}},\ \bibinfo
  {pages} {511}},\ \Eprint {https://arxiv.org/abs/1105.5723} {arXiv:1105.5723
  [hep-th]} \BibitemShut {NoStop}%
\bibitem [{\citenamefont {Kokkotas}\ and\ \citenamefont
  {Schmidt}(1999)}]{Kokkotas:1999bd}%
  \BibitemOpen
  \bibfield  {author} {\bibinfo {author} {\bibnamefont {Kokkotas},
  \bibfnamefont {K.~D.}}, and\ \bibinfo {author} {\bibfnamefont {B.~G.}\
  \bibnamefont {Schmidt}}} (\bibinfo {year} {1999}),\ \href
  {https://doi.org/10.12942/lrr-1999-2} {\bibfield  {journal} {\bibinfo
  {journal} {Living Rev. Rel.}\ }\textbf {\bibinfo {volume} {2}},\ \bibinfo
  {pages} {2}},\ \Eprint {https://arxiv.org/abs/gr-qc/9909058}
  {arXiv:gr-qc/9909058} \BibitemShut {NoStop}%
\bibitem [{\citenamefont {Konno}\ \emph {et~al.}(2009)\citenamefont {Konno},
  \citenamefont {Matsuyama},\ and\ \citenamefont {Tanda}}]{Konno:2009kg}%
  \BibitemOpen
  \bibfield  {author} {\bibinfo {author} {\bibnamefont {Konno}, \bibfnamefont
  {K.}}, \bibinfo {author} {\bibfnamefont {T.}~\bibnamefont {Matsuyama}}, and\
  \bibinfo {author} {\bibfnamefont {S.}~\bibnamefont {Tanda}}} (\bibinfo {year}
  {2009}),\ \href {https://doi.org/10.1143/PTP.122.561} {\bibfield  {journal}
  {\bibinfo  {journal} {Prog. Theor. Phys.}\ }\textbf {\bibinfo {volume}
  {122}},\ \bibinfo {pages} {561}},\ \Eprint {https://arxiv.org/abs/0902.4767}
  {arXiv:0902.4767 [gr-qc]} \BibitemShut {NoStop}%
\bibitem [{\citenamefont {Kov\'acs}(2019)}]{Kovacs:2019jqj}%
  \BibitemOpen
  \bibfield  {author} {\bibinfo {author} {\bibnamefont {Kov\'acs},
  \bibfnamefont {A.~D.}}} (\bibinfo {year} {2019}),\ \href
  {https://doi.org/10.1103/PhysRevD.100.024005} {\bibfield  {journal} {\bibinfo
   {journal} {Phys. Rev. D}\ }\textbf {\bibinfo {volume} {100}}~(\bibinfo
  {number} {2}),\ \bibinfo {pages} {024005}},\ \Eprint
  {https://arxiv.org/abs/1904.00963} {arXiv:1904.00963 [gr-qc]} \BibitemShut
  {NoStop}%
\bibitem [{\citenamefont {Kov\'acs}\ and\ \citenamefont
  {Reall}(2020{\natexlab{a}})}]{Kovacs:2020ywu}%
  \BibitemOpen
  \bibfield  {author} {\bibinfo {author} {\bibnamefont {Kov\'acs},
  \bibfnamefont {A.~D.}}, and\ \bibinfo {author} {\bibfnamefont {H.~S.}\
  \bibnamefont {Reall}}} (\bibinfo {year} {2020}{\natexlab{a}}),\ \href
  {https://doi.org/10.1103/PhysRevD.101.124003} {\bibfield  {journal} {\bibinfo
   {journal} {Phys. Rev. D}\ }\textbf {\bibinfo {volume} {101}}~(\bibinfo
  {number} {12}),\ \bibinfo {pages} {124003}},\ \Eprint
  {https://arxiv.org/abs/2003.08398} {arXiv:2003.08398 [gr-qc]} \BibitemShut
  {NoStop}%
\bibitem [{\citenamefont {Kov\'acs}\ and\ \citenamefont
  {Reall}(2020{\natexlab{b}})}]{Kovacs:2020pns}%
  \BibitemOpen
  \bibfield  {author} {\bibinfo {author} {\bibnamefont {Kov\'acs},
  \bibfnamefont {A.~D.}}, and\ \bibinfo {author} {\bibfnamefont {H.~S.}\
  \bibnamefont {Reall}}} (\bibinfo {year} {2020}{\natexlab{b}}),\ \href
  {https://doi.org/10.1103/PhysRevLett.124.221101} {\bibfield  {journal}
  {\bibinfo  {journal} {Phys. Rev. Lett.}\ }\textbf {\bibinfo {volume}
  {124}}~(\bibinfo {number} {22}),\ \bibinfo {pages} {221101}},\ \Eprint
  {https://arxiv.org/abs/2003.04327} {arXiv:2003.04327 [gr-qc]} \BibitemShut
  {NoStop}%
\bibitem [{\citenamefont {Krall}\ \emph {et~al.}(2020)\citenamefont {Krall},
  \citenamefont {Coates},\ and\ \citenamefont {Kokkotas}}]{Krall:2020kto}%
  \BibitemOpen
  \bibfield  {author} {\bibinfo {author} {\bibnamefont {Krall}, \bibfnamefont
  {V.}}, \bibinfo {author} {\bibfnamefont {A.}~\bibnamefont {Coates}}, and\
  \bibinfo {author} {\bibfnamefont {K.~D.}\ \bibnamefont {Kokkotas}}} (\bibinfo
  {year} {2020}),\ \href {https://doi.org/10.1103/PhysRevD.102.124065}
  {\bibfield  {journal} {\bibinfo  {journal} {Phys. Rev. D}\ }\textbf {\bibinfo
  {volume} {102}}~(\bibinfo {number} {12}),\ \bibinfo {pages} {124065}},\
  \Eprint {https://arxiv.org/abs/2012.03710} {arXiv:2012.03710 [gr-qc]}
  \BibitemShut {NoStop}%
\bibitem [{\citenamefont {Kramer}\ \emph {et~al.}(2021)\citenamefont {Kramer}
  \emph {et~al.}}]{Kramer:2021jcw}%
  \BibitemOpen
  \bibfield  {author} {\bibinfo {author} {\bibnamefont {Kramer}, \bibfnamefont
  {M.}},  \emph {et~al.}} (\bibinfo {year} {2021}),\ \href
  {https://doi.org/10.1103/PhysRevX.11.041050} {\bibfield  {journal} {\bibinfo
  {journal} {Phys. Rev. X}\ }\textbf {\bibinfo {volume} {11}}~(\bibinfo
  {number} {4}),\ \bibinfo {pages} {041050}},\ \Eprint
  {https://arxiv.org/abs/2112.06795} {arXiv:2112.06795 [astro-ph.HE]}
  \BibitemShut {NoStop}%
\bibitem [{\citenamefont {Kr\"uger}\ and\ \citenamefont
  {Doneva}(2021)}]{Kruger:2021yay}%
  \BibitemOpen
  \bibfield  {author} {\bibinfo {author} {\bibnamefont {Kr\"uger},
  \bibfnamefont {C.~J.}}, and\ \bibinfo {author} {\bibfnamefont {D.~D.}\
  \bibnamefont {Doneva}}} (\bibinfo {year} {2021}),\ \href
  {https://doi.org/10.1103/PhysRevD.103.124034} {\bibfield  {journal} {\bibinfo
   {journal} {Phys. Rev. D}\ }\textbf {\bibinfo {volume} {103}}~(\bibinfo
  {number} {12}),\ \bibinfo {pages} {124034}},\ \Eprint
  {https://arxiv.org/abs/2102.11698} {arXiv:2102.11698 [gr-qc]} \BibitemShut
  {NoStop}%
\bibitem [{\citenamefont {Kr\"uger}\ and\ \citenamefont
  {Kokkotas}(2020{\natexlab{a}})}]{Kruger:2020ykw}%
  \BibitemOpen
  \bibfield  {author} {\bibinfo {author} {\bibnamefont {Kr\"uger},
  \bibfnamefont {C.~J.}}, and\ \bibinfo {author} {\bibfnamefont {K.~D.}\
  \bibnamefont {Kokkotas}}} (\bibinfo {year} {2020}{\natexlab{a}}),\ \href
  {https://doi.org/10.1103/PhysRevD.102.064026} {\bibfield  {journal} {\bibinfo
   {journal} {Phys. Rev. D}\ }\textbf {\bibinfo {volume} {102}}~(\bibinfo
  {number} {6}),\ \bibinfo {pages} {064026}},\ \Eprint
  {https://arxiv.org/abs/2008.04127} {arXiv:2008.04127 [gr-qc]} \BibitemShut
  {NoStop}%
\bibitem [{\citenamefont {Kr\"uger}\ and\ \citenamefont
  {Kokkotas}(2020{\natexlab{b}})}]{Kruger:2019zuz}%
  \BibitemOpen
  \bibfield  {author} {\bibinfo {author} {\bibnamefont {Kr\"uger},
  \bibfnamefont {C.~J.}}, and\ \bibinfo {author} {\bibfnamefont {K.~D.}\
  \bibnamefont {Kokkotas}}} (\bibinfo {year} {2020}{\natexlab{b}}),\ \href
  {https://doi.org/10.1103/PhysRevLett.125.111106} {\bibfield  {journal}
  {\bibinfo  {journal} {Phys. Rev. Lett.}\ }\textbf {\bibinfo {volume}
  {125}}~(\bibinfo {number} {11}),\ \bibinfo {pages} {111106}},\ \Eprint
  {https://arxiv.org/abs/1910.08370} {arXiv:1910.08370 [gr-qc]} \BibitemShut
  {NoStop}%
\bibitem [{\citenamefont {Kuan}\ \emph
  {et~al.}(2021{\natexlab{a}})\citenamefont {Kuan}, \citenamefont {Doneva},\
  and\ \citenamefont {Yazadjiev}}]{Kuan:2021lol}%
  \BibitemOpen
  \bibfield  {author} {\bibinfo {author} {\bibnamefont {Kuan}, \bibfnamefont
  {H.-J.}}, \bibinfo {author} {\bibfnamefont {D.~D.}\ \bibnamefont {Doneva}},
  and\ \bibinfo {author} {\bibfnamefont {S.~S.}\ \bibnamefont {Yazadjiev}}}
  (\bibinfo {year} {2021}{\natexlab{a}}),\ \href
  {https://doi.org/10.1103/PhysRevLett.127.161103} {\bibfield  {journal}
  {\bibinfo  {journal} {Phys. Rev. Lett.}\ }\textbf {\bibinfo {volume}
  {127}}~(\bibinfo {number} {16}),\ \bibinfo {pages} {161103}},\ \Eprint
  {https://arxiv.org/abs/2103.11999} {arXiv:2103.11999 [gr-qc]} \BibitemShut
  {NoStop}%
\bibitem [{\citenamefont {Kuan}\ \emph
  {et~al.}(2021{\natexlab{b}})\citenamefont {Kuan}, \citenamefont {Singh},
  \citenamefont {Doneva}, \citenamefont {Yazadjiev},\ and\ \citenamefont
  {Kokkotas}}]{Kuan:2021yih}%
  \BibitemOpen
  \bibfield  {author} {\bibinfo {author} {\bibnamefont {Kuan}, \bibfnamefont
  {H.-J.}}, \bibinfo {author} {\bibfnamefont {J.}~\bibnamefont {Singh}},
  \bibinfo {author} {\bibfnamefont {D.~D.}\ \bibnamefont {Doneva}}, \bibinfo
  {author} {\bibfnamefont {S.~S.}\ \bibnamefont {Yazadjiev}}, and\ \bibinfo
  {author} {\bibfnamefont {K.~D.}\ \bibnamefont {Kokkotas}}} (\bibinfo {year}
  {2021}{\natexlab{b}}),\ \href {https://doi.org/10.1103/PhysRevD.104.124013}
  {\bibfield  {journal} {\bibinfo  {journal} {Phys. Rev. D}\ }\textbf {\bibinfo
  {volume} {104}}~(\bibinfo {number} {12}),\ \bibinfo {pages} {124013}},\
  \Eprint {https://arxiv.org/abs/2105.08543} {arXiv:2105.08543 [gr-qc]}
  \BibitemShut {NoStop}%
\bibitem [{\citenamefont {Kuan}\ \emph {et~al.}(2022)\citenamefont {Kuan},
  \citenamefont {Suvorov}, \citenamefont {Doneva},\ and\ \citenamefont
  {Yazadjiev}}]{Kuan:2022oxs}%
  \BibitemOpen
  \bibfield  {author} {\bibinfo {author} {\bibnamefont {Kuan}, \bibfnamefont
  {H.-J.}}, \bibinfo {author} {\bibfnamefont {A.~G.}\ \bibnamefont {Suvorov}},
  \bibinfo {author} {\bibfnamefont {D.~D.}\ \bibnamefont {Doneva}}, and\
  \bibinfo {author} {\bibfnamefont {S.~S.}\ \bibnamefont {Yazadjiev}}}
  (\bibinfo {year} {2022}),\ \href
  {https://doi.org/10.1103/PhysRevLett.129.121104} {\bibfield  {journal}
  {\bibinfo  {journal} {Phys. Rev. Lett.}\ }\textbf {\bibinfo {volume}
  {129}}~(\bibinfo {number} {12}),\ \bibinfo {pages} {121104}},\ \Eprint
  {https://arxiv.org/abs/2203.03672} {arXiv:2203.03672 [gr-qc]} \BibitemShut
  {NoStop}%
\bibitem [{\citenamefont {Landulfo}\ \emph {et~al.}(2012)\citenamefont
  {Landulfo}, \citenamefont {Lima}, \citenamefont {Matsas},\ and\ \citenamefont
  {Vanzella}}]{Landulfo:2012nz}%
  \BibitemOpen
  \bibfield  {author} {\bibinfo {author} {\bibnamefont {Landulfo},
  \bibfnamefont {A.~G.~S.}}, \bibinfo {author} {\bibfnamefont {W.~C.~C.}\
  \bibnamefont {Lima}}, \bibinfo {author} {\bibfnamefont {G.~E.~A.}\
  \bibnamefont {Matsas}}, and\ \bibinfo {author} {\bibfnamefont {D.~A.~T.}\
  \bibnamefont {Vanzella}}} (\bibinfo {year} {2012}),\ \href
  {https://doi.org/10.1103/PhysRevD.86.104025} {\bibfield  {journal} {\bibinfo
  {journal} {Phys. Rev. D}\ }\textbf {\bibinfo {volume} {86}},\ \bibinfo
  {pages} {104025}},\ \Eprint {https://arxiv.org/abs/1204.3654}
  {arXiv:1204.3654 [gr-qc]} \BibitemShut {NoStop}%
\bibitem [{\citenamefont {Landulfo}\ \emph {et~al.}(2015)\citenamefont
  {Landulfo}, \citenamefont {Lima}, \citenamefont {Matsas},\ and\ \citenamefont
  {Vanzella}}]{Landulfo:2014wra}%
  \BibitemOpen
  \bibfield  {author} {\bibinfo {author} {\bibnamefont {Landulfo},
  \bibfnamefont {A.~G.~S.}}, \bibinfo {author} {\bibfnamefont {W.~C.~C.}\
  \bibnamefont {Lima}}, \bibinfo {author} {\bibfnamefont {G.~E.~A.}\
  \bibnamefont {Matsas}}, and\ \bibinfo {author} {\bibfnamefont {D.~A.~T.}\
  \bibnamefont {Vanzella}}} (\bibinfo {year} {2015}),\ \href
  {https://doi.org/10.1103/PhysRevD.91.024011} {\bibfield  {journal} {\bibinfo
  {journal} {Phys. Rev. D}\ }\textbf {\bibinfo {volume} {91}}~(\bibinfo
  {number} {2}),\ \bibinfo {pages} {024011}},\ \Eprint
  {https://arxiv.org/abs/1410.2274} {arXiv:1410.2274 [gr-qc]} \BibitemShut
  {NoStop}%
\bibitem [{\citenamefont {Lang}(2014)}]{Lang:2013fna}%
  \BibitemOpen
  \bibfield  {author} {\bibinfo {author} {\bibnamefont {Lang}, \bibfnamefont
  {R.~N.}}} (\bibinfo {year} {2014}),\ \href
  {https://doi.org/10.1103/PhysRevD.89.084014} {\bibfield  {journal} {\bibinfo
  {journal} {Phys. Rev. D}\ }\textbf {\bibinfo {volume} {89}}~(\bibinfo
  {number} {8}),\ \bibinfo {pages} {084014}},\ \Eprint
  {https://arxiv.org/abs/1310.3320} {arXiv:1310.3320 [gr-qc]} \BibitemShut
  {NoStop}%
\bibitem [{\citenamefont {Langlois}\ \emph {et~al.}(2022)\citenamefont
  {Langlois}, \citenamefont {Noui},\ and\ \citenamefont
  {Roussille}}]{Langlois:2022eta}%
  \BibitemOpen
  \bibfield  {author} {\bibinfo {author} {\bibnamefont {Langlois},
  \bibfnamefont {D.}}, \bibinfo {author} {\bibfnamefont {K.}~\bibnamefont
  {Noui}}, and\ \bibinfo {author} {\bibfnamefont {H.}~\bibnamefont
  {Roussille}}} (\bibinfo {year} {2022}),\ \href
  {https://doi.org/10.1088/1475-7516/2022/09/019} {\bibfield  {journal}
  {\bibinfo  {journal} {JCAP}\ }\textbf {\bibinfo {volume} {09}},\ \bibinfo
  {pages} {019}},\ \Eprint {https://arxiv.org/abs/2204.04107} {arXiv:2204.04107
  [gr-qc]} \BibitemShut {NoStop}%
\bibitem [{\citenamefont {Lattimer}\ and\ \citenamefont
  {Prakash}(2016)}]{Lattimer:2015nhk}%
  \BibitemOpen
  \bibfield  {author} {\bibinfo {author} {\bibnamefont {Lattimer},
  \bibfnamefont {J.~M.}}, and\ \bibinfo {author} {\bibfnamefont
  {M.}~\bibnamefont {Prakash}}} (\bibinfo {year} {2016}),\ \href
  {https://doi.org/10.1016/j.physrep.2015.12.005} {\bibfield  {journal}
  {\bibinfo  {journal} {Phys. Rept.}\ }\textbf {\bibinfo {volume} {621}},\
  \bibinfo {pages} {127}},\ \Eprint {https://arxiv.org/abs/1512.07820}
  {arXiv:1512.07820 [astro-ph.SR]} \BibitemShut {NoStop}%
\bibitem [{\citenamefont {Lattimer}\ and\ \citenamefont
  {Schutz}(2005)}]{Lattimer:2004nj}%
  \BibitemOpen
  \bibfield  {author} {\bibinfo {author} {\bibnamefont {Lattimer},
  \bibfnamefont {J.~M.}}, and\ \bibinfo {author} {\bibfnamefont {B.~F.}\
  \bibnamefont {Schutz}}} (\bibinfo {year} {2005}),\ \href
  {https://doi.org/10.1086/431543} {\bibfield  {journal} {\bibinfo  {journal}
  {Astrophys. J.}\ }\textbf {\bibinfo {volume} {629}},\ \bibinfo {pages}
  {979}},\ \Eprint {https://arxiv.org/abs/astro-ph/0411470}
  {arXiv:astro-ph/0411470} \BibitemShut {NoStop}%
\bibitem [{\citenamefont {Lee}(1974)}]{Lee:1974pt}%
  \BibitemOpen
  \bibfield  {author} {\bibinfo {author} {\bibnamefont {Lee}, \bibfnamefont
  {D.~L.}}} (\bibinfo {year} {1974}),\ \href
  {https://doi.org/10.1103/PhysRevD.10.2374} {\bibfield  {journal} {\bibinfo
  {journal} {Phys. Rev. D}\ }\textbf {\bibinfo {volume} {10}},\ \bibinfo
  {pages} {2374}}\BibitemShut {NoStop}%
\bibitem [{\citenamefont {Liebling}\ and\ \citenamefont
  {Palenzuela}(2023)}]{Liebling:2012fv}%
  \BibitemOpen
  \bibfield  {author} {\bibinfo {author} {\bibnamefont {Liebling},
  \bibfnamefont {S.~L.}}, and\ \bibinfo {author} {\bibfnamefont
  {C.}~\bibnamefont {Palenzuela}}} (\bibinfo {year} {2023}),\ \href
  {https://doi.org/10.1007/s41114-023-00043-4} {\bibfield  {journal} {\bibinfo
  {journal} {Living Rev. Rel.}\ }\textbf {\bibinfo {volume} {26}},\ \bibinfo
  {pages} {1}},\ \Eprint {https://arxiv.org/abs/1202.5809} {arXiv:1202.5809
  [gr-qc]} \BibitemShut {NoStop}%
\bibitem [{\citenamefont {Lima}\ \emph {et~al.}(2010)\citenamefont {Lima},
  \citenamefont {Matsas},\ and\ \citenamefont {Vanzella}}]{Lima:2010na}%
  \BibitemOpen
  \bibfield  {author} {\bibinfo {author} {\bibnamefont {Lima}, \bibfnamefont
  {W.~C.~C.}}, \bibinfo {author} {\bibfnamefont {G.~E.~A.}\ \bibnamefont
  {Matsas}}, and\ \bibinfo {author} {\bibfnamefont {D.~A.~T.}\ \bibnamefont
  {Vanzella}}} (\bibinfo {year} {2010}),\ \href
  {https://doi.org/10.1103/PhysRevLett.105.151102} {\bibfield  {journal}
  {\bibinfo  {journal} {Phys. Rev. Lett.}\ }\textbf {\bibinfo {volume} {105}},\
  \bibinfo {pages} {151102}},\ \Eprint {https://arxiv.org/abs/1009.1771}
  {arXiv:1009.1771 [gr-qc]} \BibitemShut {NoStop}%
\bibitem [{\citenamefont {Lima}\ \emph {et~al.}(2013)\citenamefont {Lima},
  \citenamefont {Mendes}, \citenamefont {Matsas},\ and\ \citenamefont
  {Vanzella}}]{Lima:2013uya}%
  \BibitemOpen
  \bibfield  {author} {\bibinfo {author} {\bibnamefont {Lima}, \bibfnamefont
  {W.~C.~C.}}, \bibinfo {author} {\bibfnamefont {R.~F.~P.}\ \bibnamefont
  {Mendes}}, \bibinfo {author} {\bibfnamefont {G.~E.~A.}\ \bibnamefont
  {Matsas}}, and\ \bibinfo {author} {\bibfnamefont {D.~A.~T.}\ \bibnamefont
  {Vanzella}}} (\bibinfo {year} {2013}),\ \href
  {https://doi.org/10.1103/PhysRevD.87.104039} {\bibfield  {journal} {\bibinfo
  {journal} {Phys. Rev. D}\ }\textbf {\bibinfo {volume} {87}}~(\bibinfo
  {number} {10}),\ \bibinfo {pages} {104039}},\ \Eprint
  {https://arxiv.org/abs/1304.0582} {arXiv:1304.0582 [gr-qc]} \BibitemShut
  {NoStop}%
\bibitem [{\citenamefont {Lima}\ and\ \citenamefont
  {Vanzella}(2010)}]{Lima:2010xw}%
  \BibitemOpen
  \bibfield  {author} {\bibinfo {author} {\bibnamefont {Lima}, \bibfnamefont
  {W.~C.~C.}}, and\ \bibinfo {author} {\bibfnamefont {D.~A.~T.}\ \bibnamefont
  {Vanzella}}} (\bibinfo {year} {2010}),\ \href
  {https://doi.org/10.1103/PhysRevLett.104.161102} {\bibfield  {journal}
  {\bibinfo  {journal} {Phys. Rev. Lett.}\ }\textbf {\bibinfo {volume} {104}},\
  \bibinfo {pages} {161102}},\ \Eprint {https://arxiv.org/abs/1003.3421}
  {arXiv:1003.3421 [gr-qc]} \BibitemShut {NoStop}%
\bibitem [{\citenamefont {Lindblom}(2010)}]{Lindblom:2010bb}%
  \BibitemOpen
  \bibfield  {author} {\bibinfo {author} {\bibnamefont {Lindblom},
  \bibfnamefont {L.}}} (\bibinfo {year} {2010}),\ \href
  {https://doi.org/10.1103/PhysRevD.82.103011} {\bibfield  {journal} {\bibinfo
  {journal} {Phys. Rev. D}\ }\textbf {\bibinfo {volume} {82}},\ \bibinfo
  {pages} {103011}},\ \Eprint {https://arxiv.org/abs/1009.0738}
  {arXiv:1009.0738 [astro-ph.HE]} \BibitemShut {NoStop}%
\bibitem [{\citenamefont {Liu}\ \emph {et~al.}(2023{\natexlab{a}})\citenamefont
  {Liu}, \citenamefont {Zhang}, \citenamefont {Chen}, \citenamefont {Cao},
  \citenamefont {Tian},\ and\ \citenamefont {Wang}}]{Liu:2022fxy}%
  \BibitemOpen
  \bibfield  {author} {\bibinfo {author} {\bibnamefont {Liu}, \bibfnamefont
  {Y.}}, \bibinfo {author} {\bibfnamefont {C.-Y.}\ \bibnamefont {Zhang}},
  \bibinfo {author} {\bibfnamefont {Q.}~\bibnamefont {Chen}}, \bibinfo {author}
  {\bibfnamefont {Z.}~\bibnamefont {Cao}}, \bibinfo {author} {\bibfnamefont
  {Y.}~\bibnamefont {Tian}}, and\ \bibinfo {author} {\bibfnamefont
  {B.}~\bibnamefont {Wang}}} (\bibinfo {year} {2023}{\natexlab{a}}),\ \href
  {https://doi.org/10.1007/s11433-023-2160-1} {\bibfield  {journal} {\bibinfo
  {journal} {Sci. China Phys. Mech. Astron.}\ }\textbf {\bibinfo {volume}
  {66}}~(\bibinfo {number} {10}),\ \bibinfo {pages} {100412}},\ \Eprint
  {https://arxiv.org/abs/2208.07548} {arXiv:2208.07548 [gr-qc]} \BibitemShut
  {NoStop}%
\bibitem [{\citenamefont {Liu}\ \emph {et~al.}(2023{\natexlab{b}})\citenamefont
  {Liu}, \citenamefont {Zhang}, \citenamefont {Qian}, \citenamefont {Lin},\
  and\ \citenamefont {Wang}}]{Liu:2022eri}%
  \BibitemOpen
  \bibfield  {author} {\bibinfo {author} {\bibnamefont {Liu}, \bibfnamefont
  {Y.}}, \bibinfo {author} {\bibfnamefont {C.-Y.}\ \bibnamefont {Zhang}},
  \bibinfo {author} {\bibfnamefont {W.-L.}\ \bibnamefont {Qian}}, \bibinfo
  {author} {\bibfnamefont {K.}~\bibnamefont {Lin}}, and\ \bibinfo {author}
  {\bibfnamefont {B.}~\bibnamefont {Wang}}} (\bibinfo {year}
  {2023}{\natexlab{b}}),\ \href {https://doi.org/10.1007/JHEP01(2023)074}
  {\bibfield  {journal} {\bibinfo  {journal} {JHEP}\ }\textbf {\bibinfo
  {volume} {01}},\ \bibinfo {pages} {074}},\ \Eprint
  {https://arxiv.org/abs/2206.05012} {arXiv:2206.05012 [gr-qc]} \BibitemShut
  {NoStop}%
\bibitem [{\citenamefont {Luo}\ \emph {et~al.}(2022)\citenamefont {Luo},
  \citenamefont {Zhang}, \citenamefont {Liu}, \citenamefont {Niu},\ and\
  \citenamefont {Wang}}]{Luo:2022roz}%
  \BibitemOpen
  \bibfield  {author} {\bibinfo {author} {\bibnamefont {Luo}, \bibfnamefont
  {W.-K.}}, \bibinfo {author} {\bibfnamefont {C.-Y.}\ \bibnamefont {Zhang}},
  \bibinfo {author} {\bibfnamefont {P.}~\bibnamefont {Liu}}, \bibinfo {author}
  {\bibfnamefont {C.}~\bibnamefont {Niu}}, and\ \bibinfo {author}
  {\bibfnamefont {B.}~\bibnamefont {Wang}}} (\bibinfo {year} {2022}),\ \href
  {https://doi.org/10.1103/PhysRevD.106.064036} {\bibfield  {journal} {\bibinfo
   {journal} {Phys. Rev. D}\ }\textbf {\bibinfo {volume} {106}}~(\bibinfo
  {number} {6}),\ \bibinfo {pages} {064036}},\ \Eprint
  {https://arxiv.org/abs/2206.05690} {arXiv:2206.05690 [gr-qc]} \BibitemShut
  {NoStop}%
\bibitem [{\citenamefont {Macedo}(2020)}]{Macedo:2020tbm}%
  \BibitemOpen
  \bibfield  {author} {\bibinfo {author} {\bibnamefont {Macedo}, \bibfnamefont
  {C.~F.~B.}}} (\bibinfo {year} {2020}),\ \href
  {https://doi.org/10.1142/S0218271820410060} {\bibfield  {journal} {\bibinfo
  {journal} {Int. J. Mod. Phys. D}\ }\textbf {\bibinfo {volume} {29}}~(\bibinfo
  {number} {11}),\ \bibinfo {pages} {2041006}},\ \Eprint
  {https://arxiv.org/abs/2002.12719} {arXiv:2002.12719 [gr-qc]} \BibitemShut
  {NoStop}%
\bibitem [{\citenamefont {Macedo}\ \emph {et~al.}(2019)\citenamefont {Macedo},
  \citenamefont {Sakstein}, \citenamefont {Berti}, \citenamefont {Gualtieri},
  \citenamefont {Silva},\ and\ \citenamefont {Sotiriou}}]{Macedo:2019sem}%
  \BibitemOpen
  \bibfield  {author} {\bibinfo {author} {\bibnamefont {Macedo}, \bibfnamefont
  {C.~F.~B.}}, \bibinfo {author} {\bibfnamefont {J.}~\bibnamefont {Sakstein}},
  \bibinfo {author} {\bibfnamefont {E.}~\bibnamefont {Berti}}, \bibinfo
  {author} {\bibfnamefont {L.}~\bibnamefont {Gualtieri}}, \bibinfo {author}
  {\bibfnamefont {H.~O.}\ \bibnamefont {Silva}}, and\ \bibinfo {author}
  {\bibfnamefont {T.~P.}\ \bibnamefont {Sotiriou}}} (\bibinfo {year} {2019}),\
  \href {https://doi.org/10.1103/PhysRevD.99.104041} {\bibfield  {journal}
  {\bibinfo  {journal} {Phys. Rev. D}\ }\textbf {\bibinfo {volume}
  {99}}~(\bibinfo {number} {10}),\ \bibinfo {pages} {104041}},\ \Eprint
  {https://arxiv.org/abs/1903.06784} {arXiv:1903.06784 [gr-qc]} \BibitemShut
  {NoStop}%
\bibitem [{\citenamefont {Maione}\ \emph {et~al.}(2016)\citenamefont {Maione},
  \citenamefont {De~Pietri}, \citenamefont {Feo},\ and\ \citenamefont
  {L\"offler}}]{Maione:2016zqz}%
  \BibitemOpen
  \bibfield  {author} {\bibinfo {author} {\bibnamefont {Maione}, \bibfnamefont
  {F.}}, \bibinfo {author} {\bibfnamefont {R.}~\bibnamefont {De~Pietri}},
  \bibinfo {author} {\bibfnamefont {A.}~\bibnamefont {Feo}}, and\ \bibinfo
  {author} {\bibfnamefont {F.}~\bibnamefont {L\"offler}}} (\bibinfo {year}
  {2016}),\ \href {https://doi.org/10.1088/0264-9381/33/17/175009} {\bibfield
  {journal} {\bibinfo  {journal} {Class. Quant. Grav.}\ }\textbf {\bibinfo
  {volume} {33}}~(\bibinfo {number} {17}),\ \bibinfo {pages} {175009}},\
  \Eprint {https://arxiv.org/abs/1605.03424} {arXiv:1605.03424 [gr-qc]}
  \BibitemShut {NoStop}%
\bibitem [{\citenamefont {Maselli}\ \emph {et~al.}(2020)\citenamefont
  {Maselli}, \citenamefont {Franchini}, \citenamefont {Gualtieri},\ and\
  \citenamefont {Sotiriou}}]{Maselli:2020zgv}%
  \BibitemOpen
  \bibfield  {author} {\bibinfo {author} {\bibnamefont {Maselli}, \bibfnamefont
  {A.}}, \bibinfo {author} {\bibfnamefont {N.}~\bibnamefont {Franchini}},
  \bibinfo {author} {\bibfnamefont {L.}~\bibnamefont {Gualtieri}}, and\
  \bibinfo {author} {\bibfnamefont {T.~P.}\ \bibnamefont {Sotiriou}}} (\bibinfo
  {year} {2020}),\ \href {https://doi.org/10.1103/PhysRevLett.125.141101}
  {\bibfield  {journal} {\bibinfo  {journal} {Phys. Rev. Lett.}\ }\textbf
  {\bibinfo {volume} {125}}~(\bibinfo {number} {14}),\ \bibinfo {pages}
  {141101}},\ \Eprint {https://arxiv.org/abs/2004.11895} {arXiv:2004.11895
  [gr-qc]} \BibitemShut {NoStop}%
\bibitem [{\citenamefont {Maselli}\ \emph {et~al.}(2022)\citenamefont
  {Maselli}, \citenamefont {Franchini}, \citenamefont {Gualtieri},
  \citenamefont {Sotiriou}, \citenamefont {Barsanti},\ and\ \citenamefont
  {Pani}}]{Maselli:2021men}%
  \BibitemOpen
  \bibfield  {author} {\bibinfo {author} {\bibnamefont {Maselli}, \bibfnamefont
  {A.}}, \bibinfo {author} {\bibfnamefont {N.}~\bibnamefont {Franchini}},
  \bibinfo {author} {\bibfnamefont {L.}~\bibnamefont {Gualtieri}}, \bibinfo
  {author} {\bibfnamefont {T.~P.}\ \bibnamefont {Sotiriou}}, \bibinfo {author}
  {\bibfnamefont {S.}~\bibnamefont {Barsanti}}, and\ \bibinfo {author}
  {\bibfnamefont {P.}~\bibnamefont {Pani}}} (\bibinfo {year} {2022}),\ \href
  {https://doi.org/10.1038/s41550-021-01589-5} {\bibfield  {journal} {\bibinfo
  {journal} {Nature Astron.}\ }\textbf {\bibinfo {volume} {6}}~(\bibinfo
  {number} {4}),\ \bibinfo {pages} {464}},\ \Eprint
  {https://arxiv.org/abs/2106.11325} {arXiv:2106.11325 [gr-qc]} \BibitemShut
  {NoStop}%
\bibitem [{\citenamefont {Mayo}\ and\ \citenamefont
  {Bekenstein}(1996)}]{Mayo:1996mv}%
  \BibitemOpen
  \bibfield  {author} {\bibinfo {author} {\bibnamefont {Mayo}, \bibfnamefont
  {A.~E.}}, and\ \bibinfo {author} {\bibfnamefont {J.~D.}\ \bibnamefont
  {Bekenstein}}} (\bibinfo {year} {1996}),\ \href
  {https://doi.org/10.1103/PhysRevD.54.5059} {\bibfield  {journal} {\bibinfo
  {journal} {Phys. Rev. D}\ }\textbf {\bibinfo {volume} {54}},\ \bibinfo
  {pages} {5059}},\ \Eprint {https://arxiv.org/abs/gr-qc/9602057}
  {arXiv:gr-qc/9602057} \BibitemShut {NoStop}%
\bibitem [{\citenamefont {{McDermott}}\ \emph {et~al.}(1983)\citenamefont
  {{McDermott}}, \citenamefont {{van Horn}},\ and\ \citenamefont
  {{Scholl}}}]{McDermott:1983ApJ}%
  \BibitemOpen
  \bibfield  {author} {\bibinfo {author} {\bibnamefont {{McDermott}},
  \bibfnamefont {P.~N.}}, \bibinfo {author} {\bibfnamefont {H.~M.}\
  \bibnamefont {{van Horn}}}, and\ \bibinfo {author} {\bibfnamefont {J.~F.}\
  \bibnamefont {{Scholl}}}} (\bibinfo {year} {1983}),\ \href
  {https://doi.org/10.1086/161006} {\bibfield  {journal} {\bibinfo  {journal}
  {ApJ}\ }\textbf {\bibinfo {volume} {268}},\ \bibinfo {pages}
  {837}}\BibitemShut {NoStop}%
\bibitem [{\citenamefont {Mendes}(2015)}]{Mendes:2014ufa}%
  \BibitemOpen
  \bibfield  {author} {\bibinfo {author} {\bibnamefont {Mendes}, \bibfnamefont
  {R.~F.}}} (\bibinfo {year} {2015}),\ \href
  {https://doi.org/10.1103/PhysRevD.91.064024} {\bibfield  {journal} {\bibinfo
  {journal} {Phys. Rev. D}\ }\textbf {\bibinfo {volume} {91}}~(\bibinfo
  {number} {6}),\ \bibinfo {pages} {064024}},\ \Eprint
  {https://arxiv.org/abs/1412.6789} {arXiv:1412.6789 [gr-qc]} \BibitemShut
  {NoStop}%
\bibitem [{\citenamefont {Mendes}\ and\ \citenamefont
  {Ortiz}(2016)}]{Mendes:2016fby}%
  \BibitemOpen
  \bibfield  {author} {\bibinfo {author} {\bibnamefont {Mendes}, \bibfnamefont
  {R.~F.}}, and\ \bibinfo {author} {\bibfnamefont {N.}~\bibnamefont {Ortiz}}}
  (\bibinfo {year} {2016}),\ \href {https://doi.org/10.1103/PhysRevD.93.124035}
  {\bibfield  {journal} {\bibinfo  {journal} {Phys. Rev. D}\ }\textbf {\bibinfo
  {volume} {93}}~(\bibinfo {number} {12}),\ \bibinfo {pages} {124035}},\
  \Eprint {https://arxiv.org/abs/1604.04175} {arXiv:1604.04175 [gr-qc]}
  \BibitemShut {NoStop}%
\bibitem [{\citenamefont {Mendes}\ and\ \citenamefont
  {Ortiz}(2018)}]{Mendes:2018qwo}%
  \BibitemOpen
  \bibfield  {author} {\bibinfo {author} {\bibnamefont {Mendes}, \bibfnamefont
  {R.~F.}}, and\ \bibinfo {author} {\bibfnamefont {N.}~\bibnamefont {Ortiz}}}
  (\bibinfo {year} {2018}),\ \href
  {https://doi.org/10.1103/PhysRevLett.120.201104} {\bibfield  {journal}
  {\bibinfo  {journal} {Phys. Rev. Lett.}\ }\textbf {\bibinfo {volume}
  {120}}~(\bibinfo {number} {20}),\ \bibinfo {pages} {201104}},\ \Eprint
  {https://arxiv.org/abs/1802.07847} {arXiv:1802.07847 [gr-qc]} \BibitemShut
  {NoStop}%
\bibitem [{\citenamefont {Mendes}\ and\ \citenamefont
  {Ottoni}(2019)}]{Mendes:2019zpw}%
  \BibitemOpen
  \bibfield  {author} {\bibinfo {author} {\bibnamefont {Mendes}, \bibfnamefont
  {R.~F.}}, and\ \bibinfo {author} {\bibfnamefont {T.}~\bibnamefont {Ottoni}}}
  (\bibinfo {year} {2019}),\ \href {https://doi.org/10.1103/PhysRevD.99.124003}
  {\bibfield  {journal} {\bibinfo  {journal} {Phys. Rev. D}\ }\textbf {\bibinfo
  {volume} {99}}~(\bibinfo {number} {12}),\ \bibinfo {pages} {124003}},\
  \Eprint {https://arxiv.org/abs/1903.11638} {arXiv:1903.11638 [gr-qc]}
  \BibitemShut {NoStop}%
\bibitem [{\citenamefont {Mendes}\ \emph
  {et~al.}(2014{\natexlab{a}})\citenamefont {Mendes}, \citenamefont {Matsas},\
  and\ \citenamefont {Vanzella}}]{Mendes:2014vna}%
  \BibitemOpen
  \bibfield  {author} {\bibinfo {author} {\bibnamefont {Mendes}, \bibfnamefont
  {R.~F.~P.}}, \bibinfo {author} {\bibfnamefont {G.~E.~A.}\ \bibnamefont
  {Matsas}}, and\ \bibinfo {author} {\bibfnamefont {D.~A.~T.}\ \bibnamefont
  {Vanzella}}} (\bibinfo {year} {2014}{\natexlab{a}}),\ \href
  {https://doi.org/10.1103/PhysRevD.90.044053} {\bibfield  {journal} {\bibinfo
  {journal} {Phys. Rev. D}\ }\textbf {\bibinfo {volume} {90}}~(\bibinfo
  {number} {4}),\ \bibinfo {pages} {044053}},\ \Eprint
  {https://arxiv.org/abs/1407.6405} {arXiv:1407.6405 [gr-qc]} \BibitemShut
  {NoStop}%
\bibitem [{\citenamefont {Mendes}\ \emph
  {et~al.}(2014{\natexlab{b}})\citenamefont {Mendes}, \citenamefont {Matsas},\
  and\ \citenamefont {Vanzella}}]{Mendes:2013ija}%
  \BibitemOpen
  \bibfield  {author} {\bibinfo {author} {\bibnamefont {Mendes}, \bibfnamefont
  {R.~F.~P.}}, \bibinfo {author} {\bibfnamefont {G.~E.~A.}\ \bibnamefont
  {Matsas}}, and\ \bibinfo {author} {\bibfnamefont {D.~A.~T.}\ \bibnamefont
  {Vanzella}}} (\bibinfo {year} {2014}{\natexlab{b}}),\ \href
  {https://doi.org/10.1103/PhysRevD.89.047503} {\bibfield  {journal} {\bibinfo
  {journal} {Phys. Rev. D}\ }\textbf {\bibinfo {volume} {89}}~(\bibinfo
  {number} {4}),\ \bibinfo {pages} {047503}},\ \Eprint
  {https://arxiv.org/abs/1310.2185} {arXiv:1310.2185 [gr-qc]} \BibitemShut
  {NoStop}%
\bibitem [{\citenamefont {Miller}\ \emph {et~al.}(2019)\citenamefont {Miller}
  \emph {et~al.}}]{Miller:2019cac}%
  \BibitemOpen
  \bibfield  {author} {\bibinfo {author} {\bibnamefont {Miller}, \bibfnamefont
  {M.~C.}},  \emph {et~al.}} (\bibinfo {year} {2019}),\ \href
  {https://doi.org/10.3847/2041-8213/ab50c5} {\bibfield  {journal} {\bibinfo
  {journal} {Astrophys. J. Lett.}\ }\textbf {\bibinfo {volume} {887}}~(\bibinfo
  {number} {1}),\ \bibinfo {pages} {L24}},\ \Eprint
  {https://arxiv.org/abs/1912.05705} {arXiv:1912.05705 [astro-ph.HE]}
  \BibitemShut {NoStop}%
\bibitem [{\citenamefont
  {Minamitsuji}(2020{\natexlab{a}})}]{Minamitsuji:2020pak}%
  \BibitemOpen
  \bibfield  {author} {\bibinfo {author} {\bibnamefont {Minamitsuji},
  \bibfnamefont {M.}}} (\bibinfo {year} {2020}{\natexlab{a}}),\ \href
  {https://doi.org/10.1103/PhysRevD.101.104044} {\bibfield  {journal} {\bibinfo
   {journal} {Phys. Rev. D}\ }\textbf {\bibinfo {volume} {101}}~(\bibinfo
  {number} {10}),\ \bibinfo {pages} {104044}},\ \Eprint
  {https://arxiv.org/abs/2003.11885} {arXiv:2003.11885 [gr-qc]} \BibitemShut
  {NoStop}%
\bibitem [{\citenamefont
  {Minamitsuji}(2020{\natexlab{b}})}]{Minamitsuji:2020hpl}%
  \BibitemOpen
  \bibfield  {author} {\bibinfo {author} {\bibnamefont {Minamitsuji},
  \bibfnamefont {M.}}} (\bibinfo {year} {2020}{\natexlab{b}}),\ \href
  {https://doi.org/10.1103/PhysRevD.102.044048} {\bibfield  {journal} {\bibinfo
   {journal} {Phys. Rev. D}\ }\textbf {\bibinfo {volume} {102}}~(\bibinfo
  {number} {4}),\ \bibinfo {pages} {044048}},\ \Eprint
  {https://arxiv.org/abs/2008.12758} {arXiv:2008.12758 [gr-qc]} \BibitemShut
  {NoStop}%
\bibitem [{\citenamefont {Minamitsuji}(2021)}]{Minamitsuji:2021rtw}%
  \BibitemOpen
  \bibfield  {author} {\bibinfo {author} {\bibnamefont {Minamitsuji},
  \bibfnamefont {M.}}} (\bibinfo {year} {2021}),\ \href
  {https://doi.org/10.1103/PhysRevD.103.084002} {\bibfield  {journal} {\bibinfo
   {journal} {Phys. Rev. D}\ }\textbf {\bibinfo {volume} {103}}~(\bibinfo
  {number} {8}),\ \bibinfo {pages} {084002}},\ \Eprint
  {https://arxiv.org/abs/2104.03660} {arXiv:2104.03660 [gr-qc]} \BibitemShut
  {NoStop}%
\bibitem [{\citenamefont {Minamitsuji}\ and\ \citenamefont
  {Ikeda}(2019{\natexlab{a}})}]{Minamitsuji:2018xde}%
  \BibitemOpen
  \bibfield  {author} {\bibinfo {author} {\bibnamefont {Minamitsuji},
  \bibfnamefont {M.}}, and\ \bibinfo {author} {\bibfnamefont {T.}~\bibnamefont
  {Ikeda}}} (\bibinfo {year} {2019}{\natexlab{a}}),\ \href
  {https://doi.org/10.1103/PhysRevD.99.044017} {\bibfield  {journal} {\bibinfo
  {journal} {Phys. Rev. D}\ }\textbf {\bibinfo {volume} {99}}~(\bibinfo
  {number} {4}),\ \bibinfo {pages} {044017}},\ \Eprint
  {https://arxiv.org/abs/1812.03551} {arXiv:1812.03551 [gr-qc]} \BibitemShut
  {NoStop}%
\bibitem [{\citenamefont {Minamitsuji}\ and\ \citenamefont
  {Ikeda}(2019{\natexlab{b}})}]{Minamitsuji:2019iwp}%
  \BibitemOpen
  \bibfield  {author} {\bibinfo {author} {\bibnamefont {Minamitsuji},
  \bibfnamefont {M.}}, and\ \bibinfo {author} {\bibfnamefont {T.}~\bibnamefont
  {Ikeda}}} (\bibinfo {year} {2019}{\natexlab{b}}),\ \href
  {https://doi.org/10.1103/PhysRevD.99.104069} {\bibfield  {journal} {\bibinfo
  {journal} {Phys. Rev. D}\ }\textbf {\bibinfo {volume} {99}}~(\bibinfo
  {number} {10}),\ \bibinfo {pages} {104069}},\ \Eprint
  {https://arxiv.org/abs/1904.06572} {arXiv:1904.06572 [gr-qc]} \BibitemShut
  {NoStop}%
\bibitem [{\citenamefont {Minamitsuji}\ and\ \citenamefont
  {Silva}(2016)}]{Minamitsuji:2016hkk}%
  \BibitemOpen
  \bibfield  {author} {\bibinfo {author} {\bibnamefont {Minamitsuji},
  \bibfnamefont {M.}}, and\ \bibinfo {author} {\bibfnamefont {H.~O.}\
  \bibnamefont {Silva}}} (\bibinfo {year} {2016}),\ \href
  {https://doi.org/10.1103/PhysRevD.93.124041} {\bibfield  {journal} {\bibinfo
  {journal} {Phys. Rev. D}\ }\textbf {\bibinfo {volume} {93}}~(\bibinfo
  {number} {12}),\ \bibinfo {pages} {124041}},\ \Eprint
  {https://arxiv.org/abs/1604.07742} {arXiv:1604.07742 [gr-qc]} \BibitemShut
  {NoStop}%
\bibitem [{\citenamefont {Minamitsuji}\ and\ \citenamefont
  {Tsujikawa}(2023)}]{Minamitsuji:2022qku}%
  \BibitemOpen
  \bibfield  {author} {\bibinfo {author} {\bibnamefont {Minamitsuji},
  \bibfnamefont {M.}}, and\ \bibinfo {author} {\bibfnamefont {S.}~\bibnamefont
  {Tsujikawa}}} (\bibinfo {year} {2023}),\ \href
  {https://doi.org/10.1016/j.physletb.2023.137869} {\bibfield  {journal}
  {\bibinfo  {journal} {Phys. Lett. B}\ }\textbf {\bibinfo {volume} {840}},\
  \bibinfo {pages} {137869}},\ \Eprint {https://arxiv.org/abs/2208.08107}
  {arXiv:2208.08107 [gr-qc]} \BibitemShut {NoStop}%
\bibitem [{\citenamefont {Mirshekari}\ and\ \citenamefont
  {Will}(2013)}]{Mirshekari:2013vb}%
  \BibitemOpen
  \bibfield  {author} {\bibinfo {author} {\bibnamefont {Mirshekari},
  \bibfnamefont {S.}}, and\ \bibinfo {author} {\bibfnamefont {C.~M.}\
  \bibnamefont {Will}}} (\bibinfo {year} {2013}),\ \href
  {https://doi.org/10.1103/PhysRevD.87.084070} {\bibfield  {journal} {\bibinfo
  {journal} {Phys. Rev. D}\ }\textbf {\bibinfo {volume} {87}}~(\bibinfo
  {number} {8}),\ \bibinfo {pages} {084070}},\ \Eprint
  {https://arxiv.org/abs/1301.4680} {arXiv:1301.4680 [gr-qc]} \BibitemShut
  {NoStop}%
\bibitem [{\citenamefont {Morisaki}\ and\ \citenamefont
  {Suyama}(2017)}]{Morisaki:2017nit}%
  \BibitemOpen
  \bibfield  {author} {\bibinfo {author} {\bibnamefont {Morisaki},
  \bibfnamefont {S.}}, and\ \bibinfo {author} {\bibfnamefont {T.}~\bibnamefont
  {Suyama}}} (\bibinfo {year} {2017}),\ \href
  {https://doi.org/10.1103/PhysRevD.96.084026} {\bibfield  {journal} {\bibinfo
  {journal} {Phys. Rev. D}\ }\textbf {\bibinfo {volume} {96}}~(\bibinfo
  {number} {8}),\ \bibinfo {pages} {084026}},\ \Eprint
  {https://arxiv.org/abs/1707.02809} {arXiv:1707.02809 [gr-qc]} \BibitemShut
  {NoStop}%
\bibitem [{\citenamefont {Motohashi}\ and\ \citenamefont
  {Minamitsuji}(2018)}]{Motohashi:2018wdq}%
  \BibitemOpen
  \bibfield  {author} {\bibinfo {author} {\bibnamefont {Motohashi},
  \bibfnamefont {H.}}, and\ \bibinfo {author} {\bibfnamefont {M.}~\bibnamefont
  {Minamitsuji}}} (\bibinfo {year} {2018}),\ \href
  {https://doi.org/10.1016/j.physletb.2018.04.041} {\bibfield  {journal}
  {\bibinfo  {journal} {Phys. Lett. B}\ }\textbf {\bibinfo {volume} {781}},\
  \bibinfo {pages} {728}},\ \Eprint {https://arxiv.org/abs/1804.01731}
  {arXiv:1804.01731 [gr-qc]} \BibitemShut {NoStop}%
\bibitem [{\citenamefont {Motohashi}\ and\ \citenamefont
  {Suyama}(2012)}]{Motohashi:2011ds}%
  \BibitemOpen
  \bibfield  {author} {\bibinfo {author} {\bibnamefont {Motohashi},
  \bibfnamefont {H.}}, and\ \bibinfo {author} {\bibfnamefont {T.}~\bibnamefont
  {Suyama}}} (\bibinfo {year} {2012}),\ \href
  {https://doi.org/10.1103/PhysRevD.85.044054} {\bibfield  {journal} {\bibinfo
  {journal} {Phys. Rev. D}\ }\textbf {\bibinfo {volume} {85}},\ \bibinfo
  {pages} {044054}},\ \Eprint {https://arxiv.org/abs/1110.6241}
  {arXiv:1110.6241 [gr-qc]} \BibitemShut {NoStop}%
\bibitem [{\citenamefont {Mou}\ and\ \citenamefont
  {Zhang}(2022)}]{Mou:2022hqb}%
  \BibitemOpen
  \bibfield  {author} {\bibinfo {author} {\bibnamefont {Mou}, \bibfnamefont
  {Z.-G.}}, and\ \bibinfo {author} {\bibfnamefont {H.-Y.}\ \bibnamefont
  {Zhang}}} (\bibinfo {year} {2022}),\ \href
  {https://doi.org/10.1103/PhysRevLett.129.151101} {\bibfield  {journal}
  {\bibinfo  {journal} {Phys. Rev. Lett.}\ }\textbf {\bibinfo {volume}
  {129}}~(\bibinfo {number} {15}),\ \bibinfo {pages} {151101}},\ \Eprint
  {https://arxiv.org/abs/2204.11324} {arXiv:2204.11324 [hep-th]} \BibitemShut
  {NoStop}%
\bibitem [{\citenamefont {M\"uther}\ \emph {et~al.}(1987)\citenamefont
  {M\"uther}, \citenamefont {Prakash},\ and\ \citenamefont
  {Ainsworth}}]{Muther:1987xaa}%
  \BibitemOpen
  \bibfield  {author} {\bibinfo {author} {\bibnamefont {M\"uther},
  \bibfnamefont {H.}}, \bibinfo {author} {\bibfnamefont {M.}~\bibnamefont
  {Prakash}}, and\ \bibinfo {author} {\bibfnamefont {T.~L.}\ \bibnamefont
  {Ainsworth}}} (\bibinfo {year} {1987}),\ \href
  {https://doi.org/10.1016/0370-2693(87)91611-X} {\bibfield  {journal}
  {\bibinfo  {journal} {Phys. Lett. B}\ }\textbf {\bibinfo {volume} {199}},\
  \bibinfo {pages} {469}}\BibitemShut {NoStop}%
\bibitem [{\citenamefont {Myung}\ and\ \citenamefont
  {Zou}(2018)}]{Myung:2018iyq}%
  \BibitemOpen
  \bibfield  {author} {\bibinfo {author} {\bibnamefont {Myung}, \bibfnamefont
  {Y.~S.}}, and\ \bibinfo {author} {\bibfnamefont {D.-C.}\ \bibnamefont {Zou}}}
  (\bibinfo {year} {2018}),\ \href {https://doi.org/10.1103/PhysRevD.98.024030}
  {\bibfield  {journal} {\bibinfo  {journal} {Phys. Rev. D}\ }\textbf {\bibinfo
  {volume} {98}}~(\bibinfo {number} {2}),\ \bibinfo {pages} {024030}},\ \Eprint
  {https://arxiv.org/abs/1805.05023} {arXiv:1805.05023 [gr-qc]} \BibitemShut
  {NoStop}%
\bibitem [{\citenamefont {Myung}\ and\ \citenamefont
  {Zou}(2019{\natexlab{a}})}]{Myung:2019wvb}%
  \BibitemOpen
  \bibfield  {author} {\bibinfo {author} {\bibnamefont {Myung}, \bibfnamefont
  {Y.~S.}}, and\ \bibinfo {author} {\bibfnamefont {D.-C.}\ \bibnamefont {Zou}}}
  (\bibinfo {year} {2019}{\natexlab{a}}),\ \href
  {https://doi.org/10.1142/S0218271819501141} {\bibfield  {journal} {\bibinfo
  {journal} {Int. J. Mod. Phys. D}\ }\textbf {\bibinfo {volume} {28}}~(\bibinfo
  {number} {09}),\ \bibinfo {pages} {1950114}},\ \Eprint
  {https://arxiv.org/abs/1903.08312} {arXiv:1903.08312 [gr-qc]} \BibitemShut
  {NoStop}%
\bibitem [{\citenamefont {Myung}\ and\ \citenamefont
  {Zou}(2019{\natexlab{b}})}]{Myung:2018vug}%
  \BibitemOpen
  \bibfield  {author} {\bibinfo {author} {\bibnamefont {Myung}, \bibfnamefont
  {Y.~S.}}, and\ \bibinfo {author} {\bibfnamefont {D.-C.}\ \bibnamefont {Zou}}}
  (\bibinfo {year} {2019}{\natexlab{b}}),\ \href
  {https://doi.org/10.1140/epjc/s10052-019-6792-6} {\bibfield  {journal}
  {\bibinfo  {journal} {Eur. Phys. J. C}\ }\textbf {\bibinfo {volume}
  {79}}~(\bibinfo {number} {3}),\ \bibinfo {pages} {273}},\ \Eprint
  {https://arxiv.org/abs/1808.02609} {arXiv:1808.02609 [gr-qc]} \BibitemShut
  {NoStop}%
\bibitem [{\citenamefont {Myung}\ and\ \citenamefont
  {Zou}(2019{\natexlab{c}})}]{Myung:2019oua}%
  \BibitemOpen
  \bibfield  {author} {\bibinfo {author} {\bibnamefont {Myung}, \bibfnamefont
  {Y.~S.}}, and\ \bibinfo {author} {\bibfnamefont {D.-C.}\ \bibnamefont {Zou}}}
  (\bibinfo {year} {2019}{\natexlab{c}}),\ \href
  {https://doi.org/10.1140/epjc/s10052-019-7176-7} {\bibfield  {journal}
  {\bibinfo  {journal} {Eur. Phys. J. C}\ }\textbf {\bibinfo {volume}
  {79}}~(\bibinfo {number} {8}),\ \bibinfo {pages} {641}},\ \Eprint
  {https://arxiv.org/abs/1904.09864} {arXiv:1904.09864 [gr-qc]} \BibitemShut
  {NoStop}%
\bibitem [{\citenamefont {Myung}\ and\ \citenamefont
  {Zou}(2021)}]{Myung:2020etf}%
  \BibitemOpen
  \bibfield  {author} {\bibinfo {author} {\bibnamefont {Myung}, \bibfnamefont
  {Y.~S.}}, and\ \bibinfo {author} {\bibfnamefont {D.-C.}\ \bibnamefont {Zou}}}
  (\bibinfo {year} {2021}),\ \href
  {https://doi.org/10.1016/j.physletb.2021.136081} {\bibfield  {journal}
  {\bibinfo  {journal} {Phys. Lett. B}\ }\textbf {\bibinfo {volume} {814}},\
  \bibinfo {pages} {136081}},\ \Eprint {https://arxiv.org/abs/2012.02375}
  {arXiv:2012.02375 [gr-qc]} \BibitemShut {NoStop}%
\bibitem [{\citenamefont {Nelmes}\ and\ \citenamefont
  {Piette}(2012)}]{Nelmes:2012uf}%
  \BibitemOpen
  \bibfield  {author} {\bibinfo {author} {\bibnamefont {Nelmes}, \bibfnamefont
  {S.~G.}}, and\ \bibinfo {author} {\bibfnamefont {B.~M. A.~G.}\ \bibnamefont
  {Piette}}} (\bibinfo {year} {2012}),\ \href
  {https://doi.org/10.1103/PhysRevD.85.123004} {\bibfield  {journal} {\bibinfo
  {journal} {Phys. Rev. D}\ }\textbf {\bibinfo {volume} {85}},\ \bibinfo
  {pages} {123004}},\ \Eprint {https://arxiv.org/abs/1204.0910}
  {arXiv:1204.0910 [astro-ph.SR]} \BibitemShut {NoStop}%
\bibitem [{\citenamefont {Niu}\ \emph {et~al.}(2022)\citenamefont {Niu},
  \citenamefont {Xiong}, \citenamefont {Liu}, \citenamefont {Zhang},\ and\
  \citenamefont {Wang}}]{Niu:2022zlf}%
  \BibitemOpen
  \bibfield  {author} {\bibinfo {author} {\bibnamefont {Niu}, \bibfnamefont
  {C.}}, \bibinfo {author} {\bibfnamefont {W.}~\bibnamefont {Xiong}}, \bibinfo
  {author} {\bibfnamefont {P.}~\bibnamefont {Liu}}, \bibinfo {author}
  {\bibfnamefont {C.-Y.}\ \bibnamefont {Zhang}}, and\ \bibinfo {author}
  {\bibfnamefont {B.}~\bibnamefont {Wang}}} (\bibinfo {year} {2022}),\
  \href@noop {} {\ }\Eprint {https://arxiv.org/abs/2209.12117}
  {arXiv:2209.12117 [gr-qc]} \BibitemShut {NoStop}%
\bibitem [{\citenamefont {Niu}\ \emph {et~al.}(2021)\citenamefont {Niu},
  \citenamefont {Zhang}, \citenamefont {Wang},\ and\ \citenamefont
  {Zhao}}]{Niu:2021nic}%
  \BibitemOpen
  \bibfield  {author} {\bibinfo {author} {\bibnamefont {Niu}, \bibfnamefont
  {R.}}, \bibinfo {author} {\bibfnamefont {X.}~\bibnamefont {Zhang}}, \bibinfo
  {author} {\bibfnamefont {B.}~\bibnamefont {Wang}}, and\ \bibinfo {author}
  {\bibfnamefont {W.}~\bibnamefont {Zhao}}} (\bibinfo {year} {2021}),\ \href
  {https://doi.org/10.3847/1538-4357/ac1d4f} {\bibfield  {journal} {\bibinfo
  {journal} {Astrophys. J.}\ }\textbf {\bibinfo {volume} {921}}~(\bibinfo
  {number} {2}),\ \bibinfo {pages} {149}},\ \Eprint
  {https://arxiv.org/abs/2105.13644} {arXiv:2105.13644 [gr-qc]} \BibitemShut
  {NoStop}%
\bibitem [{\citenamefont {Novak}(1998{\natexlab{a}})}]{Novak:1998rk}%
  \BibitemOpen
  \bibfield  {author} {\bibinfo {author} {\bibnamefont {Novak}, \bibfnamefont
  {J.}}} (\bibinfo {year} {1998}{\natexlab{a}}),\ \href
  {https://doi.org/10.1103/PhysRevD.58.064019} {\bibfield  {journal} {\bibinfo
  {journal} {Phys. Rev. D}\ }\textbf {\bibinfo {volume} {58}},\ \bibinfo
  {pages} {064019}},\ \Eprint {https://arxiv.org/abs/gr-qc/9806022}
  {arXiv:gr-qc/9806022} \BibitemShut {NoStop}%
\bibitem [{\citenamefont {Novak}(1998{\natexlab{b}})}]{Novak:1997hw}%
  \BibitemOpen
  \bibfield  {author} {\bibinfo {author} {\bibnamefont {Novak}, \bibfnamefont
  {J.}}} (\bibinfo {year} {1998}{\natexlab{b}}),\ \href
  {https://doi.org/10.1103/PhysRevD.57.4789} {\bibfield  {journal} {\bibinfo
  {journal} {Phys. Rev. D}\ }\textbf {\bibinfo {volume} {57}},\ \bibinfo
  {pages} {4789}},\ \Eprint {https://arxiv.org/abs/gr-qc/9707041}
  {arXiv:gr-qc/9707041} \BibitemShut {NoStop}%
\bibitem [{\citenamefont {Novak}\ and\ \citenamefont
  {Ibanez}(2000)}]{Novak:1999jg}%
  \BibitemOpen
  \bibfield  {author} {\bibinfo {author} {\bibnamefont {Novak}, \bibfnamefont
  {J.}}, and\ \bibinfo {author} {\bibfnamefont {J.~M.}\ \bibnamefont {Ibanez}}}
  (\bibinfo {year} {2000}),\ \href {https://doi.org/10.1086/308627} {\bibfield
  {journal} {\bibinfo  {journal} {Astrophys. J.}\ }\textbf {\bibinfo {volume}
  {533}},\ \bibinfo {pages} {392}},\ \Eprint
  {https://arxiv.org/abs/astro-ph/9911298} {arXiv:astro-ph/9911298}
  \BibitemShut {NoStop}%
\bibitem [{\citenamefont {Ofengeim}(2020)}]{Ofengeim:2020zuc}%
  \BibitemOpen
  \bibfield  {author} {\bibinfo {author} {\bibnamefont {Ofengeim},
  \bibfnamefont {D.~D.}}} (\bibinfo {year} {2020}),\ \href
  {https://doi.org/10.1103/PhysRevD.101.103029} {\bibfield  {journal} {\bibinfo
   {journal} {Phys. Rev. D}\ }\textbf {\bibinfo {volume} {101}}~(\bibinfo
  {number} {10}),\ \bibinfo {pages} {103029}},\ \Eprint
  {https://arxiv.org/abs/2005.03549} {arXiv:2005.03549 [astro-ph.HE]}
  \BibitemShut {NoStop}%
\bibitem [{\citenamefont {Oliveira}\ and\ \citenamefont
  {Pombo}(2021)}]{Oliveira:2020dru}%
  \BibitemOpen
  \bibfield  {author} {\bibinfo {author} {\bibnamefont {Oliveira},
  \bibfnamefont {J.~a. M.~S.}}, and\ \bibinfo {author} {\bibfnamefont {A.~M.}\
  \bibnamefont {Pombo}}} (\bibinfo {year} {2021}),\ \href
  {https://doi.org/10.1103/PhysRevD.103.044004} {\bibfield  {journal} {\bibinfo
   {journal} {Phys. Rev. D}\ }\textbf {\bibinfo {volume} {103}}~(\bibinfo
  {number} {4}),\ \bibinfo {pages} {044004}},\ \Eprint
  {https://arxiv.org/abs/2012.07869} {arXiv:2012.07869 [gr-qc]} \BibitemShut
  {NoStop}%
\bibitem [{\citenamefont {Ozel}\ \emph {et~al.}(2016)\citenamefont {Ozel},
  \citenamefont {Psaltis}, \citenamefont {Guver}, \citenamefont {Baym},
  \citenamefont {Heinke},\ and\ \citenamefont {Guillot}}]{Ozel:2015fia}%
  \BibitemOpen
  \bibfield  {author} {\bibinfo {author} {\bibnamefont {Ozel}, \bibfnamefont
  {F.}}, \bibinfo {author} {\bibfnamefont {D.}~\bibnamefont {Psaltis}},
  \bibinfo {author} {\bibfnamefont {T.}~\bibnamefont {Guver}}, \bibinfo
  {author} {\bibfnamefont {G.}~\bibnamefont {Baym}}, \bibinfo {author}
  {\bibfnamefont {C.}~\bibnamefont {Heinke}}, and\ \bibinfo {author}
  {\bibfnamefont {S.}~\bibnamefont {Guillot}}} (\bibinfo {year} {2016}),\ \href
  {https://doi.org/10.3847/0004-637X/820/1/28} {\bibfield  {journal} {\bibinfo
  {journal} {Astrophys. J.}\ }\textbf {\bibinfo {volume} {820}}~(\bibinfo
  {number} {1}),\ \bibinfo {pages} {28}},\ \Eprint
  {https://arxiv.org/abs/1505.05155} {arXiv:1505.05155 [astro-ph.HE]}
  \BibitemShut {NoStop}%
\bibitem [{\citenamefont {Palenzuela}\ \emph {et~al.}(2014)\citenamefont
  {Palenzuela}, \citenamefont {Barausse}, \citenamefont {Ponce},\ and\
  \citenamefont {Lehner}}]{Palenzuela:2013hsa}%
  \BibitemOpen
  \bibfield  {author} {\bibinfo {author} {\bibnamefont {Palenzuela},
  \bibfnamefont {C.}}, \bibinfo {author} {\bibfnamefont {E.}~\bibnamefont
  {Barausse}}, \bibinfo {author} {\bibfnamefont {M.}~\bibnamefont {Ponce}},
  and\ \bibinfo {author} {\bibfnamefont {L.}~\bibnamefont {Lehner}}} (\bibinfo
  {year} {2014}),\ \href {https://doi.org/10.1103/PhysRevD.89.044024}
  {\bibfield  {journal} {\bibinfo  {journal} {Phys. Rev. D}\ }\textbf {\bibinfo
  {volume} {89}}~(\bibinfo {number} {4}),\ \bibinfo {pages} {044024}},\ \Eprint
  {https://arxiv.org/abs/1310.4481} {arXiv:1310.4481 [gr-qc]} \BibitemShut
  {NoStop}%
\bibitem [{\citenamefont {Palenzuela}\ and\ \citenamefont
  {Liebling}(2016)}]{Palenzuela:2015ima}%
  \BibitemOpen
  \bibfield  {author} {\bibinfo {author} {\bibnamefont {Palenzuela},
  \bibfnamefont {C.}}, and\ \bibinfo {author} {\bibfnamefont {S.~L.}\
  \bibnamefont {Liebling}}} (\bibinfo {year} {2016}),\ \href
  {https://doi.org/10.1103/PhysRevD.93.044009} {\bibfield  {journal} {\bibinfo
  {journal} {Phys. Rev. D}\ }\textbf {\bibinfo {volume} {93}}~(\bibinfo
  {number} {4}),\ \bibinfo {pages} {044009}},\ \Eprint
  {https://arxiv.org/abs/1510.03471} {arXiv:1510.03471 [gr-qc]} \BibitemShut
  {NoStop}%
\bibitem [{\citenamefont {Pani}\ and\ \citenamefont
  {Berti}(2014)}]{Pani:2014jra}%
  \BibitemOpen
  \bibfield  {author} {\bibinfo {author} {\bibnamefont {Pani}, \bibfnamefont
  {P.}}, and\ \bibinfo {author} {\bibfnamefont {E.}~\bibnamefont {Berti}}}
  (\bibinfo {year} {2014}),\ \href {https://doi.org/10.1103/PhysRevD.90.024025}
  {\bibfield  {journal} {\bibinfo  {journal} {Phys. Rev. D}\ }\textbf {\bibinfo
  {volume} {90}}~(\bibinfo {number} {2}),\ \bibinfo {pages} {024025}},\ \Eprint
  {https://arxiv.org/abs/1405.4547} {arXiv:1405.4547 [gr-qc]} \BibitemShut
  {NoStop}%
\bibitem [{\citenamefont {Pani}\ \emph
  {et~al.}(2011{\natexlab{a}})\citenamefont {Pani}, \citenamefont {Berti},
  \citenamefont {Cardoso},\ and\ \citenamefont {Read}}]{Pani:2011xm}%
  \BibitemOpen
  \bibfield  {author} {\bibinfo {author} {\bibnamefont {Pani}, \bibfnamefont
  {P.}}, \bibinfo {author} {\bibfnamefont {E.}~\bibnamefont {Berti}}, \bibinfo
  {author} {\bibfnamefont {V.}~\bibnamefont {Cardoso}}, and\ \bibinfo {author}
  {\bibfnamefont {J.}~\bibnamefont {Read}}} (\bibinfo {year}
  {2011}{\natexlab{a}}),\ \href {https://doi.org/10.1103/PhysRevD.84.104035}
  {\bibfield  {journal} {\bibinfo  {journal} {Phys. Rev. D}\ }\textbf {\bibinfo
  {volume} {84}},\ \bibinfo {pages} {104035}},\ \Eprint
  {https://arxiv.org/abs/1109.0928} {arXiv:1109.0928 [gr-qc]} \BibitemShut
  {NoStop}%
\bibitem [{\citenamefont {Pani}\ \emph
  {et~al.}(2011{\natexlab{b}})\citenamefont {Pani}, \citenamefont {Cardoso},
  \citenamefont {Berti}, \citenamefont {Read},\ and\ \citenamefont
  {Salgado}}]{Pani:2010vc}%
  \BibitemOpen
  \bibfield  {author} {\bibinfo {author} {\bibnamefont {Pani}, \bibfnamefont
  {P.}}, \bibinfo {author} {\bibfnamefont {V.}~\bibnamefont {Cardoso}},
  \bibinfo {author} {\bibfnamefont {E.}~\bibnamefont {Berti}}, \bibinfo
  {author} {\bibfnamefont {J.}~\bibnamefont {Read}}, and\ \bibinfo {author}
  {\bibfnamefont {M.}~\bibnamefont {Salgado}}} (\bibinfo {year}
  {2011}{\natexlab{b}}),\ \href {https://doi.org/10.1103/PhysRevD.83.081501}
  {\bibfield  {journal} {\bibinfo  {journal} {Phys. Rev. D}\ }\textbf {\bibinfo
  {volume} {83}},\ \bibinfo {pages} {081501}},\ \Eprint
  {https://arxiv.org/abs/1012.1343} {arXiv:1012.1343 [gr-qc]} \BibitemShut
  {NoStop}%
\bibitem [{\citenamefont {Papallo}(2017)}]{Papallo:2017ddx}%
  \BibitemOpen
  \bibfield  {author} {\bibinfo {author} {\bibnamefont {Papallo}, \bibfnamefont
  {G.}}} (\bibinfo {year} {2017}),\ \href
  {https://doi.org/10.1103/PhysRevD.96.124036} {\bibfield  {journal} {\bibinfo
  {journal} {Phys. Rev. D}\ }\textbf {\bibinfo {volume} {96}}~(\bibinfo
  {number} {12}),\ \bibinfo {pages} {124036}},\ \Eprint
  {https://arxiv.org/abs/1710.10155} {arXiv:1710.10155 [gr-qc]} \BibitemShut
  {NoStop}%
\bibitem [{\citenamefont {Pappas}\ and\ \citenamefont
  {Apostolatos}(2014)}]{Pappas:2013naa}%
  \BibitemOpen
  \bibfield  {author} {\bibinfo {author} {\bibnamefont {Pappas}, \bibfnamefont
  {G.}}, and\ \bibinfo {author} {\bibfnamefont {T.~A.}\ \bibnamefont
  {Apostolatos}}} (\bibinfo {year} {2014}),\ \href
  {https://doi.org/10.1103/PhysRevLett.112.121101} {\bibfield  {journal}
  {\bibinfo  {journal} {Phys. Rev. Lett.}\ }\textbf {\bibinfo {volume} {112}},\
  \bibinfo {pages} {121101}},\ \Eprint {https://arxiv.org/abs/1311.5508}
  {arXiv:1311.5508 [gr-qc]} \BibitemShut {NoStop}%
\bibitem [{\citenamefont {Pappas}\ \emph {et~al.}(2019)\citenamefont {Pappas},
  \citenamefont {Doneva}, \citenamefont {Sotiriou}, \citenamefont {Yazadjiev},\
  and\ \citenamefont {Kokkotas}}]{Pappas:2018csu}%
  \BibitemOpen
  \bibfield  {author} {\bibinfo {author} {\bibnamefont {Pappas}, \bibfnamefont
  {G.}}, \bibinfo {author} {\bibfnamefont {D.~D.}\ \bibnamefont {Doneva}},
  \bibinfo {author} {\bibfnamefont {T.~P.}\ \bibnamefont {Sotiriou}}, \bibinfo
  {author} {\bibfnamefont {S.~S.}\ \bibnamefont {Yazadjiev}}, and\ \bibinfo
  {author} {\bibfnamefont {K.~D.}\ \bibnamefont {Kokkotas}}} (\bibinfo {year}
  {2019}),\ \href {https://doi.org/10.1103/PhysRevD.99.104014} {\bibfield
  {journal} {\bibinfo  {journal} {Phys. Rev. D}\ }\textbf {\bibinfo {volume}
  {99}}~(\bibinfo {number} {10}),\ \bibinfo {pages} {104014}},\ \Eprint
  {https://arxiv.org/abs/1812.01117} {arXiv:1812.01117 [gr-qc]} \BibitemShut
  {NoStop}%
\bibitem [{\citenamefont {Pappas}\ and\ \citenamefont
  {Sotiriou}(2015)}]{Pappas:2014gca}%
  \BibitemOpen
  \bibfield  {author} {\bibinfo {author} {\bibnamefont {Pappas}, \bibfnamefont
  {G.}}, and\ \bibinfo {author} {\bibfnamefont {T.~P.}\ \bibnamefont
  {Sotiriou}}} (\bibinfo {year} {2015}),\ \href
  {https://doi.org/10.1103/PhysRevD.91.044011} {\bibfield  {journal} {\bibinfo
  {journal} {Phys. Rev. D}\ }\textbf {\bibinfo {volume} {91}}~(\bibinfo
  {number} {4}),\ \bibinfo {pages} {044011}},\ \Eprint
  {https://arxiv.org/abs/1412.3494} {arXiv:1412.3494 [gr-qc]} \BibitemShut
  {NoStop}%
\bibitem [{\citenamefont {Paschalidis}\ and\ \citenamefont
  {Stergioulas}(2017)}]{Paschalidis:2016vmz}%
  \BibitemOpen
  \bibfield  {author} {\bibinfo {author} {\bibnamefont {Paschalidis},
  \bibfnamefont {V.}}, and\ \bibinfo {author} {\bibfnamefont {N.}~\bibnamefont
  {Stergioulas}}} (\bibinfo {year} {2017}),\ \href
  {https://doi.org/10.1007/s41114-017-0008-x} {\bibfield  {journal} {\bibinfo
  {journal} {Living Rev. Rel.}\ }\textbf {\bibinfo {volume} {20}}~(\bibinfo
  {number} {1}),\ \bibinfo {pages} {7}},\ \Eprint
  {https://arxiv.org/abs/1612.03050} {arXiv:1612.03050 [astro-ph.HE]}
  \BibitemShut {NoStop}%
\bibitem [{\citenamefont {Podkowka}\ \emph {et~al.}(2018)\citenamefont
  {Podkowka}, \citenamefont {Mendes},\ and\ \citenamefont
  {Poisson}}]{Podkowka:2018gib}%
  \BibitemOpen
  \bibfield  {author} {\bibinfo {author} {\bibnamefont {Podkowka},
  \bibfnamefont {D.~M.}}, \bibinfo {author} {\bibfnamefont {R.~F.~P.}\
  \bibnamefont {Mendes}}, and\ \bibinfo {author} {\bibfnamefont
  {E.}~\bibnamefont {Poisson}}} (\bibinfo {year} {2018}),\ \href
  {https://doi.org/10.1103/PhysRevD.98.064057} {\bibfield  {journal} {\bibinfo
  {journal} {Phys. Rev. D}\ }\textbf {\bibinfo {volume} {98}}~(\bibinfo
  {number} {6}),\ \bibinfo {pages} {064057}},\ \Eprint
  {https://arxiv.org/abs/1807.01565} {arXiv:1807.01565 [gr-qc]} \BibitemShut
  {NoStop}%
\bibitem [{\citenamefont {Ponce}\ \emph {et~al.}(2015)\citenamefont {Ponce},
  \citenamefont {Palenzuela}, \citenamefont {Barausse},\ and\ \citenamefont
  {Lehner}}]{Ponce:2014hha}%
  \BibitemOpen
  \bibfield  {author} {\bibinfo {author} {\bibnamefont {Ponce}, \bibfnamefont
  {M.}}, \bibinfo {author} {\bibfnamefont {C.}~\bibnamefont {Palenzuela}},
  \bibinfo {author} {\bibfnamefont {E.}~\bibnamefont {Barausse}}, and\ \bibinfo
  {author} {\bibfnamefont {L.}~\bibnamefont {Lehner}}} (\bibinfo {year}
  {2015}),\ \href {https://doi.org/10.1103/PhysRevD.91.084038} {\bibfield
  {journal} {\bibinfo  {journal} {Phys. Rev. D}\ }\textbf {\bibinfo {volume}
  {91}}~(\bibinfo {number} {8}),\ \bibinfo {pages} {084038}},\ \Eprint
  {https://arxiv.org/abs/1410.0638} {arXiv:1410.0638 [gr-qc]} \BibitemShut
  {NoStop}%
\bibitem [{\citenamefont {Popchev}(2015)}]{Popchev2015}%
  \BibitemOpen
  \bibfield  {author} {\bibinfo {author} {\bibnamefont {Popchev}, \bibfnamefont
  {D.}}} (\bibinfo {year} {2015}),\ \emph {\bibinfo {title} {{Bifurcation of
  neutron star solutions in scalar-tensor theories of gravity}}},\ \href@noop
  {} {Master's thesis}\ (\bibinfo  {school} {University of Sofia})\BibitemShut
  {NoStop}%
\bibitem [{\citenamefont {Popchev}\ \emph {et~al.}(2019)\citenamefont
  {Popchev}, \citenamefont {Staykov}, \citenamefont {Doneva},\ and\
  \citenamefont {Yazadjiev}}]{Popchev:2018fwu}%
  \BibitemOpen
  \bibfield  {author} {\bibinfo {author} {\bibnamefont {Popchev}, \bibfnamefont
  {D.}}, \bibinfo {author} {\bibfnamefont {K.~V.}\ \bibnamefont {Staykov}},
  \bibinfo {author} {\bibfnamefont {D.~D.}\ \bibnamefont {Doneva}}, and\
  \bibinfo {author} {\bibfnamefont {S.~S.}\ \bibnamefont {Yazadjiev}}}
  (\bibinfo {year} {2019}),\ \href
  {https://doi.org/10.1140/epjc/s10052-019-6691-x} {\bibfield  {journal}
  {\bibinfo  {journal} {Eur. Phys. J. C}\ }\textbf {\bibinfo {volume}
  {79}}~(\bibinfo {number} {2}),\ \bibinfo {pages} {178}},\ \Eprint
  {https://arxiv.org/abs/1812.00347} {arXiv:1812.00347 [gr-qc]} \BibitemShut
  {NoStop}%
\bibitem [{\citenamefont {Pretorius}(2005)}]{Pretorius:2005gq}%
  \BibitemOpen
  \bibfield  {author} {\bibinfo {author} {\bibnamefont {Pretorius},
  \bibfnamefont {F.}}} (\bibinfo {year} {2005}),\ \href
  {https://doi.org/10.1103/PhysRevLett.95.121101} {\bibfield  {journal}
  {\bibinfo  {journal} {Phys. Rev. Lett.}\ }\textbf {\bibinfo {volume} {95}},\
  \bibinfo {pages} {121101}},\ \Eprint {https://arxiv.org/abs/gr-qc/0507014}
  {arXiv:gr-qc/0507014} \BibitemShut {NoStop}%
\bibitem [{\citenamefont {Proca}(1936)}]{Proca:1936fbw}%
  \BibitemOpen
  \bibfield  {author} {\bibinfo {author} {\bibnamefont {Proca}, \bibfnamefont
  {A.}}} (\bibinfo {year} {1936}),\ \href
  {https://doi.org/10.1051/jphysrad:0193600708034700} {\bibfield  {journal}
  {\bibinfo  {journal} {J. Phys. Radium}\ }\textbf {\bibinfo {volume} {7}},\
  \bibinfo {pages} {347}}\BibitemShut {NoStop}%
\bibitem [{\citenamefont {Ramazano\u{g}lu}(2017)}]{Ramazanoglu:2017xbl}%
  \BibitemOpen
  \bibfield  {author} {\bibinfo {author} {\bibnamefont {Ramazano\u{g}lu},
  \bibfnamefont {F.~M.}}} (\bibinfo {year} {2017}),\ \href
  {https://doi.org/10.1103/PhysRevD.96.064009} {\bibfield  {journal} {\bibinfo
  {journal} {Phys. Rev. D}\ }\textbf {\bibinfo {volume} {96}}~(\bibinfo
  {number} {6}),\ \bibinfo {pages} {064009}},\ \Eprint
  {https://arxiv.org/abs/1706.01056} {arXiv:1706.01056 [gr-qc]} \BibitemShut
  {NoStop}%
\bibitem [{\citenamefont
  {Ramazano\u{g}lu}(2018{\natexlab{a}})}]{Ramazanoglu:2017yun}%
  \BibitemOpen
  \bibfield  {author} {\bibinfo {author} {\bibnamefont {Ramazano\u{g}lu},
  \bibfnamefont {F.~M.}}} (\bibinfo {year} {2018}{\natexlab{a}}),\ \href
  {https://doi.org/10.1103/PhysRevD.97.024008} {\bibfield  {journal} {\bibinfo
  {journal} {Phys. Rev. D}\ }\textbf {\bibinfo {volume} {97}}~(\bibinfo
  {number} {2}),\ \bibinfo {pages} {024008}},\ \Eprint
  {https://arxiv.org/abs/1710.00863} {arXiv:1710.00863 [gr-qc]} \BibitemShut
  {NoStop}%
\bibitem [{\citenamefont
  {Ramazano\u{g}lu}(2018{\natexlab{b}})}]{Ramazanoglu:2018tig}%
  \BibitemOpen
  \bibfield  {author} {\bibinfo {author} {\bibnamefont {Ramazano\u{g}lu},
  \bibfnamefont {F.~M.}}} (\bibinfo {year} {2018}{\natexlab{b}}),\ \href
  {https://doi.org/10.1103/PhysRevD.98.044013} {\bibfield  {journal} {\bibinfo
  {journal} {Phys. Rev. D}\ }\textbf {\bibinfo {volume} {98}}~(\bibinfo
  {number} {4}),\ \bibinfo {pages} {044013}},\ \Eprint
  {https://arxiv.org/abs/1804.03158} {arXiv:1804.03158 [gr-qc]} \BibitemShut
  {NoStop}%
\bibitem [{\citenamefont
  {Ramazano\u{g}lu}(2018{\natexlab{c}})}]{Ramazanoglu:2018hwk}%
  \BibitemOpen
  \bibfield  {author} {\bibinfo {author} {\bibnamefont {Ramazano\u{g}lu},
  \bibfnamefont {F.~M.}}} (\bibinfo {year} {2018}{\natexlab{c}}),\ \href
  {https://doi.org/10.1103/PhysRevD.98.044011} {\bibfield  {journal} {\bibinfo
  {journal} {Phys. Rev. D}\ }\textbf {\bibinfo {volume} {98}}~(\bibinfo
  {number} {4}),\ \bibinfo {pages} {044011}},\ \Eprint
  {https://arxiv.org/abs/1804.00594} {arXiv:1804.00594 [gr-qc]} \BibitemShut
  {NoStop}%
\bibitem [{\citenamefont
  {Ramazano\u{g}lu}(2019{\natexlab{a}})}]{Ramazanoglu:2019tyi}%
  \BibitemOpen
  \bibfield  {author} {\bibinfo {author} {\bibnamefont {Ramazano\u{g}lu},
  \bibfnamefont {F.~M.}}} (\bibinfo {year} {2019}{\natexlab{a}}),\ \href
  {https://doi.org/10.1103/PhysRevD.99.044003} {\bibfield  {journal} {\bibinfo
  {journal} {Phys. Rev. D}\ }\textbf {\bibinfo {volume} {99}}~(\bibinfo
  {number} {4}),\ \bibinfo {pages} {044003}},\ \Eprint
  {https://arxiv.org/abs/1901.00194} {arXiv:1901.00194 [gr-qc]} \BibitemShut
  {NoStop}%
\bibitem [{\citenamefont
  {Ramazano\u{g}lu}(2019{\natexlab{b}})}]{Ramazanoglu:2019gbz}%
  \BibitemOpen
  \bibfield  {author} {\bibinfo {author} {\bibnamefont {Ramazano\u{g}lu},
  \bibfnamefont {F.~M.}}} (\bibinfo {year} {2019}{\natexlab{b}}),\ \href
  {https://doi.org/10.1103/PhysRevD.99.084015} {\bibfield  {journal} {\bibinfo
  {journal} {Phys. Rev. D}\ }\textbf {\bibinfo {volume} {99}}~(\bibinfo
  {number} {8}),\ \bibinfo {pages} {084015}},\ \Eprint
  {https://arxiv.org/abs/1901.10009} {arXiv:1901.10009 [gr-qc]} \BibitemShut
  {NoStop}%
\bibitem [{\citenamefont
  {Ramazano\u{g}lu}(2019{\natexlab{c}})}]{Ramazanoglu:2019jfy}%
  \BibitemOpen
  \bibfield  {author} {\bibinfo {author} {\bibnamefont {Ramazano\u{g}lu},
  \bibfnamefont {F.~M.}}} (\bibinfo {year} {2019}{\natexlab{c}}),\ \href
  {https://doi.org/10.3906/fiz-1908-8} {\bibfield  {journal} {\bibinfo
  {journal} {Turk. J. Phys.}\ }\textbf {\bibinfo {volume} {43}}~(\bibinfo
  {number} {6}),\ \bibinfo {pages} {586}}\BibitemShut {NoStop}%
\bibitem [{\citenamefont {Ramazano\u{g}lu}\ and\ \citenamefont
  {Pretorius}(2016)}]{Ramazanoglu:2016kul}%
  \BibitemOpen
  \bibfield  {author} {\bibinfo {author} {\bibnamefont {Ramazano\u{g}lu},
  \bibfnamefont {F.~M.}}, and\ \bibinfo {author} {\bibfnamefont
  {F.}~\bibnamefont {Pretorius}}} (\bibinfo {year} {2016}),\ \href
  {https://doi.org/10.1103/PhysRevD.93.064005} {\bibfield  {journal} {\bibinfo
  {journal} {Phys. Rev. D}\ }\textbf {\bibinfo {volume} {93}}~(\bibinfo
  {number} {6}),\ \bibinfo {pages} {064005}},\ \Eprint
  {https://arxiv.org/abs/1601.07475} {arXiv:1601.07475 [gr-qc]} \BibitemShut
  {NoStop}%
\bibitem [{\citenamefont {Ramazano\u{g}lu}\ and\ \citenamefont
  {\"Unl\"ut\"urk}(2019)}]{Ramazanoglu:2019jrr}%
  \BibitemOpen
  \bibfield  {author} {\bibinfo {author} {\bibnamefont {Ramazano\u{g}lu},
  \bibfnamefont {F.~M.}}, and\ \bibinfo {author} {\bibfnamefont {K.~I.}\
  \bibnamefont {\"Unl\"ut\"urk}}} (\bibinfo {year} {2019}),\ \href
  {https://doi.org/10.1103/PhysRevD.100.084026} {\bibfield  {journal} {\bibinfo
   {journal} {Phys. Rev. D}\ }\textbf {\bibinfo {volume} {100}}~(\bibinfo
  {number} {8}),\ \bibinfo {pages} {084026}},\ \Eprint
  {https://arxiv.org/abs/1910.02801} {arXiv:1910.02801 [gr-qc]} \BibitemShut
  {NoStop}%
\bibitem [{\citenamefont {Regge}\ and\ \citenamefont
  {Wheeler}(1957)}]{Regge:1957td}%
  \BibitemOpen
  \bibfield  {author} {\bibinfo {author} {\bibnamefont {Regge}, \bibfnamefont
  {T.}}, and\ \bibinfo {author} {\bibfnamefont {J.~A.}\ \bibnamefont
  {Wheeler}}} (\bibinfo {year} {1957}),\ \href
  {https://doi.org/10.1103/PhysRev.108.1063} {\bibfield  {journal} {\bibinfo
  {journal} {Phys. Rev.}\ }\textbf {\bibinfo {volume} {108}},\ \bibinfo {pages}
  {1063}}\BibitemShut {NoStop}%
\bibitem [{\citenamefont {Rezzolla}\ and\ \citenamefont
  {Takami}(2016)}]{Rezzolla:2016nxn}%
  \BibitemOpen
  \bibfield  {author} {\bibinfo {author} {\bibnamefont {Rezzolla},
  \bibfnamefont {L.}}, and\ \bibinfo {author} {\bibfnamefont {K.}~\bibnamefont
  {Takami}}} (\bibinfo {year} {2016}),\ \href
  {https://doi.org/10.1103/PhysRevD.93.124051} {\bibfield  {journal} {\bibinfo
  {journal} {Phys. Rev. D}\ }\textbf {\bibinfo {volume} {93}}~(\bibinfo
  {number} {12}),\ \bibinfo {pages} {124051}},\ \Eprint
  {https://arxiv.org/abs/1604.00246} {arXiv:1604.00246 [gr-qc]} \BibitemShut
  {NoStop}%
\bibitem [{\citenamefont {de~Rham}(2014)}]{deRham:2014zqa}%
  \BibitemOpen
  \bibfield  {author} {\bibinfo {author} {\bibnamefont {de~Rham}, \bibfnamefont
  {C.}}} (\bibinfo {year} {2014}),\ \href {https://doi.org/10.12942/lrr-2014-7}
  {\bibfield  {journal} {\bibinfo  {journal} {Living Rev. Rel.}\ }\textbf
  {\bibinfo {volume} {17}},\ \bibinfo {pages} {7}},\ \Eprint
  {https://arxiv.org/abs/1401.4173} {arXiv:1401.4173 [hep-th]} \BibitemShut
  {NoStop}%
\bibitem [{\citenamefont {de~Rham}\ \emph {et~al.}(2011)\citenamefont
  {de~Rham}, \citenamefont {Gabadadze},\ and\ \citenamefont
  {Tolley}}]{deRham:2010kj}%
  \BibitemOpen
  \bibfield  {author} {\bibinfo {author} {\bibnamefont {de~Rham}, \bibfnamefont
  {C.}}, \bibinfo {author} {\bibfnamefont {G.}~\bibnamefont {Gabadadze}}, and\
  \bibinfo {author} {\bibfnamefont {A.~J.}\ \bibnamefont {Tolley}}} (\bibinfo
  {year} {2011}),\ \href {https://doi.org/10.1103/PhysRevLett.106.231101}
  {\bibfield  {journal} {\bibinfo  {journal} {Phys. Rev. Lett.}\ }\textbf
  {\bibinfo {volume} {106}},\ \bibinfo {pages} {231101}},\ \Eprint
  {https://arxiv.org/abs/1011.1232} {arXiv:1011.1232 [hep-th]} \BibitemShut
  {NoStop}%
\bibitem [{\citenamefont {Rhoades}\ and\ \citenamefont
  {Ruffini}(1974)}]{Rhoades:1974fn}%
  \BibitemOpen
  \bibfield  {author} {\bibinfo {author} {\bibnamefont {Rhoades}, \bibfnamefont
  {C.~E., Jr.}}, and\ \bibinfo {author} {\bibfnamefont {R.}~\bibnamefont
  {Ruffini}}} (\bibinfo {year} {1974}),\ \href
  {https://doi.org/10.1103/PhysRevLett.32.324} {\bibfield  {journal} {\bibinfo
  {journal} {Phys. Rev. Lett.}\ }\textbf {\bibinfo {volume} {32}},\ \bibinfo
  {pages} {324}}\BibitemShut {NoStop}%
\bibitem [{\citenamefont {Ribeiro}\ and\ \citenamefont
  {Vanzella}(2020)}]{Ribeiro:2019yzv}%
  \BibitemOpen
  \bibfield  {author} {\bibinfo {author} {\bibnamefont {Ribeiro}, \bibfnamefont
  {C.~C.~H.}}, and\ \bibinfo {author} {\bibfnamefont {D.~A.~T.}\ \bibnamefont
  {Vanzella}}} (\bibinfo {year} {2020}),\ \href
  {https://doi.org/10.1103/PhysRevResearch.2.013281} {\bibfield  {journal}
  {\bibinfo  {journal} {Phys. Rev. Res.}\ }\textbf {\bibinfo {volume}
  {2}}~(\bibinfo {number} {1}),\ \bibinfo {pages} {013281}},\ \Eprint
  {https://arxiv.org/abs/1912.01971} {arXiv:1912.01971 [gr-qc]} \BibitemShut
  {NoStop}%
\bibitem [{\citenamefont {Riley}\ \emph {et~al.}(2019)\citenamefont {Riley}
  \emph {et~al.}}]{Riley:2019yda}%
  \BibitemOpen
  \bibfield  {author} {\bibinfo {author} {\bibnamefont {Riley}, \bibfnamefont
  {T.~E.}},  \emph {et~al.}} (\bibinfo {year} {2019}),\ \href
  {https://doi.org/10.3847/2041-8213/ab481c} {\bibfield  {journal} {\bibinfo
  {journal} {Astrophys. J. Lett.}\ }\textbf {\bibinfo {volume} {887}}~(\bibinfo
  {number} {1}),\ \bibinfo {pages} {L21}},\ \Eprint
  {https://arxiv.org/abs/1912.05702} {arXiv:1912.05702 [astro-ph.HE]}
  \BibitemShut {NoStop}%
\bibitem [{\citenamefont {Ripley}(2022)}]{Ripley:2022cdh}%
  \BibitemOpen
  \bibfield  {author} {\bibinfo {author} {\bibnamefont {Ripley}, \bibfnamefont
  {J.~L.}}} (\bibinfo {year} {2022}),\ \href
  {https://doi.org/10.1142/S0218271822300178} {\bibfield  {journal} {\bibinfo
  {journal} {Int. J. Mod. Phys. D}\ }\textbf {\bibinfo {volume} {31}}~(\bibinfo
  {number} {13}),\ \bibinfo {pages} {2230017}},\ \Eprint
  {https://arxiv.org/abs/2207.13074} {arXiv:2207.13074 [gr-qc]} \BibitemShut
  {NoStop}%
\bibitem [{\citenamefont {Ripley}\ and\ \citenamefont
  {Pretorius}(2020)}]{Ripley:2020vpk}%
  \BibitemOpen
  \bibfield  {author} {\bibinfo {author} {\bibnamefont {Ripley}, \bibfnamefont
  {J.~L.}}, and\ \bibinfo {author} {\bibfnamefont {F.}~\bibnamefont
  {Pretorius}}} (\bibinfo {year} {2020}),\ \href
  {https://doi.org/10.1088/1361-6382/ab9bbb} {\bibfield  {journal} {\bibinfo
  {journal} {Class. Quant. Grav.}\ }\textbf {\bibinfo {volume} {37}}~(\bibinfo
  {number} {15}),\ \bibinfo {pages} {155003}},\ \Eprint
  {https://arxiv.org/abs/2005.05417} {arXiv:2005.05417 [gr-qc]} \BibitemShut
  {NoStop}%
\bibitem [{\citenamefont {Rosca-Mead}\ \emph {et~al.}(2019)\citenamefont
  {Rosca-Mead}, \citenamefont {Moore}, \citenamefont {Agathos},\ and\
  \citenamefont {Sperhake}}]{Rosca-Mead:2019seq}%
  \BibitemOpen
  \bibfield  {author} {\bibinfo {author} {\bibnamefont {Rosca-Mead},
  \bibfnamefont {R.}}, \bibinfo {author} {\bibfnamefont {C.~J.}\ \bibnamefont
  {Moore}}, \bibinfo {author} {\bibfnamefont {M.}~\bibnamefont {Agathos}}, and\
  \bibinfo {author} {\bibfnamefont {U.}~\bibnamefont {Sperhake}}} (\bibinfo
  {year} {2019}),\ \href {https://doi.org/10.1088/1361-6382/ab256f} {\bibfield
  {journal} {\bibinfo  {journal} {Class. Quant. Grav.}\ }\textbf {\bibinfo
  {volume} {36}}~(\bibinfo {number} {13}),\ \bibinfo {pages} {134003}},\
  \Eprint {https://arxiv.org/abs/1903.09704} {arXiv:1903.09704 [gr-qc]}
  \BibitemShut {NoStop}%
\bibitem [{\citenamefont {Rosca-Mead}\ \emph
  {et~al.}(2020{\natexlab{a}})\citenamefont {Rosca-Mead}, \citenamefont
  {Moore}, \citenamefont {Sperhake}, \citenamefont {Agathos},\ and\
  \citenamefont {Gerosa}}]{Rosca-Mead:2020bzt}%
  \BibitemOpen
  \bibfield  {author} {\bibinfo {author} {\bibnamefont {Rosca-Mead},
  \bibfnamefont {R.}}, \bibinfo {author} {\bibfnamefont {C.~J.}\ \bibnamefont
  {Moore}}, \bibinfo {author} {\bibfnamefont {U.}~\bibnamefont {Sperhake}},
  \bibinfo {author} {\bibfnamefont {M.}~\bibnamefont {Agathos}}, and\ \bibinfo
  {author} {\bibfnamefont {D.}~\bibnamefont {Gerosa}}} (\bibinfo {year}
  {2020}{\natexlab{a}}),\ \href {https://doi.org/10.3390/sym12091384}
  {\bibfield  {journal} {\bibinfo  {journal} {Symmetry}\ }\textbf {\bibinfo
  {volume} {12}}~(\bibinfo {number} {9}),\ \bibinfo {pages} {1384}},\ \Eprint
  {https://arxiv.org/abs/2007.14429} {arXiv:2007.14429 [gr-qc]} \BibitemShut
  {NoStop}%
\bibitem [{\citenamefont {Rosca-Mead}\ \emph
  {et~al.}(2020{\natexlab{b}})\citenamefont {Rosca-Mead}, \citenamefont
  {Sperhake}, \citenamefont {Moore}, \citenamefont {Agathos}, \citenamefont
  {Gerosa},\ and\ \citenamefont {Ott}}]{Rosca-Mead:2020ehn}%
  \BibitemOpen
  \bibfield  {author} {\bibinfo {author} {\bibnamefont {Rosca-Mead},
  \bibfnamefont {R.}}, \bibinfo {author} {\bibfnamefont {U.}~\bibnamefont
  {Sperhake}}, \bibinfo {author} {\bibfnamefont {C.~J.}\ \bibnamefont {Moore}},
  \bibinfo {author} {\bibfnamefont {M.}~\bibnamefont {Agathos}}, \bibinfo
  {author} {\bibfnamefont {D.}~\bibnamefont {Gerosa}}, and\ \bibinfo {author}
  {\bibfnamefont {C.~D.}\ \bibnamefont {Ott}}} (\bibinfo {year}
  {2020}{\natexlab{b}}),\ \href {https://doi.org/10.1103/PhysRevD.102.044010}
  {\bibfield  {journal} {\bibinfo  {journal} {Phys. Rev. D}\ }\textbf {\bibinfo
  {volume} {102}}~(\bibinfo {number} {4}),\ \bibinfo {pages} {044010}},\
  \Eprint {https://arxiv.org/abs/2005.09728} {arXiv:2005.09728 [gr-qc]}
  \BibitemShut {NoStop}%
\bibitem [{\citenamefont {Ruegg}\ and\ \citenamefont
  {Ruiz-Altaba}(2004)}]{Ruegg:2003ps}%
  \BibitemOpen
  \bibfield  {author} {\bibinfo {author} {\bibnamefont {Ruegg}, \bibfnamefont
  {H.}}, and\ \bibinfo {author} {\bibfnamefont {M.}~\bibnamefont
  {Ruiz-Altaba}}} (\bibinfo {year} {2004}),\ \href
  {https://doi.org/10.1142/S0217751X04019755} {\bibfield  {journal} {\bibinfo
  {journal} {Int. J. Mod. Phys. A}\ }\textbf {\bibinfo {volume} {19}},\
  \bibinfo {pages} {3265}},\ \Eprint {https://arxiv.org/abs/hep-th/0304245}
  {arXiv:hep-th/0304245} \BibitemShut {NoStop}%
\bibitem [{\citenamefont {Ruiz}\ \emph {et~al.}(2012)\citenamefont {Ruiz},
  \citenamefont {Degollado}, \citenamefont {Alcubierre}, \citenamefont
  {Nunez},\ and\ \citenamefont {Salgado}}]{Ruiz:2012jt}%
  \BibitemOpen
  \bibfield  {author} {\bibinfo {author} {\bibnamefont {Ruiz}, \bibfnamefont
  {M.}}, \bibinfo {author} {\bibfnamefont {J.~C.}\ \bibnamefont {Degollado}},
  \bibinfo {author} {\bibfnamefont {M.}~\bibnamefont {Alcubierre}}, \bibinfo
  {author} {\bibfnamefont {D.}~\bibnamefont {Nunez}}, and\ \bibinfo {author}
  {\bibfnamefont {M.}~\bibnamefont {Salgado}}} (\bibinfo {year} {2012}),\ \href
  {https://doi.org/10.1103/PhysRevD.86.104044} {\bibfield  {journal} {\bibinfo
  {journal} {Phys. Rev. D}\ }\textbf {\bibinfo {volume} {86}},\ \bibinfo
  {pages} {104044}},\ \Eprint {https://arxiv.org/abs/1207.6142}
  {arXiv:1207.6142 [gr-qc]} \BibitemShut {NoStop}%
\bibitem [{\citenamefont {Saffer}\ \emph {et~al.}(2019)\citenamefont {Saffer},
  \citenamefont {Silva},\ and\ \citenamefont {Yunes}}]{Saffer:2019hqn}%
  \BibitemOpen
  \bibfield  {author} {\bibinfo {author} {\bibnamefont {Saffer}, \bibfnamefont
  {A.}}, \bibinfo {author} {\bibfnamefont {H.~O.}\ \bibnamefont {Silva}}, and\
  \bibinfo {author} {\bibfnamefont {N.}~\bibnamefont {Yunes}}} (\bibinfo {year}
  {2019}),\ \href {https://doi.org/10.1103/PhysRevD.100.044030} {\bibfield
  {journal} {\bibinfo  {journal} {Phys. Rev. D}\ }\textbf {\bibinfo {volume}
  {100}}~(\bibinfo {number} {4}),\ \bibinfo {pages} {044030}},\ \Eprint
  {https://arxiv.org/abs/1903.07779} {arXiv:1903.07779 [gr-qc]} \BibitemShut
  {NoStop}%
\bibitem [{\citenamefont {Salgado}(2006)}]{Salgado:2005hx}%
  \BibitemOpen
  \bibfield  {author} {\bibinfo {author} {\bibnamefont {Salgado}, \bibfnamefont
  {M.}}} (\bibinfo {year} {2006}),\ \href
  {https://doi.org/10.1088/0264-9381/23/14/010} {\bibfield  {journal} {\bibinfo
   {journal} {Class. Quant. Grav.}\ }\textbf {\bibinfo {volume} {23}},\
  \bibinfo {pages} {4719}},\ \Eprint {https://arxiv.org/abs/gr-qc/0509001}
  {arXiv:gr-qc/0509001} \BibitemShut {NoStop}%
\bibitem [{\citenamefont {Salgado}\ \emph {et~al.}(2008)\citenamefont
  {Salgado}, \citenamefont {Martinez-del Rio}, \citenamefont {Alcubierre},\
  and\ \citenamefont {Nunez}}]{Salgado:2008xh}%
  \BibitemOpen
  \bibfield  {author} {\bibinfo {author} {\bibnamefont {Salgado}, \bibfnamefont
  {M.}}, \bibinfo {author} {\bibfnamefont {D.}~\bibnamefont {Martinez-del
  Rio}}, \bibinfo {author} {\bibfnamefont {M.}~\bibnamefont {Alcubierre}}, and\
  \bibinfo {author} {\bibfnamefont {D.}~\bibnamefont {Nunez}}} (\bibinfo {year}
  {2008}),\ \href {https://doi.org/10.1103/PhysRevD.77.104010} {\bibfield
  {journal} {\bibinfo  {journal} {Phys. Rev. D}\ }\textbf {\bibinfo {volume}
  {77}},\ \bibinfo {pages} {104010}},\ \Eprint
  {https://arxiv.org/abs/0801.2372} {arXiv:0801.2372 [gr-qc]} \BibitemShut
  {NoStop}%
\bibitem [{\citenamefont {Salgado}\ \emph {et~al.}(1998)\citenamefont
  {Salgado}, \citenamefont {Sudarsky},\ and\ \citenamefont
  {Nucamendi}}]{Salgado:1998sg}%
  \BibitemOpen
  \bibfield  {author} {\bibinfo {author} {\bibnamefont {Salgado}, \bibfnamefont
  {M.}}, \bibinfo {author} {\bibfnamefont {D.}~\bibnamefont {Sudarsky}}, and\
  \bibinfo {author} {\bibfnamefont {U.}~\bibnamefont {Nucamendi}}} (\bibinfo
  {year} {1998}),\ \href {https://doi.org/10.1103/PhysRevD.58.124003}
  {\bibfield  {journal} {\bibinfo  {journal} {Phys. Rev. D}\ }\textbf {\bibinfo
  {volume} {58}},\ \bibinfo {pages} {124003}},\ \Eprint
  {https://arxiv.org/abs/gr-qc/9806070} {arXiv:gr-qc/9806070} \BibitemShut
  {NoStop}%
\bibitem [{\citenamefont {Salgado}\ \emph {et~al.}(2004)\citenamefont
  {Salgado}, \citenamefont {Sudarsky},\ and\ \citenamefont
  {Nucamendi}}]{Salgado:2004tg}%
  \BibitemOpen
  \bibfield  {author} {\bibinfo {author} {\bibnamefont {Salgado}, \bibfnamefont
  {M.}}, \bibinfo {author} {\bibfnamefont {D.}~\bibnamefont {Sudarsky}}, and\
  \bibinfo {author} {\bibfnamefont {U.}~\bibnamefont {Nucamendi}}} (\bibinfo
  {year} {2004}),\ \href {https://doi.org/10.1103/PhysRevD.70.084027}
  {\bibfield  {journal} {\bibinfo  {journal} {Phys. Rev. D}\ }\textbf {\bibinfo
  {volume} {70}},\ \bibinfo {pages} {084027}},\ \Eprint
  {https://arxiv.org/abs/gr-qc/0402126} {arXiv:gr-qc/0402126} \BibitemShut
  {NoStop}%
\bibitem [{\citenamefont {Sampson}\ \emph {et~al.}(2014)\citenamefont
  {Sampson}, \citenamefont {Yunes}, \citenamefont {Cornish}, \citenamefont
  {Ponce}, \citenamefont {Barausse}, \citenamefont {Klein}, \citenamefont
  {Palenzuela},\ and\ \citenamefont {Lehner}}]{Sampson:2014qqa}%
  \BibitemOpen
  \bibfield  {author} {\bibinfo {author} {\bibnamefont {Sampson}, \bibfnamefont
  {L.}}, \bibinfo {author} {\bibfnamefont {N.}~\bibnamefont {Yunes}}, \bibinfo
  {author} {\bibfnamefont {N.}~\bibnamefont {Cornish}}, \bibinfo {author}
  {\bibfnamefont {M.}~\bibnamefont {Ponce}}, \bibinfo {author} {\bibfnamefont
  {E.}~\bibnamefont {Barausse}}, \bibinfo {author} {\bibfnamefont
  {A.}~\bibnamefont {Klein}}, \bibinfo {author} {\bibfnamefont
  {C.}~\bibnamefont {Palenzuela}}, and\ \bibinfo {author} {\bibfnamefont
  {L.}~\bibnamefont {Lehner}}} (\bibinfo {year} {2014}),\ \href
  {https://doi.org/10.1103/PhysRevD.90.124091} {\bibfield  {journal} {\bibinfo
  {journal} {Phys. Rev. D}\ }\textbf {\bibinfo {volume} {90}}~(\bibinfo
  {number} {12}),\ \bibinfo {pages} {124091}},\ \Eprint
  {https://arxiv.org/abs/1407.7038} {arXiv:1407.7038 [gr-qc]} \BibitemShut
  {NoStop}%
\bibitem [{\citenamefont {Sanchis-Gual}\ \emph
  {et~al.}(2016{\natexlab{a}})\citenamefont {Sanchis-Gual}, \citenamefont
  {Degollado}, \citenamefont {Herdeiro}, \citenamefont {Font},\ and\
  \citenamefont {Montero}}]{Sanchis-Gual:2016tcm}%
  \BibitemOpen
  \bibfield  {author} {\bibinfo {author} {\bibnamefont {Sanchis-Gual},
  \bibfnamefont {N.}}, \bibinfo {author} {\bibfnamefont {J.~C.}\ \bibnamefont
  {Degollado}}, \bibinfo {author} {\bibfnamefont {C.}~\bibnamefont {Herdeiro}},
  \bibinfo {author} {\bibfnamefont {J.~A.}\ \bibnamefont {Font}}, and\ \bibinfo
  {author} {\bibfnamefont {P.~J.}\ \bibnamefont {Montero}}} (\bibinfo {year}
  {2016}{\natexlab{a}}),\ \href {https://doi.org/10.1103/PhysRevD.94.044061}
  {\bibfield  {journal} {\bibinfo  {journal} {Phys. Rev. D}\ }\textbf {\bibinfo
  {volume} {94}}~(\bibinfo {number} {4}),\ \bibinfo {pages} {044061}},\ \Eprint
  {https://arxiv.org/abs/1607.06304} {arXiv:1607.06304 [gr-qc]} \BibitemShut
  {NoStop}%
\bibitem [{\citenamefont {Sanchis-Gual}\ \emph
  {et~al.}(2016{\natexlab{b}})\citenamefont {Sanchis-Gual}, \citenamefont
  {Degollado}, \citenamefont {Montero}, \citenamefont {Font},\ and\
  \citenamefont {Herdeiro}}]{Sanchis-Gual:2015lje}%
  \BibitemOpen
  \bibfield  {author} {\bibinfo {author} {\bibnamefont {Sanchis-Gual},
  \bibfnamefont {N.}}, \bibinfo {author} {\bibfnamefont {J.~C.}\ \bibnamefont
  {Degollado}}, \bibinfo {author} {\bibfnamefont {P.~J.}\ \bibnamefont
  {Montero}}, \bibinfo {author} {\bibfnamefont {J.~A.}\ \bibnamefont {Font}},
  and\ \bibinfo {author} {\bibfnamefont {C.}~\bibnamefont {Herdeiro}}}
  (\bibinfo {year} {2016}{\natexlab{b}}),\ \href
  {https://doi.org/10.1103/PhysRevLett.116.141101} {\bibfield  {journal}
  {\bibinfo  {journal} {Phys. Rev. Lett.}\ }\textbf {\bibinfo {volume}
  {116}}~(\bibinfo {number} {14}),\ \bibinfo {pages} {141101}},\ \Eprint
  {https://arxiv.org/abs/1512.05358} {arXiv:1512.05358 [gr-qc]} \BibitemShut
  {NoStop}%
\bibitem [{\citenamefont {Santiago}\ \emph {et~al.}(2016)\citenamefont
  {Santiago}, \citenamefont {Landulfo}, \citenamefont {Lima}, \citenamefont
  {Matsas}, \citenamefont {Mendes},\ and\ \citenamefont
  {Vanzella}}]{Santiago:2015bve}%
  \BibitemOpen
  \bibfield  {author} {\bibinfo {author} {\bibnamefont {Santiago},
  \bibfnamefont {J.}}, \bibinfo {author} {\bibfnamefont {A.~G.~S.}\
  \bibnamefont {Landulfo}}, \bibinfo {author} {\bibfnamefont {W.~C.~C.}\
  \bibnamefont {Lima}}, \bibinfo {author} {\bibfnamefont {G.~E.~A.}\
  \bibnamefont {Matsas}}, \bibinfo {author} {\bibfnamefont {R.~F.~P.}\
  \bibnamefont {Mendes}}, and\ \bibinfo {author} {\bibfnamefont {D.~A.~T.}\
  \bibnamefont {Vanzella}}} (\bibinfo {year} {2016}),\ \href
  {https://doi.org/10.1103/PhysRevD.93.024043} {\bibfield  {journal} {\bibinfo
  {journal} {Phys. Rev. D}\ }\textbf {\bibinfo {volume} {93}}~(\bibinfo
  {number} {2}),\ \bibinfo {pages} {024043}},\ \Eprint
  {https://arxiv.org/abs/1512.02120} {arXiv:1512.02120 [gr-qc]} \BibitemShut
  {NoStop}%
\bibitem [{\citenamefont {Sathyaprakash}\ \emph {et~al.}(2019)\citenamefont
  {Sathyaprakash} \emph {et~al.}}]{Sathyaprakash:2019yqt}%
  \BibitemOpen
  \bibfield  {author} {\bibinfo {author} {\bibnamefont {Sathyaprakash},
  \bibfnamefont {B.~S.}},  \emph {et~al.}} (\bibinfo {year} {2019}),\
  \href@noop {} {\enquote {\bibinfo {title} {{Extreme Gravity and Fundamental
  Physics}},}\ }\Eprint {https://arxiv.org/abs/1903.09221} {arXiv:1903.09221
  [astro-ph.HE]} \BibitemShut {NoStop}%
\bibitem [{\citenamefont {Scheel}\ \emph
  {et~al.}(1995{\natexlab{a}})\citenamefont {Scheel}, \citenamefont {Shapiro},\
  and\ \citenamefont {Teukolsky}}]{Scheel:1994yr}%
  \BibitemOpen
  \bibfield  {author} {\bibinfo {author} {\bibnamefont {Scheel}, \bibfnamefont
  {M.~A.}}, \bibinfo {author} {\bibfnamefont {S.~L.}\ \bibnamefont {Shapiro}},
  and\ \bibinfo {author} {\bibfnamefont {S.~A.}\ \bibnamefont {Teukolsky}}}
  (\bibinfo {year} {1995}{\natexlab{a}}),\ \href
  {https://doi.org/10.1103/PhysRevD.51.4208} {\bibfield  {journal} {\bibinfo
  {journal} {Phys. Rev. D}\ }\textbf {\bibinfo {volume} {51}},\ \bibinfo
  {pages} {4208}},\ \Eprint {https://arxiv.org/abs/gr-qc/9411025}
  {arXiv:gr-qc/9411025} \BibitemShut {NoStop}%
\bibitem [{\citenamefont {Scheel}\ \emph
  {et~al.}(1995{\natexlab{b}})\citenamefont {Scheel}, \citenamefont {Shapiro},\
  and\ \citenamefont {Teukolsky}}]{Scheel:1994yn}%
  \BibitemOpen
  \bibfield  {author} {\bibinfo {author} {\bibnamefont {Scheel}, \bibfnamefont
  {M.~A.}}, \bibinfo {author} {\bibfnamefont {S.~L.}\ \bibnamefont {Shapiro}},
  and\ \bibinfo {author} {\bibfnamefont {S.~A.}\ \bibnamefont {Teukolsky}}}
  (\bibinfo {year} {1995}{\natexlab{b}}),\ \href
  {https://doi.org/10.1103/PhysRevD.51.4236} {\bibfield  {journal} {\bibinfo
  {journal} {Phys. Rev. D}\ }\textbf {\bibinfo {volume} {51}},\ \bibinfo
  {pages} {4236}},\ \Eprint {https://arxiv.org/abs/gr-qc/9411026}
  {arXiv:gr-qc/9411026} \BibitemShut {NoStop}%
\bibitem [{\citenamefont {Sen}\ and\ \citenamefont
  {Banerjee}(2001)}]{Sen:1998ftn}%
  \BibitemOpen
  \bibfield  {author} {\bibinfo {author} {\bibnamefont {Sen}, \bibfnamefont
  {S.}}, and\ \bibinfo {author} {\bibfnamefont {N.}~\bibnamefont {Banerjee}}}
  (\bibinfo {year} {2001}),\ \href {https://doi.org/10.1007/s12043-001-0098-5}
  {\bibfield  {journal} {\bibinfo  {journal} {Pramana}\ }\textbf {\bibinfo
  {volume} {56}},\ \bibinfo {pages} {487}},\ \Eprint
  {https://arxiv.org/abs/gr-qc/9809064} {arXiv:gr-qc/9809064} \BibitemShut
  {NoStop}%
\bibitem [{\citenamefont {Sennett}\ and\ \citenamefont
  {Buonanno}(2016)}]{Sennett:2016rwa}%
  \BibitemOpen
  \bibfield  {author} {\bibinfo {author} {\bibnamefont {Sennett}, \bibfnamefont
  {N.}}, and\ \bibinfo {author} {\bibfnamefont {A.}~\bibnamefont {Buonanno}}}
  (\bibinfo {year} {2016}),\ \href {https://doi.org/10.1103/PhysRevD.93.124004}
  {\bibfield  {journal} {\bibinfo  {journal} {Phys. Rev. D}\ }\textbf {\bibinfo
  {volume} {93}}~(\bibinfo {number} {12}),\ \bibinfo {pages} {124004}},\
  \Eprint {https://arxiv.org/abs/1603.03300} {arXiv:1603.03300 [gr-qc]}
  \BibitemShut {NoStop}%
\bibitem [{\citenamefont {Sennett}\ \emph {et~al.}(2017)\citenamefont
  {Sennett}, \citenamefont {Shao},\ and\ \citenamefont
  {Steinhoff}}]{Sennett:2017lcx}%
  \BibitemOpen
  \bibfield  {author} {\bibinfo {author} {\bibnamefont {Sennett}, \bibfnamefont
  {N.}}, \bibinfo {author} {\bibfnamefont {L.}~\bibnamefont {Shao}}, and\
  \bibinfo {author} {\bibfnamefont {J.}~\bibnamefont {Steinhoff}}} (\bibinfo
  {year} {2017}),\ \href {https://doi.org/10.1103/PhysRevD.96.084019}
  {\bibfield  {journal} {\bibinfo  {journal} {Phys. Rev. D}\ }\textbf {\bibinfo
  {volume} {96}}~(\bibinfo {number} {8}),\ \bibinfo {pages} {084019}},\ \Eprint
  {https://arxiv.org/abs/1708.08285} {arXiv:1708.08285 [gr-qc]} \BibitemShut
  {NoStop}%
\bibitem [{\citenamefont {Shao}\ \emph {et~al.}(2017)\citenamefont {Shao},
  \citenamefont {Sennett}, \citenamefont {Buonanno}, \citenamefont {Kramer},\
  and\ \citenamefont {Wex}}]{Shao:2017gwu}%
  \BibitemOpen
  \bibfield  {author} {\bibinfo {author} {\bibnamefont {Shao}, \bibfnamefont
  {L.}}, \bibinfo {author} {\bibfnamefont {N.}~\bibnamefont {Sennett}},
  \bibinfo {author} {\bibfnamefont {A.}~\bibnamefont {Buonanno}}, \bibinfo
  {author} {\bibfnamefont {M.}~\bibnamefont {Kramer}}, and\ \bibinfo {author}
  {\bibfnamefont {N.}~\bibnamefont {Wex}}} (\bibinfo {year} {2017}),\ \href
  {https://doi.org/10.1103/PhysRevX.7.041025} {\bibfield  {journal} {\bibinfo
  {journal} {Phys. Rev. X}\ }\textbf {\bibinfo {volume} {7}}~(\bibinfo {number}
  {4}),\ \bibinfo {pages} {041025}},\ \Eprint
  {https://arxiv.org/abs/1704.07561} {arXiv:1704.07561 [gr-qc]} \BibitemShut
  {NoStop}%
\bibitem [{\citenamefont {Shibata}\ \emph {et~al.}(2014)\citenamefont
  {Shibata}, \citenamefont {Taniguchi}, \citenamefont {Okawa},\ and\
  \citenamefont {Buonanno}}]{Shibata:2013pra}%
  \BibitemOpen
  \bibfield  {author} {\bibinfo {author} {\bibnamefont {Shibata}, \bibfnamefont
  {M.}}, \bibinfo {author} {\bibfnamefont {K.}~\bibnamefont {Taniguchi}},
  \bibinfo {author} {\bibfnamefont {H.}~\bibnamefont {Okawa}}, and\ \bibinfo
  {author} {\bibfnamefont {A.}~\bibnamefont {Buonanno}}} (\bibinfo {year}
  {2014}),\ \href {https://doi.org/10.1103/PhysRevD.89.084005} {\bibfield
  {journal} {\bibinfo  {journal} {Phys. Rev. D}\ }\textbf {\bibinfo {volume}
  {89}}~(\bibinfo {number} {8}),\ \bibinfo {pages} {084005}},\ \Eprint
  {https://arxiv.org/abs/1310.0627} {arXiv:1310.0627 [gr-qc]} \BibitemShut
  {NoStop}%
\bibitem [{\citenamefont {Shiralilou}\ \emph {et~al.}(2021)\citenamefont
  {Shiralilou}, \citenamefont {Hinderer}, \citenamefont {Nissanke},
  \citenamefont {Ortiz},\ and\ \citenamefont {Witek}}]{Shiralilou:2020gah}%
  \BibitemOpen
  \bibfield  {author} {\bibinfo {author} {\bibnamefont {Shiralilou},
  \bibfnamefont {B.}}, \bibinfo {author} {\bibfnamefont {T.}~\bibnamefont
  {Hinderer}}, \bibinfo {author} {\bibfnamefont {S.}~\bibnamefont {Nissanke}},
  \bibinfo {author} {\bibfnamefont {N.}~\bibnamefont {Ortiz}}, and\ \bibinfo
  {author} {\bibfnamefont {H.}~\bibnamefont {Witek}}} (\bibinfo {year}
  {2021}),\ \href {https://doi.org/10.1103/PhysRevD.103.L121503} {\bibfield
  {journal} {\bibinfo  {journal} {Phys. Rev. D}\ }\textbf {\bibinfo {volume}
  {103}}~(\bibinfo {number} {12}),\ \bibinfo {pages} {L121503}},\ \Eprint
  {https://arxiv.org/abs/2012.09162} {arXiv:2012.09162 [gr-qc]} \BibitemShut
  {NoStop}%
\bibitem [{\citenamefont {Shiralilou}\ \emph {et~al.}(2022)\citenamefont
  {Shiralilou}, \citenamefont {Hinderer}, \citenamefont {Nissanke},
  \citenamefont {Ortiz},\ and\ \citenamefont {Witek}}]{Shiralilou:2021mfl}%
  \BibitemOpen
  \bibfield  {author} {\bibinfo {author} {\bibnamefont {Shiralilou},
  \bibfnamefont {B.}}, \bibinfo {author} {\bibfnamefont {T.}~\bibnamefont
  {Hinderer}}, \bibinfo {author} {\bibfnamefont {S.~M.}\ \bibnamefont
  {Nissanke}}, \bibinfo {author} {\bibfnamefont {N.}~\bibnamefont {Ortiz}},
  and\ \bibinfo {author} {\bibfnamefont {H.}~\bibnamefont {Witek}}} (\bibinfo
  {year} {2022}),\ \href {https://doi.org/10.1088/1361-6382/ac4196} {\bibfield
  {journal} {\bibinfo  {journal} {Class. Quant. Grav.}\ }\textbf {\bibinfo
  {volume} {39}}~(\bibinfo {number} {3}),\ \bibinfo {pages} {035002}},\ \Eprint
  {https://arxiv.org/abs/2105.13972} {arXiv:2105.13972 [gr-qc]} \BibitemShut
  {NoStop}%
\bibitem [{\citenamefont {Silva}\ \emph {et~al.}(2022)\citenamefont {Silva},
  \citenamefont {Coates}, \citenamefont {Ramazano\u{g}lu},\ and\ \citenamefont
  {Sotiriou}}]{Silva:2021jya}%
  \BibitemOpen
  \bibfield  {author} {\bibinfo {author} {\bibnamefont {Silva}, \bibfnamefont
  {H.~O.}}, \bibinfo {author} {\bibfnamefont {A.}~\bibnamefont {Coates}},
  \bibinfo {author} {\bibfnamefont {F.~M.}\ \bibnamefont {Ramazano\u{g}lu}},
  and\ \bibinfo {author} {\bibfnamefont {T.~P.}\ \bibnamefont {Sotiriou}}}
  (\bibinfo {year} {2022}),\ \href
  {https://doi.org/10.1103/PhysRevD.105.024046} {\bibfield  {journal} {\bibinfo
   {journal} {Phys. Rev. D}\ }\textbf {\bibinfo {volume} {105}}~(\bibinfo
  {number} {2}),\ \bibinfo {pages} {024046}},\ \Eprint
  {https://arxiv.org/abs/2110.04594} {arXiv:2110.04594 [gr-qc]} \BibitemShut
  {NoStop}%
\bibitem [{\citenamefont {Silva}\ \emph
  {et~al.}(2021{\natexlab{a}})\citenamefont {Silva}, \citenamefont {Holgado},
  \citenamefont {C\'ardenas-Avenda\~no},\ and\ \citenamefont
  {Yunes}}]{Silva:2020acr}%
  \BibitemOpen
  \bibfield  {author} {\bibinfo {author} {\bibnamefont {Silva}, \bibfnamefont
  {H.~O.}}, \bibinfo {author} {\bibfnamefont {A.~M.}\ \bibnamefont {Holgado}},
  \bibinfo {author} {\bibfnamefont {A.}~\bibnamefont {C\'ardenas-Avenda\~no}},
  and\ \bibinfo {author} {\bibfnamefont {N.}~\bibnamefont {Yunes}}} (\bibinfo
  {year} {2021}{\natexlab{a}}),\ \href
  {https://doi.org/10.1103/PhysRevLett.126.181101} {\bibfield  {journal}
  {\bibinfo  {journal} {Phys. Rev. Lett.}\ }\textbf {\bibinfo {volume}
  {126}}~(\bibinfo {number} {18}),\ \bibinfo {pages} {181101}},\ \Eprint
  {https://arxiv.org/abs/2004.01253} {arXiv:2004.01253 [gr-qc]} \BibitemShut
  {NoStop}%
\bibitem [{\citenamefont {Silva}\ \emph {et~al.}(2015)\citenamefont {Silva},
  \citenamefont {Macedo}, \citenamefont {Berti},\ and\ \citenamefont
  {Crispino}}]{Silva:2014fca}%
  \BibitemOpen
  \bibfield  {author} {\bibinfo {author} {\bibnamefont {Silva}, \bibfnamefont
  {H.~O.}}, \bibinfo {author} {\bibfnamefont {C.~F.}\ \bibnamefont {Macedo}},
  \bibinfo {author} {\bibfnamefont {E.}~\bibnamefont {Berti}}, and\ \bibinfo
  {author} {\bibfnamefont {L.~C.}\ \bibnamefont {Crispino}}} (\bibinfo {year}
  {2015}),\ \href {https://doi.org/10.1088/0264-9381/32/14/145008} {\bibfield
  {journal} {\bibinfo  {journal} {Class. Quant. Grav.}\ }\textbf {\bibinfo
  {volume} {32}},\ \bibinfo {pages} {145008}},\ \Eprint
  {https://arxiv.org/abs/1411.6286} {arXiv:1411.6286 [gr-qc]} \BibitemShut
  {NoStop}%
\bibitem [{\citenamefont {Silva}\ \emph {et~al.}(2019)\citenamefont {Silva},
  \citenamefont {Macedo}, \citenamefont {Sotiriou}, \citenamefont {Gualtieri},
  \citenamefont {Sakstein},\ and\ \citenamefont {Berti}}]{Silva:2018qhn}%
  \BibitemOpen
  \bibfield  {author} {\bibinfo {author} {\bibnamefont {Silva}, \bibfnamefont
  {H.~O.}}, \bibinfo {author} {\bibfnamefont {C.~F.~B.}\ \bibnamefont
  {Macedo}}, \bibinfo {author} {\bibfnamefont {T.~P.}\ \bibnamefont
  {Sotiriou}}, \bibinfo {author} {\bibfnamefont {L.}~\bibnamefont {Gualtieri}},
  \bibinfo {author} {\bibfnamefont {J.}~\bibnamefont {Sakstein}}, and\ \bibinfo
  {author} {\bibfnamefont {E.}~\bibnamefont {Berti}}} (\bibinfo {year}
  {2019}),\ \href {https://doi.org/10.1103/PhysRevD.99.064011} {\bibfield
  {journal} {\bibinfo  {journal} {Phys. Rev. D}\ }\textbf {\bibinfo {volume}
  {99}}~(\bibinfo {number} {6}),\ \bibinfo {pages} {064011}},\ \Eprint
  {https://arxiv.org/abs/1812.05590} {arXiv:1812.05590 [gr-qc]} \BibitemShut
  {NoStop}%
\bibitem [{\citenamefont {Silva}\ \emph {et~al.}(2018)\citenamefont {Silva},
  \citenamefont {Sakstein}, \citenamefont {Gualtieri}, \citenamefont
  {Sotiriou},\ and\ \citenamefont {Berti}}]{Silva:2017uqg}%
  \BibitemOpen
  \bibfield  {author} {\bibinfo {author} {\bibnamefont {Silva}, \bibfnamefont
  {H.~O.}}, \bibinfo {author} {\bibfnamefont {J.}~\bibnamefont {Sakstein}},
  \bibinfo {author} {\bibfnamefont {L.}~\bibnamefont {Gualtieri}}, \bibinfo
  {author} {\bibfnamefont {T.~P.}\ \bibnamefont {Sotiriou}}, and\ \bibinfo
  {author} {\bibfnamefont {E.}~\bibnamefont {Berti}}} (\bibinfo {year}
  {2018}),\ \href {https://doi.org/10.1103/PhysRevLett.120.131104} {\bibfield
  {journal} {\bibinfo  {journal} {Phys. Rev. Lett.}\ }\textbf {\bibinfo
  {volume} {120}}~(\bibinfo {number} {13}),\ \bibinfo {pages} {131104}},\
  \Eprint {https://arxiv.org/abs/1711.02080} {arXiv:1711.02080 [gr-qc]}
  \BibitemShut {NoStop}%
\bibitem [{\citenamefont {Silva}\ \emph {et~al.}(2014)\citenamefont {Silva},
  \citenamefont {Sotani}, \citenamefont {Berti},\ and\ \citenamefont
  {Horbatsch}}]{Silva:2014ora}%
  \BibitemOpen
  \bibfield  {author} {\bibinfo {author} {\bibnamefont {Silva}, \bibfnamefont
  {H.~O.}}, \bibinfo {author} {\bibfnamefont {H.}~\bibnamefont {Sotani}},
  \bibinfo {author} {\bibfnamefont {E.}~\bibnamefont {Berti}}, and\ \bibinfo
  {author} {\bibfnamefont {M.}~\bibnamefont {Horbatsch}}} (\bibinfo {year}
  {2014}),\ \href {https://doi.org/10.1103/PhysRevD.90.124044} {\bibfield
  {journal} {\bibinfo  {journal} {Phys. Rev. D}\ }\textbf {\bibinfo {volume}
  {90}}~(\bibinfo {number} {12}),\ \bibinfo {pages} {124044}},\ \Eprint
  {https://arxiv.org/abs/1410.2511} {arXiv:1410.2511 [gr-qc]} \BibitemShut
  {NoStop}%
\bibitem [{\citenamefont {Silva}\ \emph
  {et~al.}(2021{\natexlab{b}})\citenamefont {Silva}, \citenamefont {Witek},
  \citenamefont {Elley},\ and\ \citenamefont {Yunes}}]{Silva:2020omi}%
  \BibitemOpen
  \bibfield  {author} {\bibinfo {author} {\bibnamefont {Silva}, \bibfnamefont
  {H.~O.}}, \bibinfo {author} {\bibfnamefont {H.}~\bibnamefont {Witek}},
  \bibinfo {author} {\bibfnamefont {M.}~\bibnamefont {Elley}}, and\ \bibinfo
  {author} {\bibfnamefont {N.}~\bibnamefont {Yunes}}} (\bibinfo {year}
  {2021}{\natexlab{b}}),\ \href
  {https://doi.org/10.1103/PhysRevLett.127.031101} {\bibfield  {journal}
  {\bibinfo  {journal} {Phys. Rev. Lett.}\ }\textbf {\bibinfo {volume}
  {127}}~(\bibinfo {number} {3}),\ \bibinfo {pages} {031101}},\ \Eprint
  {https://arxiv.org/abs/2012.10436} {arXiv:2012.10436 [gr-qc]} \BibitemShut
  {NoStop}%
\bibitem [{\citenamefont {Silva}\ and\ \citenamefont
  {Yunes}(2019{\natexlab{a}})}]{Silva:2019leq}%
  \BibitemOpen
  \bibfield  {author} {\bibinfo {author} {\bibnamefont {Silva}, \bibfnamefont
  {H.~O.}}, and\ \bibinfo {author} {\bibfnamefont {N.}~\bibnamefont {Yunes}}}
  (\bibinfo {year} {2019}{\natexlab{a}}),\ \href
  {https://doi.org/10.1088/1361-6382/ab3560} {\bibfield  {journal} {\bibinfo
  {journal} {Class. Quant. Grav.}\ }\textbf {\bibinfo {volume} {36}}~(\bibinfo
  {number} {17}),\ \bibinfo {pages} {17LT01}},\ \Eprint
  {https://arxiv.org/abs/1902.10269} {arXiv:1902.10269 [gr-qc]} \BibitemShut
  {NoStop}%
\bibitem [{\citenamefont {Silva}\ and\ \citenamefont
  {Yunes}(2019{\natexlab{b}})}]{Silva:2018yxz}%
  \BibitemOpen
  \bibfield  {author} {\bibinfo {author} {\bibnamefont {Silva}, \bibfnamefont
  {H.~O.}}, and\ \bibinfo {author} {\bibfnamefont {N.}~\bibnamefont {Yunes}}}
  (\bibinfo {year} {2019}{\natexlab{b}}),\ \href
  {https://doi.org/10.1103/PhysRevD.99.044034} {\bibfield  {journal} {\bibinfo
  {journal} {Phys. Rev. D}\ }\textbf {\bibinfo {volume} {99}}~(\bibinfo
  {number} {4}),\ \bibinfo {pages} {044034}},\ \Eprint
  {https://arxiv.org/abs/1808.04391} {arXiv:1808.04391 [gr-qc]} \BibitemShut
  {NoStop}%
\bibitem [{\citenamefont {Soldateschi}\ \emph {et~al.}(2020)\citenamefont
  {Soldateschi}, \citenamefont {Bucciantini},\ and\ \citenamefont
  {Del~Zanna}}]{Soldateschi:2020hju}%
  \BibitemOpen
  \bibfield  {author} {\bibinfo {author} {\bibnamefont {Soldateschi},
  \bibfnamefont {J.}}, \bibinfo {author} {\bibfnamefont {N.}~\bibnamefont
  {Bucciantini}}, and\ \bibinfo {author} {\bibfnamefont {L.}~\bibnamefont
  {Del~Zanna}}} (\bibinfo {year} {2020}),\ \href
  {https://doi.org/10.1051/0004-6361/202037918} {\bibfield  {journal} {\bibinfo
   {journal} {Astron. Astrophys.}\ }\textbf {\bibinfo {volume} {640}},\
  \bibinfo {pages} {A44}},\ \Eprint {https://arxiv.org/abs/2005.12758}
  {arXiv:2005.12758 [astro-ph.HE]} \BibitemShut {NoStop}%
\bibitem [{\citenamefont {Soldateschi}\ \emph {et~al.}(2021)\citenamefont
  {Soldateschi}, \citenamefont {Bucciantini},\ and\ \citenamefont
  {Del~Zanna}}]{Soldateschi:2020zxb}%
  \BibitemOpen
  \bibfield  {author} {\bibinfo {author} {\bibnamefont {Soldateschi},
  \bibfnamefont {J.}}, \bibinfo {author} {\bibfnamefont {N.}~\bibnamefont
  {Bucciantini}}, and\ \bibinfo {author} {\bibfnamefont {L.}~\bibnamefont
  {Del~Zanna}}} (\bibinfo {year} {2021}),\ \href
  {https://doi.org/10.1051/0004-6361/202038826} {\bibfield  {journal} {\bibinfo
   {journal} {Astron. Astrophys.}\ }\textbf {\bibinfo {volume} {645}},\
  \bibinfo {pages} {A39}},\ \Eprint {https://arxiv.org/abs/2010.14833}
  {arXiv:2010.14833 [astro-ph.HE]} \BibitemShut {NoStop}%
\bibitem [{\citenamefont {Sotani}(2012)}]{Sotani:2012eb}%
  \BibitemOpen
  \bibfield  {author} {\bibinfo {author} {\bibnamefont {Sotani}, \bibfnamefont
  {H.}}} (\bibinfo {year} {2012}),\ \href
  {https://doi.org/10.1103/PhysRevD.86.124036} {\bibfield  {journal} {\bibinfo
  {journal} {Phys. Rev. D}\ }\textbf {\bibinfo {volume} {86}},\ \bibinfo
  {pages} {124036}},\ \Eprint {https://arxiv.org/abs/1211.6986}
  {arXiv:1211.6986 [astro-ph.HE]} \BibitemShut {NoStop}%
\bibitem [{\citenamefont {Sotani}(2014)}]{Sotani:2014tua}%
  \BibitemOpen
  \bibfield  {author} {\bibinfo {author} {\bibnamefont {Sotani}, \bibfnamefont
  {H.}}} (\bibinfo {year} {2014}),\ \href
  {https://doi.org/10.1103/PhysRevD.89.064031} {\bibfield  {journal} {\bibinfo
  {journal} {Phys. Rev. D}\ }\textbf {\bibinfo {volume} {89}}~(\bibinfo
  {number} {6}),\ \bibinfo {pages} {064031}},\ \Eprint
  {https://arxiv.org/abs/1402.5699} {arXiv:1402.5699 [astro-ph.HE]}
  \BibitemShut {NoStop}%
\bibitem [{\citenamefont {Sotani}(2017)}]{Sotani:2017rrt}%
  \BibitemOpen
  \bibfield  {author} {\bibinfo {author} {\bibnamefont {Sotani}, \bibfnamefont
  {H.}}} (\bibinfo {year} {2017}),\ \href
  {https://doi.org/10.1103/PhysRevD.96.104010} {\bibfield  {journal} {\bibinfo
  {journal} {Phys. Rev. D}\ }\textbf {\bibinfo {volume} {96}}~(\bibinfo
  {number} {10}),\ \bibinfo {pages} {104010}},\ \Eprint
  {https://arxiv.org/abs/1710.10596} {arXiv:1710.10596 [astro-ph.HE]}
  \BibitemShut {NoStop}%
\bibitem [{\citenamefont {Sotani}\ and\ \citenamefont
  {Kokkotas}(2004)}]{Sotani:2004rq}%
  \BibitemOpen
  \bibfield  {author} {\bibinfo {author} {\bibnamefont {Sotani}, \bibfnamefont
  {H.}}, and\ \bibinfo {author} {\bibfnamefont {K.~D.}\ \bibnamefont
  {Kokkotas}}} (\bibinfo {year} {2004}),\ \href
  {https://doi.org/10.1103/PhysRevD.70.084026} {\bibfield  {journal} {\bibinfo
  {journal} {Phys. Rev. D}\ }\textbf {\bibinfo {volume} {70}},\ \bibinfo
  {pages} {084026}},\ \Eprint {https://arxiv.org/abs/gr-qc/0409066}
  {arXiv:gr-qc/0409066} \BibitemShut {NoStop}%
\bibitem [{\citenamefont {Sotani}\ and\ \citenamefont
  {Kokkotas}(2005)}]{Sotani:2005qx}%
  \BibitemOpen
  \bibfield  {author} {\bibinfo {author} {\bibnamefont {Sotani}, \bibfnamefont
  {H.}}, and\ \bibinfo {author} {\bibfnamefont {K.~D.}\ \bibnamefont
  {Kokkotas}}} (\bibinfo {year} {2005}),\ \href
  {https://doi.org/10.1103/PhysRevD.71.124038} {\bibfield  {journal} {\bibinfo
  {journal} {Phys. Rev. D}\ }\textbf {\bibinfo {volume} {71}},\ \bibinfo
  {pages} {124038}},\ \Eprint {https://arxiv.org/abs/gr-qc/0506060}
  {arXiv:gr-qc/0506060} \BibitemShut {NoStop}%
\bibitem [{\citenamefont {Sotani}\ and\ \citenamefont
  {Kokkotas}(2017)}]{Sotani:2017pfj}%
  \BibitemOpen
  \bibfield  {author} {\bibinfo {author} {\bibnamefont {Sotani}, \bibfnamefont
  {H.}}, and\ \bibinfo {author} {\bibfnamefont {K.~D.}\ \bibnamefont
  {Kokkotas}}} (\bibinfo {year} {2017}),\ \href
  {https://doi.org/10.1103/PhysRevD.95.044032} {\bibfield  {journal} {\bibinfo
  {journal} {Phys. Rev. D}\ }\textbf {\bibinfo {volume} {95}}~(\bibinfo
  {number} {4}),\ \bibinfo {pages} {044032}},\ \Eprint
  {https://arxiv.org/abs/1702.00874} {arXiv:1702.00874 [gr-qc]} \BibitemShut
  {NoStop}%
\bibitem [{\citenamefont {Sotiriou}\ and\ \citenamefont
  {Faraoni}(2012)}]{Sotiriou:2011dz}%
  \BibitemOpen
  \bibfield  {author} {\bibinfo {author} {\bibnamefont {Sotiriou},
  \bibfnamefont {T.~P.}}, and\ \bibinfo {author} {\bibfnamefont
  {V.}~\bibnamefont {Faraoni}}} (\bibinfo {year} {2012}),\ \href
  {https://doi.org/10.1103/PhysRevLett.108.081103} {\bibfield  {journal}
  {\bibinfo  {journal} {Phys. Rev. Lett.}\ }\textbf {\bibinfo {volume} {108}},\
  \bibinfo {pages} {081103}},\ \Eprint {https://arxiv.org/abs/1109.6324}
  {arXiv:1109.6324 [gr-qc]} \BibitemShut {NoStop}%
\bibitem [{\citenamefont {Sotiriou}\ and\ \citenamefont
  {Zhou}(2014{\natexlab{a}})}]{Sotiriou:2013qea}%
  \BibitemOpen
  \bibfield  {author} {\bibinfo {author} {\bibnamefont {Sotiriou},
  \bibfnamefont {T.~P.}}, and\ \bibinfo {author} {\bibfnamefont {S.-Y.}\
  \bibnamefont {Zhou}}} (\bibinfo {year} {2014}{\natexlab{a}}),\ \href
  {https://doi.org/10.1103/PhysRevLett.112.251102} {\bibfield  {journal}
  {\bibinfo  {journal} {Phys. Rev. Lett.}\ }\textbf {\bibinfo {volume} {112}},\
  \bibinfo {pages} {251102}},\ \Eprint {https://arxiv.org/abs/1312.3622}
  {arXiv:1312.3622 [gr-qc]} \BibitemShut {NoStop}%
\bibitem [{\citenamefont {Sotiriou}\ and\ \citenamefont
  {Zhou}(2014{\natexlab{b}})}]{Sotiriou:2014pfa}%
  \BibitemOpen
  \bibfield  {author} {\bibinfo {author} {\bibnamefont {Sotiriou},
  \bibfnamefont {T.~P.}}, and\ \bibinfo {author} {\bibfnamefont {S.-Y.}\
  \bibnamefont {Zhou}}} (\bibinfo {year} {2014}{\natexlab{b}}),\ \href
  {https://doi.org/10.1103/PhysRevD.90.124063} {\bibfield  {journal} {\bibinfo
  {journal} {Phys. Rev. D}\ }\textbf {\bibinfo {volume} {90}},\ \bibinfo
  {pages} {124063}},\ \Eprint {https://arxiv.org/abs/1408.1698}
  {arXiv:1408.1698 [gr-qc]} \BibitemShut {NoStop}%
\bibitem [{\citenamefont {Sperhake}\ \emph {et~al.}(2017)\citenamefont
  {Sperhake}, \citenamefont {Moore}, \citenamefont {Rosca}, \citenamefont
  {Agathos}, \citenamefont {Gerosa},\ and\ \citenamefont
  {Ott}}]{Sperhake:2017itk}%
  \BibitemOpen
  \bibfield  {author} {\bibinfo {author} {\bibnamefont {Sperhake},
  \bibfnamefont {U.}}, \bibinfo {author} {\bibfnamefont {C.~J.}\ \bibnamefont
  {Moore}}, \bibinfo {author} {\bibfnamefont {R.}~\bibnamefont {Rosca}},
  \bibinfo {author} {\bibfnamefont {M.}~\bibnamefont {Agathos}}, \bibinfo
  {author} {\bibfnamefont {D.}~\bibnamefont {Gerosa}}, and\ \bibinfo {author}
  {\bibfnamefont {C.~D.}\ \bibnamefont {Ott}}} (\bibinfo {year} {2017}),\ \href
  {https://doi.org/10.1103/PhysRevLett.119.201103} {\bibfield  {journal}
  {\bibinfo  {journal} {Phys. Rev. Lett.}\ }\textbf {\bibinfo {volume}
  {119}}~(\bibinfo {number} {20}),\ \bibinfo {pages} {201103}},\ \Eprint
  {https://arxiv.org/abs/1708.03651} {arXiv:1708.03651 [gr-qc]} \BibitemShut
  {NoStop}%
\bibitem [{\citenamefont {Staykov}\ and\ \citenamefont
  {Doneva}(2022)}]{Staykov:2022uwq}%
  \BibitemOpen
  \bibfield  {author} {\bibinfo {author} {\bibnamefont {Staykov}, \bibfnamefont
  {K.~V.}}, and\ \bibinfo {author} {\bibfnamefont {D.~D.}\ \bibnamefont
  {Doneva}}} (\bibinfo {year} {2022}),\ \href
  {https://doi.org/10.1103/PhysRevD.106.104064} {\bibfield  {journal} {\bibinfo
   {journal} {Phys. Rev. D}\ }\textbf {\bibinfo {volume} {106}}~(\bibinfo
  {number} {10}),\ \bibinfo {pages} {104064}},\ \Eprint
  {https://arxiv.org/abs/2209.01038} {arXiv:2209.01038 [gr-qc]} \BibitemShut
  {NoStop}%
\bibitem [{\citenamefont {Staykov}\ \emph {et~al.}(2019)\citenamefont
  {Staykov}, \citenamefont {Doneva},\ and\ \citenamefont
  {Yazadjiev}}]{Staykov:2019pwj}%
  \BibitemOpen
  \bibfield  {author} {\bibinfo {author} {\bibnamefont {Staykov}, \bibfnamefont
  {K.~V.}}, \bibinfo {author} {\bibfnamefont {D.~D.}\ \bibnamefont {Doneva}},
  and\ \bibinfo {author} {\bibfnamefont {S.~S.}\ \bibnamefont {Yazadjiev}}}
  (\bibinfo {year} {2019}),\ \href {https://doi.org/10.1007/s10509-019-3666-1}
  {\bibfield  {journal} {\bibinfo  {journal} {Astrophys. Space Sci.}\ }\textbf
  {\bibinfo {volume} {364}}~(\bibinfo {number} {10}),\ \bibinfo {pages}
  {178}},\ \Eprint {https://arxiv.org/abs/1902.09208} {arXiv:1902.09208
  [gr-qc]} \BibitemShut {NoStop}%
\bibitem [{\citenamefont {Staykov}\ \emph {et~al.}(2015)\citenamefont
  {Staykov}, \citenamefont {Doneva}, \citenamefont {Yazadjiev},\ and\
  \citenamefont {Kokkotas}}]{Staykov:2015cfa}%
  \BibitemOpen
  \bibfield  {author} {\bibinfo {author} {\bibnamefont {Staykov}, \bibfnamefont
  {K.~V.}}, \bibinfo {author} {\bibfnamefont {D.~D.}\ \bibnamefont {Doneva}},
  \bibinfo {author} {\bibfnamefont {S.~S.}\ \bibnamefont {Yazadjiev}}, and\
  \bibinfo {author} {\bibfnamefont {K.~D.}\ \bibnamefont {Kokkotas}}} (\bibinfo
  {year} {2015}),\ \href {https://doi.org/10.1103/PhysRevD.92.043009}
  {\bibfield  {journal} {\bibinfo  {journal} {Phys. Rev. D}\ }\textbf {\bibinfo
  {volume} {92}}~(\bibinfo {number} {4}),\ \bibinfo {pages} {043009}},\ \Eprint
  {https://arxiv.org/abs/1503.04711} {arXiv:1503.04711 [gr-qc]} \BibitemShut
  {NoStop}%
\bibitem [{\citenamefont {Staykov}\ \emph {et~al.}(2018)\citenamefont
  {Staykov}, \citenamefont {Popchev}, \citenamefont {Doneva},\ and\
  \citenamefont {Yazadjiev}}]{Staykov:2018hhc}%
  \BibitemOpen
  \bibfield  {author} {\bibinfo {author} {\bibnamefont {Staykov}, \bibfnamefont
  {K.~V.}}, \bibinfo {author} {\bibfnamefont {D.}~\bibnamefont {Popchev}},
  \bibinfo {author} {\bibfnamefont {D.~D.}\ \bibnamefont {Doneva}}, and\
  \bibinfo {author} {\bibfnamefont {S.~S.}\ \bibnamefont {Yazadjiev}}}
  (\bibinfo {year} {2018}),\ \href
  {https://doi.org/10.1140/epjc/s10052-018-6064-x} {\bibfield  {journal}
  {\bibinfo  {journal} {Eur. Phys. J. C}\ }\textbf {\bibinfo {volume}
  {78}}~(\bibinfo {number} {7}),\ \bibinfo {pages} {586}},\ \Eprint
  {https://arxiv.org/abs/1805.07818} {arXiv:1805.07818 [gr-qc]} \BibitemShut
  {NoStop}%
\bibitem [{\citenamefont {Stefanov}\ \emph
  {et~al.}(2007{\natexlab{a}})\citenamefont {Stefanov}, \citenamefont
  {Yazadjiev},\ and\ \citenamefont {Todorov}}]{Stefanov:2007qw}%
  \BibitemOpen
  \bibfield  {author} {\bibinfo {author} {\bibnamefont {Stefanov},
  \bibfnamefont {I.~Z.}}, \bibinfo {author} {\bibfnamefont {S.~S.}\
  \bibnamefont {Yazadjiev}}, and\ \bibinfo {author} {\bibfnamefont {M.~D.}\
  \bibnamefont {Todorov}}} (\bibinfo {year} {2007}{\natexlab{a}}),\ \href
  {https://doi.org/10.1103/PhysRevD.75.084036} {\bibfield  {journal} {\bibinfo
  {journal} {Phys. Rev. D}\ }\textbf {\bibinfo {volume} {75}},\ \bibinfo
  {pages} {084036}},\ \Eprint {https://arxiv.org/abs/0704.3784}
  {arXiv:0704.3784 [gr-qc]} \BibitemShut {NoStop}%
\bibitem [{\citenamefont {Stefanov}\ \emph
  {et~al.}(2007{\natexlab{b}})\citenamefont {Stefanov}, \citenamefont
  {Yazadjiev},\ and\ \citenamefont {Todorov}}]{Stefanov:2007bn}%
  \BibitemOpen
  \bibfield  {author} {\bibinfo {author} {\bibnamefont {Stefanov},
  \bibfnamefont {I.~Z.}}, \bibinfo {author} {\bibfnamefont {S.~S.}\
  \bibnamefont {Yazadjiev}}, and\ \bibinfo {author} {\bibfnamefont {M.~D.}\
  \bibnamefont {Todorov}}} (\bibinfo {year} {2007}{\natexlab{b}}),\ \href
  {https://doi.org/10.1142/S0217732307023560} {\bibfield  {journal} {\bibinfo
  {journal} {Mod. Phys. Lett. A}\ }\textbf {\bibinfo {volume} {22}},\ \bibinfo
  {pages} {1217}},\ \Eprint {https://arxiv.org/abs/0708.3203} {arXiv:0708.3203
  [gr-qc]} \BibitemShut {NoStop}%
\bibitem [{\citenamefont {Stefanov}\ \emph {et~al.}(2008)\citenamefont
  {Stefanov}, \citenamefont {Yazadjiev},\ and\ \citenamefont
  {Todorov}}]{Stefanov:2007eq}%
  \BibitemOpen
  \bibfield  {author} {\bibinfo {author} {\bibnamefont {Stefanov},
  \bibfnamefont {I.~Z.}}, \bibinfo {author} {\bibfnamefont {S.~S.}\
  \bibnamefont {Yazadjiev}}, and\ \bibinfo {author} {\bibfnamefont {M.~D.}\
  \bibnamefont {Todorov}}} (\bibinfo {year} {2008}),\ \href
  {https://doi.org/10.1142/S0217732308028351} {\bibfield  {journal} {\bibinfo
  {journal} {Mod. Phys. Lett. A}\ }\textbf {\bibinfo {volume} {23}},\ \bibinfo
  {pages} {2915}},\ \Eprint {https://arxiv.org/abs/0708.4141} {arXiv:0708.4141
  [gr-qc]} \BibitemShut {NoStop}%
\bibitem [{\citenamefont {Strohmayer}\ and\ \citenamefont
  {Watts}(2005)}]{Strohmayer:2005ks}%
  \BibitemOpen
  \bibfield  {author} {\bibinfo {author} {\bibnamefont {Strohmayer},
  \bibfnamefont {T.~E.}}, and\ \bibinfo {author} {\bibfnamefont {A.~L.}\
  \bibnamefont {Watts}}} (\bibinfo {year} {2005}),\ \href
  {https://doi.org/10.1086/497911} {\bibfield  {journal} {\bibinfo  {journal}
  {Astrophys. J. Lett.}\ }\textbf {\bibinfo {volume} {632}},\ \bibinfo {pages}
  {L111}},\ \Eprint {https://arxiv.org/abs/astro-ph/0508206}
  {arXiv:astro-ph/0508206} \BibitemShut {NoStop}%
\bibitem [{\citenamefont {Strohmayer}\ and\ \citenamefont
  {Watts}(2006)}]{Strohmayer:2006py}%
  \BibitemOpen
  \bibfield  {author} {\bibinfo {author} {\bibnamefont {Strohmayer},
  \bibfnamefont {T.~E.}}, and\ \bibinfo {author} {\bibfnamefont {A.~L.}\
  \bibnamefont {Watts}}} (\bibinfo {year} {2006}),\ \href
  {https://doi.org/10.1086/508703} {\bibfield  {journal} {\bibinfo  {journal}
  {Astrophys. J.}\ }\textbf {\bibinfo {volume} {653}},\ \bibinfo {pages}
  {593}},\ \Eprint {https://arxiv.org/abs/astro-ph/0608463}
  {arXiv:astro-ph/0608463} \BibitemShut {NoStop}%
\bibitem [{\citenamefont {Takami}\ \emph {et~al.}(2014)\citenamefont {Takami},
  \citenamefont {Rezzolla},\ and\ \citenamefont {Baiotti}}]{Takami:2014zpa}%
  \BibitemOpen
  \bibfield  {author} {\bibinfo {author} {\bibnamefont {Takami}, \bibfnamefont
  {K.}}, \bibinfo {author} {\bibfnamefont {L.}~\bibnamefont {Rezzolla}}, and\
  \bibinfo {author} {\bibfnamefont {L.}~\bibnamefont {Baiotti}}} (\bibinfo
  {year} {2014}),\ \href {https://doi.org/10.1103/PhysRevLett.113.091104}
  {\bibfield  {journal} {\bibinfo  {journal} {Phys. Rev. Lett.}\ }\textbf
  {\bibinfo {volume} {113}}~(\bibinfo {number} {9}),\ \bibinfo {pages}
  {091104}},\ \Eprint {https://arxiv.org/abs/1403.5672} {arXiv:1403.5672
  [gr-qc]} \BibitemShut {NoStop}%
\bibitem [{\citenamefont {Takami}\ \emph {et~al.}(2015)\citenamefont {Takami},
  \citenamefont {Rezzolla},\ and\ \citenamefont {Baiotti}}]{Takami:2014tva}%
  \BibitemOpen
  \bibfield  {author} {\bibinfo {author} {\bibnamefont {Takami}, \bibfnamefont
  {K.}}, \bibinfo {author} {\bibfnamefont {L.}~\bibnamefont {Rezzolla}}, and\
  \bibinfo {author} {\bibfnamefont {L.}~\bibnamefont {Baiotti}}} (\bibinfo
  {year} {2015}),\ \href {https://doi.org/10.1103/PhysRevD.91.064001}
  {\bibfield  {journal} {\bibinfo  {journal} {Phys. Rev. D}\ }\textbf {\bibinfo
  {volume} {91}}~(\bibinfo {number} {6}),\ \bibinfo {pages} {064001}},\ \Eprint
  {https://arxiv.org/abs/1412.3240} {arXiv:1412.3240 [gr-qc]} \BibitemShut
  {NoStop}%
\bibitem [{\citenamefont {Taniguchi}\ and\ \citenamefont
  {Gourgoulhon}(2003)}]{Taniguchi:2003hx}%
  \BibitemOpen
  \bibfield  {author} {\bibinfo {author} {\bibnamefont {Taniguchi},
  \bibfnamefont {K.}}, and\ \bibinfo {author} {\bibfnamefont {E.}~\bibnamefont
  {Gourgoulhon}}} (\bibinfo {year} {2003}),\ \href
  {https://doi.org/10.1103/PhysRevD.68.124025} {\bibfield  {journal} {\bibinfo
  {journal} {Phys. Rev. D}\ }\textbf {\bibinfo {volume} {68}},\ \bibinfo
  {pages} {124025}},\ \Eprint {https://arxiv.org/abs/gr-qc/0309045}
  {arXiv:gr-qc/0309045} \BibitemShut {NoStop}%
\bibitem [{\citenamefont {Taniguchi}\ and\ \citenamefont
  {Shibata}(2010)}]{Taniguchi:2010kj}%
  \BibitemOpen
  \bibfield  {author} {\bibinfo {author} {\bibnamefont {Taniguchi},
  \bibfnamefont {K.}}, and\ \bibinfo {author} {\bibfnamefont {M.}~\bibnamefont
  {Shibata}}} (\bibinfo {year} {2010}),\ \href
  {https://doi.org/10.1088/0067-0049/188/1/187} {\bibfield  {journal} {\bibinfo
   {journal} {Astrophys. J. Suppl.}\ }\textbf {\bibinfo {volume} {188}},\
  \bibinfo {pages} {187}},\ \Eprint {https://arxiv.org/abs/1005.0958}
  {arXiv:1005.0958 [astro-ph.SR]} \BibitemShut {NoStop}%
\bibitem [{\citenamefont {Taniguchi}\ \emph {et~al.}(2015)\citenamefont
  {Taniguchi}, \citenamefont {Shibata},\ and\ \citenamefont
  {Buonanno}}]{Taniguchi:2014fqa}%
  \BibitemOpen
  \bibfield  {author} {\bibinfo {author} {\bibnamefont {Taniguchi},
  \bibfnamefont {K.}}, \bibinfo {author} {\bibfnamefont {M.}~\bibnamefont
  {Shibata}}, and\ \bibinfo {author} {\bibfnamefont {A.}~\bibnamefont
  {Buonanno}}} (\bibinfo {year} {2015}),\ \href
  {https://doi.org/10.1103/PhysRevD.91.024033} {\bibfield  {journal} {\bibinfo
  {journal} {Phys. Rev. D}\ }\textbf {\bibinfo {volume} {91}}~(\bibinfo
  {number} {2}),\ \bibinfo {pages} {024033}},\ \Eprint
  {https://arxiv.org/abs/1410.0738} {arXiv:1410.0738 [gr-qc]} \BibitemShut
  {NoStop}%
\bibitem [{\citenamefont {Tasinato}(2014)}]{Tasinato:2014eka}%
  \BibitemOpen
  \bibfield  {author} {\bibinfo {author} {\bibnamefont {Tasinato},
  \bibfnamefont {G.}}} (\bibinfo {year} {2014}),\ \href
  {https://doi.org/10.1007/JHEP04(2014)067} {\bibfield  {journal} {\bibinfo
  {journal} {JHEP}\ }\textbf {\bibinfo {volume} {04}},\ \bibinfo {pages}
  {067}},\ \Eprint {https://arxiv.org/abs/1402.6450} {arXiv:1402.6450 [hep-th]}
  \BibitemShut {NoStop}%
\bibitem [{\citenamefont {Taylor}\ and\ \citenamefont
  {Weisberg}(1982)}]{Taylor:1982zz}%
  \BibitemOpen
  \bibfield  {author} {\bibinfo {author} {\bibnamefont {Taylor}, \bibfnamefont
  {J.~H.}}, and\ \bibinfo {author} {\bibfnamefont {J.~M.}\ \bibnamefont
  {Weisberg}}} (\bibinfo {year} {1982}),\ \href
  {https://doi.org/10.1086/159690} {\bibfield  {journal} {\bibinfo  {journal}
  {Astrophys. J.}\ }\textbf {\bibinfo {volume} {253}},\ \bibinfo {pages}
  {908}}\BibitemShut {NoStop}%
\bibitem [{\citenamefont {Taylor}\ \emph {et~al.}(1993)\citenamefont {Taylor},
  \citenamefont {Wolszczan},\ and\ \citenamefont {Damour}}]{Taylor:1993zz}%
  \BibitemOpen
  \bibfield  {author} {\bibinfo {author} {\bibnamefont {Taylor}, \bibfnamefont
  {J.~N.}}, \bibinfo {author} {\bibfnamefont {A.}~\bibnamefont {Wolszczan}},
  and\ \bibinfo {author} {\bibfnamefont {T.}~\bibnamefont {Damour}}} (\bibinfo
  {year} {1993}),\ \href {https://doi.org/10.1038/355132a0} {\bibfield
  {journal} {\bibinfo  {journal} {Nature}\ }\textbf {\bibinfo {volume} {355}},\
  \bibinfo {pages} {132}}\BibitemShut {NoStop}%
\bibitem [{\citenamefont {{Thorne}}\ and\ \citenamefont
  {{Campolattaro}}(1967)}]{Thorne:1967ApJ}%
  \BibitemOpen
  \bibfield  {author} {\bibinfo {author} {\bibnamefont {{Thorne}},
  \bibfnamefont {K.~S.}}, and\ \bibinfo {author} {\bibfnamefont
  {A.}~\bibnamefont {{Campolattaro}}}} (\bibinfo {year} {1967}),\ \href
  {https://doi.org/10.1086/149288} {\bibinfo  {journal} {Astrophysical
  Journal}\ ,\ \bibinfo {pages} {591}}\BibitemShut {NoStop}%
\bibitem [{\citenamefont {Tuna}\ \emph {et~al.}(2022)\citenamefont {Tuna},
  \citenamefont {\"Unl\"ut\"urk},\ and\ \citenamefont
  {Ramazano\u{g}lu}}]{Tuna:2022qqr}%
  \BibitemOpen
\bibfield  {journal} {  }\bibfield  {author} {\bibinfo {author} {\bibnamefont
  {Tuna}, \bibfnamefont {S.}}, \bibinfo {author} {\bibfnamefont {K.~I.}\
  \bibnamefont {\"Unl\"ut\"urk}}, and\ \bibinfo {author} {\bibfnamefont
  {F.~M.}\ \bibnamefont {Ramazano\u{g}lu}}} (\bibinfo {year} {2022}),\ \href
  {https://doi.org/10.1103/PhysRevD.105.124070} {\bibfield  {journal} {\bibinfo
   {journal} {Phys. Rev. D}\ }\textbf {\bibinfo {volume} {105}}~(\bibinfo
  {number} {12}),\ \bibinfo {pages} {124070}},\ \Eprint
  {https://arxiv.org/abs/2204.02138} {arXiv:2204.02138 [gr-qc]} \BibitemShut
  {NoStop}%
\bibitem [{\citenamefont {Ventagli}\ \emph {et~al.}(2021)\citenamefont
  {Ventagli}, \citenamefont {Antoniou}, \citenamefont {Leh\'ebel},\ and\
  \citenamefont {Sotiriou}}]{Ventagli:2021ubn}%
  \BibitemOpen
  \bibfield  {author} {\bibinfo {author} {\bibnamefont {Ventagli},
  \bibfnamefont {G.}}, \bibinfo {author} {\bibfnamefont {G.}~\bibnamefont
  {Antoniou}}, \bibinfo {author} {\bibfnamefont {A.}~\bibnamefont {Leh\'ebel}},
  and\ \bibinfo {author} {\bibfnamefont {T.~P.}\ \bibnamefont {Sotiriou}}}
  (\bibinfo {year} {2021}),\ \href
  {https://doi.org/10.1103/PhysRevD.104.124078} {\bibfield  {journal} {\bibinfo
   {journal} {Phys. Rev. D}\ }\textbf {\bibinfo {volume} {104}}~(\bibinfo
  {number} {12}),\ \bibinfo {pages} {124078}},\ \Eprint
  {https://arxiv.org/abs/2111.03644} {arXiv:2111.03644 [gr-qc]} \BibitemShut
  {NoStop}%
\bibitem [{\citenamefont {Ventagli}\ \emph {et~al.}(2020)\citenamefont
  {Ventagli}, \citenamefont {Leh\'ebel},\ and\ \citenamefont
  {Sotiriou}}]{Ventagli:2020rnx}%
  \BibitemOpen
  \bibfield  {author} {\bibinfo {author} {\bibnamefont {Ventagli},
  \bibfnamefont {G.}}, \bibinfo {author} {\bibfnamefont {A.}~\bibnamefont
  {Leh\'ebel}}, and\ \bibinfo {author} {\bibfnamefont {T.~P.}\ \bibnamefont
  {Sotiriou}}} (\bibinfo {year} {2020}),\ \href
  {https://doi.org/10.1103/PhysRevD.102.024050} {\bibfield  {journal} {\bibinfo
   {journal} {Phys. Rev. D}\ }\textbf {\bibinfo {volume} {102}}~(\bibinfo
  {number} {2}),\ \bibinfo {pages} {024050}},\ \Eprint
  {https://arxiv.org/abs/2006.01153} {arXiv:2006.01153 [gr-qc]} \BibitemShut
  {NoStop}%
\bibitem [{\citenamefont {Voisin}\ \emph {et~al.}(2020)\citenamefont {Voisin},
  \citenamefont {Cognard}, \citenamefont {Freire}, \citenamefont {Wex},
  \citenamefont {Guillemot}, \citenamefont {Desvignes}, \citenamefont
  {Kramer},\ and\ \citenamefont {Theureau}}]{Voisin:2020lqi}%
  \BibitemOpen
  \bibfield  {author} {\bibinfo {author} {\bibnamefont {Voisin}, \bibfnamefont
  {G.}}, \bibinfo {author} {\bibfnamefont {I.}~\bibnamefont {Cognard}},
  \bibinfo {author} {\bibfnamefont {P.~C.~C.}\ \bibnamefont {Freire}}, \bibinfo
  {author} {\bibfnamefont {N.}~\bibnamefont {Wex}}, \bibinfo {author}
  {\bibfnamefont {L.}~\bibnamefont {Guillemot}}, \bibinfo {author}
  {\bibfnamefont {G.}~\bibnamefont {Desvignes}}, \bibinfo {author}
  {\bibfnamefont {M.}~\bibnamefont {Kramer}}, and\ \bibinfo {author}
  {\bibfnamefont {G.}~\bibnamefont {Theureau}}} (\bibinfo {year} {2020}),\
  \href {https://doi.org/10.1051/0004-6361/202038104} {\bibfield  {journal}
  {\bibinfo  {journal} {Astron. Astrophys.}\ }\textbf {\bibinfo {volume}
  {638}},\ \bibinfo {pages} {A24}},\ \Eprint {https://arxiv.org/abs/2005.01388}
  {arXiv:2005.01388 [gr-qc]} \BibitemShut {NoStop}%
\bibitem [{\citenamefont {Wald}(1993)}]{Wald:1993nt}%
  \BibitemOpen
  \bibfield  {author} {\bibinfo {author} {\bibnamefont {Wald}, \bibfnamefont
  {R.~M.}}} (\bibinfo {year} {1993}),\ \href
  {https://doi.org/10.1103/PhysRevD.48.R3427} {\bibfield  {journal} {\bibinfo
  {journal} {Phys. Rev. D}\ }\textbf {\bibinfo {volume} {48}}~(\bibinfo
  {number} {8}),\ \bibinfo {pages} {R3427}},\ \Eprint
  {https://arxiv.org/abs/gr-qc/9307038} {arXiv:gr-qc/9307038} \BibitemShut
  {NoStop}%
\bibitem [{\citenamefont {Watts}\ \emph {et~al.}(2016)\citenamefont {Watts}
  \emph {et~al.}}]{Watts:2016uzu}%
  \BibitemOpen
  \bibfield  {author} {\bibinfo {author} {\bibnamefont {Watts}, \bibfnamefont
  {A.~L.}},  \emph {et~al.}} (\bibinfo {year} {2016}),\ \href
  {https://doi.org/10.1103/RevModPhys.88.021001} {\bibfield  {journal}
  {\bibinfo  {journal} {Rev. Mod. Phys.}\ }\textbf {\bibinfo {volume}
  {88}}~(\bibinfo {number} {2}),\ \bibinfo {pages} {021001}},\ \Eprint
  {https://arxiv.org/abs/1602.01081} {arXiv:1602.01081 [astro-ph.HE]}
  \BibitemShut {NoStop}%
\bibitem [{\citenamefont {Weih}\ \emph {et~al.}(2018)\citenamefont {Weih},
  \citenamefont {Most},\ and\ \citenamefont {Rezzolla}}]{Weih:2017mcw}%
  \BibitemOpen
  \bibfield  {author} {\bibinfo {author} {\bibnamefont {Weih}, \bibfnamefont
  {L.~R.}}, \bibinfo {author} {\bibfnamefont {E.~R.}\ \bibnamefont {Most}},
  and\ \bibinfo {author} {\bibfnamefont {L.}~\bibnamefont {Rezzolla}}}
  (\bibinfo {year} {2018}),\ \href {https://doi.org/10.1093/mnrasl/slx178}
  {\bibfield  {journal} {\bibinfo  {journal} {Mon. Not. Roy. Astron. Soc.}\
  }\textbf {\bibinfo {volume} {473}}~(\bibinfo {number} {1}),\ \bibinfo {pages}
  {L126}},\ \Eprint {https://arxiv.org/abs/1709.06058} {arXiv:1709.06058
  [gr-qc]} \BibitemShut {NoStop}%
\bibitem [{\citenamefont {Wex}(2014)}]{Wex:2014nva}%
  \BibitemOpen
  \bibfield  {author} {\bibinfo {author} {\bibnamefont {Wex}, \bibfnamefont
  {N.}}} (\bibinfo {year} {2014}),\ \href@noop {} {\enquote {\bibinfo {title}
  {{Testing Relativistic Gravity with Radio Pulsars}},}\ }\Eprint
  {https://arxiv.org/abs/1402.5594} {arXiv:1402.5594 [gr-qc]} \BibitemShut
  {NoStop}%
\bibitem [{\citenamefont {Whinnett}(2000)}]{Whinnett:1999sc}%
  \BibitemOpen
  \bibfield  {author} {\bibinfo {author} {\bibnamefont {Whinnett},
  \bibfnamefont {A.}}} (\bibinfo {year} {2000}),\ \href
  {https://doi.org/10.1103/PhysRevD.61.124014} {\bibfield  {journal} {\bibinfo
  {journal} {Phys. Rev. D}\ }\textbf {\bibinfo {volume} {61}},\ \bibinfo
  {pages} {124014}},\ \Eprint {https://arxiv.org/abs/gr-qc/9911052}
  {arXiv:gr-qc/9911052} \BibitemShut {NoStop}%
\bibitem [{\citenamefont {Whinnett}(1999)}]{Whinnett:1999ws}%
  \BibitemOpen
  \bibfield  {author} {\bibinfo {author} {\bibnamefont {Whinnett},
  \bibfnamefont {A.~W.}}} (\bibinfo {year} {1999}),\ \href
  {https://doi.org/10.1088/0264-9381/16/8/316} {\bibfield  {journal} {\bibinfo
  {journal} {Class. Quant. Grav.}\ }\textbf {\bibinfo {volume} {16}},\ \bibinfo
  {pages} {2797}}\BibitemShut {NoStop}%
\bibitem [{\citenamefont {Whinnett}\ and\ \citenamefont
  {Torres}(2004)}]{Whinnett:2004rr}%
  \BibitemOpen
  \bibfield  {author} {\bibinfo {author} {\bibnamefont {Whinnett},
  \bibfnamefont {A.~W.}}, and\ \bibinfo {author} {\bibfnamefont {D.~F.}\
  \bibnamefont {Torres}}} (\bibinfo {year} {2004}),\ \href
  {https://doi.org/10.1086/383145} {\bibfield  {journal} {\bibinfo  {journal}
  {Astrophys. J. Lett.}\ }\textbf {\bibinfo {volume} {603}},\ \bibinfo {pages}
  {L133}},\ \Eprint {https://arxiv.org/abs/astro-ph/0401521}
  {arXiv:astro-ph/0401521} \BibitemShut {NoStop}%
\bibitem [{\citenamefont {Will}(2014)}]{Will:2014kxa}%
  \BibitemOpen
  \bibfield  {author} {\bibinfo {author} {\bibnamefont {Will}, \bibfnamefont
  {C.~M.}}} (\bibinfo {year} {2014}),\ \href
  {https://doi.org/10.12942/lrr-2014-4} {\bibfield  {journal} {\bibinfo
  {journal} {Living Rev. Rel.}\ }\textbf {\bibinfo {volume} {17}},\ \bibinfo
  {pages} {4}},\ \Eprint {https://arxiv.org/abs/1403.7377} {arXiv:1403.7377
  [gr-qc]} \BibitemShut {NoStop}%
\bibitem [{\citenamefont {Will}(2018)}]{Will:2018bme}%
  \BibitemOpen
  \bibfield  {author} {\bibinfo {author} {\bibnamefont {Will}, \bibfnamefont
  {C.~M.}}} (\bibinfo {year} {2018}),\ \href@noop {} {\emph {\bibinfo {title}
  {{Theory and Experiment in Gravitational Physics}}}}\ (\bibinfo  {publisher}
  {Cambridge University Press})\BibitemShut {NoStop}%
\bibitem [{\citenamefont {Witek}\ \emph {et~al.}(2019)\citenamefont {Witek},
  \citenamefont {Gualtieri}, \citenamefont {Pani},\ and\ \citenamefont
  {Sotiriou}}]{Witek:2018dmd}%
  \BibitemOpen
  \bibfield  {author} {\bibinfo {author} {\bibnamefont {Witek}, \bibfnamefont
  {H.}}, \bibinfo {author} {\bibfnamefont {L.}~\bibnamefont {Gualtieri}},
  \bibinfo {author} {\bibfnamefont {P.}~\bibnamefont {Pani}}, and\ \bibinfo
  {author} {\bibfnamefont {T.~P.}\ \bibnamefont {Sotiriou}}} (\bibinfo {year}
  {2019}),\ \href {https://doi.org/10.1103/PhysRevD.99.064035} {\bibfield
  {journal} {\bibinfo  {journal} {Phys. Rev. D}\ }\textbf {\bibinfo {volume}
  {99}}~(\bibinfo {number} {6}),\ \bibinfo {pages} {064035}},\ \Eprint
  {https://arxiv.org/abs/1810.05177} {arXiv:1810.05177 [gr-qc]} \BibitemShut
  {NoStop}%
\bibitem [{\citenamefont {Wong}\ \emph {et~al.}(2022)\citenamefont {Wong},
  \citenamefont {Herdeiro},\ and\ \citenamefont {Radu}}]{Wong:2022wni}%
  \BibitemOpen
  \bibfield  {author} {\bibinfo {author} {\bibnamefont {Wong}, \bibfnamefont
  {L.~K.}}, \bibinfo {author} {\bibfnamefont {C.~A.~R.}\ \bibnamefont
  {Herdeiro}}, and\ \bibinfo {author} {\bibfnamefont {E.}~\bibnamefont {Radu}}}
  (\bibinfo {year} {2022}),\ \href
  {https://doi.org/10.1103/PhysRevD.106.024008} {\bibfield  {journal} {\bibinfo
   {journal} {Phys. Rev. D}\ }\textbf {\bibinfo {volume} {106}}~(\bibinfo
  {number} {2}),\ \bibinfo {pages} {024008}},\ \Eprint
  {https://arxiv.org/abs/2204.09038} {arXiv:2204.09038 [gr-qc]} \BibitemShut
  {NoStop}%
\bibitem [{\citenamefont {Xiong}\ \emph {et~al.}(2022)\citenamefont {Xiong},
  \citenamefont {Liu}, \citenamefont {Niu}, \citenamefont {Zhang},\ and\
  \citenamefont {Wang}}]{Xiong:2022ozw}%
  \BibitemOpen
  \bibfield  {author} {\bibinfo {author} {\bibnamefont {Xiong}, \bibfnamefont
  {W.}}, \bibinfo {author} {\bibfnamefont {P.}~\bibnamefont {Liu}}, \bibinfo
  {author} {\bibfnamefont {C.}~\bibnamefont {Niu}}, \bibinfo {author}
  {\bibfnamefont {C.-Y.}\ \bibnamefont {Zhang}}, and\ \bibinfo {author}
  {\bibfnamefont {B.}~\bibnamefont {Wang}}} (\bibinfo {year} {2022}),\ \href
  {https://doi.org/10.1088/1674-1137/ac70ad} {\bibfield  {journal} {\bibinfo
  {journal} {Chin. Phys. C}\ }\textbf {\bibinfo {volume} {46}}~(\bibinfo
  {number} {9}),\ \bibinfo {pages} {095103}},\ \Eprint
  {https://arxiv.org/abs/2205.07538} {arXiv:2205.07538 [gr-qc]} \BibitemShut
  {NoStop}%
\bibitem [{\citenamefont {Xu}\ \emph {et~al.}(2020)\citenamefont {Xu},
  \citenamefont {Gao},\ and\ \citenamefont {Shao}}]{Xu:2020vbs}%
  \BibitemOpen
  \bibfield  {author} {\bibinfo {author} {\bibnamefont {Xu}, \bibfnamefont
  {R.}}, \bibinfo {author} {\bibfnamefont {Y.}~\bibnamefont {Gao}}, and\
  \bibinfo {author} {\bibfnamefont {L.}~\bibnamefont {Shao}}} (\bibinfo {year}
  {2020}),\ \href {https://doi.org/10.1103/PhysRevD.102.064057} {\bibfield
  {journal} {\bibinfo  {journal} {Phys. Rev. D}\ }\textbf {\bibinfo {volume}
  {102}}~(\bibinfo {number} {6}),\ \bibinfo {pages} {064057}},\ \Eprint
  {https://arxiv.org/abs/2007.10080} {arXiv:2007.10080 [gr-qc]} \BibitemShut
  {NoStop}%
\bibitem [{\citenamefont {Yagi}\ \emph {et~al.}(2014)\citenamefont {Yagi},
  \citenamefont {Kyutoku}, \citenamefont {Pappas}, \citenamefont {Yunes},\ and\
  \citenamefont {Apostolatos}}]{Yagi:2014bxa}%
  \BibitemOpen
  \bibfield  {author} {\bibinfo {author} {\bibnamefont {Yagi}, \bibfnamefont
  {K.}}, \bibinfo {author} {\bibfnamefont {K.}~\bibnamefont {Kyutoku}},
  \bibinfo {author} {\bibfnamefont {G.}~\bibnamefont {Pappas}}, \bibinfo
  {author} {\bibfnamefont {N.}~\bibnamefont {Yunes}}, and\ \bibinfo {author}
  {\bibfnamefont {T.~A.}\ \bibnamefont {Apostolatos}}} (\bibinfo {year}
  {2014}),\ \href {https://doi.org/10.1103/PhysRevD.89.124013} {\bibfield
  {journal} {\bibinfo  {journal} {Phys. Rev. D}\ }\textbf {\bibinfo {volume}
  {89}}~(\bibinfo {number} {12}),\ \bibinfo {pages} {124013}},\ \Eprint
  {https://arxiv.org/abs/1403.6243} {arXiv:1403.6243 [gr-qc]} \BibitemShut
  {NoStop}%
\bibitem [{\citenamefont {Yagi}\ \emph {et~al.}(2012)\citenamefont {Yagi},
  \citenamefont {Stein}, \citenamefont {Yunes},\ and\ \citenamefont
  {Tanaka}}]{Yagi:2011xp}%
  \BibitemOpen
  \bibfield  {author} {\bibinfo {author} {\bibnamefont {Yagi}, \bibfnamefont
  {K.}}, \bibinfo {author} {\bibfnamefont {L.~C.}\ \bibnamefont {Stein}},
  \bibinfo {author} {\bibfnamefont {N.}~\bibnamefont {Yunes}}, and\ \bibinfo
  {author} {\bibfnamefont {T.}~\bibnamefont {Tanaka}}} (\bibinfo {year}
  {2012}),\ \href {https://doi.org/10.1103/PhysRevD.85.064022} {\bibfield
  {journal} {\bibinfo  {journal} {Phys. Rev. D}\ }\textbf {\bibinfo {volume}
  {85}},\ \bibinfo {pages} {064022}},\ \bibinfo {note} {[Erratum: Phys.Rev.D
  93, 029902 (2016)]},\ \Eprint {https://arxiv.org/abs/1110.5950}
  {arXiv:1110.5950 [gr-qc]} \BibitemShut {NoStop}%
\bibitem [{\citenamefont {Yagi}\ and\ \citenamefont
  {Stepniczka}(2021)}]{Yagi:2021loe}%
  \BibitemOpen
  \bibfield  {author} {\bibinfo {author} {\bibnamefont {Yagi}, \bibfnamefont
  {K.}}, and\ \bibinfo {author} {\bibfnamefont {M.}~\bibnamefont {Stepniczka}}}
  (\bibinfo {year} {2021}),\ \href
  {https://doi.org/10.1103/PhysRevD.104.044017} {\bibfield  {journal} {\bibinfo
   {journal} {Phys. Rev. D}\ }\textbf {\bibinfo {volume} {104}}~(\bibinfo
  {number} {4}),\ \bibinfo {pages} {044017}},\ \Eprint
  {https://arxiv.org/abs/2105.01614} {arXiv:2105.01614 [gr-qc]} \BibitemShut
  {NoStop}%
\bibitem [{\citenamefont {Yagi}\ and\ \citenamefont
  {Yunes}(2013{\natexlab{a}})}]{Yagi:2013bca}%
  \BibitemOpen
  \bibfield  {author} {\bibinfo {author} {\bibnamefont {Yagi}, \bibfnamefont
  {K.}}, and\ \bibinfo {author} {\bibfnamefont {N.}~\bibnamefont {Yunes}}}
  (\bibinfo {year} {2013}{\natexlab{a}}),\ \href
  {https://doi.org/10.1126/science.1236462} {\bibfield  {journal} {\bibinfo
  {journal} {Science}\ }\textbf {\bibinfo {volume} {341}},\ \bibinfo {pages}
  {365}},\ \Eprint {https://arxiv.org/abs/1302.4499} {arXiv:1302.4499 [gr-qc]}
  \BibitemShut {NoStop}%
\bibitem [{\citenamefont {Yagi}\ and\ \citenamefont
  {Yunes}(2013{\natexlab{b}})}]{Yagi:2013awa}%
  \BibitemOpen
  \bibfield  {author} {\bibinfo {author} {\bibnamefont {Yagi}, \bibfnamefont
  {K.}}, and\ \bibinfo {author} {\bibfnamefont {N.}~\bibnamefont {Yunes}}}
  (\bibinfo {year} {2013}{\natexlab{b}}),\ \href
  {https://doi.org/10.1103/PhysRevD.88.023009} {\bibfield  {journal} {\bibinfo
  {journal} {Phys. Rev. D}\ }\textbf {\bibinfo {volume} {88}}~(\bibinfo
  {number} {2}),\ \bibinfo {pages} {023009}},\ \Eprint
  {https://arxiv.org/abs/1303.1528} {arXiv:1303.1528 [gr-qc]} \BibitemShut
  {NoStop}%
\bibitem [{\citenamefont {Yagi}\ and\ \citenamefont
  {Yunes}(2017)}]{Yagi:2016bkt}%
  \BibitemOpen
  \bibfield  {author} {\bibinfo {author} {\bibnamefont {Yagi}, \bibfnamefont
  {K.}}, and\ \bibinfo {author} {\bibfnamefont {N.}~\bibnamefont {Yunes}}}
  (\bibinfo {year} {2017}),\ \href
  {https://doi.org/10.1016/j.physrep.2017.03.002} {\bibfield  {journal}
  {\bibinfo  {journal} {Phys. Rept.}\ }\textbf {\bibinfo {volume} {681}},\
  \bibinfo {pages} {1}},\ \Eprint {https://arxiv.org/abs/1608.02582}
  {arXiv:1608.02582 [gr-qc]} \BibitemShut {NoStop}%
\bibitem [{\citenamefont {Yamashita}\ \emph {et~al.}(2014)\citenamefont
  {Yamashita}, \citenamefont {De~Felice},\ and\ \citenamefont
  {Tanaka}}]{Yamashita:2014fga}%
  \BibitemOpen
  \bibfield  {author} {\bibinfo {author} {\bibnamefont {Yamashita},
  \bibfnamefont {Y.}}, \bibinfo {author} {\bibfnamefont {A.}~\bibnamefont
  {De~Felice}}, and\ \bibinfo {author} {\bibfnamefont {T.}~\bibnamefont
  {Tanaka}}} (\bibinfo {year} {2014}),\ \href
  {https://doi.org/10.1142/S0218271814430032} {\bibfield  {journal} {\bibinfo
  {journal} {Int. J. Mod. Phys. D}\ }\textbf {\bibinfo {volume} {23}},\
  \bibinfo {pages} {1443003}},\ \Eprint {https://arxiv.org/abs/1408.0487}
  {arXiv:1408.0487 [hep-th]} \BibitemShut {NoStop}%
\bibitem [{\citenamefont {Yazadjiev}(1999)}]{Yazadjiev:1999hy}%
  \BibitemOpen
  \bibfield  {author} {\bibinfo {author} {\bibnamefont {Yazadjiev},
  \bibfnamefont {S.}}} (\bibinfo {year} {1999}),\ \href
  {https://doi.org/10.1088/0264-9381/16/10/102} {\bibfield  {journal} {\bibinfo
   {journal} {Class. Quant. Grav.}\ }\textbf {\bibinfo {volume} {16}},\
  \bibinfo {pages} {L63}},\ \Eprint {https://arxiv.org/abs/gr-qc/9906038}
  {arXiv:gr-qc/9906038} \BibitemShut {NoStop}%
\bibitem [{\citenamefont {Yazadjiev}\ \emph {et~al.}(2017)\citenamefont
  {Yazadjiev}, \citenamefont {Doneva},\ and\ \citenamefont
  {Kokkotas}}]{Yazadjiev:2017vpg}%
  \BibitemOpen
  \bibfield  {author} {\bibinfo {author} {\bibnamefont {Yazadjiev},
  \bibfnamefont {S.~S.}}, \bibinfo {author} {\bibfnamefont {D.~D.}\
  \bibnamefont {Doneva}}, and\ \bibinfo {author} {\bibfnamefont {K.~D.}\
  \bibnamefont {Kokkotas}}} (\bibinfo {year} {2017}),\ \href
  {https://doi.org/10.1103/PhysRevD.96.064002} {\bibfield  {journal} {\bibinfo
  {journal} {Phys. Rev. D}\ }\textbf {\bibinfo {volume} {96}}~(\bibinfo
  {number} {6}),\ \bibinfo {pages} {064002}},\ \Eprint
  {https://arxiv.org/abs/1705.06984} {arXiv:1705.06984 [gr-qc]} \BibitemShut
  {NoStop}%
\bibitem [{\citenamefont {Yazadjiev}\ \emph {et~al.}(2016)\citenamefont
  {Yazadjiev}, \citenamefont {Doneva},\ and\ \citenamefont
  {Popchev}}]{Yazadjiev:2016pcb}%
  \BibitemOpen
  \bibfield  {author} {\bibinfo {author} {\bibnamefont {Yazadjiev},
  \bibfnamefont {S.~S.}}, \bibinfo {author} {\bibfnamefont {D.~D.}\
  \bibnamefont {Doneva}}, and\ \bibinfo {author} {\bibfnamefont
  {D.}~\bibnamefont {Popchev}}} (\bibinfo {year} {2016}),\ \href
  {https://doi.org/10.1103/PhysRevD.93.084038} {\bibfield  {journal} {\bibinfo
  {journal} {Phys. Rev. D}\ }\textbf {\bibinfo {volume} {93}}~(\bibinfo
  {number} {8}),\ \bibinfo {pages} {084038}},\ \Eprint
  {https://arxiv.org/abs/1602.04766} {arXiv:1602.04766 [gr-qc]} \BibitemShut
  {NoStop}%
\bibitem [{\citenamefont {Yunes}\ and\ \citenamefont
  {Pretorius}(2009)}]{Yunes:2009hc}%
  \BibitemOpen
  \bibfield  {author} {\bibinfo {author} {\bibnamefont {Yunes}, \bibfnamefont
  {N.}}, and\ \bibinfo {author} {\bibfnamefont {F.}~\bibnamefont {Pretorius}}}
  (\bibinfo {year} {2009}),\ \href {https://doi.org/10.1103/PhysRevD.79.084043}
  {\bibfield  {journal} {\bibinfo  {journal} {Phys. Rev. D}\ }\textbf {\bibinfo
  {volume} {79}},\ \bibinfo {pages} {084043}},\ \Eprint
  {https://arxiv.org/abs/0902.4669} {arXiv:0902.4669 [gr-qc]} \BibitemShut
  {NoStop}%
\bibitem [{\citenamefont {Yunes}\ and\ \citenamefont
  {Siemens}(2013)}]{Yunes:2013dva}%
  \BibitemOpen
  \bibfield  {author} {\bibinfo {author} {\bibnamefont {Yunes}, \bibfnamefont
  {N.}}, and\ \bibinfo {author} {\bibfnamefont {X.}~\bibnamefont {Siemens}}}
  (\bibinfo {year} {2013}),\ \href {https://doi.org/10.12942/lrr-2013-9}
  {\bibfield  {journal} {\bibinfo  {journal} {Living Rev. Rel.}\ }\textbf
  {\bibinfo {volume} {16}},\ \bibinfo {pages} {9}},\ \Eprint
  {https://arxiv.org/abs/1304.3473} {arXiv:1304.3473 [gr-qc]} \BibitemShut
  {NoStop}%
\bibitem [{\citenamefont {Yunes}\ and\ \citenamefont
  {Stein}(2011)}]{Yunes:2011we}%
  \BibitemOpen
  \bibfield  {author} {\bibinfo {author} {\bibnamefont {Yunes}, \bibfnamefont
  {N.}}, and\ \bibinfo {author} {\bibfnamefont {L.~C.}\ \bibnamefont {Stein}}}
  (\bibinfo {year} {2011}),\ \href {https://doi.org/10.1103/PhysRevD.83.104002}
  {\bibfield  {journal} {\bibinfo  {journal} {Phys. Rev. D}\ }\textbf {\bibinfo
  {volume} {83}},\ \bibinfo {pages} {104002}},\ \Eprint
  {https://arxiv.org/abs/1101.2921} {arXiv:1101.2921 [gr-qc]} \BibitemShut
  {NoStop}%
\bibitem [{\citenamefont {Zenginoglu}(2008)}]{Zenginoglu:2007jw}%
  \BibitemOpen
  \bibfield  {author} {\bibinfo {author} {\bibnamefont {Zenginoglu},
  \bibfnamefont {A.}}} (\bibinfo {year} {2008}),\ \href
  {https://doi.org/10.1088/0264-9381/25/14/145002} {\bibfield  {journal}
  {\bibinfo  {journal} {Class. Quant. Grav.}\ }\textbf {\bibinfo {volume}
  {25}},\ \bibinfo {pages} {145002}},\ \Eprint
  {https://arxiv.org/abs/0712.4333} {arXiv:0712.4333 [gr-qc]} \BibitemShut
  {NoStop}%
\bibitem [{\citenamefont {Zhang}\ \emph {et~al.}(2022)\citenamefont {Zhang},
  \citenamefont {Chen}, \citenamefont {Liu}, \citenamefont {Luo}, \citenamefont
  {Tian},\ and\ \citenamefont {Wang}}]{Zhang:2022cmu}%
  \BibitemOpen
  \bibfield  {author} {\bibinfo {author} {\bibnamefont {Zhang}, \bibfnamefont
  {C.-Y.}}, \bibinfo {author} {\bibfnamefont {Q.}~\bibnamefont {Chen}},
  \bibinfo {author} {\bibfnamefont {Y.}~\bibnamefont {Liu}}, \bibinfo {author}
  {\bibfnamefont {W.-K.}\ \bibnamefont {Luo}}, \bibinfo {author} {\bibfnamefont
  {Y.}~\bibnamefont {Tian}}, and\ \bibinfo {author} {\bibfnamefont
  {B.}~\bibnamefont {Wang}}} (\bibinfo {year} {2022}),\ \href
  {https://doi.org/10.1103/PhysRevD.106.L061501} {\bibfield  {journal}
  {\bibinfo  {journal} {Phys. Rev. D}\ }\textbf {\bibinfo {volume}
  {106}}~(\bibinfo {number} {6}),\ \bibinfo {pages} {L061501}},\ \Eprint
  {https://arxiv.org/abs/2204.09260} {arXiv:2204.09260 [gr-qc]} \BibitemShut
  {NoStop}%
\bibitem [{\citenamefont {Zhang}(2021)}]{Zhang:2021btn}%
  \BibitemOpen
  \bibfield  {author} {\bibinfo {author} {\bibnamefont {Zhang}, \bibfnamefont
  {S.-J.}}} (\bibinfo {year} {2021}),\ \href
  {https://doi.org/10.1140/epjc/s10052-021-09249-8} {\bibfield  {journal}
  {\bibinfo  {journal} {Eur. Phys. J. C}\ }\textbf {\bibinfo {volume}
  {81}}~(\bibinfo {number} {5}),\ \bibinfo {pages} {441}},\ \Eprint
  {https://arxiv.org/abs/2102.10479} {arXiv:2102.10479 [gr-qc]} \BibitemShut
  {NoStop}%
\bibitem [{\citenamefont {Zhang}\ \emph {et~al.}(2020)\citenamefont {Zhang},
  \citenamefont {Wang}, \citenamefont {Wang},\ and\ \citenamefont
  {Saavedra}}]{Zhang:2020pko}%
  \BibitemOpen
  \bibfield  {author} {\bibinfo {author} {\bibnamefont {Zhang}, \bibfnamefont
  {S.-J.}}, \bibinfo {author} {\bibfnamefont {B.}~\bibnamefont {Wang}},
  \bibinfo {author} {\bibfnamefont {A.}~\bibnamefont {Wang}}, and\ \bibinfo
  {author} {\bibfnamefont {J.~F.}\ \bibnamefont {Saavedra}}} (\bibinfo {year}
  {2020}),\ \href {https://doi.org/10.1103/PhysRevD.102.124056} {\bibfield
  {journal} {\bibinfo  {journal} {Phys. Rev. D}\ }\textbf {\bibinfo {volume}
  {102}}~(\bibinfo {number} {12}),\ \bibinfo {pages} {124056}},\ \Eprint
  {https://arxiv.org/abs/2010.05092} {arXiv:2010.05092 [gr-qc]} \BibitemShut
  {NoStop}%
\bibitem [{\citenamefont {Zhao}\ \emph {et~al.}(2022)\citenamefont {Zhao},
  \citenamefont {Freire}, \citenamefont {Kramer}, \citenamefont {Shao},\ and\
  \citenamefont {Wex}}]{Zhao:2022vig}%
  \BibitemOpen
  \bibfield  {author} {\bibinfo {author} {\bibnamefont {Zhao}, \bibfnamefont
  {J.}}, \bibinfo {author} {\bibfnamefont {P.~C.~C.}\ \bibnamefont {Freire}},
  \bibinfo {author} {\bibfnamefont {M.}~\bibnamefont {Kramer}}, \bibinfo
  {author} {\bibfnamefont {L.}~\bibnamefont {Shao}}, and\ \bibinfo {author}
  {\bibfnamefont {N.}~\bibnamefont {Wex}}} (\bibinfo {year} {2022}),\ \href
  {https://doi.org/10.1088/1361-6382/ac69a3} {\bibfield  {journal} {\bibinfo
  {journal} {Class. Quant. Grav.}\ }\textbf {\bibinfo {volume} {39}}~(\bibinfo
  {number} {11}),\ \bibinfo {pages} {11LT01}},\ \Eprint
  {https://arxiv.org/abs/2201.03771} {arXiv:2201.03771 [astro-ph.HE]}
  \BibitemShut {NoStop}%
\bibitem [{\citenamefont {Zumalac\'arregui}\ and\ \citenamefont
  {Garc\'\i{}a-Bellido}(2014)}]{Zumalacarregui:2013pma}%
  \BibitemOpen
  \bibfield  {author} {\bibinfo {author} {\bibnamefont {Zumalac\'arregui},
  \bibfnamefont {M.}}, and\ \bibinfo {author} {\bibfnamefont {J.}~\bibnamefont
  {Garc\'\i{}a-Bellido}}} (\bibinfo {year} {2014}),\ \href
  {https://doi.org/10.1103/PhysRevD.89.064046} {\bibfield  {journal} {\bibinfo
  {journal} {Phys. Rev. D}\ }\textbf {\bibinfo {volume} {89}},\ \bibinfo
  {pages} {064046}},\ \Eprint {https://arxiv.org/abs/1308.4685}
  {arXiv:1308.4685 [gr-qc]} \BibitemShut {NoStop}%
\end{thebibliography}%

\end{document}